\documentclass[a4paper, traditabstract,longauth]{aa} 
\usepackage{graphicx,amsmath}
\usepackage{epsf}
\usepackage{ifthen}
\usepackage[table,usenames,dvipsnames]{xcolor}
\usepackage{fixltx2e}

\usepackage{color}
\usepackage{epsfig}
\usepackage[breaklinks, colorlinks, citecolor=blue]{hyperref} 
\usepackage[nonamebreak]{natbib}

\usepackage[varg]{txfonts}

\usepackage{etex}
\reserveinserts{28}

\pdfoutput=1
\pdfmapfile{+txfonts.map}

\bibpunct{(}{)}{;}{a}{}{,} % to follow the A&A style

\def\setsymbol#1#2{\expandafter\def\csname #1\endcsname{#2}}
\def\getsymbol#1{\csname #1\endcsname}

\def\Planck{\textit{Planck}}

\newbox\tablebox    \newdimen\tablewidth
\def\leaderfil{\leaders\hbox to 5pt{\hss.\hss}\hfil}
\def\endPlancktable{\tablewidth=\columnwidth 
    $$\hss\copy\tablebox\hss$$
    \vskip-\lastskip\vskip -2pt}
\def\endPlancktablewide{\tablewidth=\textwidth 
    $$\hss\copy\tablebox\hss$$
    \vskip-\lastskip\vskip -2pt}
\def\tablenote#1 #2\par{\begingroup \parindent=0.8em
    \abovedisplayshortskip=0pt\belowdisplayshortskip=0pt
    \noindent
    $$\hss\vbox{\hsize\tablewidth \hangindent=\parindent \hangafter=1 \noindent
    \hbox to \parindent{$^#1$\hss}\strut#2\strut\par}\hss$$
    \endgroup}
\def\doubleline{\vskip 3pt\hrule \vskip 1.5pt \hrule \vskip 5pt}

\def\L2{\ifmmode L_2\else $L_2$\fi}

\def\DeltaT{\ifmmode \Delta T\else $\Delta T$\fi}
\def\deltat{\ifmmode \Delta t\else $\Delta t$\fi}
\def\fknee{\ifmmode f_{\rm knee}\else $f_{\rm knee}$\fi}
\def\Fmax{\ifmmode F_{\rm max}\else $F_{\rm max}$\fi}
\def\solar{\ifmmode{\rm M}_{\mathord\odot}\else${\rm M}_{\mathord\odot}$\fi}
\def\Msolar{\ifmmode{\rm M}_{\mathord\odot}\else${\rm M}_{\mathord\odot}$\fi}
\def\Lsolar{\ifmmode{\rm L}_{\mathord\odot}\else${\rm L}_{\mathord\odot}$\fi}

\def\inv{\ifmmode^{-1}\else$^{-1}$\fi}
\def\mo{\ifmmode^{-1}\else$^{-1}$\fi}
\def\sup#1{\ifmmode ^{\rm #1}\else $^{\rm #1}$\fi}
\def\expo#1{\ifmmode \times 10^{#1}\else $\times 10^{#1}$\fi}
\def\,{\thinspace}
\def\lsim{\mathrel{\raise .4ex\hbox{\rlap{$<$}\lower 1.2ex\hbox{$\sim$}}}}
\def\gsim{\mathrel{\raise .4ex\hbox{\rlap{$>$}\lower 1.2ex\hbox{$\sim$}}}}
\let\lea=\lsim

\def\simprop{\mathrel{\raise .4ex\hbox{\rlap{$\propto$}\lower 1.2ex\hbox{$\sim$}}}}
\def\deg{\ifmmode^\circ\else$^\circ$\fi}
\def\pdeg{\ifmmode $\setbox0=\hbox{$^{\circ}$}\rlap{\hskip.11\wd0 .}$^{\circ}
          \else \setbox0=\hbox{$^{\circ}$}\rlap{\hskip.11\wd0 .}$^{\circ}$\fi}
\def\arcs{\ifmmode {^{\scriptstyle\prime\prime}}
          \else $^{\scriptstyle\prime\prime}$\fi}
\def\arcm{\ifmmode {^{\scriptstyle\prime}}
          \else $^{\scriptstyle\prime}$\fi}
\newdimen\sa  \newdimen\sb
\def\parcs{\sa=.07em \sb=.03em
     \ifmmode \hbox{\rlap{.}}^{\scriptstyle\prime\kern -\sb\prime}\hbox{\kern -\sa}
     \else \rlap{.}$^{\scriptstyle\prime\kern -\sb\prime}$\kern -\sa\fi}
\def\parcm{\sa=.08em \sb=.03em
     \ifmmode \hbox{\rlap{.}\kern\sa}^{\scriptstyle\prime}\hbox{\kern-\sb}
     \else \rlap{.}\kern\sa$^{\scriptstyle\prime}$\kern-\sb\fi}
\def\ra[#1 #2 #3.#4]{#1\sup{h}#2\sup{m}#3\sup{s}\llap.#4}
\def\dec[#1 #2 #3.#4]{#1\deg#2\arcm#3\arcs\llap.#4}
\def\deco[#1 #2 #3]{#1\deg#2\arcm#3\arcs}
\def\rra[#1 #2]{#1\sup{h}#2\sup{m}}

\def\dots{\relax\ifmmode \ldots\else $\ldots$\fi}
\def\WHzsr{\ifmmode $W\,Hz\mo\,sr\mo$\else W\,Hz\mo\,sr\mo\fi}
\def\mHz{\ifmmode $\,mHz$\else \,mHz\fi}
\def\GHz{\ifmmode $\,GHz$\else \,GHz\fi}
\def\mKs{\ifmmode $\,mK\,s$^{1/2}\else \,mK\,s$^{1/2}$\fi}
\def\muKs{\ifmmode \,\mu$K\,s$^{1/2}\else \,$\mu$K\,s$^{1/2}$\fi}
\def\muKRJs{\ifmmode \,\mu$K$_{\rm RJ}$\,s$^{1/2}\else \,$\mu$K$_{\rm RJ}$\,s$^{1/2}$\fi}
\def\muKHz{\ifmmode \,\mu$K\,Hz$^{-1/2}\else \,$\mu$K\,Hz$^{-1/2}$\fi}
\def\MJysr{\ifmmode \,$MJy\,sr\mo$\else \,MJy\,sr\mo\fi}
\def\MJysrmK{\ifmmode \,$MJy\,sr\mo$\,mK$_{\rm CMB}\mo\else \,MJy\,sr\mo\,mK$_{\rm CMB}\mo$\fi}
\def\microns{\ifmmode \,\mu$m$\else \,$\mu$m\fi}

\def\muK{\ifmmode \,\mu$K$\else \,$\mu$\hbox{K}\fi}
\def\microK{\ifmmode \,\mu$K$\else \,$\mu$\hbox{K}\fi}
\def\muW{\ifmmode \,\mu$W$\else \,$\mu$\hbox{W}\fi}
\def\kms{\ifmmode $\,km\,s$^{-1}\else \,km\,s$^{-1}$\fi}
\def\kmsMpc{\ifmmode $\,\kms\,Mpc\mo$\else \,\kms\,Mpc\mo\fi}
\setsymbol{LFI:center:frequency:70GHz:units}{70.3\,GHz}
\setsymbol{LFI:center:frequency:44GHz:units}{44.1\,GHz}
\setsymbol{LFI:center:frequency:30GHz:units}{28.5\,GHz}

\setsymbol{LFI:center:frequency:70GHz}{70.3}
\setsymbol{LFI:center:frequency:44GHz}{44.1}
\setsymbol{LFI:center:frequency:30GHz}{28.5}

\setsymbol{LFI:center:frequency:LFI18:Rad:M:units}{71.7\GHz}
\setsymbol{LFI:center:frequency:LFI19:Rad:M:units}{67.5\GHz}
\setsymbol{LFI:center:frequency:LFI20:Rad:M:units}{69.2\GHz}
\setsymbol{LFI:center:frequency:LFI21:Rad:M:units}{70.4\GHz}
\setsymbol{LFI:center:frequency:LFI22:Rad:M:units}{71.5\GHz}
\setsymbol{LFI:center:frequency:LFI23:Rad:M:units}{70.8\GHz}
\setsymbol{LFI:center:frequency:LFI24:Rad:M:units}{44.4\GHz}
\setsymbol{LFI:center:frequency:LFI25:Rad:M:units}{44.0\GHz}
\setsymbol{LFI:center:frequency:LFI26:Rad:M:units}{43.9\GHz}
\setsymbol{LFI:center:frequency:LFI27:Rad:M:units}{28.3\GHz}
\setsymbol{LFI:center:frequency:LFI28:Rad:M:units}{28.8\GHz}
\setsymbol{LFI:center:frequency:LFI18:Rad:S:units}{70.1\GHz}
\setsymbol{LFI:center:frequency:LFI19:Rad:S:units}{69.6\GHz}
\setsymbol{LFI:center:frequency:LFI20:Rad:S:units}{69.5\GHz}
\setsymbol{LFI:center:frequency:LFI21:Rad:S:units}{69.5\GHz}
\setsymbol{LFI:center:frequency:LFI22:Rad:S:units}{72.8\GHz}
\setsymbol{LFI:center:frequency:LFI23:Rad:S:units}{71.3\GHz}
\setsymbol{LFI:center:frequency:LFI24:Rad:S:units}{44.1\GHz}
\setsymbol{LFI:center:frequency:LFI25:Rad:S:units}{44.1\GHz}
\setsymbol{LFI:center:frequency:LFI26:Rad:S:units}{44.1\GHz}
\setsymbol{LFI:center:frequency:LFI27:Rad:S:units}{28.5\GHz}
\setsymbol{LFI:center:frequency:LFI28:Rad:S:units}{28.2\GHz}

\setsymbol{LFI:center:frequency:LFI18:Rad:M}{71.7}
\setsymbol{LFI:center:frequency:LFI19:Rad:M}{67.5}
\setsymbol{LFI:center:frequency:LFI20:Rad:M}{69.2}
\setsymbol{LFI:center:frequency:LFI21:Rad:M}{70.4}
\setsymbol{LFI:center:frequency:LFI22:Rad:M}{71.5}
\setsymbol{LFI:center:frequency:LFI23:Rad:M}{70.8}
\setsymbol{LFI:center:frequency:LFI24:Rad:M}{44.4}
\setsymbol{LFI:center:frequency:LFI25:Rad:M}{44.0}
\setsymbol{LFI:center:frequency:LFI26:Rad:M}{43.9}
\setsymbol{LFI:center:frequency:LFI27:Rad:M}{28.3}
\setsymbol{LFI:center:frequency:LFI28:Rad:M}{28.8}
\setsymbol{LFI:center:frequency:LFI18:Rad:S}{70.1}
\setsymbol{LFI:center:frequency:LFI19:Rad:S}{69.6}
\setsymbol{LFI:center:frequency:LFI20:Rad:S}{69.5}
\setsymbol{LFI:center:frequency:LFI21:Rad:S}{69.5}
\setsymbol{LFI:center:frequency:LFI22:Rad:S}{72.8}
\setsymbol{LFI:center:frequency:LFI23:Rad:S}{71.3}
\setsymbol{LFI:center:frequency:LFI24:Rad:S}{44.1}
\setsymbol{LFI:center:frequency:LFI25:Rad:S}{44.1}
\setsymbol{LFI:center:frequency:LFI26:Rad:S}{44.1}
\setsymbol{LFI:center:frequency:LFI27:Rad:S}{28.5}
\setsymbol{LFI:center:frequency:LFI28:Rad:S}{28.2}

\setsymbol{LFI:white:noise:sensitivity:70GHz:units}{134.7\muKs}
\setsymbol{LFI:white:noise:sensitivity:44GHz:units}{164.7\muKs}
\setsymbol{LFI:white:noise:sensitivity:30GHz:units}{143.4\muKs}

\setsymbol{LFI:white:noise:sensitivity:70GHz}{134.7}
\setsymbol{LFI:white:noise:sensitivity:44GHz}{164.7}
\setsymbol{LFI:white:noise:sensitivity:30GHz}{143.4}

\setsymbol{LFI:white:noise:sensitivity:LFI18:Rad:M:units}{512.0\muKs}
\setsymbol{LFI:white:noise:sensitivity:LFI19:Rad:M:units}{581.4\muKs}
\setsymbol{LFI:white:noise:sensitivity:LFI20:Rad:M:units}{590.8\muKs}
\setsymbol{LFI:white:noise:sensitivity:LFI21:Rad:M:units}{455.2\muKs}
\setsymbol{LFI:white:noise:sensitivity:LFI22:Rad:M:units}{492.0\muKs}
\setsymbol{LFI:white:noise:sensitivity:LFI23:Rad:M:units}{507.7\muKs}
\setsymbol{LFI:white:noise:sensitivity:LFI24:Rad:M:units}{462.2\muKs}
\setsymbol{LFI:white:noise:sensitivity:LFI25:Rad:M:units}{413.6\muKs}
\setsymbol{LFI:white:noise:sensitivity:LFI26:Rad:M:units}{478.6\muKs}
\setsymbol{LFI:white:noise:sensitivity:LFI27:Rad:M:units}{277.7\muKs}
\setsymbol{LFI:white:noise:sensitivity:LFI28:Rad:M:units}{312.3\muKs}
\setsymbol{LFI:white:noise:sensitivity:LFI18:Rad:S:units}{465.7\muKs}
\setsymbol{LFI:white:noise:sensitivity:LFI19:Rad:S:units}{555.6\muKs}
\setsymbol{LFI:white:noise:sensitivity:LFI20:Rad:S:units}{623.2\muKs}
\setsymbol{LFI:white:noise:sensitivity:LFI21:Rad:S:units}{564.1\muKs}
\setsymbol{LFI:white:noise:sensitivity:LFI22:Rad:S:units}{534.4\muKs}
\setsymbol{LFI:white:noise:sensitivity:LFI23:Rad:S:units}{542.4\muKs}
\setsymbol{LFI:white:noise:sensitivity:LFI24:Rad:S:units}{399.2\muKs}
\setsymbol{LFI:white:noise:sensitivity:LFI25:Rad:S:units}{392.6\muKs}
\setsymbol{LFI:white:noise:sensitivity:LFI26:Rad:S:units}{418.6\muKs}
\setsymbol{LFI:white:noise:sensitivity:LFI27:Rad:S:units}{302.9\muKs}
\setsymbol{LFI:white:noise:sensitivity:LFI28:Rad:S:units}{285.3\muKs}

\setsymbol{LFI:white:noise:sensitivity:LFI18:Rad:M}{512.0}
\setsymbol{LFI:white:noise:sensitivity:LFI19:Rad:M}{581.4}
\setsymbol{LFI:white:noise:sensitivity:LFI20:Rad:M}{590.8}
\setsymbol{LFI:white:noise:sensitivity:LFI21:Rad:M}{455.2}
\setsymbol{LFI:white:noise:sensitivity:LFI22:Rad:M}{492.0}
\setsymbol{LFI:white:noise:sensitivity:LFI23:Rad:M}{507.7}
\setsymbol{LFI:white:noise:sensitivity:LFI24:Rad:M}{462.2}
\setsymbol{LFI:white:noise:sensitivity:LFI25:Rad:M}{413.6}
\setsymbol{LFI:white:noise:sensitivity:LFI26:Rad:M}{478.6}
\setsymbol{LFI:white:noise:sensitivity:LFI27:Rad:M}{277.7}
\setsymbol{LFI:white:noise:sensitivity:LFI28:Rad:M}{312.3}
\setsymbol{LFI:white:noise:sensitivity:LFI18:Rad:S}{465.7}
\setsymbol{LFI:white:noise:sensitivity:LFI19:Rad:S}{555.6}
\setsymbol{LFI:white:noise:sensitivity:LFI20:Rad:S}{623.2}
\setsymbol{LFI:white:noise:sensitivity:LFI21:Rad:S}{564.1}
\setsymbol{LFI:white:noise:sensitivity:LFI22:Rad:S}{534.4}
\setsymbol{LFI:white:noise:sensitivity:LFI23:Rad:S}{542.4}
\setsymbol{LFI:white:noise:sensitivity:LFI24:Rad:S}{399.2}
\setsymbol{LFI:white:noise:sensitivity:LFI25:Rad:S}{392.6}
\setsymbol{LFI:white:noise:sensitivity:LFI26:Rad:S}{418.6}
\setsymbol{LFI:white:noise:sensitivity:LFI27:Rad:S}{302.9}
\setsymbol{LFI:white:noise:sensitivity:LFI28:Rad:S}{285.3}

\setsymbol{LFI:knee:frequency:70GHz:units}{29.5\mHz}
\setsymbol{LFI:knee:frequency:44GHz:units}{56.2\mHz}
\setsymbol{LFI:knee:frequency:30GHz:units}{113.7\mHz}

\setsymbol{LFI:knee:frequency:70GHz}{29.5}
\setsymbol{LFI:knee:frequency:44GHz}{56.2}
\setsymbol{LFI:knee:frequency:30GHz}{113.7}

\setsymbol{LFI:knee:frequency:LFI18:Rad:M:units}{16.3\mHz}
\setsymbol{LFI:knee:frequency:LFI19:Rad:M:units}{15.1\mHz}
\setsymbol{LFI:knee:frequency:LFI20:Rad:M:units}{18.7\mHz}
\setsymbol{LFI:knee:frequency:LFI21:Rad:M:units}{37.2\mHz}
\setsymbol{LFI:knee:frequency:LFI22:Rad:M:units}{12.7\mHz}
\setsymbol{LFI:knee:frequency:LFI23:Rad:M:units}{34.6\mHz}
\setsymbol{LFI:knee:frequency:LFI24:Rad:M:units}{46.2\mHz}
\setsymbol{LFI:knee:frequency:LFI25:Rad:M:units}{24.9\mHz}
\setsymbol{LFI:knee:frequency:LFI26:Rad:M:units}{67.6\mHz}
\setsymbol{LFI:knee:frequency:LFI27:Rad:M:units}{187.4\mHz}
\setsymbol{LFI:knee:frequency:LFI28:Rad:M:units}{122.2\mHz}
\setsymbol{LFI:knee:frequency:LFI18:Rad:S:units}{17.7\mHz}
\setsymbol{LFI:knee:frequency:LFI19:Rad:S:units}{22.0\mHz}
\setsymbol{LFI:knee:frequency:LFI20:Rad:S:units}{8.7\mHz}
\setsymbol{LFI:knee:frequency:LFI21:Rad:S:units}{25.9\mHz}
\setsymbol{LFI:knee:frequency:LFI22:Rad:S:units}{15.8\mHz}
\setsymbol{LFI:knee:frequency:LFI23:Rad:S:units}{129.8\mHz}
\setsymbol{LFI:knee:frequency:LFI24:Rad:S:units}{100.9\mHz}
\setsymbol{LFI:knee:frequency:LFI25:Rad:S:units}{38.9\mHz}
\setsymbol{LFI:knee:frequency:LFI26:Rad:S:units}{58.9\mHz}
\setsymbol{LFI:knee:frequency:LFI27:Rad:S:units}{104.4\mHz}
\setsymbol{LFI:knee:frequency:LFI28:Rad:S:units}{40.7\mHz}

\setsymbol{LFI:knee:frequency:LFI18:Rad:M}{16.3}
\setsymbol{LFI:knee:frequency:LFI19:Rad:M}{15.1}
\setsymbol{LFI:knee:frequency:LFI20:Rad:M}{18.7}
\setsymbol{LFI:knee:frequency:LFI21:Rad:M}{37.2}
\setsymbol{LFI:knee:frequency:LFI22:Rad:M}{12.7}
\setsymbol{LFI:knee:frequency:LFI23:Rad:M}{34.6}
\setsymbol{LFI:knee:frequency:LFI24:Rad:M}{46.2}
\setsymbol{LFI:knee:frequency:LFI25:Rad:M}{24.9}
\setsymbol{LFI:knee:frequency:LFI26:Rad:M}{67.6}
\setsymbol{LFI:knee:frequency:LFI27:Rad:M}{187.4}
\setsymbol{LFI:knee:frequency:LFI28:Rad:M}{122.2}
\setsymbol{LFI:knee:frequency:LFI18:Rad:S}{17.7}
\setsymbol{LFI:knee:frequency:LFI19:Rad:S}{22.0}
\setsymbol{LFI:knee:frequency:LFI20:Rad:S}{8.7}
\setsymbol{LFI:knee:frequency:LFI21:Rad:S}{25.9}
\setsymbol{LFI:knee:frequency:LFI22:Rad:S}{15.8}
\setsymbol{LFI:knee:frequency:LFI23:Rad:S}{129.8}
\setsymbol{LFI:knee:frequency:LFI24:Rad:S}{100.9}
\setsymbol{LFI:knee:frequency:LFI25:Rad:S}{38.9}
\setsymbol{LFI:knee:frequency:LFI26:Rad:S}{58.9}
\setsymbol{LFI:knee:frequency:LFI27:Rad:S}{104.4}
\setsymbol{LFI:knee:frequency:LFI28:Rad:S}{40.7}

\setsymbol{LFI:slope:70GHz:units}{$-1.03$\mHz}
\setsymbol{LFI:slope:44GHz:units}{$-0.89$\mHz}
\setsymbol{LFI:slope:30GHz:units}{$-0.87$\mHz}

\setsymbol{LFI:slope:70GHz}{$-1.03$}
\setsymbol{LFI:slope:44GHz}{$-0.89$}
\setsymbol{LFI:slope:30GHz}{$-0.87$}

\setsymbol{LFI:slope:LFI18:Rad:M:units}{$-1.04$\mHz}
\setsymbol{LFI:slope:LFI19:Rad:M:units}{$-1.09$\mHz}
\setsymbol{LFI:slope:LFI20:Rad:M:units}{$-0.69$\mHz}
\setsymbol{LFI:slope:LFI21:Rad:M:units}{$-1.56$\mHz}
\setsymbol{LFI:slope:LFI22:Rad:M:units}{$-1.01$\mHz}
\setsymbol{LFI:slope:LFI23:Rad:M:units}{$-0.96$\mHz}
\setsymbol{LFI:slope:LFI24:Rad:M:units}{$-0.83$\mHz}
\setsymbol{LFI:slope:LFI25:Rad:M:units}{$-0.91$\mHz}
\setsymbol{LFI:slope:LFI26:Rad:M:units}{$-0.95$\mHz}
\setsymbol{LFI:slope:LFI27:Rad:M:units}{$-0.87$\mHz}
\setsymbol{LFI:slope:LFI28:Rad:M:units}{$-0.88$\mHz}
\setsymbol{LFI:slope:LFI18:Rad:S:units}{$-1.15$\mHz}
\setsymbol{LFI:slope:LFI19:Rad:S:units}{$-1.00$\mHz}
\setsymbol{LFI:slope:LFI20:Rad:S:units}{$-0.95$\mHz}
\setsymbol{LFI:slope:LFI21:Rad:S:units}{$-0.92$\mHz}
\setsymbol{LFI:slope:LFI22:Rad:S:units}{$-1.01$\mHz}
\setsymbol{LFI:slope:LFI23:Rad:S:units}{$-0.95$\mHz}
\setsymbol{LFI:slope:LFI24:Rad:S:units}{$-0.73$\mHz}
\setsymbol{LFI:slope:LFI25:Rad:S:units}{$-1.16$\mHz}
\setsymbol{LFI:slope:LFI26:Rad:S:units}{$-0.79$\mHz}
\setsymbol{LFI:slope:LFI27:Rad:S:units}{$-0.82$\mHz}
\setsymbol{LFI:slope:LFI28:Rad:S:units}{$-0.91$\mHz}

\setsymbol{LFI:slope:LFI18:Rad:M}{$-1.04$}
\setsymbol{LFI:slope:LFI19:Rad:M}{$-1.09$}
\setsymbol{LFI:slope:LFI20:Rad:M}{$-0.69$}
\setsymbol{LFI:slope:LFI21:Rad:M}{$-1.56$}
\setsymbol{LFI:slope:LFI22:Rad:M}{$-1.01$}
\setsymbol{LFI:slope:LFI23:Rad:M}{$-0.96$}
\setsymbol{LFI:slope:LFI24:Rad:M}{$-0.83$}
\setsymbol{LFI:slope:LFI25:Rad:M}{$-0.91$}
\setsymbol{LFI:slope:LFI26:Rad:M}{$-0.95$}
\setsymbol{LFI:slope:LFI27:Rad:M}{$-0.87$}
\setsymbol{LFI:slope:LFI28:Rad:M}{$-0.88$}
\setsymbol{LFI:slope:LFI18:Rad:S}{$-1.15$}
\setsymbol{LFI:slope:LFI19:Rad:S}{$-1.00$}
\setsymbol{LFI:slope:LFI20:Rad:S}{$-0.95$}
\setsymbol{LFI:slope:LFI21:Rad:S}{$-0.92$}
\setsymbol{LFI:slope:LFI22:Rad:S}{$-1.01$}
\setsymbol{LFI:slope:LFI23:Rad:S}{$-0.95$}
\setsymbol{LFI:slope:LFI24:Rad:S}{$-0.73$}
\setsymbol{LFI:slope:LFI25:Rad:S}{$-1.16$}
\setsymbol{LFI:slope:LFI26:Rad:S}{$-0.79$}
\setsymbol{LFI:slope:LFI27:Rad:S}{$-0.82$}
\setsymbol{LFI:slope:LFI28:Rad:S}{$-0.91$}

\setsymbol{LFI:FWHM:70GHz:units}{13\parcm01}
\setsymbol{LFI:FWHM:44GHz:units}{27\parcm92}
\setsymbol{LFI:FWHM:30GHz:units}{32\parcm65}

\setsymbol{LFI:FWHM:70GHz}{13.01}
\setsymbol{LFI:FWHM:44GHz}{27.92}
\setsymbol{LFI:FWHM:30GHz}{32.65}

\setsymbol{LFI:FWHM:LFI18:units}{13\parcm39}
\setsymbol{LFI:FWHM:LFI19:units}{13\parcm01}
\setsymbol{LFI:FWHM:LFI20:units}{12\parcm75}
\setsymbol{LFI:FWHM:LFI21:units}{12\parcm74}
\setsymbol{LFI:FWHM:LFI22:units}{12\parcm87}
\setsymbol{LFI:FWHM:LFI23:units}{13\parcm27}
\setsymbol{LFI:FWHM:LFI24:units}{22\parcm98}
\setsymbol{LFI:FWHM:LFI25:units}{30\parcm46}
\setsymbol{LFI:FWHM:LFI26:units}{30\parcm31}
\setsymbol{LFI:FWHM:LFI27:units}{32\parcm65}
\setsymbol{LFI:FWHM:LFI28:units}{32\parcm66}

\setsymbol{LFI:FWHM:LFI18}{13.39}
\setsymbol{LFI:FWHM:LFI19}{13.01}
\setsymbol{LFI:FWHM:LFI20}{12.75}
\setsymbol{LFI:FWHM:LFI21}{12.74}
\setsymbol{LFI:FWHM:LFI22}{12.87}
\setsymbol{LFI:FWHM:LFI23}{13.27}
\setsymbol{LFI:FWHM:LFI24}{22.98}
\setsymbol{LFI:FWHM:LFI25}{30.46}
\setsymbol{LFI:FWHM:LFI26}{30.31}
\setsymbol{LFI:FWHM:LFI27}{32.65}
\setsymbol{LFI:FWHM:LFI28}{32.66}

\setsymbol{LFI:FWHM:uncertainty:LFI18:units}{0.170\arcm}
\setsymbol{LFI:FWHM:uncertainty:LFI19:units}{0.174\arcm}
\setsymbol{LFI:FWHM:uncertainty:LFI20:units}{0.170\arcm}
\setsymbol{LFI:FWHM:uncertainty:LFI21:units}{0.156\arcm}
\setsymbol{LFI:FWHM:uncertainty:LFI22:units}{0.164\arcm}
\setsymbol{LFI:FWHM:uncertainty:LFI23:units}{0.171\arcm}
\setsymbol{LFI:FWHM:uncertainty:LFI24:units}{0.652\arcm}
\setsymbol{LFI:FWHM:uncertainty:LFI25:units}{1.075\arcm}
\setsymbol{LFI:FWHM:uncertainty:LFI26:units}{1.131\arcm}
\setsymbol{LFI:FWHM:uncertainty:LFI27:units}{1.266\arcm}
\setsymbol{LFI:FWHM:uncertainty:LFI28:units}{1.287\arcm}

\setsymbol{LFI:FWHM:uncertainty:LFI18}{0.170}
\setsymbol{LFI:FWHM:uncertainty:LFI19}{0.174}
\setsymbol{LFI:FWHM:uncertainty:LFI20}{0.170}
\setsymbol{LFI:FWHM:uncertainty:LFI21}{0.156}
\setsymbol{LFI:FWHM:uncertainty:LFI22}{0.164}
\setsymbol{LFI:FWHM:uncertainty:LFI23}{0.171}
\setsymbol{LFI:FWHM:uncertainty:LFI24}{0.652}
\setsymbol{LFI:FWHM:uncertainty:LFI25}{1.075}
\setsymbol{LFI:FWHM:uncertainty:LFI26}{1.131}
\setsymbol{LFI:FWHM:uncertainty:LFI27}{1.266}
\setsymbol{LFI:FWHM:uncertainty:LFI28}{1.287}

\setsymbol{HFI:center:frequency:100GHz:units}{100\,GHz}
\setsymbol{HFI:center:frequency:143GHz:units}{143\,GHz}
\setsymbol{HFI:center:frequency:217GHz:units}{217\,GHz}
\setsymbol{HFI:center:frequency:353GHz:units}{353\,GHz}
\setsymbol{HFI:center:frequency:545GHz:units}{545\,GHz}
\setsymbol{HFI:center:frequency:857GHz:units}{857\,GHz}

\setsymbol{HFI:center:frequency:100GHz}{100}
\setsymbol{HFI:center:frequency:143GHz}{143}
\setsymbol{HFI:center:frequency:217GHz}{217}
\setsymbol{HFI:center:frequency:353GHz}{353}
\setsymbol{HFI:center:frequency:545GHz}{545}
\setsymbol{HFI:center:frequency:857GHz}{857}

\setsymbol{HFI:Ndetectors:100GHz}{8}
\setsymbol{HFI:Ndetectors:143GHz}{11}
\setsymbol{HFI:Ndetectors:217GHz}{12}
\setsymbol{HFI:Ndetectors:353GHz}{12}
\setsymbol{HFI:Ndetectors:545GHz}{3}
\setsymbol{HFI:Ndetectors:857GHz}{4}

\setsymbol{HFI:FWHM:Maps:100GHz:units}{9\parcm88}
\setsymbol{HFI:FWHM:Maps:143GHz:units}{7\parcm18}
\setsymbol{HFI:FWHM:Maps:217GHz:units}{4\parcm87}
\setsymbol{HFI:FWHM:Maps:353GHz:units}{4\parcm65}
\setsymbol{HFI:FWHM:Maps:545GHz:units}{4\parcm72}
\setsymbol{HFI:FWHM:Maps:857GHz:units}{4\parcm39}
\setsymbol{HFI:FWHM:Maps:100GHz}{9.88}
\setsymbol{HFI:FWHM:Maps:143GHz}{7.18}
\setsymbol{HFI:FWHM:Maps:217GHz}{4.87}
\setsymbol{HFI:FWHM:Maps:353GHz}{4.65}
\setsymbol{HFI:FWHM:Maps:545GHz}{4.72}
\setsymbol{HFI:FWHM:Maps:857GHz}{4.39}

\setsymbol{HFI:beam:ellipticity:Maps:100GHz}{1.15}
\setsymbol{HFI:beam:ellipticity:Maps:143GHz}{1.01}
\setsymbol{HFI:beam:ellipticity:Maps:217GHz}{1.06}
\setsymbol{HFI:beam:ellipticity:Maps:353GHz}{1.05}
\setsymbol{HFI:beam:ellipticity:Maps:545GHz}{1.14}
\setsymbol{HFI:beam:ellipticity:Maps:857GHz}{1.19}

\setsymbol{HFI:FWHM:Mars:100GHz:units}{9\parcm37}
\setsymbol{HFI:FWHM:Mars:143GHz:units}{7\parcm04}
\setsymbol{HFI:FWHM:Mars:217GHz:units}{4\parcm68}
\setsymbol{HFI:FWHM:Mars:353GHz:units}{4\parcm43}
\setsymbol{HFI:FWHM:Mars:545GHz:units}{3\parcm80}
\setsymbol{HFI:FWHM:Mars:857GHz:units}{3\parcm67}

\setsymbol{HFI:FWHM:Mars:100GHz}{9.37}
\setsymbol{HFI:FWHM:Mars:143GHz}{7.04}
\setsymbol{HFI:FWHM:Mars:217GHz}{4.68}
\setsymbol{HFI:FWHM:Mars:353GHz}{4.43}
\setsymbol{HFI:FWHM:Mars:545GHz}{3.80}
\setsymbol{HFI:FWHM:Mars:857GHz}{3.67}

\setsymbol{HFI:beam:ellipticity:Mars:100GHz}{1.18}
\setsymbol{HFI:beam:ellipticity:Mars:143GHz}{1.03}
\setsymbol{HFI:beam:ellipticity:Mars:217GHz}{1.14}
\setsymbol{HFI:beam:ellipticity:Mars:353GHz}{1.09}
\setsymbol{HFI:beam:ellipticity:Mars:545GHz}{1.25}
\setsymbol{HFI:beam:ellipticity:Mars:857GHz}{1.03}

\setsymbol{HFI:CMB:relative:calibration:100GHz}{$\lsim 1\%$}
\setsymbol{HFI:CMB:relative:calibration:143GHz}{$\lsim 1\%$}
\setsymbol{HFI:CMB:relative:calibration:217GHz}{$\lsim 1\%$}
\setsymbol{HFI:CMB:relative:calibration:353GHz}{$\lsim 1\%$}
\setsymbol{HFI:CMB:relative:calibration:545GHz}{}
\setsymbol{HFI:CMB:relative:calibration:857GHz}{}

\setsymbol{HFI:CMB:absolute:calibration:100GHz}{$\lsim 2\%$}
\setsymbol{HFI:CMB:absolute:calibration:143GHz}{$\lsim 2\%$}
\setsymbol{HFI:CMB:absolute:calibration:217GHz}{$\lsim 2\%$}
\setsymbol{HFI:CMB:absolute:calibration:353GHz}{$\lsim 2\%$}
\setsymbol{HFI:CMB:absolute:calibration:545GHz}{}
\setsymbol{HFI:CMB:absolute:calibration:857GHz}{}

\setsymbol{HFI:FIRAS:gain:calibration:accuracy:statistical:100GHz}{}
\setsymbol{HFI:FIRAS:gain:calibration:accuracy:statistical:143GHz}{}
\setsymbol{HFI:FIRAS:gain:calibration:accuracy:statistical:217GHz}{}
\setsymbol{HFI:FIRAS:gain:calibration:accuracy:statistical:353GHz}{2.5\%}
\setsymbol{HFI:FIRAS:gain:calibration:accuracy:statistical:545GHz}{1\%}
\setsymbol{HFI:FIRAS:gain:calibration:accuracy:statistical:857GHz}{0.5\%}

\setsymbol{HFI:FIRAS:gain:calibration:accuracy:systematic:100GHz}{}
\setsymbol{HFI:FIRAS:gain:calibration:accuracy:systematic:143GHz}{}
\setsymbol{HFI:FIRAS:gain:calibration:accuracy:systematic:217GHz}{}
\setsymbol{HFI:FIRAS:gain:calibration:accuracy:systematic:353GHz}{}
\setsymbol{HFI:FIRAS:gain:calibration:accuracy:systematic:545GHz}{7\%}
\setsymbol{HFI:FIRAS:gain:calibration:accuracy:systematic:857GHz}{7\%}

\setsymbol{HFI:FIRAS:zero:point:accuracy:100GHz:units}{0.8\MJysr}
\setsymbol{HFI:FIRAS:zero:point:accuracy:143GHz:units}{}
\setsymbol{HFI:FIRAS:zero:point:accuracy:217GHz:units}{}
\setsymbol{HFI:FIRAS:zero:point:accuracy:353GHz:units}{1.4\MJysr}
\setsymbol{HFI:FIRAS:zero:point:accuracy:545GHz:units}{2.2\MJysr}
\setsymbol{HFI:FIRAS:zero:point:accuracy:857GHz:units}{1.7\MJysr}

\setsymbol{HFI:FIRAS:zero:point:accuracy:100GHz}{0.8}
\setsymbol{HFI:FIRAS:zero:point:accuracy:143GHz}{}
\setsymbol{HFI:FIRAS:zero:point:accuracy:217GHz}{}
\setsymbol{HFI:FIRAS:zero:point:accuracy:353GHz}{1.4}
\setsymbol{HFI:FIRAS:zero:point:accuracy:545GHz}{2.2}
\setsymbol{HFI:FIRAS:zero:point:accuracy:857GHz}{1.7}

\setsymbol{HFI:unit:conversion:100GHz:units}{0.2415\MJysrmK}
\setsymbol{HFI:unit:conversion:143GHz:units}{0.3694\MJysrmK}
\setsymbol{HFI:unit:conversion:217GHz:units}{0.4811\MJysrmK}
\setsymbol{HFI:unit:conversion:353GHz:units}{0.2883\MJysrmK}
\setsymbol{HFI:unit:conversion:545GHz:units}{0.05826\MJysrmK}
\setsymbol{HFI:unit:conversion:857GHz:units}{0.002238\MJysrmK}

\setsymbol{HFI:unit:conversion:100GHz}{0.2415}
\setsymbol{HFI:unit:conversion:143GHz}{0.3694}
\setsymbol{HFI:unit:conversion:217GHz}{0.4811}
\setsymbol{HFI:unit:conversion:353GHz}{0.2883}
\setsymbol{HFI:unit:conversion:545GHz}{0.05826}
\setsymbol{HFI:unit:conversion:857GHz}{0.002238}

\setsymbol{HFI:colour:correction:alpha=-2:V1.01:100GHz}{0.9893}
\setsymbol{HFI:colour:correction:alpha=-2:V1.01:143GHz}{0.9759}
\setsymbol{HFI:colour:correction:alpha=-2:V1.01:217GHz}{1.0007}
\setsymbol{HFI:colour:correction:alpha=-2:V1.01:353GHz}{1.0028}
\setsymbol{HFI:colour:correction:alpha=-2:V1.01:545GHz}{1.0019}
\setsymbol{HFI:colour:correction:alpha=-2:V1.01:857GHz}{0.9889}

\setsymbol{HFI:colour:correction:alpha=0:V1.01:100GHz}{1.0008}
\setsymbol{HFI:colour:correction:alpha=0:V1.01:143GHz}{1.0148}
\setsymbol{HFI:colour:correction:alpha=0:V1.01:217GHz}{0.9909}
\setsymbol{HFI:colour:correction:alpha=0:V1.01:353GHz}{0.9888}
\setsymbol{HFI:colour:correction:alpha=0:V1.01:545GHz}{0.9878}
\setsymbol{HFI:colour:correction:alpha=0:V1.01:857GHz}{1.0014}

\newcommand{\fnl}{f_{\rm{NL}}}

\newcommand{\taunl}{\tau_{\rm{NL}}}
\newcommand{\gnl}{g_{\rm{NL}}}
\newcommand{\htaunl}{\hat{\tau}_{\rm NL}}
\newcommand{\barmodulation}{\bar{f}}
\newcommand{\hatmodulation}{\hat{f}}
\newcommand{\mfmodulation}{\barmodulation^{\rm MF}}

\providecommand{\fsky}{f_{\rm sky}}

\providecommand{\planck}{\textit{Planck}}

\newcommand{\covinvT}{\bar{T}}
\newcommand{\Ltemp}{\tilde{T}}
\newcommand{\lensedC}{\tilde{C}}
\newcommand{\Lmin}{{L_{\rm min}}}
\newcommand{\Lmax}{{L_{\rm max}}}

\providecommand{\alt}{\lea}

\providecommand{\LCDM}{{$\rm{\Lambda CDM}$}}

\newcommand\ba{\begin{eqnarray}}
\newcommand\ea{\end{eqnarray}}
\newcommand\bea{\begin{eqnarray}}
\newcommand\eea{\end{eqnarray}}

\newcommand\be{\begin{equation}}
\newcommand\ee{\end{equation}}

\providecommand{\cov}{\text{cov}}

\newcommand{\ud}{{\rm d}}

\newcommand{\boldvec}[1]{{{\vec{#1}}}}

\newcommand{\vn}{\boldvec{n}}

\newcommand{\vv}{\boldvec{v}}

\newcommand{\vx}{\boldvec{x}}

\newcommand{\clf}{\mathcal{F}}

\newcommand{\clo}{\mathcal{O}}

\newcommand{\vnhat}{\hat{\vn}}

\def\Commander{\texttt{Commander}}
\def\Ruler{\texttt{Ruler}}
\def\CR{\texttt{C-R}}
\def\NILC{\texttt{NILC}}
\def\SMICA{\texttt{SMICA}}
\def\SEVEM{\texttt{SEVEM}}

\def\px{\approx}

\def\={\nonumber &=}
\def\nn{\nonumber}
\def\({\left(}
\def\){\right)}
\def\[{\left[}
\def\]{\right]}
\def\<{\left\langle}
\def\>{\right\rangle}

\def\curl{\mathcal}

\def\eq{\begin{eqnarray}}
\def\qe{\end{eqnarray}}

\def\and{\quad \mbox{and} \quad}

\def\fnl{f_\textrm{NL}}

\def\Fnl{ F_\textrm {NL}}
\def\barFnl{ \bar F_\textrm {NL}}

\def\fnllocal{f_\textrm {NL}^\textrm{local}}

\def\bfnl{\kern2pt\overline{\kern-2ptf}_\textrm{NL}}

\def\lmax{\ell_\textrm{max}}

\def\blll{b_{\ell_1\ell_2\ell_3}}

\def\hlll{h_{\ell_1\ell_2\ell_3}}

\def\th{\textrm{th}}
\def\exp{\textrm{exp}}

\def\kall{k_1,k_2,k_3}

\def\alm{a_{\ell m}}
\def\almone{a_{\ell_1m_1}}
\def\almtwo{a_{\ell_2m_2}}
\def\almthree{a_{\ell_3m_3}}

\def\Ylm{Y_{\ell m}}

\def\Vtetra{{{\cal V}_{\cal T}}}

\def\barQ{\kern2pt\overline{\kern-2pt\curl{Q}}}

\def\bargamma{\kern2pt\overline{\kern-2pt\gamma}}
\def\barzeta{\kern2pt\overline{\kern-2pt\zeta}}

\def\barR{\kern2pt\overline{\kern-2pt\curl{R}}}

\def\nmax{n_\textrm{max}}

\def\aQ{\alpha^{\scriptscriptstyle{\cal Q}}}
\def\aQn{\aQ_n}

\def\bbQ{\bar{\beta}^{\scriptscriptstyle{\cal Q}}}
\def\bbQn{\bbQ_n}

\def\alll{\ell_1 \ell_2 \ell_3}

\def\llist{\ell_1,\ell_2,\ell_3}
\def\klist{k_1,k_2,k_3}
\def\eqref#1{(\ref{#1})}

\def\leaderfi1{\leaders\hbox to 5pt{\hss.\hss}\hfil}

\def\setsize{\csname @setfontsize\endcsname \setsize}

\begin{document}
\title{ \textit{Planck} 2013 results. XXIV. Constraints on primordial non-Gaussianity}

%This author list corresponds to \title{Author list for SVN P09a\_Fnl\_Bispectrum\_Trispectrum, Proj. Ref. 4\_2: Constraints on F\_NL based on Planck data}
%Prepared by R. Leonardi (rleonardi@sciops.esa.int), ESAC/ESA
%This version is from Thu Sep 05 13:37:08 2013 CET
%\subtitle{There are 239 co-authors in this list}
\author{\small
Planck Collaboration:
P.~A.~R.~Ade\inst{87}
\and
N.~Aghanim\inst{60}
\and
C.~Armitage-Caplan\inst{93}
\and
M.~Arnaud\inst{73}
\and
M.~Ashdown\inst{70, 6}
\and
F.~Atrio-Barandela\inst{18}
\and
J.~Aumont\inst{60}
\and
C.~Baccigalupi\inst{86}
\and
A.~J.~Banday\inst{96, 9}
\and
R.~B.~Barreiro\inst{67}
% ...
}
\institute{\small
APC, AstroParticule et Cosmologie, Universit\'{e} Paris Diderot, CNRS/IN2P3, CEA/lrfu, Observatoire de Paris, Sorbonne Paris Cit\'{e}, 10, rue Alice Domon et L\'{e}onie Duquet, 75205 Paris Cedex 13, France\\
\and
Aalto University Mets\"{a}hovi Radio Observatory, Mets\"{a}hovintie 114, FIN-02540 Kylm\"{a}l\"{a}, Finland\\
\and
African Institute for Mathematical Sciences, 6-8 Melrose Road, Muizenberg, Cape Town, South Africa\\
\and
Agenzia Spaziale Italiana Science Data Center, Via del Politecnico snc, 00133, Roma, Italy\\
\and
Agenzia Spaziale Italiana, Viale Liegi 26, Roma, Italy\\
\and
Astrophysics Group, Cavendish Laboratory, University of Cambridge, J J Thomson Avenue, Cambridge CB3 0HE, U.K.\\
\and
Astrophysics \& Cosmology Research Unit, School of Mathematics, Statistics \& Computer Science, University of KwaZulu-Natal, Westville Campus, Private Bag X54001, Durban 4000, South Africa\\
\and
CITA, University of Toronto, 60 St. George St., Toronto, ON M5S 3H8, Canada\\
\and
CNRS, IRAP, 9 Av. colonel Roche, BP 44346, F-31028 Toulouse cedex 4, France\\
\and
California Institute of Technology, Pasadena, California, U.S.A.\\
\and
Centre for Theoretical Cosmology, DAMTP, University of Cambridge, Wilberforce Road, Cambridge CB3 0WA, U.K.\\
\and
Centro de Estudios de F\'{i}sica del Cosmos de Arag\'{o}n (CEFCA), Plaza San Juan, 1, planta 2, E-44001, Teruel, Spain\\
\and
Computational Cosmology Center, Lawrence Berkeley National Laboratory, Berkeley, California, U.S.A.\\
\and
Consejo Superior de Investigaciones Cient\'{\i}ficas (CSIC), Madrid, Spain\\
\and
DSM/Irfu/SPP, CEA-Saclay, F-91191 Gif-sur-Yvette Cedex, France\\
\and
DTU Space, National Space Institute, Technical University of Denmark, Elektrovej 327, DK-2800 Kgs. Lyngby, Denmark\\
\and
D\'{e}partement de Physique Th\'{e}orique, Universit\'{e} de Gen\`{e}ve, 24, Quai E. Ansermet,1211 Gen\`{e}ve 4, Switzerland\\
\and
Departamento de F\'{\i}sica Fundamental, Facultad de Ciencias, Universidad de Salamanca, 37008 Salamanca, Spain\\
\and
Departamento de F\'{\i}sica, Universidad de Oviedo, Avda. Calvo Sotelo s/n, Oviedo, Spain\\
\and
Department of Astronomy and Astrophysics, University of Toronto, 50 Saint George Street, Toronto, Ontario, Canada\\
\and
Department of Astrophysics/IMAPP, Radboud University Nijmegen, P.O. Box 9010, 6500 GL Nijmegen, The Netherlands\\
\and
Department of Electrical Engineering and Computer Sciences, University of California, Berkeley, California, U.S.A.\\
\and
Department of Physics \& Astronomy, University of British Columbia, 6224 Agricultural Road, Vancouver, British Columbia, Canada\\
\and
Department of Physics and Astronomy, Dana and David Dornsife College of Letter, Arts and Sciences, University of Southern California, Los Angeles, CA 90089, U.S.A.\\
\and
Department of Physics and Astronomy, University College London, London WC1E 6BT, U.K.\\
\and
Department of Physics and Astronomy, University of Sussex, Brighton BN1 9QH, U.K.\\
\and
Department of Physics, Florida State University, Keen Physics Building, 77 Chieftan Way, Tallahassee, Florida, U.S.A.\\
\and
Department of Physics, Gustaf H\"{a}llstr\"{o}min katu 2a, University of Helsinki, Helsinki, Finland\\
\and
Department of Physics, Princeton University, Princeton, New Jersey, U.S.A.\\
\and
Department of Physics, University of California, Berkeley, California, U.S.A.\\
\and
Department of Physics, University of California, One Shields Avenue, Davis, California, U.S.A.\\
\and
Department of Physics, University of California, Santa Barbara, California, U.S.A.\\
\and
Department of Physics, University of Illinois at Urbana-Champaign, 1110 West Green Street, Urbana, Illinois, U.S.A.\\
\and
Dipartimento di Fisica e Astronomia G. Galilei, Universit\`{a} degli Studi di Padova, via Marzolo 8, 35131 Padova, Italy\\
\and
Dipartimento di Fisica e Scienze della Terra, Universit\`{a} di Ferrara, Via Saragat 1, 44122 Ferrara, Italy\\
\and
Dipartimento di Fisica, Universit\`{a} La Sapienza, P. le A. Moro 2, Roma, Italy\\
\and
Dipartimento di Fisica, Universit\`{a} degli Studi di Milano, Via Celoria, 16, Milano, Italy\\
\and
Dipartimento di Fisica, Universit\`{a} degli Studi di Trieste, via A. Valerio 2, Trieste, Italy\\
\and
Dipartimento di Fisica, Universit\`{a} di Roma Tor Vergata, Via della Ricerca Scientifica, 1, Roma, Italy\\
\and
Dipartimento di Matematica, Universit\`{a} di Roma Tor Vergata, Via della Ricerca Scientifica, 1, Roma, Italy\\
\and
Discovery Center, Niels Bohr Institute, Blegdamsvej 17, Copenhagen, Denmark\\
\and
Dpto. Astrof\'{i}sica, Universidad de La Laguna (ULL), E-38206 La Laguna, Tenerife, Spain\\
\and
European Space Agency, ESAC, Planck Science Office, Camino bajo del Castillo, s/n, Urbanizaci\'{o}n Villafranca del Castillo, Villanueva de la Ca\~{n}ada, Madrid, Spain\\
\and
European Space Agency, ESTEC, Keplerlaan 1, 2201 AZ Noordwijk, The Netherlands\\
\and
Helsinki Institute of Physics, Gustaf H\"{a}llstr\"{o}min katu 2, University of Helsinki, Helsinki, Finland\\
\and
INAF - Osservatorio Astronomico di Padova, Vicolo dell'Osservatorio 5, Padova, Italy\\
\and
INAF - Osservatorio Astronomico di Roma, via di Frascati 33, Monte Porzio Catone, Italy\\
\and
INAF - Osservatorio Astronomico di Trieste, Via G.B. Tiepolo 11, Trieste, Italy\\
\and
INAF Istituto di Radioastronomia, Via P. Gobetti 101, 40129 Bologna, Italy\\
\and
INAF/IASF Bologna, Via Gobetti 101, Bologna, Italy\\
\and
INAF/IASF Milano, Via E. Bassini 15, Milano, Italy\\
\and
INFN, Sezione di Bologna, Via Irnerio 46, I-40126, Bologna, Italy\\
\and
INFN, Sezione di Roma 1, Universit\`{a} di Roma Sapienza, Piazzale Aldo Moro 2, 00185, Roma, Italy\\
\and
IPAG: Institut de Plan\'{e}tologie et d'Astrophysique de Grenoble, Universit\'{e} Joseph Fourier, Grenoble 1 / CNRS-INSU, UMR 5274, Grenoble, F-38041, France\\
\and
IUCAA, Post Bag 4, Ganeshkhind, Pune University Campus, Pune 411 007, India\\
\and
Imperial College London, Astrophysics group, Blackett Laboratory, Prince Consort Road, London, SW7 2AZ, U.K.\\
\and
Infrared Processing and Analysis Center, California Institute of Technology, Pasadena, CA 91125, U.S.A.\\
\and
Institut N\'{e}el, CNRS, Universit\'{e} Joseph Fourier Grenoble I, 25 rue des Martyrs, Grenoble, France\\
\and
Institut Universitaire de France, 103, bd Saint-Michel, 75005, Paris, France\\
\and
Institut d'Astrophysique Spatiale, CNRS (UMR8617) Universit\'{e} Paris-Sud 11, B\^{a}timent 121, Orsay, France\\
\and
Institut d'Astrophysique de Paris, CNRS (UMR7095), 98 bis Boulevard Arago, F-75014, Paris, France\\
\and
Institute for Space Sciences, Bucharest-Magurale, Romania\\
\and
Institute of Astronomy and Astrophysics, Academia Sinica, Taipei, Taiwan\\
\and
Institute of Astronomy, University of Cambridge, Madingley Road, Cambridge CB3 0HA, U.K.\\
\and
Institute of Theoretical Astrophysics, University of Oslo, Blindern, Oslo, Norway\\
\and
Instituto de Astrof\'{\i}sica de Canarias, C/V\'{\i}a L\'{a}ctea s/n, La Laguna, Tenerife, Spain\\
\and
Instituto de F\'{\i}sica de Cantabria (CSIC-Universidad de Cantabria), Avda. de los Castros s/n, Santander, Spain\\
\and
Jet Propulsion Laboratory, California Institute of Technology, 4800 Oak Grove Drive, Pasadena, California, U.S.A.\\
\and
Jodrell Bank Centre for Astrophysics, Alan Turing Building, School of Physics and Astronomy, The University of Manchester, Oxford Road, Manchester, M13 9PL, U.K.\\
\and
Kavli Institute for Cosmology Cambridge, Madingley Road, Cambridge, CB3 0HA, U.K.\\
\and
LAL, Universit\'{e} Paris-Sud, CNRS/IN2P3, Orsay, France\\
\and
LERMA, CNRS, Observatoire de Paris, 61 Avenue de l'Observatoire, Paris, France\\
\and
Laboratoire AIM, IRFU/Service d'Astrophysique - CEA/DSM - CNRS - Universit\'{e} Paris Diderot, B\^{a}t. 709, CEA-Saclay, F-91191 Gif-sur-Yvette Cedex, France\\
\and
Laboratoire Traitement et Communication de l'Information, CNRS (UMR 5141) and T\'{e}l\'{e}com ParisTech, 46 rue Barrault F-75634 Paris Cedex 13, France\\
\and
Laboratoire de Physique Subatomique et de Cosmologie, Universit\'{e} Joseph Fourier Grenoble I, CNRS/IN2P3, Institut National Polytechnique de Grenoble, 53 rue des Martyrs, 38026 Grenoble cedex, France\\
\and
Laboratoire de Physique Th\'{e}orique, Universit\'{e} Paris-Sud 11 \& CNRS, B\^{a}timent 210, 91405 Orsay, France\\
\and
Lawrence Berkeley National Laboratory, Berkeley, California, U.S.A.\\
\and
Max-Planck-Institut f\"{u}r Astrophysik, Karl-Schwarzschild-Str. 1, 85741 Garching, Germany\\
\and
McGill Physics, Ernest Rutherford Physics Building, McGill University, 3600 rue University, Montr\'{e}al, QC, H3A 2T8, Canada\\
\and
MilliLab, VTT Technical Research Centre of Finland, Tietotie 3, Espoo, Finland\\
\and
National University of Ireland, Department of Experimental Physics, Maynooth, Co. Kildare, Ireland\\
\and
Niels Bohr Institute, Blegdamsvej 17, Copenhagen, Denmark\\
\and
Observational Cosmology, Mail Stop 367-17, California Institute of Technology, Pasadena, CA, 91125, U.S.A.\\
\and
Optical Science Laboratory, University College London, Gower Street, London, U.K.\\
\and
SB-ITP-LPPC, EPFL, CH-1015, Lausanne, Switzerland\\
\and
SISSA, Astrophysics Sector, via Bonomea 265, 34136, Trieste, Italy\\
\and
School of Physics and Astronomy, Cardiff University, Queens Buildings, The Parade, Cardiff, CF24 3AA, U.K.\\
\and
School of Physics and Astronomy, University of Nottingham, Nottingham NG7 2RD, U.K.\\
\and
Space Research Institute (IKI), Russian Academy of Sciences, Profsoyuznaya Str, 84/32, Moscow, 117997, Russia\\
\and
Space Sciences Laboratory, University of California, Berkeley, California, U.S.A.\\
\and
Special Astrophysical Observatory, Russian Academy of Sciences, Nizhnij Arkhyz, Zelenchukskiy region, Karachai-Cherkessian Republic, 369167, Russia\\
\and
Stanford University, Dept of Physics, Varian Physics Bldg, 382 Via Pueblo Mall, Stanford, California, U.S.A.\\
\and
Sub-Department of Astrophysics, University of Oxford, Keble Road, Oxford OX1 3RH, U.K.\\
\and
Theory Division, PH-TH, CERN, CH-1211, Geneva 23, Switzerland\\
\and
UPMC Univ Paris 06, UMR7095, 98 bis Boulevard Arago, F-75014, Paris, France\\
\and
Universit\'{e} de Toulouse, UPS-OMP, IRAP, F-31028 Toulouse cedex 4, France\\
% ...
}

\date{Received xxxx, Accepted xxxxx}

%%%%%%%%%%%%%%% abstract %%%%%%%%%%%%%%%

 \abstract {
 The  \textit{Planck}  nominal mission cosmic microwave background (CMB) maps yield unprecedented  constraints on  primordial non-Gaussianity (NG).
Using three optimal bispectrum estimators, separable template-fitting (KSW), binned, and modal, we obtain consistent values for the primordial local, equilateral, and orthogonal bispectrum amplitudes, quoting as our final result  $\fnllocal=2.7 \pm 5.8$, $f_\textrm{NL}^\textrm{equil}= - 42 \pm 75$, and $f_\textrm{NL}^\textrm{ortho}= - 25 \pm 39$ ($68\%$ CL statistical). Non-Gaussianity is detected in the data; using skew-$C_\ell$ statistics we find a nonzero bispectrum from residual point sources, and the Integrated-Sachs-Wolfe-lensing bispectrum at a level expected in the \LCDM\ scenario. The results are based on  comprehensive cross-validation of these estimators on Gaussian and non-Gaussian simulations, are stable across component separation techniques, pass an extensive suite of tests, and are confirmed by skew-$C_\ell$, wavelet bispectrum and Minkowski functional estimators. Beyond estimates of individual shape amplitudes, we present model-independent, three-dimensional reconstructions of the \textit{Planck} CMB bispectrum and thus derive constraints on early-Universe scenarios that generate primordial NG, including general single-field models of inflation, excited initial states (non-Bunch-Davies vacua), and directionally-dependent vector models. We provide an initial survey of scale-dependent feature and resonance models.  These results bound both general single-field and multi-field model parameter ranges, such as the speed of sound, $c_{\textrm{s}} \geq 0.02$ ($95\%$ CL), in an effective field theory parametrization, and  the curvaton decay fraction $r_\mathrm{D} \geq 0.15$ ($95\%$ CL).   The \textit{Planck} data significantly limit 
 the viable parameter space of the ekpyrotic/cyclic scenarios. The amplitude of the four-point function in the local model   $\tau_{\mathrm{NL}} < 2800$ ($95\%$ CL). Taken together, these constraints represent the highest precision tests to date of physical mechanisms for the origin of cosmic structure. 
}

\keywords{cosmology: cosmic background radiation -- cosmology: observations -- cosmology: theory -- cosmology: early Universe -- cosmology: inflation -- methods: data analysis}
%\keywords{Primordial non-Gaussianity -- \textit{Planck}  }

\authorrunning{Planck Collaboration}
\titlerunning{ \textit{Planck} 2013 Results. XXIV. Constraints on primordial NG}

\maketitle
\clearpage

%%%%%%%%%%%%%%% Section 1 %%%%%%%%%%%%%%%
\section {Introduction}
\label{intro}

This paper, one of a set associated with the 2013 release of data from the \Planck\footnote{\Planck\ (\url{http://www.esa.int/Planck}) is a project of the European Space Agency (ESA) with instruments provided by two scientific consortia funded by ESA member states (in particular the lead countries France and Italy), with contributions from NASA (USA) and telescope reflectors provided by a collaboration between ESA and a scientific consortium led and funded by Denmark.} mission~\citep{planck2013-p01}, describes the constraints on primordial non-Gaussianity (NG) obtained using the cosmic microwave background (CMB) maps derived from the data acquired by \textit{Planck} during its nominal operations period, i.e., between 12 August 2009 and 27 November 2010.
 
Primordial NG is one of the most informative fingerprints of the origin of structure in the Universe, probing physics at extremely high energy scales inaccessible to laboratory experiments. Possible departures from a purely Gaussian distribution of the CMB anisotropies provide powerful observational access to this extreme physics~(\citealt{1987PhLB..197...66A,1990PhRvD..42.3936S,1993ApJ...403L...1F,1994ApJ...430..447G,2000MNRAS.313..141V,2000MNRAS.313..323G,2000PhRvD..61f3504W,2001PhRvD..63f3002K,2003NuPhB.667..119A,2003JHEP...05..013M,2004JCAP...08..009B}; for recent reviews~\citealt{Bartolo:2004if},~\citealt{2010AdAst2010E..73L},~\citealt{2010AdAst2010E..72C},~\citealt{2010CQGra..27l4010K},~\citealt{2010AdAst2010E..71Y}).  A robust detection of primordial NG -- or a strong constraint on it -- discriminates among competing mechanisms for the generation of the cosmological perturbations in the early Universe. Different inflationary models, firmly rooted in modern theoretical particle physics, predict different {\it amplitudes}, {\it shapes}, and {\it scale dependence} of NG. As a result, primordial NG is complementary to the scalar-spectral index of curvature perturbations and the tensor-to-scalar amplitude ratio, distinguishing between inflationary models that are degenerate on the basis of their power spectra alone. Even in the simplest models of inflation, consisting of a single slowly-rolling scalar field, a small (but calculable) level of NG is predicted~(\citealt{2003NuPhB.667..119A,2003JHEP...05..013M}); this is undetectable in present-quality CMB and large-scale structure measurements. However, as demonstrated by a large body of work in recent years, extending this simplest paradigm will generically lead to detectable levels of NG in CMB anisotropies. Critically, a robust detection of primordial NG would rule out \emph{all} canonical single-field slow-roll models of inflation, pointing to physics beyond the simplest ``textbook'' picture of inflation. Conversely, significant improvements in the constraints on primordial NG strongly limit extensions to the simplest paradigm, thus providing powerful clues to the physical mechanism that generated cosmic structure.

If the primordial fluctuations are Gaussian-distributed, then they are completely characterised by  their two-point correlation function, or equivalently, their power spectrum. If they are non-Gaussian, there is additional statistical information in the higher-order correlation functions, which is not captured by the two-point correlation function. In particular, the 3-point correlation function, or its Fourier counterpart, the {\it bispectrum}, is important because it is the lowest-order statistic that can distinguish between Gaussian and non-Gaussian perturbations.  One of the main goals of this paper is to constrain the amplitude and shape of primordial NG using the angular bispectrum of the CMB anisotropies. The CMB angular bispectrum is related to the primordial bispectrum defined by
\begin{equation}
\label{bispectrumPhi}
\langle  \Phi({\vec k}_1) \Phi({\vec k}_2) \Phi({\vec k}_3) \rangle= (2 \pi)^3 \delta^{(3)}({\vec k}_1+{\vec k}_2+{\vec k}_3) B_{\rm \Phi}(k_1,k_2,k_3).
\end{equation}
Here we define the potential $\Phi$ in terms of the comoving curvature perturbation $\zeta$ on super-horizon scales by $\Phi \equiv (3/5) \zeta$. In matter
domination, on super-horizon scales, $\Phi$ is equivalent to Bardeen's gauge-invariant gravitational potential (\citealt{1980PhRvD..22.1882B}), and we adopt this notation for historical consistency. 
The bispectrum  $B_{\Phi}(k_1,k_2,k_3)$ measures the correlation among three perturbation modes. Assuming translational and rotational invariance, it depends only on the magnitudes of the three wavevectors. In general the bispectrum can be written as 
\begin{equation}
\label{amplitudeandF}
B_{\Phi}(k_1,k_2,k_3)= f_{\rm NL} F(k_1,k_2,k_3) \, .
\end{equation}
Here, $f_{\rm NL}$ is the so-called ``nonlinearity parameter'' ~(\citealt{1994ApJ...430..447G,2000PhRvD..61f3504W,2001PhRvD..63f3002K,2004JCAP...08..009B}), a dimensionless parameter measuring the amplitude of NG. 
The bispectrum is measured by sampling triangles in Fourier space. The dependence of the function $F(k_1,k_2,k_3)$ on the type of triangle (i.e., the configuration) formed by the three wavevectors describes the \emph{shape} of the bispectrum~(\citealt{2004JCAP...08..009B}), which encodes much physical information. It can also encode the scale dependence, i.e., the \emph{running}, of the bispectrum~(\citealt{2005PhRvD..72l3518C}).\footnote{Specifically, one can define the shape of the bispectrum as the dependence of $F(k_1,k_2,k_3) (k_1 k_2 k_3)^2$ on the ratios of momenta, e.g., $(k_2/k_1)$ and $(k_3/k_1)$, once the overall scale of the triangle $K=k_1+k_2+k_3$ is fixed. The scale dependence of the bispectrum can be characterized by the dependence of  $F(k_1,k_2,k_3) (k_1 k_2 k_3)^2$ on the overall scale $K$, once the ratios $(k_2/k_1)$ and $(k_3/k_1)$ are fixed (see, e.g.,~\citealt{2010AdAst2010E..72C}).} Different NG shapes are linked to distinctive physical mechanisms that can generate such non-Gaussian fingerprints in the early Universe. For example, 
the so-called ``local'' NG~(\citealt{1994ApJ...430..447G,2000MNRAS.313..141V,2000PhRvD..61f3504W,2001PhRvD..63f3002K}) is characterized by a signal that is maximal for ``squeezed'' triangles with $k_1 \ll k_2\simeq k_3$~(or permutations; \citealt{2003JHEP...05..013M}) which occurs, in general, when the primordial NG is generated on super-horizon scales. Conversely, ``equilateral'' NG~(\citealt{2004JCAP...08..009B}) peaks for  equilateral configurations $k_1 \approx k_2 \approx k_3$, due to correlations between fluctuation modes that are of comparable wavelengths, which can occur if the  three perturbation modes mostly interact when they cross the horizon approximately at the same time. Other relevant shapes include the so-called ``folded'' (or flattened) NG~(\citealt{2007JCAP...01..002C}), which is due to correlations between perturbation modes that are enhanced for  
$k_1 + k_2 \approx k_3$, or the ``orthogonal'' NG~(\citealt{2010JCAP...01..028S}) that generates a signal with a positive peak at the equilateral configuration and a negative peak at the folded configuration.

We now sketch how non-Gaussian information in the initial conditions is transferred to observable quantities (in this instance, the CMB anisotropies) in the context of inflation. Primordial perturbations in the inflaton field(s) $\phi({\vec x},t)=\phi_0(t)+\delta \phi({\vec x},t)$ (where $\delta \phi$ denotes quantum fluctuations about the background value $\phi_0(t)$) can be characterized by the comoving curvature perturbation $\zeta$,  since this is conserved on super-horizon scales for adiabatic perturbations. The inflaton fluctuations $\delta \phi$ (in the flat gauge) induce a curvature perturbation\footnote{For the curvature perturbation, we follow the notation and sign conventions of \cite{2011ApJS..192...18K}. $\zeta$ is also sometimes denoted ${\mathcal R}$ (see e.g.,~\citealt{1997RvMP...69..373L},~\citealt{1999PhR...314....1L} and references therein), while the comoving curvature perturbation ${\mathcal R}$ as defined, e.g., in \cite{2009PhR...475....1M} is such that ${\mathcal R}=-\zeta$.} $\zeta=-(H/\dot{\phi_0})\, \delta \phi$ at linear order; however, nonlinearities induce corrections to this relation.  The primordial NG in the curvature perturbation $\zeta$ is intrinsically nonlinear, so that its contribution to the CMB anisotropies is transferred linearly at leading order.  In particular, at the linear level, the curvature perturbation $\zeta$ is related to Bardeen's gravitational potential $\Phi$ during the matter-dominated epoch by $\Phi=(3/5) \zeta$ and $\Delta T/T\sim g\, \zeta$, where $g$ is the linear radiation transfer function; thus, any primordial NG will be transferred to the CMB even at linear order. For example, in the large-angular scale limit, the linear Sachs-Wolfe effect reads $\DeltaT /T=-\Phi/3=- \zeta/5$. Further, any other field excited during the inflationary phase which develops quantum fluctuations contributing to the  primordial curvature perturbation -- whether or not it is driving inflation -- can leave its non-Gaussian imprint in the CMB anisotropies. 

Thus the bispectrum of Eq.~(\ref{bispectrumPhi}) measures the fundamental (self-) interactions of the scalar field(s) involved in the inflationary phase and/or generating the primordial curvature perturbation, as well as measuring nonlinear processes occurring during or after inflation. It therefore brings insights into the fundamental physics behind inflation, possibly allowing for the first time a reconstruction of the inflationary Lagrangian itself.  For example, in a large class of inflationary models which involve 
additional light field(s) different from the inflaton, the super-horizon evolution of the fluctuations in the additional field(s) and their transfer to the adiabatic curvature perturbations can generate a large primordial NG of the local type. This is the case for curvaton-type models~(\citealt{1997PhRvD..56..535L,2002PhLB..524....5L,2003PhRvD..67b3503L}) where the late-time decay of a scalar field, belonging to the non-inflationary sector of the theory, induces curvature perturbations; models where the curvature perturbation is generated by the local fluctuations of the inflaton's coupling to matter during the reheating phase~(\citealt{2003astro.ph..3614K,2004PhRvD..69h3505D}); and multi-field models of inflation~(see, e.g.,~\citealt{2002PhRvD..65j3505B},~\citealt{2002PhRvD..66j3506B},~\citealt{2006JCAP...05..019V},~\citealt{2006PhRvD..73h3522R,2007PhRvD..76h3512R,2005PhRvL..95l1302L},~\citealt{2010AdAst2010E..76B}). Since the nonlinear processes take place on super-horizon scales, the form of NG is local in real space and thus, in Fourier space, the bispectrum correlates large and small Fourier modes.
``Equilateral''  NG~(\citealt{2004JCAP...08..009B}) is a generic feature of  single-field models with a non-canonical kinetic term, which can also generate the ``orthogonal'' type of NG~(\citealt{2010JCAP...01..028S}). In general, these models are characterized by higher-derivative interactions of the inflaton field. The correlation between the fluctuation modes is suppressed when one of the modes is on super-horizon scales, because the derivative terms are redshifted away, so that the correlation is maximal for three modes of comparable wavelengths that cross the horizon at the same time. An example of ``folded'' NG is the one generated in a class of single-field models with non-Bunch-Davies vacuum~(\citealt{2007JCAP...01..002C,2008JCAP...05..001H}).  Indeed, these and other types of primordial NG can also be produced in other models, and we refer to Sect.~\ref{SectionII} for more details.  All these models can easily yield primordial NG with an amplitude much bigger than the one predicted in the standard models of single-field slow-roll inflation, for which the NG amplitude turns out to be proportional to the usual slow-roll parameters $f_{\rm NL} \sim {\cal O}({\epsilon, \eta})$~(\citealt{2003NuPhB.667..119A,2003JHEP...05..013M}).

Given that a robust detection of primordial NG would represent a breakthrough in the understanding of the physics governing the Universe during its very first stages, it is crucial that all sources of contamination are sufficiently understood to firmly control their effects. In particular, any nonlinearity in the post-inflationary Universe can introduce NG into perturbations that were initially Gaussian. Therefore, one must ensure that a primordial origin is not ascribed to a non-primordial contaminant; however, estimators of (primordial) NG from CMB data will also typically be sensitive to such contaminating signals. Potential non-primordial sources of NG can be classified into four broad categories: instrumental systematic effects (see e.g.,~\citealt{2009ApJ...706.1226D}); residual foregrounds and point sources; secondary CMB anisotropies, such as the Sunyaev-Zeldovich (SZ) effect~\citep{1969Ap&SS...4..301Z}, gravitational lensing~(see~\citealt{2006PhR...429....1L} for a review), the Integrated Sachs-Wolfe (ISW) effect~\citep{Sachs:1967er} or the Rees-Sciama effect~\citep{1968Natur.217..511R}; and effects arising from nonlinear (second-order) perturbations in the Boltzmann equations (due to the nonlinear nature of General Relativity and the nonlinear dynamics of the photon-baryon fluid at recombination). Among the secondary anisotropies, the cross-correlation of the ISW/Rees-Sciama and lensing~(\citealt{1999PhRvD..59j3002G}) produces the dominant contamination to the (local) primordial NG. The impact is mainly on the local type of primordial NG, because the ISW-lensing correlation couples the large-scale gravitational potential fluctuations sourcing the ISW effect with the small-scale lensing effects of the CMB, thus producing a bispectrum which peaks on the squeezed configurations, as for the local shape. Detailed analyses have shown that the ISW-lensing bispectrum can introduce a bias to local primordial NG, while the bias to equilateral primordial NG is negligible (see~\citealt{2008PhRvD..77j7305S},~\citealt{2011MNRAS.417....2S},~\citealt{2009PhRvD..80h3004H},~\citealt{2011JCAP...03..018L},~\citealt{2009PhRvD..80l3007M},~\citealt{2012arXiv1204.3789J},~\citealt{2012arXiv1204.5018L},~\citealt{2013arXiv1303.1722M}). In our analysis we have carefully accounted for this effect (we report the values of the ISW-lensing bias in Sect.~\ref{sec:lensingISW}, and demonstrate the detection of the effect with skew-$C_\ell$s), as well as validating our results through an extensive suite of simulations and null tests in order to quantify the effects of systematic effects and diffuse and point-source foregrounds. 
Finally, a consistent treatment of weak NG in the CMB must account for additional contributions that arise at the nonlinear (second-order) level both in the gravitational perturbations after inflation ends, and for the evolution of the CMB anisotropies at second-order in perturbation theory at large and small angular scales. It has been shown that these second-order CMB effects yield negligible contamination to primordial NG for \textit{Planck}-quality data~(\citealt{2004JCAP...01..003B,2004PhRvD..70h3532C,2005JCAP...08..010B,2009JCAP...08..029B,2009JCAP...05..014N,2009JCAP...09..038S,2009PhRvD..79b3501K,2009JCAP...03..017B,2010PhRvD..81j3518K,2010AdAst2010E..75B,2011JCAP...11..025C,2012JCAP...02..017B,2012arXiv1212.3573H,2012arXiv1212.6968S,2013arXiv1302.0832P}).

Previous constraints on various shapes of primordial NG come from the \textit{WMAP}-9 data~(\citealt{2012arXiv1212.5225B}). For the local shape they find $f_{\rm NL}^{\rm local} = 37 \pm 20$ ($68 \%$ CL). For equilateral-type NG, they obtain $f_{\rm NL}^{\rm equil} = 51\pm 136$ ($68 \%$ CL), while for the orthogonal shape $f_{\rm NL}^{\rm ortho}= - 245\pm100$ ($68 \%$ CL). 
Other analyses employing different estimators give compatible constraints. 
Limits on other shapes, such as e.g. flattened and feature models, have also been obtained~(\citealt{2010arXiv1006.1642F}). 

Before concluding this section let us point out the connection between the analyses presented here and in the companion paper~\cite{planck2013-p09} on the statistical and isotropy properties of the CMB.  
Statistical anisotropy and NG are essentially two alternative 
descriptions of the same phenomenon on the sky \citep{1997PhRvD..56.4578F}. 
Specifically any Gaussian but statistically anisotropic model becomes, after 
averaging over the possible (a priori unknown) orientations of the 
anisotropy, a statistically isotropic non-Gaussian model. For example local 
NG can be generated by large-scale field fluctuations that 
couple to the small-scale power. For the given fixed realization of 
large-scale modes that we see, the small-scale anisotropies look anisotropic 
on the sky, and it is equally valid to describe this as a Gaussian anisotropic 
model (assuming the large-scale modes are Gaussian). In this paper we mostly 
focus on the non-Gaussian interpretation of various physically motivated 
models, although it is useful to bear both perspectives in mind, in 
particular when considering what forms of non-primordial signal might cause 
contamination.  \cite{planck2013-p09} consider a broad class of more general 
phenomenological forms of anisotropy, which are complementary to the analysis 
presented here.

This paper is organized as follows. In  Sect.~\ref{SectionII}, we present models generating primordial NG that have been tested in this paper. Section~\ref{sec:se} summarizes the statistical estimators used to constrain 
the CMB bispectrum from \textit{Planck} data and the methods for the reconstruction of the CMB bispectrum. Section~\ref{sec:CMBt} summarizes the statistical estimator used to constrain the CMB trispectrum. In Sect.~\ref{sec:npNG}, we discuss the non-primordial contributions to the CMB bispectrum and trispectrum, including foreground residuals after component separation and focusing on the $f_{\rm NL}$ bias induced by the ISW-lensing bispectrum. Section~\ref{sec:Validation} describes an extensive suite of tests performed on realistic simulations to validate the different estimator pipelines, and compare their performance. Using simulations, we also quantify the impact on $f_{\rm NL}$ of using a variety of component-separation techniques. Section~\ref{sec:Results} contains our main results: we present constraints on $f_{\rm NL}$ for the local, equilateral, and orthogonal bispectra, and a selected set of other bispectrum shapes; we show a reconstruction of the CMB bispectrum, and give limits on the CMB trispectrum. In Sect.~\ref{Sec_valid_data} we validate these results by performing a series of null tests on the data to assess the robustness of our results. We also evaluate the impact of the \Planck\ data processing on the primordial NG signal.
In Sect.~\ref{sec:Implications}, we discuss the main implications of \textit{Planck}'s constraints on primordial NG for early Universe models. We conclude in Sect.~\ref{sec:Conc}. The realistic \Planck\ simulations used in various steps of the analysis and validation tests are described in Appendix~\ref{sec:FFP6}.  Appendix~\ref{sec:AA} contains a derivation of the expected scatter between $f_{\rm NL}$ results on the same map from different estimators used in the validation tests of Sect.~\ref{sec:Validation}, while Appendix~\ref{sec:AB} presents  a comparison of constraints on some selected non-standard bispectrum shapes using different foreground-cleaned maps.

%%%%%%%%%%%%%%% Section 2 %%%%%%%%%%%%%%%
\section{Inflationary models for primordial non-Gaussianity}
\label{SectionII}

There is a simple reason why standard single-field models of slow-roll inflation predict a tiny level of NG, of the order of the usual slow-roll parameters $f_{\rm NL}\sim{\cal O}({\epsilon, \eta})$:\footnote{This has been shown in the pioneering research which demonstrated that perturbations produced in single-field models of slow-roll inflation are characterized by a low-amplitude NG~(\citealt{1990PhRvD..42.3936S,1993ApJ...403L...1F,1994ApJ...430..447G}). Later \cite{2003NuPhB.667..119A} and~\cite{2003JHEP...05..013M} obtained a complete quantitative prediction for the nonlinearity parameter in single-field slow-roll inflation models, also showing that the predicted NG is characterized by a shape dependence which is more complex than suggested by previous results expressed in terms of the simple parameterization $\Phi({\vec x})=\Phi_{\rm L}({\vec x})+f_{\rm NL}\Phi^2_{\rm L}({\vec x})$~(\citealt{1994ApJ...430..447G,2000MNRAS.313..141V,2000PhRvD..61f3504W,2001PhRvD..63f3002K}), where $\Phi_{\rm L}$ is the linear gravitational potential.}  in order to achieve an accelerated period of expansion, the inflaton potential must be very flat, thus suppressing the inflaton (self-)interactions and any sources of nonlinearity, and leaving only its weak gravitational interactions as the main source of NG. This fact leads to a clear distinction between the simplest models of inflation, and scenarios where a significant amplitude of NG can be generated (e.g.,~\citealt{2010CQGra..27l4010K}), as follows. The simplest inflationary models are based on a set of minimal conditions: (i)~a single weakly-coupled neutral single scalar field (the \textit{inflaton}, which drives inflation and generates the curvature perturbations);  (ii)~with a canonical kinetic term; (iii) ~slowly rolling down its (featureless) potential; (iv)~initially lying in a Bunch-Davies (ground) vacuum state. In the last few years, an important theoretical realization has taken place: a detectable amplitude of NG with specific triangular configurations (corresponding broadly to well-motivated classes of
physical models) can be generated if any one of the above conditions is violated (\citealt{Bartolo:2004if,2010AdAst2010E..73L,2010AdAst2010E..72C,2010CQGra..27l4010K,2010AdAst2010E..71Y}):  
\begin{itemize}
\item ``local''  NG, where the signal peaks in ``squeezed''  triangles ($k_1 \ll k_2\simeq k_3$) (e.g., multi-field models of inflation); 
\item ``equilateral'' NG, peaking for $k_1 \approx k_2 \approx k_3$. Examples of this class include single-field models with non-canonical kinetic term~(\citealt{2007JCAP...01..002C}), such as $k$-inflation~(\citealt{1999PhLB..458..209A,2007JCAP...01..002C}) or Dirac-Born-Infield (DBI)  inflation~(\citealt{2004PhRvD..70j3505S,2004PhRvD..70l3505A}), models characterized by more general higher-derivative interactions of the inflaton field, such as ghost inflation~(\citealt{2004JCAP...04..001A}), and models arising from effective field theories~(\citealt{2008JHEP...03..014C}); 
\item ``folded'' (or flattened) NG.
Examples of this class include: single-field models with non-Bunch-Davies vacuum~(\citealt{2007JCAP...01..002C,2008JCAP...05..001H}) and models with general higher-derivative interactions~(\citealt{2010JCAP...01..028S,2010JCAP...08..008B}); 
\item ``orthogonal'' NG
which is generated, e.g., in single-field models of inflation with a non-canonical kinetic term~(\citealt{2010JCAP...01..028S}), or with general higher-derivative interactions. 
\end{itemize}
All these models naturally predict values of $| f_{\rm NL} | \gg  1$. A detection of such a signal would rule out the simplest models of single-field inflation, which, obeying \emph{all} the conditions above, are characterized by weak gravitational interactions with $| f_{\rm NL}| \ll 1$. 

The above scheme provides a general classification of inflationary models in terms of the corresponding NG shapes, which we adopt for the data analysis presented in this paper: 
\begin{enumerate}
\item ``general'' single-field inflationary models (tested using the equilateral, orthogonal and folded shapes);
\item multi-field models of inflation (tested using the local shape).
\end{enumerate}
In each class, there exist specific realizations of inflationary models which are characterized by the same underlying physical mechanism, generating a specific NG shape. We will investigate these classes of inflationary models by constraining the corresponding NG content, focusing on amplitudes and shapes. We also perform a survey of non-standard models giving rise to alternative specific shapes of NG. Different NG shapes are observationally distinguishable if their cross-correlation is sufficiently low; almost all of the shapes analysed in this paper are highly orthogonal to each other (e.g., \citealt{2004JCAP...08..009B,2007PhRvD..76h3523F}).

There are exceptional cases which evade this classification: for example, some exotic non-local single-field theories of inflation produce local NG~(\citealt{2008JCAP...06..030B}), while some multi-field models can produce equilateral NG, e.g., if some particle production mechanism is present (examples include trapped inflation~\citealt{2009PhRvD..80f3533G},  and some models of axion inflation \citealt{2011PhRvL.106r1301B, 2011JCAP...04..009B,2012PhRvD..85b3525B}). Another example arises in a class of multi-field models where the second scalar field is not light, but has a mass $m \approx H$, of the order of the Hubble rate during inflation. Then NG with an intermediate shape, interpolating between local and equilateral, can be produced -- ``quasi-single field'' models of inflation~\citep{2010PhRvD..81f3511C,2010JCAP...04..027C} -- for which the NG shape is similar to the so-called constant NG of~\cite{2007PhRvD..76h3523F}. Furthermore, there is the possibility of a superposition of shapes (and/or running of NG), generated if different mechanisms sourcing NG act simultaneously during the inflationary evolution. For example, in multi-field DBI inflation, equilateral NG is generated at horizon crossing from the higher-derivative interactions of the scalar fields, and it adds to the local NG arising from the super-horizon nonlinear evolution (e.g.,~\citealt{2008PhRvL.101f1301L,2008PhRvD..78f3523L,2008JCAP...08..015A,2009JCAP...10..012R}). 

In the following subsections, we discuss each of these possibilities in turn. The reader already familiar with this background material may skip to Sect.~\ref{sec:se}.

%%%%%%%%%%%%%%%%%%%%%%%%%%%%%%%%%%%%%%%%%%%%%%%%%%%%

\subsection{General single-field models of inflation}

Typically in models with a non-standard kinetic term (or more general higher-derivative interactions), inflaton perturbations propagate with an effective sound speed $c_s $ which can be smaller than the speed of light, and this results in a contribution to the NG amplitude $f_{\rm NL} \sim c_{\rm s}^{-2}$ in the limit $c_{\rm s} \ll 1$. For example, models with a non-standard kinetic term are described by an inflaton Lagrangian ${\mathcal L}=P(X, \phi)$, where $X=g^{\mu \nu} \partial_\mu \phi \, \partial_\nu \phi$, with at most one derivative on $\phi$, and the sound speed is $c_{\rm s} ^2=(\partial P/\partial X)/(\partial P/\partial X+2 X (\partial^2 P/\partial X^2))$.

In this case, two interaction terms give the dominant contribution to primordial NG, one of the type $(\dot{\delta \phi})^3$ and the other of the type $\dot{\delta \phi} (\nabla \delta \phi)^2$, which arise from expanding the $P(X,\phi)$ Lagrangian. Each of these two interaction terms generates a bispectrum with a shape similar to the equilateral type, with the second inflaton interaction yielding a nonlinearity parameter $f_{\rm NL} \approx  c_{\rm s}^{-2}$, independent of the amplitude of the other bispectrum. Equilateral NG is usually generated by derivative interactions of the inflaton field; derivative terms are suppressed when one perturbation mode is frozen on super-horizon scales during inflation, and the other two are still crossing the horizon, so that the correlation between the three perturbation modes will be suppressed, while it is maximal when all the three modes cross the horizon at the same time, which happens for $k_1\approx k_2\approx k_3$. 
 
The equilateral type NG is well approximated by the template~(\citealt{2006JCAP...05..004C})
\begin{eqnarray}
\label{equilateralBis}
\nonumber
& &	B_{\Phi}^{\rm equil}(k_1,k_2,k_3)= 6A^2 f_{\rm NL}^{\rm equil}\\
\nonumber
& \times& \left\{
-\frac1{k^{4-n_{\rm s}}_1k^{4-n_{\rm s}}_2}-\frac1{k^{4-n_{\rm s}}_2k^{4-n_{\rm s}}_3}
-\frac1{k^{4-n_{\rm s}}_3k^{4-n_{\rm s}}_1} 
%\right. \\
%\nonumber
%& &
-\frac2{(k_1k_2k_3)^{2(4-n_{\rm s})/3}}
\right. \\
%\nonumber
& &
\left.+\left[\frac1{k^{(4-n_{\rm s})/3}_1k^{2(4-n_{\rm s})/3}_2k^{4-n_{\rm s}}_3}
%\right.\\
%& &\left.
+\mbox{(5 permutations)}\right]\right\}\, ,
\end{eqnarray}
where $P_{\Phi}(k)=A/k^{4-n_{\rm s}}$ is the power spectrum of Bardeen's gravitational potential with normalization $A^2$ and scalar spectral index $n_{\rm s}$.
For example, the models introduced in the string theory framework based on the DBI action~(\citealt{2004PhRvD..70j3505S,2004PhRvD..70l3505A}) can be described within the $P(X, \phi)$-class, and they give rise to an equilateral NG with an overall amplitude $f^{\rm equil}_{\rm NL}=-(35/108)c_{\rm s}^{-2}$ for $c_{\rm s} \ll 1$, which turns out typically to be $f^{\rm equil}_{\rm NL} <-5$.
\footnote{An effectively single-field model with a non-standard kinetic term and a reduced sound speed for the adiabatic perturbation modes might also arise in coupled multi-field systems, where the heavy fields are integrated out: see 
discussions in, e.g.,~\cite{2010PhRvD..81d3502T,2011JCAP...01..030A,2011PhRvD..84j3509S}.}

The equilateral shape emerges also in models characterized by more general higher-derivative interactions, such as ghost inflation~(\citealt{2004JCAP...04..001A}) or models within effective field theories of inflation~(\citealt{2008JHEP...03..014C,2010JCAP...01..028S,2010JCAP...08..008B}).
  
Taken individually, each higher-derivative interaction of the inflaton field generically gives rise to a bispectrum with a shape which is similar -- but not identical to -- the equilateral form (an example is provided by the two interaction terms discussed above for an inflaton with a non-standard kinetic term).  Therefore it has been shown, using an effective field theory approach to inflationary perturbations, that it is possible to build a combination of the corresponding similar equilateral shapes to generate a bispectrum that is orthogonal to the equilateral one, the so-called ``orthogonal'' shape. This can be approximated by the template~(\citealt{2010JCAP...01..028S})

\begin{eqnarray}\label{orthogonalBis}
\nonumber
& &	B^{\rm ortho}_{\Phi}(k_1,k_2,k_3)= 6A^2 f_{\rm NL}^{\rm ortho}\\
\nonumber
& \times& \left\{
-\frac3{k^{4-n_{\rm s}}_1k^{4-n_{\rm s}}_2}-\frac3{k^{4-n_{\rm s}}_2k^{4-n_{\rm s}}_3}
-\frac3{k^{4-n_{\rm s}}_3k^{4-n_{\rm s}}_1}\right.
%\\
%\nonumber
%& &
-\frac8{(k_1k_2k_3)^{2(4-n_{\rm s})/3}}
\\
%\nonumber
& &
\left.+\left[\frac3{k^{(4-n_{\rm s})/3}_1k^{2(4-n_{\rm s})/3}_2k^{4-n_{\rm s}}_3}
%\\
%& &
+\mbox{(5 perm.)}\right]\right\}.
\end{eqnarray}
The orthogonal bispectrum can also arise as the predominant shape in some inflationary realizations of Galileon inflation~(\citealt{2011JCAP...11..042R}). 

\medskip

\noindent {\it Non-separable single-field bispectrum shapes}: While most single-field inflation bispectra can be well-characterized by the equilateral and orthogonal shapes, we note that these are separable ans\"atze which only approximate the contributions from two leading order terms in the cubic Lagrangian.   In an effective field theory approach these correspond to two shapes which can be associated directly with the  
inflaton field interactions $\dot \pi (\partial_i\pi)^2$ and $\dot \pi^3$ (in the language of the effective field theory of inflation the inflaton scalar degree of freedom $\pi$ is related to the comoving curvature perturbation as 
$\zeta = -H \pi$). They are, respectively (\citealt{2010JCAP...01..028S}, see also  \citealt{2007JCAP...01..002C}~\footnote{Notice that the two shapes~(\ref{EFT1Bis}) and~(\ref{EFT2Bis}) correspond to 
a linear combination of the two shapes found in~\cite{2007JCAP...01..002C}.})
\eq\label{EFT1Bis}
&&B_\Phi^{\rm EFT1}(\kall) = \frac{6A ^2\fnl^{\rm EFT1}}{(k_1k_2k_3)^3}\frac{(-9/17)}{(k_1 + k_2 + k_3)^3}~\times\nn\\
&&\qquad\left\{\sum_i k_i^6 + \sum_{i \neq j}\[3 k_i^5 k_j - k_i^4k_j^2 - 3 k_i^3 k_j^3\]\right. \\
&&\qquad\left.+ \sum_{i \neq j \neq l}\[3 k_i^4 k_j k_l - 9 k_i^3 k_j^2 k_l- 2k_i^2 k_j ^2k_l^2\]\right\}\,,\nn\\
\label{EFT2Bis}
&&B_\Phi^{\rm EFT2}(\kall) = \frac{6A ^2\fnl^{\rm EFT2}}{k_1 k_2 k_3 \,} \frac{27}{(k_1+k_2+k_3)^3}\,.
\qe
These shapes 
differ from equilateral in the flattened or collinear limit. DBI inflation gives a closely related shape of particular interest phenomenologically \citep{2004PhRvD..70l3505A},
\eq\label{dbiBis}
 &&B_\Phi^{\rm DBI}(\kall) = \frac{6A ^2\fnl^{\rm DBI}}{(k_1 k_2 k_3)^3}\,\frac{(-3/7)}{(k_1+k_2+k_3)^2}\;\times\\ 
&&~\left\{\sum_i k_i^5 + \sum_{i \neq j}\[2 k_i^4 k_j - 3 k_i^3 k_j^2\] 
+ \sum_{i \neq j \neq l}\[k_i^3 k_j k_l - 4 k_i^2 k_j^2 k_l\]\right\}.\nn
\qe
For brevity, we have given the scale-invariant form of the shape functions, without the mild power spectrum running.   There are also sub-leading order terms which give rise to additional non-separable shapes, but these are expected to be much smaller without special fine-tuning. 

%%%%%%%%%%%%%%%%%%%%%%%%%%%%%%%%%%%%%%%%%%%%%%%%%%%%

\subsection{Multi-field models}
This class of models generally includes an additional light scalar field (or more fields) during inflation, which can be different from the inflaton, and whose fluctuations contribute to the final primordial curvature perturbation of the gravitational potential. It could be the case of inflation driven by several scalar fields -- ``multiple-field inflation'' -- or the one where the inflaton drives the accelerated expansion, while other scalar fields remain subdominant during inflation. This encompasses, for instance, a large class of multi-field models which leads to non-Gaussian isocurvature perturbations (for earlier works, see e.g.,~\citealt{1997PhRvD..56..535L},~\citealt{1997ApJ...483L...1P},~\citealt{1997PhRvD..55.7415B}). More importantly, such models can also lead to cross-correlated and non-Gaussian adiabatic and isocurvature modes, where NG is first generated by large nonlinearities in some scalar (possibly non-inflatonic) sector of the theory, and then efficiently transferred to the inflaton adiabatic sector(s) through the cross-correlation of adiabatic and isocurvature perturbations\footnote{This may happen, for instance, if the inflaton field is coupled to the other scalar degrees of freedom, as expected on particle physics grounds. These scalar degrees
of freedom may have large self-interactions, so that their quantum fluctuations are intrinsically non-Gaussian, because, unlike the inflaton case, the self-interaction strength in such an extra scalar
sector does not suffer from the usual slow-roll conditions.}~(\citealt{2002PhRvD..65j3505B,2002PhRvD..66j3506B,2006JCAP...05..019V,2006PhRvD..73h3522R,2007PhRvD..76h3512R,2005PhRvL..95l1302L,Tzavara:2010ge}; for a review on NG from multiple-field inflation models, see,~\citealt{2010AdAst2010E..76B}). Another interesting possibility is the curvaton model~(\citealt{1990PhRvD..42..313M,2002NuPhB.626..395E,2002PhLB..524....5L,2001PhLB..522..215M}), where a second light scalar field, subdominant during inflation, decays after inflation ends, producing the primordial density perturbations which can be characterized by a high NG level~(e.g., \citealt{2002PhLB..524....5L,2003PhRvD..67b3503L,2004PhRvD..69d3503B}).  NG in the curvature perturbation can be generated at the end of inflation, e.g., due to the nonlinear dynamics of (p)reheating (e.g.,~\citealt{2005PhRvL..94p1301E,2008PhRvL.100d1302C,2006PhRvD..73j6012B}; see also~\citealt{2009PhRvL.103g1301B}) or, as in modulated (p)reheating and modulated hybrid inflation, due to local fluctuations in the decay rate/interactions of the inflaton field~(\citealt{2003astro.ph..3614K,2004PhRvD..69h3505D,2004PhRvD..69b3505D,2004PhRvD..70h3004B,Zaldarriaga:2003my,2005JCAP...11..006L,2005PhRvD..72l3516S,2006PhRvL..97l1301L,2006PhRvD..73b3522K,2012JCAP...05..039C}). The common feature of all these models is that a large NG in the curvature perturbation can be produced via both a transfer of super-horizon non-Gaussian isocurvature perturbations in the second field (not necessarily the inflaton) to the adiabatic density perturbations, and via additional nonlinearities in the transfer mechanism. Since, typically, this process occurs on super-horizon scales, the form of NG is local in real space. Being local in real space, the bispectrum correlates large and small scale Fourier modes. The local bispectrum is given by~(\citealt{1993ApJ...403L...1F,1994ApJ...430..447G,2000MNRAS.313..141V,2000PhRvD..61f3504W,2001PhRvD..63f3002K})
\begin{eqnarray}
\label{localBis}
B^{\rm local}_{\Phi}(k_1,k_2,k_3)&=&2 f^{\rm local}_{\rm NL} \Big[ P_{\Phi}(k_1) P_{\Phi}(k_2)+P_{\Phi}(k_1) P_{\Phi}(k_3) \nonumber\\
&+&P_{\Phi}(k_2) P_{\Phi}(k_3) \Big] \nonumber \\
&=& 2 A^2  f^{\rm local}_{\rm NL} \left[  \frac{1}{k_1^{4-n_{\rm s}}k_2^{4-n_{\rm s}}} +{\rm cycl.} \right]\, .
\end{eqnarray}
Most of the signal-to-noise ratio in fact peaks in the squeezed configurations ($k_1 \ll k_2\simeq k_3$) 
\begin{equation}
B^{\rm local}_{\Phi}(k_1\rightarrow 0,k_2,k_3) \rightarrow 4  f^{\rm local}_{\rm NL} P_{\Phi}(k_1) P_{\Phi}(k_2)\, .
\end{equation}
The typical example of a curvature perturbation that generates the bispectrum of Eq.~(\ref{localBis}) is the standard local form for the gravitational potential~(\citealt{1990NuPhB.335..197H,1991ASPC...15..339K,1990PhRvD..42.3936S,1994ApJ...430..447G,2000MNRAS.313..141V,2000PhRvD..61f3504W,2001PhRvD..63f3002K})
\begin{equation}
\label{PotentialLocal}
\Phi({\vec x})=\Phi_L({\vec x})+f^{\rm local}_{\rm NL} (\Phi_L^2({\vec x})-\langle  \Phi_L^2({\vec x}) \rangle)\, ,  
\end{equation}
where $\Phi_L({\vec x})$ is the linear Gaussian gravitational potential and $f^{\rm local}_{\rm NL}$ is the amplitude of a quadratic nonlinear correction (though this is not the only possibility: e.g., the gravitational potential produced in multiple-field inflation models generally cannot be reduced to the Eq.~(\ref{PotentialLocal})). For example, in the (simplest) adiabatic curvaton models, the NG amplitude turns out to be~\citep{2004PhRvD..69d3503B,2004PhRvL..93w1301B} $f_{\rm NL}^{\rm local}=(5/4r_\mathrm{D})-5r_\mathrm{D}/6-5/3$, for a quadratic potential of the curvaton field~(\citealt{2002PhLB..524....5L,2003PhRvD..67b3503L,2005PhRvL..95l1302L,2006JCAP...09..008M,2006PhRvD..74j3003S}), where $r_{\rm D}=[3\rho_{\rm curvaton}/(3 \rho_{\rm curvaton}+4\rho_{\rm radiation})]_{\rm D}$  is the ``curvaton decay fraction''  evaluated at the epoch of the curvaton decay in the sudden decay approximation. Therefore, for $r_\mathrm{D} \ll 1$, a high level of NG is imprinted. 

There exists a clear distinction between multi-field and single-field models of inflation that can be probed via a consistency condition~(\citealt{2003JHEP...05..013M,2004JCAP...10..006C,2007JCAP...01..002C,2010AdAst2010E..72C}):
in the squeezed limit, single-field models predict a bispectrum 
\begin{equation}
\label{cr}
B^{\rm single-field}_{\Phi}(k_1 \rightarrow 0,k_2,k_3=k_2) \rightarrow \frac{5}{3} (1-n_{\rm s}) P_{\Phi}(k_1) P_{\Phi}(k_2)\, ,
\end{equation}
and thus $f_{\rm NL}\sim{\mathcal O}(n_{\rm s}-1)$ in the squeezed limit, in a model-independent sense (i.e., not only for standard single-field models). 
This means that a significant detection of local NG (in the squeezed limit) would rule out a very large class of single-field models of inflation (not just the simplest ones). Although based on very general conditions, the consistency condition of Eq.~(\ref{cr}) can be violated in some well-motivated inflationary settings (we refer the reader to~\cite{2010AdAst2010E..72C,2013arXiv1301.5699C} and references therein for more details). 

\medskip
\noindent {\it Quasi-single field inflation:}
\label{sec:qsf}
Quasi-single field inflation has an extra field (or fields) with mass $m$ close to the Hubble parameter $H$ during inflation; these models evolve quiescently, producing a calculable non-Gaussian signature~(\citealt{2010JCAP...04..027C}).  The resulting one-parameter bispectrum smoothly interpolates between local and equilateral models, though in a non-trivial manner:
\eq\label{qsiBis}
%B_\Phi^{\rm QSI}(\kall) = \frac{6A ^2\fnl^{\rm QSI} }{(k_1 k_2 k_3)^{3/2}}\frac{3^{3/2}N_\nu[8 k_1 k_2 k_3/(k_1+k_2+k_3)^3]}{N_\nu[8/27](k_1+k_2+k_3)^{3/2}}\,
B_\Phi^{\rm QSI}(\kall) = \frac{\sqrt{972}A ^2\fnl^{\rm QSI} }{(k_1 k_2 k_3)^{3/2}}\frac{N_\nu[8 k_1 k_2 k_3/(k_1+k_2+k_3)^3]}{N_\nu[8/27](k_1+k_2+k_3)^{3/2}}\, 
\qe
where $\nu =(9/4-m^2/H^2)^{1/2}$ and $N_\nu$ is the Neumann function of order $\nu$.   Quasi-single field models can also produce an essentially ``constant'' bispectrum defined by $B^{\rm const}(\kall) = 6A ^2\fnl^{\rm const}/ (k_1k_2k_3)^2$.  The constant model is the  simplest possible non-zero primordial shape, with all its late-time CMB structure simply reflecting the behaviour of the transfer functions.   

\medskip
\noindent {\it Alternatives to inflation:}
Local NG can also be generated in some alternative scenarios to inflation, for instance in cyclic/ekpyrotic models~(for a review, see~\citealt{2010AdAst2010E..67L}),  due to the same basic curvaton mechanism described above. In this case, typical values of the nonlinearity parameter can easily reach $|f_{\rm NL}^{\rm local}| > 10$. 

%%%%%%%%%%%%%%%%%%%%%%%%%%%%%%%%%%%%%%%%%%%%%%%%%%%%

\subsection{Non-standard models giving rise to alternative specific forms of NG}
\label{New}
\noindent{\it Non-Bunch-Davies vacuum and higher-derivative interactions:} 
Another interesting bispectrum shape is the folded one, which peaks in flattened configurations. To facilitate data analyses, the flat shape has been usually parametrized  by the template~(\citealt{Meerburg:2009ys})
\begin{eqnarray}
\label{flatBis}
\nonumber
& &	B^{\rm flat}_{\Phi}(k_1,k_2,k_3)= 6A^2 f_{\rm NL}^{\rm flat}\\
\nonumber
& \times& \left\{
\frac1{k^{4-n_{\rm s}}_1k^{4-n_{\rm s}}_2}+\frac1{k^{4-n_{\rm s}}_2k^{4-n_{\rm s}}_3}
+\frac1{k^{4-n_{\rm s}}_3k^{4-n_{\rm s}}_1}\right.
%\\
%\nonumber
%& &
+\frac3{(k_1k_2k_3)^{2(4-n_{\rm s})/3}}
\\ & &
\left. -\left[\frac1{k^{(4-n_{\rm s})/3}_1k^{2(4-n_{\rm s})/3}_2k^{4-n_{\rm s}}_3}
%\\ & &\left.\left. 
+\mbox{(5 perm.)}\right]\right\}.
\end{eqnarray}
The initial quantum state of the inflaton is usually specified by requiring that, at asymptotically early times and short distances, its fluctuations behave as in flat space. Deviations from this standard ``Bunch-Davies'' vacuum can result in interesting features in the bispectrum. Models with an initial non-Bunch-Davies vacuum state~(\citealt{2007JCAP...01..002C,2008JCAP...05..001H,Meerburg:2009ys,2011JCAP...03..025A}) can generate sizeable NG similar to this type. NG highly correlated with such a template can be produced in single-field models of inflation from higher-derivative interactions~(\citealt{2010JCAP...08..008B}), and in models where a ``Galilean'' symmetry is imposed~(\citealt{2011JCAP...02..006C}). In both cases, cubic inflaton interactions with two derivatives of the inflaton field arise.  Single-field inflation models with a small sound speed, studied in \cite{2010JCAP...01..028S}, can generate the flat shape, as a result of a linear combination of the orthogonal and equilateral shapes. In fact, from a simple parametrization point of view, the flat shape can be always written as  $F_{\rm flat}(k_1,k_2,k_3)=[F_{\rm equil}(k_1,k_2,k_3)-F_{\rm ortho}(k_1,k_2,k_3)]/2$~(\citealt{2010JCAP...01..028S}). Despite this, we provide constraints also on the amplitude of the flat bispectrum 
shape of Eq.~(\ref{flatBis}).

For models with excited (i.e., non-Bunch-Davies) initial states, the resulting NG shapes are model-dependent, but they are usually characterized by the importance of flattened or collinear triangles, with $k_3 \approx k_1+k_2$ along the edges of the tetrapyd.
We will denote the original flattened bispectrum shape, given in Eq.~(6.2) and (6.3) of~\cite{2007JCAP...01..002C}, by $B_\Phi^{\rm NBD}$;  it is generically much more flattened than the ``flat'' model of Eq.~\eqref{flatBis}.
Although this shape was derived specifically for power-law $k$-inflation, it encapsulates several different shapes, with amplitudes which can vary between different phenomenological models.  These shapes are also typically oscillatory, being regularized by a cutoff scale $k_{\rm c}$ giving the oscillation period; this cutoff $k_{\rm c} \approx (c_{\rm s}\tau_{\rm c})^{-1}$  is determined by the (finite) time $\tau_{\rm c}$ in the past when the non-Bunch-Davies component was initially excited.  For excited canonical single-field inflation, the two leading order shapes can be described \citep{Agullo:2011xv} by the ansatz
\eq\label{NBD2Bis}
&&B_\Phi^{\rm NBD\it i} =\frac{2A ^2\fnl^{\rm NBD\it i}}{(k_1k_2k_3)^3} ~\Bigg \{ \,f_i(\klist)~\times\\
&&~~~\left.\frac{1-\cos [(k_2+k_3-k_1)/k_{\rm c}]}{k_2+k_3-k_1}+ \hbox{2 perm.}\right\}\,,\nn
\qe
where $f_1(\klist) = k_1^2(k_2^2+k_3^2)/2$ is dominated by squeezed configurations, $f_2(\klist) = k_2^2k_3^2$ has a flattened shape, and $i=1,2$.  Note that for all oscillatory shapes, the relevant bispectrum equation defines the normalisation of $\fnl$.  The flattened signal is most easily enhanced in the limit of small sound speed $c_{\rm s}$, for which a regularized ansatz is given by \citep{2010AdAst2010E..72C}
\eq\label{NBD3Bis}
B_\Phi^{\rm NBD3} = \frac{2A ^2\fnl^{\rm NBD3}}{k_1k_2k_3} \left[ \frac{k_1+k_2-k_3}{(k_{\rm c}+k_1+k_2-k_3)^4} + \hbox{2 perm.}\right]\,.
\qe

\medskip
\noindent {\it Scale-dependent feature and resonant models:}  
Oscillating bispectra can be generated from violation of a smooth slow-roll evolution (``feature'' or ``resonant'' NG).  These models have the distinctive property of a strong running NG, which breaks approximate scale-invariance. A sharp feature in the inflaton potential forces the inflaton field away from the attractor solution, and causes oscillations as it relaxes back; these oscillations can appear in the bispectrum
\citep{2000PhRvD..61f3504W,2007JCAP...06..023C,2008JCAP...04..010C}, as well as the power spectrum and other correlators.   An analytic form for the oscillatory bispectrum for these feature models is \citep{2008JCAP...04..010C}
\begin{align}
\label{featureBis}
B_\Phi^{\rm feat}(\kall) = \frac{6A ^2\fnl^{\rm feat} }{(k_1 k_2 k_3)^2}\sin\[\frac{2 \pi(k_1+k_2+k_3)}{3k_{\rm c}} + \phi\]\,,
\end{align}
where $\phi$ is a phase factor and $k_{\rm c}$ is a scale associated with the feature, which is linked in turn to an effective multipole periodicity $\ell_{\rm c}$ of the CMB bispectrum.
Typically, these oscillations will decay with an envelope of the form $ \exp [-(k_1+k_2+k_3)/mk_{\rm c}]$ for a model-dependent parameter $m$.  

Closely related ``resonant" bispectra can be created by periodic features superimposed on a smooth inflation potential (\citealt{2008JCAP...04..010C,2011JCAP...01..017F}); these induce small periodic features in the background evolution, with which the quantum inflaton fluctuations can resonate while still inside the horizon.  Resonant models are particularly relevant in the context of axion inflation models~(e.g.,~\citealt{2010JCAP...06..009F,2011JCAP...01..017F,2012PhRvD..85b3525B}).  These mechanisms also create oscillatory behaviour in the bispectrum, but with a more constant amplitude and a wavelength that becomes logarithmically stretched. Here, the resonant oscillations for most models can be represented in the form  
\eq\label{resonantBis}
B_\Phi^{\rm res}(\kall) = \frac{6A ^2\fnl^{\rm res} }{(k_1 k_2 k_3)^2}\sin\[C\ln({k_1+k_2+k_3}) + \phi\]\,, 
\qe
where the constant $C=1/\ln (3k_{\rm c})$ and $\phi$ is a phase.

Finally, we note that periodic features in the inflationary potential can excite the vacuum state, as well as perturbing the background inflation trajectory (\citealt{Chen:2010bka}).  Such models offer the intriguing possibility of combining the  flattened non-Bunch-Davies shape with periodic oscillations:
\eq\label{resNBDBis}
&&B_\Phi^{\rm resNBD}(\kall) = \frac{2A ^2\fnl^{\rm resNBD} }{(k_1 k_2 k_3)^2}\Big\{\exp[- k_{\rm c}^{3/5}(k_2+k_3-k_1)/2k_1] \nn \\
&&~~~~~~~\times \sin[ k_{\rm c} ((k_2+k_3-k_1)/2k_1 + \ln k_1) + \phi] + 2\; \hbox{perm.} \Big\}\,.
\qe
This ansatz represents the dominant folded resonant contribution in inflationary models with non-canonical kinetic terms, which competes with resonant (Eq.~\eqref{resonantBis}) and equilateral (Eq.~\eqref{equilateralBis}) contributions; however, for slow-roll single-field inflation, there are additional terms. 
\medskip

\noindent {\it Directional dependence motivated by gauge fields:}  Additional variations of the bispectrum shape have been proposed for models with vector fields, which can have an additional directional dependence through the parameter $\mu_{12} =\hat {\vec k}_1\cdot\hat{\vec k}_2$ where $\hat{\vec k} = {\vec k}/k$.    For example, primordial magnetic fields sourcing curvature perturbations can cause a dependence on  both  $\mu$ and $\mu^2$~\citep{Shiraishi:2012rm}, and a coupling between the inflaton $\phi$  and the gauge field strength $F^2$ can yield a $\mu^2$ dependence (\citealt{Barnaby:2012tk,Bartolo:2012sd}). We can parameterize these shapes as variations on the local shape, following ~\cite{Shiraishi:2013vja}, as 
\eq\label{vectorBis}
B_\Phi(\kall) = \sum_L c_L [P_L(\mu_{12})P_{\Phi}(k_1)P_{\Phi}(k_2)+\hbox{2\,perm}],
\qe
where $P_L(\mu)$ is the Legendre polynomial with $P_0=1, \;P_1=\mu$ and $P_2=\textstyle {\frac{1}{2} (3 \mu^2-1)}$.   For example, for $L=1$ we have the shape 
\eq\label{eq:dbiS}
 B_\Phi^{L=1}(\kall) = \frac{2A ^2\fnl^{L=1}}{(k_1 k_2 k_3)^2}\,\left[\frac{k_3^2}{k_1^2k_2^2}(k_1^2+k_2^2-k_3^2)  + \hbox{2\,perm.}\right].
 \qe
Also the recently introduced ``solid inflation'' model~\citep{2012arXiv1210.0569E} generates bispectra similar to Eq.~(\ref{vectorBis}).
Here and in the following the nonlinearity parameters $f_{\rm NL}^L$ are related to the $c_L$ coefficients by $c_0=2 f_{\rm NL}^{L=0}$, $c_1=-4 f_{\rm NL}^{L=1}$, and $c_2=-16 f_{\rm NL}^{L=2}$.
The $L=1,2$ shapes exhibit sharp variations in the flattened limit for e.g., $k_1+k_2\approx k_3$, while in the squeezed limit, $L=1$ is suppressed whereas $L=2$ grows like the local bispectrum shape (i.e., the $L=0$ case).   Whether or not the underlying gauge field models prove robust, this directional dependence on the wave vectors is a generic feature which yields distinct bispectrum families, deserving closer study.

\medskip
\noindent {\it Warm inflation:}
In warm inflation \citep{1995PhRvL..75.3218B}, where dissipative effects are important, a non-Gaussian signal can be generated~(e.g.,~\citealt{2007JCAP...04..007M}) that peaks in the squeezed limit -- but with a more complex shape than the local one -- and exhibiting a low cross-correlation 
with the other shapes (see references in \citealt{2010AdAst2010E..73L}). 

%%%%%%%%%%%%%%%%%%%%%%%%%%%%%%%%%%%%%%%%%%%%%%%%%%%%

\subsection{Higher-order non-Gaussianity: the trispectrum}
\label{trispectrum}
The connected four-point functions of CMB anisotropies (or the harmonic counterpart, the so-called trispectrum) can also provide crucial information about  the mechanism
that gave rise to the primordial curvature perturbations~\citep{2002PhRvD..66f3008O}. The primordial trispectrum is usually characterised by two
amplitudes $\tau_{\rm NL}$ and $g_{\rm NL}$: $\tau_{\rm NL}$ is most often related to $f^{2}_{\rm NL}$-type contributions, while $g_{\rm NL}$ is the amplitude of intrinsic cubic
nonlinearities in the primordial gravitational potential (corresponding, in terms of field interactions, to a scalar-exchange and to a contact interaction term, respectively). They correspond to 'soft' limits of the full four-point function, with respectively the diagonal and one side of the general wavevector trapezoid being much smaller than the others. In the CMB maps they appear respectively approximately as a spatial variation in amplitude of the small-scale fluctuations, and a spatial variations in the value of $f_{\rm NL}$ correlated with the large-scale temperature. In addition to possible primordial signals that are the focus of this paper there is also expected to be a large lensing trispectrum (of very different shape), discussed in detail in~\cite{planck2013-p12}.

The simplest local trispectrum is given by
\begin{eqnarray}
& &\langle  \Phi({\vec k}_1)  \Phi({\vec k}_2)   \Phi({\vec k}_3)   \Phi({\vec k}_4) \rangle= (2 \pi)^3 \delta^{(3)}({\vec k}_1+{\vec k}_2+{\vec k}_3+{\vec k}_4) \nonumber \\
&\times& \Bigg \{ \frac{25}{9}   \tau_{\rm NL} \[P_{\Phi}(k_1) P_{\Phi}(k_2)  P_{\Phi}(k_{13}) +(11\, {\rm perm.}) \]    \nonumber \\
&& + 6g_{\rm NL} \[P_{\Phi}(k_1) P_{\Phi}(k_2)  P_{\Phi}(k_{3}) +(3\, {\rm perm.}) \]  \Bigg\} \, ,
\end{eqnarray}
where $k_{ij}\equiv |{\vec k}_i+{\vec k}_j|$.
Previous constraints on $\tau_{\rm NL}$ and $g_{\rm NL}$ have been derived, e.g., by \cite{2010PhRvD..81l3007S} who obtained $-7.4 \times 10^5 < g_{\rm NL} < 8.2 \times 10^5$ and $-0.6 \times 10^4 < \tau_{\rm NL} < 3.3 \times 10^4$ (at $95 \%$ CL)
analysing \textit{WMAP}-5 data; for the same datasets \cite{2010arXiv1012.6039F} obtained $-5.4 \times 10^5 < g_{\rm NL} < 8.6 \times 10^5$ ($68\%$ CL).
This kind of trispectrum typically arises in multi-field inflationary models where large NG arise from the conversion of isocurvature perturbations on superhorizon scales. If the curvature perturbation is the standard local form, in real space one has $\Phi({\vec x})=\Phi_L({\vec x})+f^{\rm local}_{\rm NL} (\Phi_L^2({\vec x})-\langle \Phi_L^2\rangle)+ g_{\rm NL} \Phi_L^3({\vec x})$. In this case, $\tau_{\rm NL}=(6 f^{\rm local}_{\rm NL}/5)^2$; however, in general the trispectrum amplitude can be larger.

The trispectrum is a complementary observable to the CMB bispectrum as it can further distinguish different inflationary scenarios. This is because
the same interactions that lead to the bispectrum might be responsible also for a large trispectrum, so that the different NG parameters can be related to each other in a well-defined way within specific models. If there is a non-zero squeezed-shape bispectrum there must necessarily be a trispectrum, with $\tau_{\rm NL} \geq (6 f^{\rm local}_{\rm NL}/5)^2$~\citep{2008PhRvD..77b3505S,2010JCAP...12..030S,2011PhRvL.106y1301S,2012arXiv1201.4048S,Lewis:2011au,2011PhRvL.107s1301S,2012JCAP...11..047A,2012arXiv1205.1523K}. In the simplest inflationary scenarios the prediction would be $\tau_{\rm NL} = (6 f^{\rm local}_{\rm NL}/5)^2$, but larger values would indicate more complicated dynamics. Several inflationary scenarios
have been found in which the bispectrum is suppressed,
thus leaving the trispectrum as the largest higher-order correlator
in the data. A detection of a large trispectrum and a negligible
bispectrum would be a smoking gun for these models. This is the case, for example, of certain curvaton and multi-field field models of inflation
\citep{2006PhRvD..74l3519B,2006PhRvD..74j3003S,2010AdAst2010E..76B}, which for particular parameter choice can produce a significant $\taunl$ and $\gnl$ and small $\fnl$. Large trispectra are also possible in single-field models of inflation with higher-derivative interactions (see, e.g.,~\citealt{2009JCAP...08..008C,2009PhRvD..80d3527A,2011JCAP...01..003S,2010JCAP...09..035B}), but these would be suppressed in the squeezed limit since they are generated by derivative interactions at horizon-crossing, and hence only project weakly onto the local shapes. These equilateral trispectra arise can be well-described by some template forms~\citep{2010arXiv1012.6039F}. Naturally, higher-order correlations could also be considered, but are not directly studied in this paper.

%%%%%%%%%%%%%%% Section 3 %%%%%%%%%%%%%%%
\section{Statistical estimation of the CMB bispectrum}
\label{sec:se}

In this Section, we review the statistical techniques that we use to estimate the nonlinearity parameter $f_{\textrm{NL}}$. We begin by fixing some notation and describing the CMB angular bispectrum in Sect.~\ref{Sec_angbisp}. We then introduce in Sect.~\ref{Sec_optimalest} the  optimal $\fnl$ bispectrum estimator. From Sect.~\ref{Sec_KSW} onwards we describe in detail the different implementations of the optimal estimator that were developed and applied to {\it Planck} data.

%%%%%%%%%%%%%%%%%%%%%%%%%%%%%%%%%%%%%%%%%%%%%%%%%%%%
 
\subsection{The CMB angular bispectrum}\label{Sec_angbisp}
\begin{figure}[!t]
\centering
\includegraphics[width=\hsize]{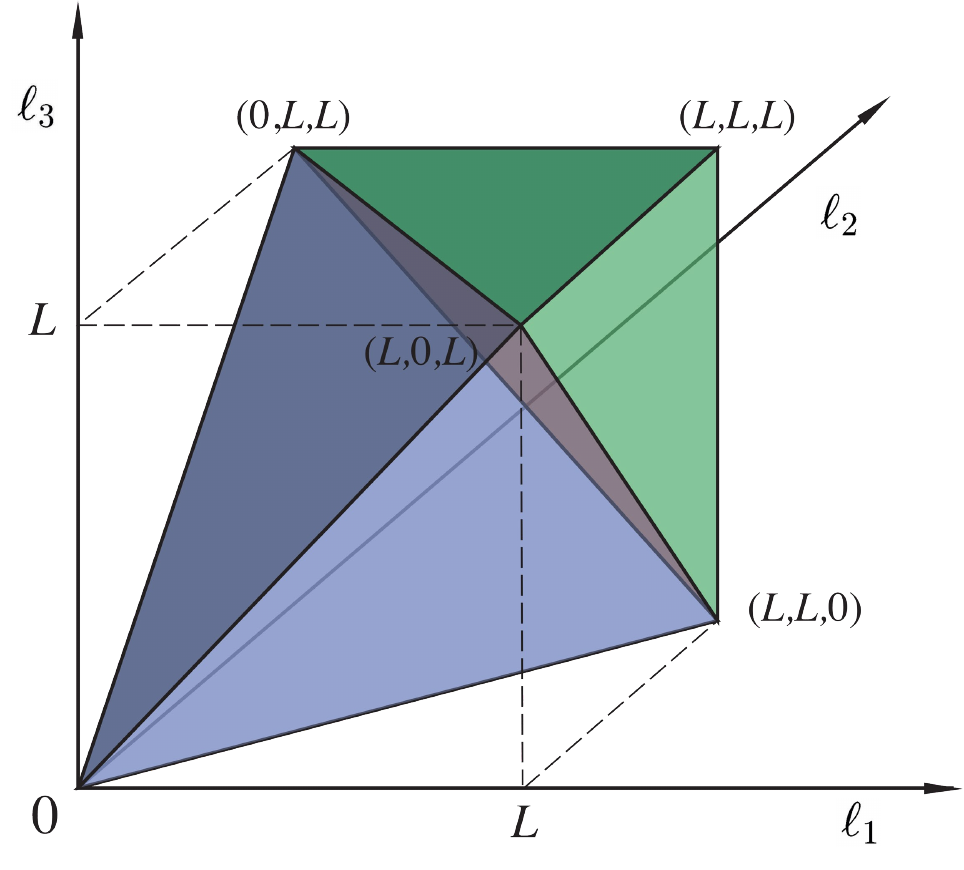}
\caption[Tetrahedral domain]{\small  Permitted observational domain of Eq.~(\ref{eq:tetrapydl}) for the CMB bispectrum $\blll$.  Allowed multipole values $(\ell_1,\,\ell_2,\,\ell_3)$ lie inside the shaded  ``tetrapyd'' region (tetrahedron$+$pyramid), satisfying both the triangle condition and the experimental resolution $\ell <L $$\,\equiv\,$$\lmax$.}
\label{fig:tetrapyd}
\end{figure}

Temperature anisotropies are represented using the $a_{\ell m}$ coefficients of a spherical harmonic decomposition of the CMB map, 
\begin{equation}
\label{eq:alm}
\frac{\Delta
T}{T}(\vnhat) = \sum_{\ell m} a_{\ell m} Y_{\ell m}(\vnhat)\,;
\end{equation}
we write $C_{\ell}=\langle |a_{\ell m}|^{2}\rangle$ for the
angular power spectrum and $\hat{C}_{\ell} = (2\ell+1)^{-1}\sum _m
|a_{\ell m}|^{2}$ for the corresponding (ideal) estimator; hats ``$\,\hat{\,}\,$'' denote estimated quantities.
The CMB angular bispectrum is the three-point correlator of the $a_{\ell m}$:
\begin{equation}
B_{\ell_1 \ell_2 \ell_3}^{m_1 m_2 m_3} \equiv\langle a_{\ell_1 m_1}a_{\ell_2 m_2}a_{\ell_3 m_3}\rangle .
\end{equation}
If the CMB sky is rotationally invariant, the angular bispectrum can be factorized as follows:
\begin{equation}\label{eq:Bred}
\langle a_{\ell_1 m_1}a_{\ell_2 m_2}a_{\ell_3 m_3}\rangle = 
\curl{G}^{\ell_1 \ell_2 \ell_3}_{m_1 m_2 m_3} \,b_{\ell_1 \ell_2
\ell_3}\, ,
\end{equation} 
where $b_{\ell_1 \ell_2\ell_3}$ is the so called {\em reduced bispectrum}, and $\curl{G}^{\ell_1 \ell_2 \ell_3}_{m_1 m_2 m_3}$ 
 is the Gaunt integral, defined as:
\eq\label{eq:Gaunt}
 \curl{G}^{\ell_1 \ell_2 \ell_3}_{m_1 m_2 m_3} &\equiv\int Y_{\ell_1 m_1}(\vnhat) \, Y_{\ell_2 m_2}(\vnhat) \, Y_{\ell_3 m_3}(\vnhat) \, d^2\vnhat \cr
 &= \hlll \( \begin{array}{ccc} \ell_1 & \ell_2 & \ell_3 \\ m_1 & m_2 & m_3 \end{array} \)\,,
\qe
where $h_{\ell_1 \ell_2
\ell_3}$ is a geometrical factor,
\eq\label{eq:hdefn} h_{\ell_1 \ell_2
\ell_3} =
\sqrt{\frac{(2\ell_1+1)(2\ell_2+1)(2\ell_3+1)}{4\pi}} \(
\begin{array}{ccc} \ell_1 & \ell_2 & \ell_3 \\ 0 & 0 & 0 \end{array} \)\, .
\qe
The Wigner-$3j$ symbol in parentheses enforces rotational symmetry, and allows 
us to restrict attention to a tetrahedral domain of multipole triplets
$\{\ell_1,\ell_2,\ell_3\}$, satisfying both a triangle condition and a limit
given by some maximum resolution $\ell_\mathrm{max}$ (the latter being defined by the finite angular 
resolution of the experiment under study).  This
three-dimensional domain $\Vtetra$ of allowed multipoles, sometimes referred to 
 in the following as a ``tetrapyd'', is
illustrated in Fig.~\ref{fig:tetrapyd} and it is explicitly
defined by
\eq \label{eq:tetrapydl} \nonumber
&&\mbox{Triangle condition:}~~ \ell_1 \leq \ell_2+\ell_3  ~~\hbox{for}~~ \ell_1 \geq \ell_2,\,\ell_3,  +\hbox{perms.},\\
&&\mbox{Parity condition:} ~~\quad \ell_1+ \ell_2+
\ell_3= 2n\, ,~~~n\in\mathbb{N}\,,\\
&&\mbox{Resolution:} ~~\quad \qquad~ \ell_1, \ell_2,
\ell_3  \leq \ell_{\mathrm{max}}\,,\quad \ell_1, \ell_2,
\ell_3 \in  \mathbb{N}\,.\nonumber
\qe
Here, $\Vtetra$ is the isotropic subset of the
full space of bispectra, denoted by ${\cal V}$.

One can also define an alternative rotationally-invariant reduced bispectrum
$B_{\alll}$ in the following way:
\begin{equation}\label{eq:Bav}
 B_{\alll} \equiv h_{\alll} \sum_{m_1 m_2 m_3} \( \begin{array}{ccc} \ell_1 & \ell_2 & \ell_3 \\ m_1 & m_2 & m_3 \end{array} \)  B_{\alll}^{m_1 m_2 m_3} \; .
\end{equation}
Note that this $B_{\alll}$ is equal to $h_{\alll}$ times the angle-averaged
bispectrum as defined in the literature.
From Eqs.~(\ref{eq:Bred}) and~(\ref{eq:Gaunt}), and the fact that the
sum over all $m_i$ of the Wigner-$3j$ symbol squared is equal to 1,
it is easy to see that $B_{\alll}$ is related to the reduced bispectrum by
\begin{equation}\label{eq:red2av}
B_{\alll} = h_{\alll}^2 b_{\alll} \; .
\end{equation}
The interest in this bispectrum $B_{\alll}$ is that it can be estimated directly 
from maximally-filtered maps of the data:
\begin{equation}\label{eq:integral_Bav}
\hat{B}_{\alll} = \int d^2\vnhat T_{\ell_1} (\vnhat) T_{\ell_2}(\vnhat) T_{\ell_3} (\vnhat) \; ,
\end{equation}
where the filtered maps $T_\ell(\vnhat)$ are defined as:
\begin{equation}
T_{\ell}(\vnhat) \equiv \sum_m \alm \Ylm(\vnhat) \; .
\end{equation}
This can be seen by replacing the $B_{\alll}^{m_1 m_2 m_3}$ in Eq.~(\ref{eq:Bav}) by
its estimate $\almone \almtwo \almthree$ and then 
using Eq.~(\ref{eq:Gaunt}) to rewrite the Wigner symbol in terms of a Gaunt 
integral, which in its turn is expressed as an integral over the product of 
three spherical harmonics.

%%%%%%%%%%%%%%%%%%%%%%%%%%%%%%%%%%%%%%%%%%%%%%%%%%%%

%%%%%%%%%%%%%%%%%%%%%%%%%%%%%%%%%%%%%%%%%%%%%%%%%%%%

\subsection{CMB bispectrum estimators}\label{Sec_optimalest}

The full bispectrum for a high-resolution map cannot be
evaluated explicitly because of the sheer number of operations involved, $O(\ell^5_
{\rm max})$, as well as
the fact that the signal will be too weak to measure in individual multipoles with any significance.
Instead, we essentially use a least-squares fit to compare the bispectrum of the observed CMB multipoles 
with a particular theoretical bispectrum $b_{\alll}$. We then extract an overall ``amplitude parameter'' $\fnl$ for that specific template, after defining a suitable normalization convention so that we can write $b_{\alll} = \fnl b_{\alll}^{{\rm th}}$, where $b_{\alll}^{{\rm th}}$ is defined as the value of the theoretical bispectrum ansatz for $\fnl=1$. 

Optimal 3-point estimators, introduced by  \cite{1998MNRAS.299..805H}  (see also \citealt{2000PhRvD..62j3004G}),  are those which saturate the Cram\' er-Rao bound.  Taking into account the fact that instrument noise and masking can break rotational invariance, it has been shown that the general optimal $\fnl$ estimator can be written as \citep{2005PhRvD..72d3003B,2006JCAP...05..004C,2010JCAP...01..028S,2013arXiv1301.6017V}:
\begin{align}\label{eq:optimalestimator}
\hat{f}_{\textrm{NL}} =& \frac{1}{N} \sum_{\ell_i,m_i}\curl{G}^{\,\,\ell_1\; \ell_2\; \ell_3}_{m_1 m_2 m_3 } b^{\rm th}_{\ell_1 \ell_2 \ell_3}\\
&\times\, \big[C^{-1}_{\ell_1 m_1, \ell_1' m_1'} a_{\ell_1'm_1'}\, C^{-1}_{\ell_2 m_2, \ell_2' m_2'} a_{\ell_2'm_2'}\, 
C^{-1}_{\ell_3 m_3, \ell_3' m_3'}a_{\ell_3'm_3'} \nonumber  \\
&- 3 \,C^{-1}_{\ell_1 m_1, \ell_2 m_2} C^{-1}_{\ell_3 m_3,\ell_3' m_3'} a_{\ell_3'm_3'} \big]\,,\nonumber
\end{align}
where $C^{-1}$ is the inverse of the covariance matrix $C_{\ell_1 m_1, \ell_2 m_2} \equiv \langle \almone \almtwo \rangle$ and $N$ is a suitable normalization chosen to produce unit response to $ b_{\ell_1 \ell_2 \ell_3}^{\rm th}$.
 
In the expression of the optimal estimator above we note the presence of two contributions, one (hereafter defined the ``cubic term" of the estimator) is cubic in the observed $\alm$ and correlates the bispectrum of the data to the theoretical fitting template $b_{\ell_1 \ell_2 \ell_3}^{\rm th}$, while the other is linear in the observed $\alm$ (hereafter, the ``linear term"), which is zero on average. In the rotationally-invariant case the linear term is proportional to the monopole in the map, which has been set to zero, so in this case the estimator simply reduces to the cubic term. However, when rotational invariance is broken by realistic experimental features such as a Galactic mask or an anisotropic noise distribution, the linear term has an important effect on the estimator variance.
 In this case, the coupling between different $\ell$ would in fact produce a spurious increase in the error bars (coupling of Fourier modes due to statistical anisotropy can be ``misinterpreted" by the estimator as NG). The linear term correlates the observed $\alm$ to the  power spectrum anisotropies and removes this effect, thus restoring optimality \citep{2006JCAP...05..004C,Yadav:2007ny,Yadav:2007rk}.

The actual problem with Eq.~(\ref{eq:optimalestimator}) is that its direct implementation to get an optimal $\fnl$ estimator  would require measurement of all the bispectrum configurations from the data. As already mentioned at the beginning of this section, the computational cost of this would scale like $\ell_{\rm max}^5$ and be totally prohibitive for  high-resolution CMB experiments. Even taking into account
the constraints imposed by isotropy, the number of multipole triples $\{\ell_{1},\ell_{2},\ell_{3}\}$ is of the order of $10^9$ at
{\it Planck} resolution, and the number of different observed bispectrum configurations $\hat{b}^{\ell_1 \ell_2 \ell_3}_{m_1 m_2 m_3}$ is of the order of $10^{15}$. For each of them, costly numerical evaluation of the Wigner symbol is also required. This is completely out of reach of existing supercomputers. It is then necessary to find numerical solutions that circumvent this 
problem and in the following subsections we will show how the different estimators used for the $\fnl$ {\it Planck} data analysis   address this challenge.  Before entering into a more accurate description of these different methods, we would like however to stress again that they are all going to be different implementations of the optimal $\fnl$ estimator defined by Eq.~(\ref{eq:optimalestimator}); therefore they are  conceptually equivalent and expected to produce $\fnl$ results that are in very tight agreement. This will later on allow for stringent validation tests based on comparing different pipelines. On the other hand, it will soon become clear that the different approaches that we are going to discuss also open up a range of additional applications beyond simple $\fnl$ estimation for standard bispectra. Such applications include, for example, full bispectrum reconstruction (in a suitably smoothed domain), tests of directional dependence of $\fnl$, and other ways to reduce the amount of data, going beyond simple single-number $\fnl$ estimation. So different methods will also provide a vast range of complementary information.

Another important preliminary point, to notice before discussing different techniques, is that none of the estimators in the following sections implement exactly Eq.~(\ref{eq:optimalestimator}), but a slightly modified version of it. In Eq.~(\ref{eq:optimalestimator}) the CMB multipoles always appear weighted by the inverse of the full covariance matrix. Inverse 
covariance filtering of CMB data at the high angular resolutions achieved by experiments like {\it WMAP} and {\it Planck} is another very challenging numerical issue, which was fully addressed only recently \citep{2009JCAP...09..006S,2011ApJS..192...18K,2013A&A...549A.111E}. For our analyses we developed two independent inverse-covariance filtering pipelines. The former is based on an extension to {\it Planck} resolution of the algorithm used for {\it WMAP} analysis \citep{2009JCAP...09..006S,2011ApJS..192...18K}; the latter is based on the algorithm described in \cite{2013A&A...549A.111E}.
However, detailed comparisons interestingly showed that our estimators perform {\em equally} well (i.e.,  they saturate the Cram\' er-Rao bound) if we approximate the covariance matrix as diagonal in the filtering procedure {\em and}  we apply a simple diffusive inpainting procedure to the masked areas of the input CMB maps. A more detailed description of our inpainting and Wiener filtering algorithms can be found in Sect.~\ref{Sec_Wiener}. 

In the diagonal covariance approximation, the minimum variance estimator is obtained by making the replacement $(C^{-1} a)_{\ell m} \rightarrow {\alm/C_\ell}$ in the cubic term and then including the linear term that minimizes the variance for this class of cubic estimator \citep{2006JCAP...05..004C}. This procedure leads to the following expression:
\begin{align}\label{eq:optimalestimator2}
\hat{f}_{\textrm{NL}} =& \frac{1}{N} \sum_{\ell_i,m_i}\curl{G}^{\,\,\ell_1\; \ell_2\; \ell_3}_{m_1 m_2 m_3 } \tilde{b}^{\rm th}_{\ell_1 \ell_2 \ell_3} \times\, \nonumber \\
&\left[ \frac{\tilde{a}_{\ell_1 m_1}}{\tilde{C}_{\ell_1}}\, \frac{\tilde{a}_{\ell_2 m_2}}{\tilde{C}_{\ell_2}}\, 
\frac{\tilde{a}_{\ell_3 m_3}}{\tilde{C}_{\ell_3}}
- 6 \, \frac{\tilde{C}_{\ell_1 m_1, \ell_2 m_2}} {\tilde{C}_{\ell_1} \tilde{C}_{\ell_2}} \frac{\tilde{a}_{\ell_3 m_3}}{\tilde{C}_{\ell_3 m_3}} \right]\,,
\end{align}
where the tilde denotes the modification of $C_\ell$ and $b_{\alll}$ to incorporate instrument beam and noise effects, and indicates that the multipoles are obtained from a map that was masked and preprocessed through the inpainting procedure detailed in Sect.~\ref{Sec_Wiener}. This means that
\begin{align}
\tilde{b}_{\alll} \equiv b_{\ell_1} b_{\ell_2} b_{\ell_3} b_{\alll}  \,, \qquad
\tilde{C}_\ell \equiv  b^2_{\ell} C_{\ell} + N_{\ell} \; ,
\end{align}
where $b_\ell$ denotes the experimental beam, and $N_{\ell}$ is the noise power spectrum.
For simplicity of notation, in the following we will drop the tilde and always assume that beam, noise and inpainting 
 effects are properly included.

Using Eqs.~(\ref{eq:Bav}) and (\ref{eq:red2av}) we can rewrite Eq.~(\ref{eq:optimalestimator2}) in terms of the bispectrum $B_{\alll}$:
\begin{equation}\label{eq:optimalestaveraged}
{\hat f}_\mathrm{NL} = 
\frac{6}{N} \sum_{\ell _1 \leq \ell _2 \leq \ell_3} 
\frac{ 
B^\mathrm{th}_{\ell _1 \ell _2 \ell _3} 
\left(B^\mathrm{obs}_{\ell _1 \ell _2 \ell _3} 
- B^\mathrm{lin}_{\ell _1 \ell _2 \ell _3}\right) 
}{ 
V_{\ell _1 \ell _2 \ell _3}
}.
\end{equation}
In the above expression, $B^\mathrm{th}$ is the theoretical template for $B$ (with $\fnl=1$) and $B^{\mathrm{obs}}$ denotes the observed bispectrum (the cubic term), extracted from the (inpainted) data using Eq.~(\ref{eq:integral_Bav}). $B^\mathrm{lin}$ is the linear correction, also computed using
Eq.~(\ref{eq:integral_Bav}) by replacing two of the filtered temperature maps by 
simulated Gaussian ones and averaging over a large number of them 
(three permutations). 
The variance $V$ in the inverse-variance weights is given by
$V_{\ell_1 \ell_2 \ell_3} = g_{\ell_1 \ell_2 \ell_3} h^2_{\ell_1 \ell_2 \ell_3} 
C_{\ell_1} C_{\ell_2} C_{\ell_3}$ (remember that these should be viewed as being
the quantities with tildes, having beam and noise effects included) with
$g_{\alll}$ a permutation factor ($g_{\alll} = 6$ when all $\ell$ are equal, $g_{\alll}=2$ when two $\ell$ are equal, and $g_{\alll}=1$ otherwise). Both Eqs.~(\ref{eq:optimalestimator2}) and~(\ref{eq:optimalestaveraged}) will be used in the following. Eq.~(\ref{eq:optimalestimator2}) will provide the starting point for the KSW, skew-$C_\ell$ and modal estimators, while Eq.~(\ref{eq:optimalestaveraged}) will be the basis for the binned and wavelets estimators. 

Next, we will describe in detail the different methods, and show how they address the numerical challenge posed by the necessity to evaluate a huge number of bispectrum configurations. To summarize loosely: the KSW estimator,
the skew-$C_{\ell}$ approach and the separable modal methodology
achieve massive reductions in computational costs by exploiting structural properties of $b^\mathrm{th}$, e.g., separability.
 On the other hand, the binned bispectrum and the
wavelet approaches achieve computational gains by data compression of 
$B^\mathrm{obs}$.

\subsubsection{The KSW estimator}\label{Sec_KSW}

To understand the rationale behind the KSW estimator \citep{Komatsu:2003iq,
2003ApJS..148..119K,2010JCAP...01..028S,2006JCAP...05..004C,Yadav:2007ny,Yadav:2007rk,2008PhRvL.100r1301Y,2011MNRAS.417....2S}, 
assume that the theoretical reduced bispectrum $b^\mathrm{th}_{\ell_1 \ell_2 \ell_3}$ can be
exactly decomposed into a separable structure, e.g., there
exist some sequences of functions $\alpha(\ell,r),\,\beta(\ell,r)$ such
that we can approximate $b_{\ell_1 \ell_2 \ell_3}$ as
\eq\label{eq:presutti0}
b_{\ell_1 \ell_2 \ell_3}&\simeq& \int
\left[\beta(\ell_1,r)\beta(\ell_2,r)\alpha(\ell_3,r)+\beta(\ell_1,r)\beta(\ell_3,r)\alpha(\ell_2,r)\right. \nonumber\\ 
&&\left. +\beta(\ell_2,r)\beta(\ell_3,r)\alpha(\ell_1,r)\right]\, r^2 \, dr \, ,
\qe
where $r$ is a radial coordinate. This assumption is fulfilled in particular by the local shape \citep{2001PhRvD..63f3002K,2004JCAP...08..009B}, with $\alpha(\ell, r)$ and $\beta(\ell, r)$ 
 involving integrals of products of spherical Bessel functions and CMB radiation transfer functions.
Let us consider the optimal estimator of Eq.~($\ref{eq:optimalestimator2}$) and neglect for the moment the linear part. 
Exploiting Eq.~(\ref{eq:presutti0}) and the factorizability property of the Gaunt integral (Eq.~(\ref{eq:Gaunt})), the cubic term of 
the estimator can be written as:
\begin{align}\label{eq:presutti3}
S_{\mathrm{cub}} = \int dr\, r^2 \int\, d^2\vnhat\, A(\vnhat,r) B^2(\vnhat,r)
\end{align}
where
\begin{equation}\label{eq:presutti6}
A(\vnhat,r)=\sum_{\ell m}\frac{\alpha(\ell,r)\,a_{\ell m}Y_{\ell m}(\vnhat)}{C_{\ell}} \; ,
\end{equation}
and
\begin{equation}\label{eq:presutti5}
B(\vnhat,r)=\sum_{\ell m} \frac{\beta(\ell,r)\,a_{\ell m}Y_{\ell m}(\vnhat)}{C_{\ell}} \; .
\end{equation}
From the formulae above we see that the overall triple integral over all the configurations $\ell_1, \ell_2, \ell_3$ has been factorized into a 
 product of three separate sums over different $\ell$. This produces a massive reduction in computational time, as the problem now scales like $\ell^3_{\rm max}$ 
instead of the original $\ell^5_{\rm max}$ . Moreover, the bispectrum can be evaluated in terms of a cubic statistic in pixel space from Eq.~(\ref{eq:presutti3}), and 
 the functions $A(\vnhat,r)$, $B(\vnhat,r)$ are obtained from the observed $\alm$ by means of Fast Harmonic Transforms. 

It is easy to see that the linear term can be factorized in analogous fashion. Again considering the local shape type of decomposition of Eq.~(\ref{eq:presutti0}), it is possible to find:
\begin{align}
S_{\mathrm{lin}} = & \frac{-6}{N} \int dr \, r^2 \int d^2\vnhat \, \big[2 \left\langle A(r,\vnhat) B(r, \vnhat) \right \rangle_{\mathrm{MC}}\times \nonumber \\
& \times B(r, \vnhat)  + \left\langle B(r,\vnhat) B(r, \vnhat) \right \rangle_{\mathrm{MC}}  A(r,\vnhat) \big] \; ,
\end{align}
where $\left\langle \cdot \right\rangle_{\mathrm{MC}}$ denotes a Monte Carlo (MC) average over simulations accurately reproducing the properties of the actual data set (basically we are taking a MC approach to estimate the 
 product between the theoretical bispectrum and the $\alm$ covariance matrix appearing in the linear term expression).

The estimator can be finally expressed as a function of $S_{\mathrm{cub}}$ and $S_{\mathrm{lin}}$:
\begin{equation}
\hat{f}_{\mathrm{NL}} = \frac{S_{\mathrm{cub}} + S_{\mathrm{lin}}}{N}.
\end{equation}
Whenever it can be applied, the KSW approach makes the problem of $\fnl$ estimation computationally feasible, even at the high angular resolution of the {\it Planck} satellite. One important  caveat is that 
 factorizability of the shape, which is the starting point of the method, is not a general property of theoretical bispectrum templates. Strictly speaking, only the local shape is manifestly separable. However, a large class 
 of inflationary models can be extremely well approximated by separable equilateral and orthogonal templates \citep{2004JCAP...08..009B,2006JCAP...05..004C,2010JCAP...01..028S}.  The 
specific expressions of cubic and linear terms are of course template-dependent, but as long as the template itself is separable their structure is analogous to the example shown in this Section, i.e., 
they can be written as pixel space integrals of cubic products of suitably-filtered CMB maps (involving MC approximations of the $\alm$ covariance for the linear term). For a complete and compact summary 
 of KSW implementations for local, equilateral and orthogonal bispectra see \citet[Appendix]{2009ApJS..180..330K}.

\subsubsection{The Skew-$C_{\ell}$ Extension}
\label{skewCl}

The skew-$C_\ell$ statistics were introduced by \cite{2010MNRAS.401.2406M} to address an issue with estimators such as KSW which 
reduce the map to an estimator of $f_{\mathrm{NL}}$ for a given type of NG.  This level of data compression, to a single number, has the
disadvantage that it does not allow verification that a NG signal is of the type which has been estimated.  KSW on its own cannot tell if a measurement of $f_{\textrm{NL}}$ of given type is actually caused by NG of that type, or by contamination from some other source or sources.  The skew-$C_\ell$ statistics 
perform a less radical data compression than KSW (to a function of $\ell$), and thus retain enough information to distinguish different NG signals. The desire to find a statistic which is able to fulfil this r\^ ole, but which is still optimal, drives one to a statistic which is closely related to KSW, and indeed reduces to it when the scale-dependent information is not used.  A further advantage of the skew-$C_\ell$ is that it allows joint estimation of the level of many types of NG simultaneously.  This requires a large number of simulations for accurate estimation of their covariance matrix, and they are not used in this role in this paper. However, they do play an important part in identifying which sources of NG are clearly detected in the data, and which are not.

We define the skew-$C_\ell$ statistics by extending from KSW, as follows: from Eq.~(\ref{eq:presutti3}),
the numerator $\cal{E}$ can be rewritten as
\begin{equation}
{\cal E}=\sum_{\ell}[C^{A,B^2}_{\ell}+2C^{AB,B}_{\ell} ] \label{skew0}
\end{equation}
where
\eq\label{eq:skew1}
C^{A,B^2}_{\ell}&=&\int
\int_{S^2}\sum_{\ell_1,\ell_{2}}\sum_{m_1,m_{2},m}\left[\frac{\beta(\ell_1;r)a_{\ell_1m_1}Y_{\ell_1m_1}(\vnhat)}{C_{\ell_1}}\nonumber \right.\\
&\times&\left.\frac{\beta(\ell_2;r)a_{\ell_2m_2}Y_{\ell_2m_2}(\vnhat)}{C_{\ell_2}}\frac{\alpha(\ell;r)a_{\ell m}Y_{\ell m}(\vnhat)}{C_{\ell}} \right]r^2
d^2\vnhat dr
\qe
and
\eq\label{eq:skew3}
C^{AB,B}_{\ell}&=&\int
\int_{S^2}\sum_{\ell_1,\ell_{2}}\sum_{m_1,m_{2},m}\left[\frac{\beta(\ell_1;r)a_{\ell_1m_1}Y_{\ell_1m_1}(\vnhat)}{C_{\ell_1}}\right.\nonumber \\
&\times&\left.
\frac{\alpha(\ell_2;r)a_{\ell_2m_2}Y_{\ell_2,m_2}(\vnhat)}{C_{\ell_2}}\frac{\beta(\ell;r)a_{\ell m}Y_{\ell m}(\vnhat)}{C_{\ell}}\right]r^2
d^2\vnhat dr \, .
\qe
The skew-$C_{\ell}$ approach allows for the full implementation of the KSW
procedure, when the sum in Eq.~(\ref{skew0}) is fully evaluated; furthermore, it
allows for extra degrees of flexibility, e.g., by restricting the sum to subsets of the multipole space, which
may highlight specific features of the NG signal. Each form of NG considered has its own $\alpha,\beta$, hence its own set of skew-$C_\ell$, denoted $S_\ell \equiv C^{A,B^2}_{\ell}+2C^{AB,B}_{\ell}$, and we have chosen to illustrate here with the local form, but as with KSW the method can be extended to other separable shapes, and some skew-$C_\ell$ do not involve integrals, such as the ISW-lensing skew statistic.  Note that in this paper we do not fit the $S_\ell$ directly, but instead we estimate the NG using KSW, and then verify (or not) the nature of the NG by comparing the skew-$C_\ell$ with the theoretical expectation.  No further free parameters are introduced at this stage.  This procedure allows investigation of KSW detections of NG of a given type, assessing whether or not they are actually due to NG of that type.

\subsubsection{Separable Modal Methodology}\label{Sec_modalest}

Primordial bispectra need not be manifestly separable (like the local bispectrum), or be easily approximated by separable ad hoc templates (equilateral and orthogonal),
 so the direct KSW approach above cannot be applied generically (nor to late-time bispectra).    
However, we can employ a highly-efficient generalization by considering a complete basis of separable modes describing any late-time bispectrum 
(see \citealt{2007PhRvD..76h3523F, 2010PhRvD..82b3502F}), as applied to {\textit {WMAP}}-7 data for a wide variety of separable and non-separable bispectrum models 
\citep{2010arXiv1006.1642F}. See also \cite{planck2013-p09} and \cite{planck2013-p20}. We can achieve this by expanding the signal-to-noise-weighted bispectrum  as 
\begin{equation} \label{eq:cmbestmodes}
\frac{ \blll}{\sqrt{C_{\ell_1}C_{\ell_2}C_{\ell_3}}} \,= \sum_{i,j,k}\alpha_{ijk} Q_{ijk} (\ell_1, \ell_2, \ell_3)\,,
\end{equation}
where the (non-orthogonal) separable modes $Q_{n}$ are defined by 
\begin{align}\label{eq:separablemodes}
Q_{ijk} (\llist) = {\frac{1}{6}}&[ q_i(\ell_1)\, q_j(\ell_2)\,  q_k(\ell_3) + q_j(\ell_1)\, q_i(\ell_2)\, q_k(\ell_3) \nonumber\\
&+  \mbox{cyclic perms in \,$i,j,k$}\,]\,.
\end{align}
It is more efficient to label the basis as $Q_n$, with the subscript $n$ representing an ordering of the $\{i,j,k\}$ products (e.g., by distance $i^2+j^2+k^2$).   
The $\bar q_i(\ell)$ are any complete basis functions up to a given resolution of interest and they can be augmented with other special functions adapted to target particular bispectra. 
The modal coefficients in the bispectrum of Eq.~\eqref{eq:cmbestmodes} are given by the inner product of the weighted bispectrum with each mode
\begin{equation}\label{eq:alphacoeff}
\alpha_{n}= \sum_{p} \gamma_{np}^{\,-1}\left \langle \frac{ \blll}{\sqrt{C_{\ell_1}C_{\ell_2}C_{\ell_3}}} \,, \;Q_{p} (\llist) \right\rangle\, 
\end{equation}
where the modal transformation matrix is 
\begin{align}\label{eq:gammamatrix}
\gamma_{np} &= \langle Q_{n},\, Q_{p} \rangle  \nonumber \\
&\equiv ~{\sum_{\llist}} h_{\ell_1 \ell_2 \ell_3}^2 Q_{ijk} (\llist) \, Q_{i'j'k'} (\llist)\,.  
\end{align}
In the following, the specific basis functions $\bar q_i(\ell)$  we employ include either weighted Legendre-like polynomials or trigonometric functions. These are combined with the Sachs-Wolfe local shape and the separable ISW shape in order to obtain high correlations to all known bispectrum shapes (usually in excess of 99\%).

Substituting the separable mode expansion of Eq.~(\ref{eq:cmbestmodes}) into the estimator and exploiting the separability of the Gaunt integral (Eq.~(\ref{eq:Gaunt})), yields
\begin{align}\label{eq:cmbestsep}
{\cal E} = \frac{1}{N^2}  \sum _{n\leftrightarrow prs}\kern-6pt \alpha_n \int & d^2\vnhat\,\left[\bar M_{\{p}(\vnhat)\bar M_r(\vnhat)\bar M_{s\}}(\vnhat)\right.\nonumber\\
&\left. -~
6\left\langle\bar M^{\rm G}_{\{p}(\vnhat)\bar M^{\rm G}_r(\vnhat)\right\rangle\bar M_{s\}}(\vnhat)\right]\,.
\end{align} 
where  the $\bar M_p(\vnhat)$ represent versions of the original CMB map filtered with 
the basis function $q_p$ (and the weights $( \sqrt{C_\ell})^{-1}$), that is, 
\begin{equation}\label{eq:barmapfilter}
\bar M_p(\vnhat) = \sum_{\ell m} q_p(\ell)\frac{a_{\ell m}}{\sqrt{C_\ell}} Y_{\ell m}(\vnhat)\,.
\end{equation}
The maps  $\bar M^{\rm G}_p(\vnhat)$ incorporate the same mask 
and a realistic model of the inhomogeneous instrument noise; a large ensemble of these maps, calculated from Gaussian simulations, is used in the averaged linear
term in the estimator of Eq.~(\ref{eq:cmbestsep}), allowing for the subtraction of these important effects.  Defining the integral over
these convolved product maps as cubic and linear terms respectively, 
\begin{align}\label{eq:mapintegral}   
\beta_n{}^{\rm cub} &=  \int d^2\vnhat\, \bar M_{\{p}(\vnhat)\bar M_r(\vnhat)\bar M_{s\}}(\vnhat)\,,\\
\beta_n{}^{\rm lin} &=  \int d^2\vnhat\, \left\langle\bar M^{\rm G}_{\{p}(\vnhat)\bar M^{\rm G}_r(\vnhat)\right\rangle\bar M_{s\}}(\vnhat)\,,
\end{align}
the estimator reduces to a simple sum over the mode coefficients
\begin{equation}\label{eq:estimatorsum}
{\cal E} = \frac{1}{N^2} \sum_n \alpha_n \bar \beta_n\,,
\end{equation}
where $\bbQn \equiv  \bbQn{}^{\rm cub} - \bbQn{}^{\rm lin}$.   
The estimator sum in Eq.~(\ref{eq:estimatorsum})  is now straightforward to evaluate because of separability, since it has been reduced to 
a product of three sums over the observational maps (Eq.~(\ref{eq:cmbestsep})), followed by a single 2D integral over
all directions (Eq.~(\ref{eq:mapintegral})).   The number of operations in evaluating the estimator sum is only ${\cal{O}} (\ell^2)$.  

For the purposes of testing a wide range inflationary models, we can also define a set of primordial basis functions $\overline Q_{ink}(k_1,k_2,k_3) =  \bar q_{i}(k_1) \bar q_{j}(k_2) \bar q_{k}(k_3) +\hbox{perms.}$ for wavenumbers satisfying the triangle condition (again we will order the $\{i,j,k\}$ with $n$).  This provides a separable expansion for an arbitrary primordial shape function $S(\klist) = B(\klist)/(k_1k_2k_3)^2$, that is, 
\begin{equation}\label{primmodes}
S(\klist) = \sum_{n}\bar\alpha_{n} \overline Q_{n} (\klist)\,.
\end{equation}
Using the same transfer functions as in the KSW integral \eqref{eq:presutti3}, we can efficiently project forward each separable primordial mode $\overline Q_{n} (k_1,k_2,k_3)$ to a corresponding late-time solution $\widetilde {\cal Q}_{n}(l_1l_2l_3)$ (essentially the projected CMB bispectrum for $\overline Q_{n} (k_1,k_2,k_3)$).  By finding the inner product between these projected modes $\widetilde Q_{n}(\llist)$ and the CMB basis functions $Q_{n}(\llist)$,  we can obtain the transformation matrix  (\citealt{2010PhRvD..82b3502F,2010arXiv1012.6039F})
\begin{equation}\label{eq:transform}
\Gamma _{n p} = \sum_{r} \bar \gamma^{-1}_{np} \langle \widetilde {\cal Q}_{r}(\llist),\, Q(\llist)\rangle\,,
\end{equation}
which projects the primordial expansion coefficients $\aQn$ to late-time:
\begin{equation}
\alpha_n = \sum _{p} \Gamma_{np} \bar \alpha_p \,.
 \end{equation}
When $\Gamma_{np}$ is calculated once we can efficiently convert any given primordial bispectrum $B(\kall)$ into its late-time CMB bispectrum counterpart using Eq.~\eqref{eq:cmbestmodes}.   We can use this to extend the KSW methodology and to search for the much broader range of primordial models beyond local, equilateral and orthogonal, having validated on these standard shapes.  

\subsubsection{Binned Bispectrum}
\label{binned}
In the binned bispectrum approach~\citep{Bucher:2009nm}, the
computational gains are achieved by data compression of the observed 
$\hat{B}$. This is quite
feasible, because like the power spectrum the bispectrum is a rather
smooth function, with features on the scale of the acoustic peaks.
In this way one obtains an enormous reduction of the computational
resources needed at the cost of only a tiny increase in variance
(typically $1\%$).

More precisely, the following statistic is considered,
\begin{equation}
T_i(\vnhat) = \sum_{\ell\in\Delta_i} \sum_{m=-\ell}^{+\ell}
a_{\ell m} Y_{\ell m}(\vnhat), \label{Tmapbinned}
\end{equation}
where the $\Delta_i$ are intervals (bins) of multipole values
$[\ell_i,\ell_{i+1}-1]$, for $i=0,\ldots ,(N_\mathrm{bins}-1)$, with $\ell_0=\ell_\mathrm{min}$ and
$\ell_{N_\mathrm{bins}} = \ell_\mathrm{max}+1$, and the other bin boundaries
chosen in such a way as to minimize the variance of $\hat{f}_\mathrm{NL}$.
The binned bispectrum is then obtained by using $T_i$ instead of
$T_\ell$ in the expression for the sample bispectrum of Eq.~(\ref{eq:integral_Bav}), 
to obtain:
\begin{equation}
B_{i_1 i_2 i_3}^\mathrm{bin}= \int  T_{i_1}(\vnhat )
T_{i_2}(\vnhat ) T_{i_3}(\vnhat )d^2\vnhat . \label{Bobsbinned}
\end{equation}
The linear term $B^\mathrm{lin}$ is binned in an analogous way, and the
theoretical bispectrum template $B^\mathrm{th}$ and variance $V$ are also
binned by summing them over the values of $\ell$ inside the bin. 
Finally $\fnl$ is determined using the binned version of
Eq.~(\ref{eq:optimalestaveraged}), i.e., by replacing all quantities by
their binned equivalent and replacing the sum over $\ell$ by a sum over
bin indices $i$.
An important point is that the binned bispectrum
estimator does not mix the observed bispectrum and the theoretical
template weights until the very last step of the computation of
$\hat{f}_\mathrm{NL}$ (the final sum over bin indices).
Thus, the (binned) bispectrum of the map is also a direct output of the code. 
Moreover, one can easily study the $\ell$-dependence of the results by 
omitting bins from this final sum.

The full binned bispectrum allows one to explore the bispectral
properties of maps independent of a theoretical model. Binned
bispectra have been used to compare component separation maps and
single-frequency maps dominated by foregrounds, as presented 
in Sects.~\ref{binnedrec} and~\ref{sec:Results}. 
In its simplest implementation, which is used in this
paper, the binned estimator uses top-hat filters in harmonic space,
which makes the Gaussian noise independent between different bins;
however, slightly overlapping bins could be used to provide better
localization properties in pixel space.  In this sense the binned
estimator is related to the wavelet estimators, which we discuss
below.

\subsubsection{Wavelet $f_{\rm NL}$ estimator}

Wavelet methods are well-established in the CMB literature and have
been applied to virtually all areas of the data analysis pipeline,
including map-making and component separation, point source detection,
search for anomalies and anisotropies, cross-correlation with large
scale structure and the ISW effect, and many others (see for instance
\citealt{antoine:1998}, \citealt{martinez:2002}, \citealt{cayon2003},
\citealt{mcewen1}, \citealt{pbm06},
\citealt{starck}, \citealt{mcewen2}, \citealt{Cruz1}, \citealt{Fay08},
\citealt{Peiris}, \citealt{gm1}, \citealt{gm2}, \citealt{starck2010},
\citealt{sandro}, \citealt{fernandez2012}).  These methods have the
advantage of possessing localization properties both in real and
harmonic space, allowing the effects of masked regions and anisotropic
noise to be  dealt with efficiently. 

In terms of the current discussion, wavelets can be viewed as a way to
compress the sample bispectrum vector by a careful binning scheme in
the harmonic domain. See also \cite{planck2013-p09}. In particular, the wavelet bispectrum can be
rewritten as
\begin{equation}
q_{i j k}=\frac{1}{4\pi}\frac{1}{\sigma_i\sigma_j\sigma_k}\int d^2\vnhat \,w(R_i,\vnhat)w(R_j,\vnhat)w(R_k,\vnhat),
\label{waveneedbisp}
\end{equation}
where
\begin{equation}
w(R; {\vec b}) = \int d^2\vnhat \, f(\vnhat)\Psi(\vnhat, {\vec b}; R) = \sum_{\ell m}a_{\ell m}\omega_\ell (R)Y_{\ell m}(\vnhat).
\label{waveneedcomp}
\end{equation}
Here ${\vec b}$ is the position on the sky at which the wavelet coefficient
is evaluated and $\sigma$ is is the dispersion of the wavelet
coefficient map $w(R; {\vec b})$ for the scale $R$. The filters
$\omega_\ell (R)$ can be seen as the coefficients of the expansion
into spherical harmonic of the Spherical Mexican Hat Wavelet (SMHW) filter
\begin{equation}
\Psi({\bf x},{\bf n};R)=\frac{1}{\sqrt{2\pi}}\frac{1}{N(R)}\left[1+\left(\frac{y}{2}\right)^2\right]^2\left[2-\left(\frac{y}{R}\right)^2\right]e^{-y^2/2R^2}\, .
\end{equation}
Here $N(R)=R\sqrt{1+R^2/2+R^4/4}$ is a normalizing constant and
$y=2\tan(\theta/2)$ represents the distance between ${\bf x}$ and ${\bf n}$,
evaluated on the stereographic projection on the tangent plane at ${\bf n}$;
$\theta$ is the corresponding angular distance, evaluated on the
spherical surface. 

The implementation of the linear-term correction can proceed in
analogy with the earlier cases. However, note that, in view of the
real-space localization properties of the wavelet filters, the linear
term here is smaller than for KSW and related approaches, although not
negligible. Moreover, it can be well-approximated by a term-by-term
sample-mean subtraction for the wavelet coefficients, which allows for
a further reduction of computational costs. Further details can be
found in \citet{curto:2011a,curto:2011b,curto:2012,regan:2013} (see
also \citealt{needbisp,rudjord09,
2009MNRAS.396.1682P,2010MNRAS.402L..34P,
donzelli:2012} for related needlet-based procedures).
%

%%%%%%%%%%%%%%%%%%%%%%%%%%%%%%%%%%%%%%%%%%%%%%%%%%%%

\subsection{Wiener filtering}\label{Sec_Wiener}

As discussed in Sect.~\ref{Sec_optimalest}, the $\fnl$ bispectrum estimator 
requires, in principle, inverse covariance filtering of the data to achieve optimality 
(equivalent to Wiener filtering up to a trivial multiplication by the inverse of the signal  
 power spectrum).

We have used the iterative method of \citet{2013A&A...549A.111E} for 
Wiener filtering simulations and data. The algorithm makes use of a 
messenger field, introduced to mediate between the pixel space and 
harmonic space representation, where noise and signal properties can be 
specified most directly. Since the Wiener filter is the maximum a 
posteriori solution, we 
monitor the $\chi^2$ of the current solution as a convergence 
diagnostics. We terminate the algorithm as soon as the improvement in 
the posterior between two consecutive steps has dropped below a 
threshold of $\Delta \chi^2 \le 10^{-4}\sigma_{\chi^2}$, where 
$\sigma_{\chi^2}$ is the standard deviation of $\chi^2$-distributed 
variables with a number of degrees of freedom given by the number of 
observed pixels. Results of this method have been cross-validated using an independent
 conjugate gradient inversion algorithm with multi-grid preconditioning, originally 
 developed for the analysis of {\it WMAP} data in \cite{2009JCAP...09..006S}. Applying this estimator 
 to simulations pre-processed using the above mentioned algorithms yielded optimal error 
bars as expected.
 
However, we found  that optimal error bars could also be achieved for all shapes 
 using a much simpler diffusive inpainting pre-filtering procedure that can be described as follows: 
all masked areas of the sky (both Galactic and point sources) are filled in with an iterative scheme. 
Each pixel in the mask is filled with the average of 
all surrounding pixels, and this is repeated $2000$ times 
over all masked pixels (we checked on simulations that convergence of all $\fnl$ estimates was 
achieved with $2000$ iterations). Note that the effect of this inpainting procedure, especially visible for the Galactic mask, is effectively to apodize the mask, reproducing small-scale structure near the edges and only large-scale modes in the interior. This helps to prevent propagating any sharp-edge effects or lack of large-scale power in the interior of the mask to the unmasked regions during harmonic transforms.

Any bias and/or excess variance arising from the inpainting procedure 
were assessed through MC validation (see 
Sect.~\ref{sec:Validation}) and found to be negligible. Since the inpainting procedure 
 is particularly simple to implement, easily allows inclusion of realistic correlated-noise models 
 in the simulations, and retains optimality, we chose inpainting as our data filtering procedure 
 for the $\fnl$ analysis. 

%%%%%%%%%%%%%%%%%%%%%%%%%%%%%%%%%%%%%%%%%%%%%%%%%%%%

\subsection{CMB bispectrum reconstruction}
\label{sec:bisp_reconst}

\subsubsection{Modal bispectrum reconstruction}

Modal and related estimators can be used to reconstruct the full bispectrum from the modal coefficients $\beta_{ijk}$ obtained from map filtering with the separable basis functions $Q_{ijk}(\llist)$ (Eq.~\eqref{eq:mapintegral})  \citep{2010PhRvD..82b3502F}. It is easy to show 
that the expectation value for $\beta_{ijk}$ (or equivalently $\beta_n$) for an ensemble of non-Gaussian maps generated with a CMB bispectrum shape given by $\alpha_n$ (Eq.~\eqref{eq:alphacoeff})
(with amplitude $\fnl$) is  
\begin{equation}\label{eq:reconmodes}
\langle \beta_n\rangle = \fnl \sum_p  \gamma_{np} \alpha_p\,,
\end{equation}
where $ \gamma_{np}$ is defined in Eq.~\eqref{eq:gammamatrix}.
Hence, we expect the best fit coefficients for a particular $\alpha_n$ realization to be given by the $\beta_n$s themselves (for
a sufficiently large signal).  
Assuming that we can extract the $\beta_n$ coefficients accurately from a particular experiment, 
we can directly reconstruct the CMB bispectrum using the expansion of Eq.~(\ref{eq:separablemodes}), that is, 
\eq \label{eq:cmbrecmodes}
 \blll &=& \sqrt{C_{\ell_1}C_{\ell_2}C_{\ell_3}}\,\sum_{n,p}\gamma_{np}^{-1}\,\beta_{p}\, Q_{n} (\ell_1, \ell_2, \ell_3)\cr
 &=&\sqrt{C_{\ell_1}C_{\ell_2}C_{\ell_3}}\,\sum_{n}\beta^R_{n}\, R_{n} (\ell_1, \ell_2, \ell_3)\,,\label{eq:cmborthomodes}
\qe 
where, for convenience, we have also defined orthonormalized basis functions $R_n(\llist)$ with coefficients $\alpha^R_n$ and $\beta_n^R$ such that $\langle R_n,\,R_p\rangle = \delta_{np}$.  
This method has been validated for simulated maps, showing the accurate recovery of CMB bispectra, and it has been
applied to the {\it WMAP}-7 data to reconstruct the full 3D CMB bispectrum \citep{2010arXiv1006.1642F}.

To quantify whether or not there is a model-independent deviation from Gaussianity, we can consider the total integrated bispectrum.  
By summing over all multipoles, we can define a  total integrated nonlinearity parameter $\bar F_{\rm NL}$ which, with the orthonormal modal decomposition of Eq.~\eqref{eq:cmborthomodes} becomes \citep{2010arXiv1006.1642F}
\begin{equation}\label{eq:totalbispectrum}
\bar F_{\rm NL}^2 ~ = ~\frac{1}{N_{\rm loc}^2} \sum_{l_i}\frac{\hlll^2\blll^2 }{C_{\ell_1} C_{\ell_2} C_{\ell_3}}~=~ \frac{\sum_{n} {\beta^R_n}{}^2}{\sum_{n} {\alpha^R_n}{}^2}\,,
\end{equation}
where $N_{\rm loc}$ is the normalization for the local $\fnl=1$ model.  
The expectation value $\langle \bar F_{\rm NL}^2\rangle$ contains more than just the three-point correlator, with contributions from products of the two-point correlators and even higher-order contributions,
\begin{equation}
\label{eq:fnlsummary}
\barFnl^2 \px \frac{1}{N_{\rm loc}^2} \left[6\nmax + \sum_{n=1}^{\nmax} \left( \Fnl^2\alpha^R_n{}^2 +  \langle\beta^R_n{}^2\rangle_{\rm 6pt}  + ...\right)\right]\,.
\end{equation}
Here $\bar{F}_{\mathrm{NL}}^2$ is the full 6-point function over the unconnected
Gaussian part, i.e., the product of 3 $C_{\ell}$. So, for a Gaussian 
input this recovers an average of 1 per mode.  In the non-Gaussian 
case the leading-order contributions after the Gaussian part are the 
bispectrum squared and the product of the power spectrum and the 
trispectrum, which enter at the same order (for additional explanations see 
\citealt{2010arXiv1006.1642F}). 

\subsubsection{Smoothed binned bispectrum reconstruction}
\label{binnedrec}

As explained in Sect.~\ref{binned}, the full binned bispectrum of the maps
under study is one of the products of the binned estimator code.

Given the relatively fine binning (about 50 bins up to $\ell=2000$ or
$2500$), most of the measurement in any single bin-triplet is noise. If
combinations of maps are chosen so that the CMB primordial signal
dominates, most of this noise is Gaussian, reflecting the fact that
even when $\langle xyz\rangle =0$, for a particular statistical
realization, $xyz$ is almost certainly non-zero. If our goal is to
test whether there is any statistically significant signal, then it
makes sense to normalize by defining a new field $\mathcal{B}_{i_1 i_2 i_3}$, 
which is the binned bispectrum divided by its expected standard deviation 
(computed in the standard way assuming Gaussian statistics). The 
distribution of $\mathcal{B}_{i_1 i_2 i_3}$ in any one bin-triplet is very 
nearly Gaussian as a result of the central limit theorem, and there is 
almost no correlation between different bins.

We could study the significance of the extreme values at this fine
resolution, but it also makes sense to smooth in order to detect
features coherent over a wider range of $\ell$. If the
three-dimensional domain over which the bispectrum is defined
(consisting of those triplets in the range $[\ell_\mathrm{min},
  \ell_\mathrm{max}]$ satisfying the triangle inequality and parity
condition) did not have boundaries, the smoothing would be more
straightforward. We smooth with a Gaussian kernel of varying width
in $\Delta (\ln \ell )$, normalized so that the smoothed function in
each pixel would be normal-distributed if the input map were
Gaussian. In this way, based on the extreme values, it is possible to
decide on whether there is a statistically significant NG
signal in the map in a ``blind'' (non-parametric) way.

%%%%%%%%%%%%%%% Section 4 %%%%%%%%%%%%%%%
\section{Statistical estimation of the CMB trispectrum}
\label{sec:CMBt}

\subsection{The squeezed-diagonal trispectrum: $\tau_{\rm NL}$}
\label{sec:taunl}
The 4-point function, or equivalently the trispectrum, of the CMB, can also place interesting constraints on inflationary physics.
There are several physically interesting
``shapes'' of the trispectrum
(e.g.,~\citealt{Huang:2006eha,2006PhRvD..74l3519B,2010arXiv1012.6039F,Izumi:2011di}),
in analogy to the bispectrum case.
In the simplest non-Gaussian models, the CMB bispectrum has larger signal-to-noise
than the trispectrum, but there are examples of technically natural models in which
the trispectrum has larger signal-to-noise
(e.g.,~\citealt{2011JCAP...01..003S,Baumann:2011nk}; see also~\citealt{2010JCAP...09..035B}).
This can happen in models in which the field modulating the fluctuation amplitude is only weakly correlated to the observed large-scale curvature perturbation.

Analysis of the trispectrum is more challenging than that of the bispectrum, due
to the increased range of systematic effects and secondary signals which can contribute.
For example, gravitational lensing of the CMB generates a many-sigma contribution to
the trispectrum, though it has a distinctive anisotropic shape that differs from primordial NG modulated by scalar fields.
As an instrumental example, any mismatch between the true covariance of the observed
CMB plus noise and the covariance which is assumed in the analysis (due, for example, to
mischaracterisation of the pointing, beams, or noise properties) will generally lead
to biases in the estimated trispectrum.
Due to these challenges, we have deferred a full analysis of the primordial trispectrum
to a future paper, and here focus on the simplest squeezed shape that can provide useful constraints on primordial models, $\taunl$.

$\taunl$ is most easily understood as measuring the large-scale modulation of small-scale power. The constraints on $\fnl$ show that such a modulation must be small if correlated with the temperature. However it is possible for multi-field inflation models to produce squeezed-shape modulations which are uncorrelated with the large-scale curvature perturbations. Such models can be constrained by the trispectrum, conventionally parameterized by $\taunl$ in the squeezed-diagonal shape.

For example, consider the case where a small-scale Gaussian curvature perturbation $\zeta_0$ is modulated by another field $\phi$ so that the primordial perturbation is given by
\be
\zeta(\vx) = \zeta_0(\vx)[1+\phi(\vx)],
\ee
where $\phi(\vx)$ is a large-scale modulating field (with amplitude $\ll 1$). The large-scale modes of $\phi$ can be measured from the modulation they induce in the small-scale $\zeta$ power spectrum. If $\phi$ has a nearly scale-invariant spectrum, the nearly-white cosmic variance noise on the reconstruction dominates on small-scales, so only the very largest modes can be reconstructed \citep{Kogo:2006kh}. A reconstruction of $\phi$ is going to be limited to only very large-scale variations, in which case the scale of the variation is very large compared to the width of the last-scattering surface; i.e., in any particular direction a large-scale modulating field will modulate all small-scale perturbations through the last-scattering surface by approximately the same amount. This approximation is good at the percent level, and can readily be related to the full trispectrum estimator, as discussed in more detail in~\citet[Sect.~IV]{Pearson:2012ba}; see also \citealt{2002PhRvD..66f3008O,Munshi:2009wy,2010PhRvD..81l3007S}.

A large-scale power modulation therefore translates directly into a large-scale modulation of the small-scale CMB temperature:
\be
T(\vnhat) \approx T_{\rm g}(\vnhat)[1 + \phi(\vnhat,r_*)] \equiv  T_{\rm g}(\vnhat)[1 + f(\vnhat)],
\label{Tmodulation}
\ee
where $T_{\rm g}$ are the usual small-scale Gaussian CMB temperature anisotropies and $r_*$ is the radial distance to the last-scattering surface. We can quantify the trispectrum as a function of modulation scale by using the power spectrum of the modulation,
\be
\taunl(L) \equiv \frac{C_L^f}{C_L^{\zeta_\star}}.
\label{taunl_L}
\ee
As is conventional, we normalize relative to $C_L^{\zeta_\star}$, the power spectrum of the primordial curvature perturbation at the location of the recombination surface. The field $f$ is directly observable, but $C_L^{\zeta_\star}$ is not, since the curvature perturbation can only be constrained very indirectly on very large scales. We shall therefore give constraints on $f$, which is directly constrained by \planck, but also on $\taunl$ for comparison with the inflation literature. Note that $\taunl\sim 500$ corresponds to an $f= \clo(10^{-3})$ modulation.

A general quadratic estimator methodology for reconstructing $f$ was developed in~\cite{Hanson:2009gu}, which we broadly follow here. The structure is essentially identical to that for lensing reconstruction \citep{planck2013-p12}, where here instead of reconstructing a lensing potential (or deflection angle), we are reconstructing a scalar modulation field. The
quadratic maximum likelihood estimator for the large-scale modulation field $f$ (assuming it is small) is given by
\be
\hatmodulation_{LM}= \clf^{-1}_{LM L'M'} \left[ \barmodulation_{L'M'}  - \langle \barmodulation_{L'M'} \rangle \right]\,,
\ee
where $\barmodulation$ is a quadratic function of the filtered data that can be calculated quickly in real space:
\be
\barmodulation_{LM}
=
\int d^2\vnhat Y_{LM}^{*} \left[ \sum_{\ell_1 m_1}\covinvT_{\ell_1 m_1}^i Y_{\ell_1 m_1} \right] \left[ \sum_{\ell_2 m_2}\lensedC_{\ell_2} \covinvT_{\ell_2 m_2}^j Y_{\ell_2 m_2} \right].
\ee
Here $\covinvT^i = C^{-1} \Ltemp^i$ is an inverse-variance filtering sky map (which accounts for sky cuts and inhomogeneous noise),
and $\lensedC_{\ell_2}$ is the lensed $C_\ell$. The ``mean field"  $\mfmodulation \equiv \langle \barmodulation \rangle$ can be estimated from simulations, along with the Fisher normalization $\clf$ that is given by the covariance of $ \barmodulation  - \mfmodulation$. The $i,j$ indices are included here, since we shall be using different sky maps with independent noise to avoid noise biases at the level of modulation field reconstruction.
For low $L$ and high $\lmax$ the reconstruction noise is very nearly constant (white, because each small patch of sky gives a nearly-uncorrelated but noisy estimate of the small-scale power), and the reconstruction is very local.

In practice the inverse-variance filtering is imperfect, the noise cannot be modelled exactly, and the normalizing Fisher matrix $\clf_{\ell m \ell'm'}$ evaluated from simulations would be inaccurate. Instead we focus on $\barmodulation$ directly, which is approximately an inverse-variance weighed reconstruction of the modulation, and is manifestly very local in real space (and hence zero in the cut part of the sky). Since the reconstruction noise, which also approximately determines the normalization, is nearly white (constant in $L$), $\barmodulation  - \mfmodulation$ in real space has an expectation nearly proportional to the underlying modulation outside the mask.

We then define an estimator of the modulation power spectrum
\begin{equation}
\hat{C}_L^{f} =
k_L  \left( \frac{A_L}{2L+1} \sum_M | \barmodulation_{LM}  - \mfmodulation_{LM} |^2 -
N_{L}^{(0)}\right)
\label{CLphi}
\end{equation}
where
\begin{equation}\label{N_L0}
N_L^{(0)}=\frac{A_L}{2L+1} \sum_M \langle |\barmodulation_{LM}  - \mfmodulation_{LM}|^2\rangle_0
\end{equation}
is a noise bias for zero signal estimated from simulations. The normalization $A_L$ is the analytic ideal full-sky normalization which is very close to a constant, and $k_L $ is a calibration factor determined from with-signal simulations. On small scales $k_L\propto \fsky^{-1}$, but has some scale dependence: it increases towards $k_L \propto \fsky^{-2}$ at $L=0$ at very low $L$. We shall sometimes plot $\bar{C}_L^f\equiv k_L^{-1}\hat{C}_L^{f}$, corresponding to the uncalibrated reconstruction of the power modulation, which is very local in real space.

For each value of the modulation scale $L$, Eq.~\eqref{CLphi} defines a separate estimator for $\taunl$
\be
\htaunl(L) \equiv \frac{\hat{C}_L^{f}}{C_L^{\zeta_\star}}.
\ee
We can combine estimators from all $\ell$ by constructing
\be
 \hat{\tau}_{{\rm NL},1}  \approx  N^{-1} \sum_{L'=\Lmin}^{\Lmax} C_{L'}^{\zeta_\star} \cov^{-1}_{L'L} \hat{C}_L^f,
\label{taunlest_covweight}
\ee
where $N = \sum_{L'=\Lmin}^\Lmax C_{L'}^{\zeta_\star}\cov^{-1}_{L'L}C_{L}^{\zeta_\star}$ and $\cov_{LL'}$ is the covariance of $\hat{C}_L^f$ from simulations with $\taunl=0$. On the full-sky the estimators from each $L$ would be independent, but the mask introduces significant coupling between the very low multipoles and this form of the estimator allows us to account for this.  In the full-sky uncorrelated approximation, with a nearly scale-invariant primordial spectrum and using the whiteness of the reconstruction noise, the estimator for $\taunl$ simplifies to ~\citep{Pearson:2012ba}
\be
 \htaunl  \approx  N^{-1} \sum_{L=\Lmin}^\Lmax \frac{2L+1}{L^2(L+1)^2}\frac{\hat{C}_L^f}{C_L^{\zeta_\star}},
\label{taunlest_scaleinv}
\ee
where $N\equiv \Lmin^{-2}-(\Lmax+1)^{-2}$. This result does not require many simulations to estimate the covariance accurately for inversion, and is typically expected to give very similar results with an error bar that is less than $10\%$ larger. We calculate both as a cross-check, but report results for $\htaunl$ because it is more robust, and in our simulation results actually has slightly lower tails (though larger variance).
Mean fields and the $N^{(0)}_L$ bias are estimated in all cases from 1000 zero-$\taunl$ simulations, and the mask used retains about $70\%$ of the sky.

If there is a nearly scale-invariant signal, so $C_L^f \propto C_L^\zeta$ as expected in most multi-field inflation models,  the contributions fall rapidly $\propto 1/L^3$, as expected when measuring a scale-invariant signal that has large white reconstruction noise. The signal is therefore on very large scales, with typically half the Fisher signal in the dipole modulation and $95\%$ of the signal at $L\le 4$, justifying the squeezed approximations used. We use $\Lmax=10$ for the estimators, which includes almost all of the signal-to-noise but avoids excessive contamination with the `blue' spectrum of lensing contributions. However, due to the small number of modes involved, the posterior distributions of $\taunl$ can have quite broad tails corresponding to the finite probability that all the largest-scale modulation modes just happen to be near zero. To improve constraints on large values it can help to include a larger range of $L$, and we consider $L$ up to $\Lmax=50$, which is about the limit of where the approximations are valid.

%%%%%%%%%%%%%%% Section 5 %%%%%%%%%%%%%%%
\section{Non-primordial contributions to the CMB bispectrum and trispectrum}
\label{sec:npNG}

In this subsection we present the steps followed to account for and remove the main non-primordial contributions to CMB NG.

\subsection{Foreground subtraction}

Foreground emission signals in the microwave bands have a strong non-Gaussian signature. Therefore any residual emission in the CMB data can give a spurious apparent primordial NG detection. In our analysis we considered \Planck\ CMB foreground-cleaned maps created using several independent techniques, as described in \citet{planck2013-p06}: explicit parametrization and fitting of foregrounds in real space
(\Commander-\Ruler, \CR) \citep{Eriksen2006ApJ641,Eriksen2008ApJ676}; Spectral Matching of foregrounds implementing Independent Component Analysis (\SMICA) (\citealt{delabrouille2003,cardoso2008}); Internal Linear Combination (Needlet Internal Linear Combination, (\NILC) (\citealt{delabrouille2009}); and Internal Template Fitting (\SEVEM) \citep{fernandez2012}. These and other techniques underwent a pre-launch testing phase \citep{leach2008}.
Each method provides a \Planck\ CMB foreground-cleaned map with a confidence mask, which defines the trusted cleaned region of the sky; an estimate of the noise in the output CMB map obtained from half-ring difference maps; and an estimate of the beam transfer function of the processed map. The resolution
reaches 5 arcminutes. In addition a union of all the confidence masks, denoted as U73, is provided.
Channels from both the Low Frequency Instrument (LFI,~\citealt{planck2013-p02})
and the High Frequency Instrument (HFI,~\citealt{planck2013-p03}) of \Planck\
are used to achieve each of the reconstructed CMB templates. The validation of CMB reconstruction through component separation
is based on the inspection of several observables, as explained in detail in \citet{planck2013-p06}: the two-point correlation function and derived cosmological parameters; indicators of NG including the $f_\textrm{NL}$ results presented in the present paper; and cross-correlation with known foreground templates.
Based on various figures-of-merit, the foreground cleaning techniques performed comparably well  \citep{planck2013-p06}.

In order to test foreground residuals a battery of simulations is required. In the simulations the foreground emission was modelled with the pre-launch version of the \Planck\ Sky Model (PSM), based on observations of the emission from our own Galaxy and known extra-Galactic sources, largely in the radio and infrared bands.
The PSM is described in \citet{delabrouille2012} and includes models of CMB (including a dipole), diffuse Galactic emissions (synchrotron, free-free, thermal dust, Anomalous Microwave Emission and CO molecular lines), emission from compact objects
(thermal SZ effect, kinetic SZ effect, radio sources, infrared sources, correlated far-infrared background and ultra-compact \ion{H}{ii} regions). The sky model includes total intensity as well as polarization, which was not used in this paper.

The PSM has been used to create the sixth round of Full Focal Plane (FFP6) simulations, a set of simulations for the 2013 data release based on detailed models of the sky and instrument (e.g., noise properties, beams, satellite pointing and map-making process), consisting of both Gaussian and non-Gaussian CMB realizations. A description of the FFP6 simulations can be found in Appendix~\ref{sec:FFP6} (see \cite{planck2013-p28} for details). The FFP6 set has been used to test and validate the component separation algorithms employed in \Planck\, and to establish uncertainties on the outputs~\citep{planck2013-p06}. We will also use FFP6 simulations in the next Sections in order to validate our analysis (see Sects.~\ref{sec:Validation},~\ref{sec:Results},~\ref{Sec_valid_data}).

\subsection{The Integrated Sachs-Wolfe-lensing bispectrum}
\label{sec:lensingISW}

One of the most relevant mechanisms that can generate NG from secondary CMB anisotropies is the coupling between weak lensing and the ISW \citep{Sachs:1967er} effect. This is in fact the leading contribution to the CMB secondary bispectrum with a blackbody frequency dependence \citep{1999PhRvD..59j3002G, 2002PhRvD..65d3007V, 2005PhRvD..71j3009G}.

%%%%%%%%%%%%%%%%%%%%%%%%%%%%%%%%%%%%%%%%%%%%%%%%%%%%
\begin{table}             % table* is a two-column table.  Drop the * for one column.
\begingroup
\newdimen\tblskip \tblskip=5pt
\caption{The bias in the three primordial $f_{\rm NL}$ parameters due to the
ISW-lensing signal for the four component-separation methods.}
\label{t:dx9pred}                            % Label goes here.
\nointerlineskip
\vskip -6mm
\footnotesize
\setbox\tablebox=\vbox{
   \newdimen\digitwidth
   \setbox0=\hbox{\rm 0}
   \digitwidth=\wd0
   \catcode`*=\active
   \def*{\kern\digitwidth}
   \newdimen\signwidth
   \setbox0=\hbox{+}
   \signwidth=\wd0
   \catcode`!=\active
   \def!{\kern\signwidth}
   \newdimen\dotwidth
   \setbox0=\hbox{.}
   \dotwidth=\wd0
   \catcode`^=\active
   \def^{\kern\dotwidth}
\halign{\hbox to 1.5in{#\leaderfil}\tabskip 1em&
\hfil#\hfil&
\hfil#\hfil&
\hfil#\hfil&
\hfil#\hfil\tabskip 0pt\cr
\noalign{\doubleline\vskip 2pt}
\omit&\multispan4\hfil ISW-lensing $f_{\rm NL}$ bias\hfil\cr
\omit&\multispan4\hrulefill\cr
Shape\hfill&\SMICA&\NILC&\SEVEM&\CR\cr
\noalign{\vskip 4pt\hrule\vskip 6pt}
Local & !*7.1& !*7.0& !*7.1& !*6.0\cr
Equilateral & !*0.4& !*0.5& !*0.4& !*1.4\cr
Orthogonal & $-$22^*& $-$21^*& $-$21^*& $-$19^*\cr
\noalign{\vskip 3pt\hrule\vskip 4pt}}}
\endPlancktable                    % ends one-column \halign

\endgroup
\end{table}                        % table* is a two-column table.  Drop the * for one column.
%%%%%%%%%%%%%%%%%%%%%%%%%%%%%%%%%%%%%%%%%%%%%%%%%%%%

%%%%%%%%%%%%%%%%%%%%%%%%%%%%%%%%%%%%%%%%%%%%%%%%%%%%
\begin{table}               % table* is a two-column table.  Drop the * for one column.
\begingroup
\newdimen\tblskip \tblskip=5pt
\caption{Results for the amplitude of the ISW-lensing bispectrum from the
\SMICA, \NILC, \SEVEM, and \CR\ foreground-cleaned maps,
for the KSW, binned, and modal (polynomial) estimators; error
bars are 68\% CL\,.}                          % Caption goes here.
\label{tab:ISWL}                            % Label goes here.
\nointerlineskip
\vskip -6mm
\footnotesize
\setbox\tablebox=\vbox{
   \newdimen\digitwidth
   \setbox0=\hbox{\rm 0}
   \digitwidth=\wd0
   \catcode`*=\active
   \def*{\kern\digitwidth}
   \newdimen\signwidth
   \setbox0=\hbox{+}
   \signwidth=\wd0
   \catcode`!=\active
   \def!{\kern\signwidth}
\halign{\hbox to 0.7in{#\leaderfil}\tabskip 1em&
\hfil#\hfil&\tabskip 0.8em&
\hfil#\hfil&
\hfil#\hfil&
\hfil#\hfil\tabskip 0pt\cr
\noalign{\doubleline\vskip 2pt}
\omit&\multispan4\hfil ISW-lensing amplitude\hfil\cr
\omit&\multispan4\hrulefill\cr
Method\hfill&\SMICA&\NILC&\SEVEM&\CR\cr
\noalign{\vskip 4pt\hrule\vskip 6pt}
KSW & 0.81 $\pm$ 0.31 & 0.85 $\pm$ 0.32 & 0.68 $\pm$ 0.32 & 0.75 $\pm$ 0.32\cr
Binned & 0.91 $\pm$ 0.37 & 1.03 $\pm$ 0.37 & 0.83 $\pm$ 0.39 & 0.80 $\pm$ 0.40\cr
Modal & 0.77 $\pm$ 0.37 & 0.93 $\pm$ 0.37 & 0.60 $\pm$ 0.37 & 0.68 $\pm$ 0.39\cr
\noalign{\vskip 3pt\hrule\vskip 4pt}}}
\endPlancktable                    % ends one-column \halign
%\endPlancktablewide                 % ends two-column \halign
%\tablenote a Footnote a.\par
%\tablenote b Footnote b.\par
\endgroup
\end{table}                        % table* is a two-column table.  Drop the * for one column.
%%%%%%%%%%%%%%%%%%%%%%%%%%%%%%%%%%%%%%%%%%%%%%%%%%%%

Weak lensing of the CMB is caused by gradients in the matter gravitational potential that distorts the CMB photon geodesics. The ISW on the other hand arise because of time-varying gravitational potentials due to the linear and nonlinear growth of structure in the evolving Universe.
Both the lensing and the ISW effect are then related
to the matter gravitational potential and thus are correlated phenomena.
This gives rise to a non-vanishing 3-point correlation function. Furthermore, lensing is related to nonlinear processes which are therefore non-Gaussian.
 A detailed description of the signal, which accounts also for the contribution from the early-ISW effect, can be found in \cite{2012arXiv1204.5018L}.

The ISW-lensing bispectrum takes the form:
\begin{equation}
B_{\ell_1 \ell_2 \ell_3}^{m_1 m_2 m_3}\equiv\langle a_{\ell_1 m_1}a_{\ell_2 m_2}a_{\ell_3 m_3}\rangle=
\langle a_{\ell_1 m_1}^{\rm P}a_{\ell_2 m_2}^{\rm L}a_{\ell_3 m_3}^{\rm ISW}\rangle+ 5
\; \mbox{perm.}\,,
\end{equation}
where P, L, and ISW indicate primordial, lensing and ISW contributions respectively.
This becomes
\be \label{bis-l-rs0}
B_{\ell_1 \ell_2 \ell_3}^{m_1 m_2 m_3 \, ({\rm ISW-L})}={\cal G}^{m_1 m_2m_3}_{\ell_1\ell_2\ell_3}b^{\rm ISW-L}_{\ell_1\ell_2\ell_3},
\ee
where ${\cal G}^{m_1 m_2m_3}_{\ell_1\ell_2\ell_3}$ is the Gaunt integral and $b^{{\rm ISW-L}}_{\ell_1\ell_2\ell_3}$ is the reduced bispectrum given by
\begin{align}\label{bis-l-rs}
b_{\ell_1 \ell_2 \ell_3}^{\rm ISW-L}&=
\frac{\ell_1(\ell_1+1)-\ell_2(\ell_2+1)+\ell_3(\ell_3+1)}{2} \nonumber\\
 & \times \, \tilde{C}_{\ell_1}^{\rm TT}  C^{{\rm T} \phi}_{\ell_3}
  + (5 \mbox{ perm.})\,.
\end{align}
Here $\tilde{C}_{\ell}^{\rm TT}$ is the lensed
CMB power spectrum and $C^{{\rm T} \phi}_\ell$ is the ISW-lensing cross-power spectrum~(\citealt{2012arXiv1204.5018L,1999PhRvD..59j3002G, 2002PhRvD..65d3007V,Cooray:1999kg})
that expresses the statistical expectation of the correlation between the lensing and the ISW effect.

As shown in \cite{2009PhRvD..80h3004H},~\cite{2009PhRvD..80l3007M}, and ~\cite{2011JCAP...03..018L}, the ISW-lensing bispectrum can introduce a contamination in the constraints on primordial local NG from the CMB bispectrum. Both bispectra are maximal for squeezed or nearly squeezed configurations.
The bias on a primordial $f_{\rm NL}$ (e.g., local) due to the presence of the ISW-lensing cross correlation signal is
\be\label{eq:fnl-est}
{\Delta f}^{\rm local}_{\rm NL}= \frac{\hat{S}}{N},
\ee
with
\be
\hat{S}= \sum_{2 \leqslant \ell_1 \ell_2 \ell_3} \frac{ B^{\rm ISW-L}_{\ell_1 \ell_2 \ell_3} \, B^{\rm P}_{\ell_1 \ell_2 \ell_3} }{ V_{\ell_1 \ell_2 \ell_3} } \;, \quad N= \sum_{2 \leqslant \ell_1 \ell_2 \ell_3} \frac{ \(B^{\rm P}_{\ell_1 \ell_2 \ell_3}\)^2 }{ V_{\ell_1 \ell_2 \ell_3} },
\ee
where $B^{\rm ISW-L}$ and $B^{\rm P}$ refer respectively to the ISW-lensing and
the primordial bispectrum, and $V$ is defined below Eq.~(\ref{eq:optimalestaveraged}).

The bias in the estimation of the three primordial $f_{\rm NL}$ from
\textit{Planck} is given in Table~\ref{t:dx9pred}.
As one can see, taking into account the $f_{\rm NL}$ statistical error bars shown, e.g., in Table~\ref{Tab_KSW+SMICA},
the local shape is most affected by this bias (at the level of more than
$1\,\sigma_\mathrm{local}$), followed by the orthogonal shape (at the level of
about $0.5\,\sigma_\mathrm{ortho}$), while the equilateral shape is hardly
affected. In this paper we have taken into account the bias reported in Table~\ref{t:dx9pred} by subtracting it from the measured $f_{\rm NL}$.\footnote{See~\cite{2013arXiv1302.5799K}
for other debiasing techniques.}

The results for the amplitude of the ISW-lensing bispectrum from the different
foreground-cleaned maps are given in Table~\ref{tab:ISWL}. It should be
noted that the binned and modal estimators are less correlated to the exact
template for the ISW-lensing shape than they are for the primordial shapes,
hence their larger error bars compared to KSW (which uses the exact template
by construction~\citep{2013arXiv1303.1722M}). The conclusion is that we detect the ISW-lensing bispectrum
at a value consistent with the fiducial value of 1, at a significance level
of $2.6\,\sigma$ (taking the \SMICA-KSW value as reference).
For details about comparisons between different estimators and analysis
of the data regarding primordial shapes we refer the reader to
Sects.~\ref{sec:Validation} and Sect.~\ref{sec:Results}.

%%%%%%%%%%%%%%%%%%%%%%%%%%%%%%%%%%%%%%%%%%%%%%%%%%%%
\begin{figure}
\includegraphics[width=\hsize]{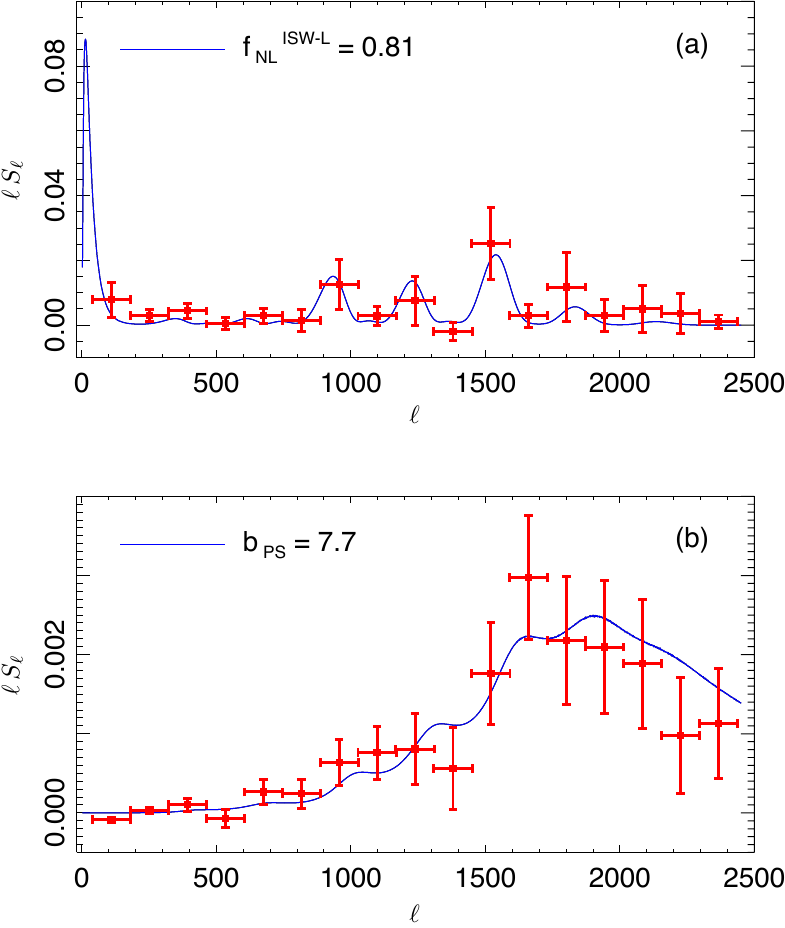}
\caption{The binned skew-$C_\ell$ statistics from the \SMICA\ map for
(a) ISW-lensing and (b) Poisson point sources. Theoretical curves are
not fitted to the data shown, but are plotted with the amplitude (the only
free parameter) determined from the KSW technique.
The Poisson point-source foreground is clearly detected, and the ISW-lensing
skew-spectrum is evident and consistent with the overall $2.6\sigma$ detection.  $b_{\textrm{ps}}$ is the Poisson point-source amplitude in dimensionless units of $10^{-29}$, and $f_{\textrm{NL}}^{\textrm{ISW-L}}$ is the ISW-lensing amplitude in units of that expected from the {\it Planck} best-fit cosmology.  Error bars come from covariance estimates from 1000 simulated maps, and the points are mildly correlated.}
\label{fig:skewcllensps}
\end{figure}
%%%%%%%%%%%%%%%%%%%%%%%%%%%%%%%%%%%%%%%%%%%%%%%%%%%%

We show for the \SMICA\ map in the top figure of Fig.~\ref{fig:skewcllensps}
the measured skew-$C_\ell$ spectrum (see Sect.~\ref{skewCl}) for optimal
detection of the ISW-lensing bispectrum,
along with the best-fitting estimates of $f_{\rm NL}$ from the KSW method for
different values of $\ell$.
It should be noted that the skew-$C_\ell$ spectrum is {\em not} a fit to
the KSW data points; its shape is fully fixed by the template under
consideration, with only the overall amplitude as a free parameter.
Hence the agreement between the curve and the points is good evidence that KSW is really
detecting the ISW-lensing effect and not some other source of NG. Note that point sources, at the level determined by their own
skew-spectrum, do not contribute significantly to the ISW-lensing statistic).
See \cite{planck2013-p12},~\cite{planck2013-p14} for further information
about the detection by \planck\ of the ISW-lensing signal.

%%%%%%%%%%%%%%%%%%%%%%%%%%%%%%%%%%%%%%%%%%%%%%%%%%%%

\subsection{Point-sources bispectrum}
\label{sec:psbisp}

Extra-Galactic point sources at \textit{Planck} frequencies are divided into two broad categories: radio sources with synchrotron and/or free-free emission; and infrared galaxies with thermal emission from dust heated by young stars.
Radio sources are dominant at central CMB frequencies up to 143~GHz, and can be considered unclustered \citep{Toffolatti1998,Gonzalez-Nuevo2005}. Hence their bispectrum is constant and is related to their number counts as
\begin{equation}
b_\mathrm{ps} = k_{\nu}^3 \int_{0}^{S_{\rm c}} S^3 \frac{dn}{dS} dS,
\end{equation}
with $S$ the flux density, $dn/dS$ the number counts per steradian, $S_{\rm c}$ the flux cut and $k_{\nu}$ the conversion factor from flux to relative temperature elevation, depending on the frequency and instrumental bandpass.

Infrared galaxies become important at higher frequencies, 217GHz and above, and are highly clustered in dark matter halos, which enhances their bispectrum on large angular scales \citep{Lacasa2012,Curto2013}. However, in the
\Planck\ context it was shown by \citet{Lacasa2013} that the IR bispectrum is more than 90\% correlated with the Poissonian template of the radio sources.
So a joint estimation of $f_{\rm NL}$ with a Poissonian bispectrum template will essentially account for the IR signal, and provide quasi-identical values compared to an analysis accounting for the IR bispectrum template.
Indeed, in our final optimal bispectrum constraints for primordial shapes, we will account for the potential contamination from point sources by jointly fitting primordial and Poisson templates to the data.

Our final measured point-source bispectrum amplitudes from the data are
reported in Table~\ref{tab:PB}. The amplitude is expressed in dimensionless
units, i.e., it has been divided by the appropriate power of the monopole
temperature $T_0$, and has been multiplied by $10^{29}$.
As shown in Sect.~\ref{Sec_deplmax}, the Poisson template is the only
one that still evolves significantly between $\ell=2000$ and $\ell=2500$.
This explains the differences between the values of the KSW and binned (that use
$\ell_\mathrm{max}=2500$) and the modal (that uses $\ell_\mathrm{max}=2000$)
estimators. It has been shown that for the same value of $\ell_\mathrm{max}$ all
three estimators agree very well.

We finally conclude from Table~\ref{tab:PB} that we detect the point-source
bispectrum with high significance in the \SMICA, \NILC, and \SEVEM\ cleaned
maps,
while it is absent from the \CR\ cleaned map. The measured skew-$C_\ell$
spectrum of the \SMICA\ map in the bottom figure of Fig.~\ref{fig:skewcllensps}
gives further evidence that the NG from foreground point sources is
convincingly detected. The only degree of freedom in this plot is
the amplitude, which is not set by a direct fit to the skew-$C_\ell$, but
rather is estimated by KSW.  As a result, the good agreement with the shape
of this skew-$C_\ell$
spectrum is powerful evidence that there is NG from point sources.
However, this still turns out to be a negligible contaminant for primordial
$f_{\rm NL}$ studies, due to the very low correlation between the Poisson
bispectrum and the primordial shapes.

%%%%%%%%%%%%%%%%%%%%%%%%%%%%%%%%%%%%%%%%%%%%%%%%%%%%
\begin{table}[tmb]                 % table* is a two-column table.  Drop the * for one column.
\begingroup
\newdimen\tblskip \tblskip=5pt
\caption{Results for the amplitude of the point source (Poisson) bispectrum
(in dimensionless units of $10^{-29}$)
from the \SMICA, \NILC, \SEVEM, and \CR\ foreground-cleaned maps,
for the KSW, binned, and modal (polynomial) estimators; error
bars are 68\% CL. Note that the KSW and binned estimators use
$\ell_\mathrm{max}=2500$, while the modal estimator has $\ell_\mathrm{max}=2000$.}
\label{tab:PB}
\nointerlineskip
\vskip -3mm
\footnotesize
\setbox\tablebox=\vbox{
   \newdimen\digitwidth
   \setbox0=\hbox{\rm 0}
   \digitwidth=\wd0
   \catcode`*=\active
   \def*{\kern\digitwidth}
   \newdimen\signwidth
   \setbox0=\hbox{+}
   \signwidth=\wd0
   \catcode`!=\active
   \def!{\kern\signwidth}
 \newdimen\dotwidth
   \setbox0=\hbox{.}
   \dotwidth=\wd0
   \catcode`;=\active
   \def;{\kern\dotwidth}
\halign{\hbox to 0.7in{#\leaderfil}\tabskip 1em&
#\hfil&\tabskip 0.8em&
#\hfil&
#\hfil&
#\hfil\tabskip 0pt\cr
\noalign{\doubleline\vskip 2pt}
\omit&\multispan{4}\hfil Point source bispectrum amplitude/$10^{-29}$\hfil\cr
\omit&\multispan4\hrulefill\cr
Method\hfill&\hfil\SMICA\hfil&\hfil\NILC\hfil&\hfil\SEVEM\hfil&\hfil\CR\hfil\cr
\noalign{\vskip 4pt\hrule\vskip 6pt}
KSW&*7.7 $\pm$ 1.5&*9.2 $\pm$ 1.7&*7.6 $\pm$ 1.7&1.1 $\pm$ 5.1\cr
Binned&*7.7 $\pm$ 1.6&*8.2 $\pm$ 1.6&*7.5 $\pm$ 1.7&0.9 $\pm$ 4.8\cr
Modal&10;* $\pm$ 3&11;* $\pm$ 3&10;* $\pm$ 3&0.5 $\pm$ 6\cr
\noalign{\vskip 3pt\hrule\vskip 4pt}}}
                      % Template goes here.
%\noalign{\doubleline}
                                    % Table headings go here.
%\noalign{\vskip 3pt\hrule\vskip 5pt}
                                    % Body of table goes here.
%\noalign{\vskip 5pt\hrule\vskip 3pt}}}
\endPlancktable                    % ends one-column \halign
%\endPlancktablewide                 % ends two-column \halign
%\tablenote a Footnote a.\par
%\tablenote b Footnote b.\par
\endgroup
\end{table}                        % table* is a two-column table.  Drop the * for one column.
%%%%%%%%%%%%%%%%%%%%%%%%%%%%%%%%%%%%%%%%%%%%%%%%%%%%

\subsection{Non-primordial contributions to the trispectrum}
\label{taunl_nonprimordial}

The main non-instrumental source of non-primordial signal is the kinematic modulation dipole due to the peculiar velocity of the earth, ${\vec{v}}$, whose magnitude is $\clo(10^{-3})$ (\citealt{Challinor:2002zh,Kosowsky:2010jm,Amendola:2010ty}). If data are used to constrain $\taunl$ using the dipole modulation (which shrinks the Fisher error by a factor of two relative to starting at $L=2$), the dipole-induced signal must be subtracted, since its modulation reconstruction has signal-to-noise larger than unity at \planck\ resolution. Confirmation that this signal is detected with the expected magnitude and direction is a good test of our methodology. The dipole signal seen by \planck\ is studied in detail in~\cite{planck2013-pipaberration}, so we only summarize the key points here.

The local Doppler effect  modulates the observed CMB temperature $T_0[1+\Delta T(\vnhat)]$ by $1+\vnhat\cdot\vv$ at leading order, so that $T(\vnhat)=T_0 [1+\Delta T(\vnhat)](1+\vnhat\cdot\vv)$. The spectrum in each direction remains a blackbody, but the relative response in the intensity $I_\nu(\nu,\vnhat)$ at the observed \planck\ frequencies is however frequency dependent. The effective thermodynamic fractional temperature anisotropy $\Delta \Theta$ at each frequency for zero peculiar velocity is defined by
\be
I_\nu(\nu,\vnhat) = I_\nu(\nu)\left[ 1 + \left.\frac{\ud \ln I_\nu}{\ud\ln T} \right|_\nu \Delta\Theta(\vnhat)\right].
\ee
With peculiar velocity the temperature $T$ depends on the second order term $\Delta T \vnhat\cdot\vv$, so expanding the Planck function to second order then gives a change in the effective small-scale temperature anisotropy from both first and second order terms in the expansion of $I_\nu$:
\bea
\Delta\Theta(\vnhat) &\rightarrow& \left[ 1 +  \vnhat\cdot\vv + T\frac{ \ud^2 I_\nu/ \ud T^2}{\ud I_\nu/\ud T} \vnhat\cdot\vv \right]\Delta\Theta(\vnhat) \nonumber \\
&=& \left(1+ \left[x\coth (x/2) -1\right]\vnhat\cdot\vv\right) \Delta\Theta(\vnhat) ,
\eea
where $x\equiv h \nu/k_{\textrm B}T_0$ (and we neglect small second-order non-modulation terms). Thus the anisotropies in the \planck\ maps have a dipolar modulation given by $1+ b_\nu\vnhat\cdot\vv$, where for the frequency bands we use $b_{143}\approx 2$, $b_{217}\approx 3$, and $\beta\equiv |\vv|= 1.23\times 10^{-3}$ in the direction of CMB dipole.
In addition our peculiar velocity induces kinetic aberration, which looks at leading order exactly like a dipole lensing convergence and only projects weakly into the power anisotropy estimator.  For constraining $\taunl$ both of the expected kinematic signals can be included in the simulations, and hence subtracted in the mean field of the modulation reconstruction.

Secondary effects are dominated by the significant and very blue lensing signal. However unlike for the $\fnl$ bispectrum lensing only overlaps with $\taunl$ at a small fraction of the error bar as long as only low modulation multipoles $L  \alt 10$ are used, where the $\taunl$ signal peaks~\citep{Pearson:2012ba}. We include lensing in the simulations, so lensing is straightforwardly accounted for in our analysis by its inclusion in the $N_L^{(0)}$ noise bias (Eq.~(\ref{N_L0})) and mean field.

A variety of instrumental effects can also give a spurious modulation signal if not modelled accurately. In particular the mean field due to anisotropic noise is very large \citep{Hanson:2009dr}. On the ultra-large scales of interest for $\taunl$, our understanding of the noise is not adequate to calculate accurately and subtract this large signal. Instead, as for the power spectrum estimation, we use cross-map estimators that have no noise mean field on average. Both the noise and most other instrumental effects such as gain variations are expected to produce a signal with approximate symmetry about the ecliptic plane. Our modulation reconstruction methodology is especially useful here, since we can easily inspect the orientation of any signal found; for example a naive treatment of the noise not using cross-maps would give a large apparent quadrupolar modulation signal aligned with the ecliptic, corresponding to percent-level misestimation of the noise mean field from inaccurate noise simulation.

Beam asymmetries are included in the simulation, as described in \cite{planck2013-p12}, but their effect is very small, since the modulation we are reconstructing is isotropic.

Since we are reconstructing a modulation of \emph{small-scale} power, the estimator is totally insensitive to smooth large-scale foregrounds. However large-scale variation in small-scale foreground power can mimic a trispectrum modulation. We project out 857~GHz as a dust template in our inverse-variance filtering procedure, as described in Appendix A of \cite{planck2013-p12}, but do not include any other foreground model in the trispectrum analysis. Any unmodelled foreground power variation would increase the $\taunl$ signal, so our modelling is sufficient to place a robust upper limit.

%%%%%%%%%%%%%%% Section 6 %%%%%%%%%%%%%%%
\section{Validation Tests}
\label{sec:Validation}

The $f_{\rm NL}$ results quoted in
this paper have all been cross-validated using multiple
bispectrum-based estimators from different groups.
Having multiple estimators was extremely useful for the
entire analysis, for two main reasons. First, it allowed
great improvement in the robustness of the final results. In the early
stages of the work the comparison between different independent
techniques helped to resolve bugs and other technical issues in the
various computer codes, while during the later stages it was very
useful to understand the data and find the optimal way of extracting
information about the various bispectrum templates.  Secondly,
besides these cross-checking purposes, different estimators provide
also interesting complementary information, going beyond simple 
$f_{\rm NL}$ estimation. For example, the binned and modal estimators provide a
reconstruction of the full bispectrum of the data (smoothed in
different domains), the skew-$C_\ell$ estimator allows monitoring of the
contribution to $f_{\rm NL}$ from different sources of NG, 
the wavelets reconstruction allows $f_{\rm NL}$ directionality
tests, and so on.

In this Section we are concerned with the first point above, that is, the use of multiple bispectrum-based pipelines as a way
to improve the robustness of the results.
For this purpose, a large amount of work was dedicated to the
development and analysis of various test maps, in order to validate
the estimators. This means not only checking that the various
estimators recover the input $f_{\rm NL}$ within the expected errors, but also that the results agree on a map-by-map basis.

The Section is split into two parts. Sect.~\ref{sec:Sec_valid_est}
shows results on a set of initially full-sky, noiseless, Gaussian CMB 
simulations, to which we add, in several steps, realistic complications, including  
primordial NG, anisotropic coloured noise, and a mask, showing the impact 
on the results at each step.
In Sect.~\ref{Sec_valid_compsep} we show our results on a set of simulations
that mimic the real data as closely as possible (except for the presence
of foreground residuals, which will be studied in Sect.~\ref{sec:Valid_FGresid}): 
no primordial NG, but NG due to the ISW-lensing 
effect; simulated instrumental effects and realistic noise; and simulations
passed through the component separation pipelines. In fact these are
the FFP6 simulations (see Appendix~\ref{sec:FFP6}) that are used to determine the error bars for the final
{\it Planck} results.

We present here only a small subset of the large number
of validation tests that were performed. For example,
we also had a number of ``blind $f_{\rm NL}$ challenges'', in which
the different groups received a simulated data set with an
unknown value of input $f_{\rm NL}$ for a given shape and they had
to report their estimated values. In addition different noise models
were tested (white vs.\ coloured and isotropic vs.\ anisotropic),
leading to the conclusion that it is important to make the noise in the
estimator calibration as realistic as possible (coloured and anisotropic).
We also tested different Galactic and point source masks, with and without inpainting, 
concluding that it is best to fill in both the point sources and the Galactic mask, using a sufficient
number of iterations in our diffusive procedure to entirely fill in the 
point source gaps, while at the same time only effectively apodizing 
the Galactic mask (no small-scale structure in its interior).
There were also various tests on FFP simulations of 
  {\it Planck} data with Gaussian or non-Gaussian CMB 
and all foregrounds, provided by the PSM (see Appendix~\ref{sec:ffp6_valid}). These simulations were tested
both before and after they had passed through the component separation
pipelines. In all comparison tests the results were
consistent with input $f_{\rm NL}$ values and differences between
estimators were consistent with theoretical expectations. 

%%%%%%%%%%%%%%%%%%%%%%%%%%%%%%%%%%%%%%%%%%%%%%%%%%%%

\subsection{Validation of estimators in the presence of primordial non-Gaussianity}
\label{sec:Sec_valid_est}

The aim of the first set of validation tests is threefold. 
First, we want to study the level of agreement from different estimators in ideal conditions (i.e., full-sky noiseless data). 
 The expected scatter between measurements is, in this 
 case, entirely due to the slightly imperfect correlation between weights of estimators that adopt different schemes to approximate the 
 primordial shape templates. For this case the scatter can be computed analytically (see Appendix~\ref{sec:AA} for details). We can then verify that our 
 results in ideal conditions match theoretical expectations. This is done in Sect.~\ref{Sec_valid_ideal}.
Second, we want to make sure that the estimators are unbiased and correctly recover $\fnl$ in input for local, equilateral,  
and orthogonal shapes. This is done in Sects.~\ref{Sec_valid_nomask} and~\ref{Sec_valid_mask}, where a superposition 
of local, equilateral and orthogonal bispectra is included in the simulations and the three $\fnl$ values are estimated 
both independently and jointly. Finally we want to understand how much the agreement between pipelines in ideal conditions is degraded 
when we include a realistic correlated noise component and a sky cut, thus requiring the introduction of a linear term in the estimators 
in order to account for off-diagonal covariance terms introduced by the breaking of rotational invariance. Since we want to study the impact 
of adding noise and masking separately, we will first work on a set of full-sky maps with noise in Sect.~\ref{Sec_valid_nomask}, and then add a 
mask in Sect.~\ref{Sec_valid_mask}.

The tests were applied to the KSW, binned and 
 modal estimators. These are the bispectrum pipelines used to analyse {\it Planck} data in Sect.~\ref{sec:Results}. Our goal for this set of tests is not so much to attain the tightest possible agreement between methods, as it is to address the points summarized in the above paragraph. For this reason the estimator implementations used in this specific Section were slightly less accurate but faster to compute than those adopted for the final data analysis of Sect.~\ref{sec:Results}. The primary difference with respect to the main analysis is that a smaller number of simulations was used to calibrate the linear term (80--100 in these tests, as against 200 or more for the full analysis). For the modal estimator we also use a faster expansion with a smaller number of modes: 300 here versus 600 in the high accuracy version of the pipeline\footnote{While most of the modal results in this paper come from the most accurate 600 modes pipeline, a few computationally intensive data validation tests of Sect.~\ref{Sec_valid_data} also use the fast 300 modes version; therefore the results in this Section also provide a direct validation of the fast modal pipeline.} used in Sect.~\ref{sec:Results}. Even with many fewer modes, the modal estimator is still quite accurate: the correlation coefficient for the modal expansion of the local template is $0.95$, while for the equilateral and orthogonal shapes it is $0.98$.   

%%%%%%%%%%%%%%%%%%%%%%%%%%%%%%%%%%%%%%%%%%%%%%%%%%%%

\subsubsection{Ideal Gaussian simulations}
\label{Sec_valid_ideal}

As a basis for the other tests we start with the
ideal case, a set of 96 simulations of a full-sky Gaussian CMB, 
with a Gaussian beam with FWHM 5 arcmin and without any noise, cut off at 
$\ell_\mathrm{max}=2000$ in our analyses. The independent Fisher matrix 
error bars in that case are $4.2$ for local NG, $56$ for equilateral, and 
$28$ for orthogonal.

Note that this test does not make sense for all estimators, and hence
results are not included for all of them. For example, for the binned
estimator the optimal binning depends on the noise. While this dependence
is not very strong, the difference between no noise and {\it Planck} noise is
sufficiently large that a completely different binning would have to be
used just for this test, going against the purpose of this Section to
validate the estimators as used for the data analysis.\footnote{While the 
binning with 48 bins and $\ell_\mathrm{max}=2000$ used in the validation tests
of Sects.~\ref{Sec_valid_nomask} and \ref{Sec_valid_mask} is also slightly
different from the binning used for the data analysis with 51 bins and
$\ell_\mathrm{max}=2500$, these differences are very small and the binnings have
very similar correlation coefficients of 0.99 or more for local and equilateral shapes,
and 0.95 for orthogonal.}

The purpose here is mostly aimed at checking consistency with the following formula
(derived in Appendix~\ref{sec:AA}) for the expected scatter
(standard deviation) between $f_{\rm NL}$ results of the same map from
an exact and an approximate estimator: 
\begin{equation}\label{eqn:expectation}
\sigma_{\delta f_{\rm NL}} = \Delta_{\rm th} \frac{\sqrt{1-r^2}}{r} \; .
\end{equation}
Here $\Delta_{\rm th}$ is the standard deviation of the exact
estimator and $r$ is the correlation coefficient that gives the correlation
of the approximate bispectrum template with the exact one, defined as
\begin{equation}\label{eqn:corr}
 r \equiv \frac{  
\sum_{\ell_1 \leq \ell_2 \leq \ell_3} \frac{ B^{t\rm h}_{\ell_1 \ell_2 \ell_3} B^{\rm exp}_{\ell_1 \ell_2 \ell_3}}{g_{\ell_1 \ell_2 \ell_3} C_{\ell_1} C_{\ell_2} C_{\ell_3}}}
{\sqrt{\sum_{\ell_1 \leq \ell_2 \leq \ell_3} \frac{ (B^{\rm th}_{\ell_1 \ell_2 \ell_3})^2}{g_{\ell_1 \ell_2 \ell_3} C_{\ell_1} C_{\ell_2} C_{\ell_3}}
\sum_{\ell_1 \leq \ell_2 \leq \ell_3} \frac{ (B^{\rm exp}_{\ell_1 \ell_2 \ell_3})^2}{g_{\ell_1 \ell_2 \ell_3} C_{\ell_1} C_{\ell_2} C_{\ell_3}}} } \; ,
\end{equation}
where the label ``th'' denotes the initial bispectrum shape to fit to
the data, and ``exp'' is the approximate expanded one.
Note that this formula has been obtained under the simplifying assumptions
of Gaussianity, full-sky coverage and homogeneous noise. For
applications dealing with more realistic cases we might expect the
scatter to become larger, while remaining
qualitatively consistent.

The results averaged over the whole set of maps are given in 
Table~\ref{Tab_valid_ideal} for the KSW and modal estimators individually, 
as well as for their difference. 
The plane wave modal expansion implemented here achieves about 98\% 
correlation with the separable shapes used by KSW. According to the formula 
above we then expect a standard deviation of map-by-map differences of order 
$0.2 \Delta_{f_\mathrm{NL}}$ for a given shape, where $\Delta_{f_\mathrm{NL}}$ is the 
corresponding $f_\mathrm{NL}$ error bar. Looking at the left-hand side of 
Table~\ref{Tab_valid_ideal}, we see that the error bars are 4 for local NG, 
50 for equilateral, and 30 for orthogonal. So we predict a standard 
deviation of map-by-map differences of 0.8, 10 and 6 for local, equilateral, 
and orthogonal NG, respectively. As one can see from the ``Modal-KSW" column,
the measurements are in excellent agreement with the theoretical expectation. 
%

%%%%%%%%%%%%%%%%%%%%%%%%%%%%%%%%%%%%%%%%%%%%%%%%%%%%
\begin{table}[tmb]                 % table* is a two-column table.  Drop the * for one column.
\begingroup
\newdimen\tblskip \tblskip=5pt
\caption{Results for $f_\mathrm{NL}$ for the set of ideal Gaussian 
simulations described in Sect.~\ref{Sec_valid_ideal} for the 
KSW and modal estimators and for their difference, assuming all 
shapes to be independent.}
\label{Tab_valid_ideal}
\nointerlineskip
\vskip -6mm
\footnotesize
\setbox\tablebox=\vbox{
   \newdimen\digitwidth 
   \setbox0=\hbox{\rm 0} 
   \digitwidth=\wd0 
   \catcode`*=\active 
   \def*{\kern\digitwidth}
   \newdimen\signwidth 
   \setbox0=\hbox{$-$} 
   \signwidth=\wd0 
   \catcode`!=\active 
   \def!{\kern\signwidth}
 \newdimen\dotwidth 
   \setbox0=\hbox{.} 
   \dotwidth=\wd0 
   \catcode`^=\active 
   \def^{\kern\dotwidth}
\halign{\hbox to 1in{#\leaderfil}\tabskip 1em&
\hfil#\hfil\tabskip 0.8em&
\hfil#\hfil&
%\hbox to 0.2in{#\leaderfil}&
\hfil#\hfil \tabskip 0pt\cr
\noalign{\doubleline\vskip 2pt}
\omit&\multispan3\hfil $f_{\rm NL}$\hfil\cr
\omit&\multispan3\hrulefill\cr
Shape\hfill&KSW&Modal&Modal $-$ KSW\cr
\noalign{\vskip 4pt\hrule\vskip 6pt}
Local&$-$0.5 $\pm$ *4.1&$-$0.5 $\pm$ *4.1&$-$0.05 $\pm$ 0.63\cr
Equilateral&!2.2 $\pm$ 48^*&!1.3 $\pm$ 48^*&$-$0.9 $\pm$ 8.9\cr
Orthogonal&$-$1.1 $\pm$ 29^*&$-$1.0 $\pm$ 30^*&!0.1 $\pm$ 6.5\cr
\noalign{\vskip 3pt\hrule\vskip 4pt}}}
\endPlancktable                    % ends one-column \halign
%\endPlancktablewide                 % ends two-column \halign
\endgroup
\end{table}                         % table* is a two-column table.  Drop the * for one column.
%%%%%%%%%%%%%%%%%%%%%%%%%%%%%%%%%%%%%%%%%%%%%%%%%%%%

%%%%%%%%%%%%%%%%%%%%%%%%%%%%%%%%%%%%%%%%%%%%%%%%%%%%

\subsubsection{Non-Gaussian simulations with realistic noise}
\label{Sec_valid_nomask}
%
%%%%%%%%%%%%%%%%%%%%%%%%%%%%%%%%%%%%%%%%%%%%%%%%%%%%
\begin{table*}[tmb]                 % table* is a two-column table.  Drop the * for one column.
\begingroup
\newdimen\tblskip \tblskip=5pt
\caption{Results from the different estimators for $f_\mathrm{NL}$ for the set 
of full-sky non-Gaussian simulations described in Sect.~\ref{Sec_valid_nomask}.
The simulations have $f_{\rm NL}=12, 35, -22$ respectively.
Both the results for the estimators individually and for the differences with
KSW are given.}
\label{Tab_valid_nomask}
\nointerlineskip
\vskip -3mm
\footnotesize
\setbox\tablebox=\vbox{
   \newdimen\digitwidth 
   \setbox0=\hbox{\rm 0} 
   \digitwidth=\wd0 
   \catcode`*=\active 
   \def*{\kern\digitwidth}
   \newdimen\signwidth 
   \setbox0=\hbox{+} 
   \signwidth=\wd0 
   \catcode`!=\active 
   \def!{\kern\signwidth}
   \newdimen\dotwidth 
   \setbox0=\hbox{.} 
   \dotwidth=\wd0 
   \catcode`^=\active 
   \def^{\kern\dotwidth}
\halign{\hbox to 1.5in{#\leaderfil}\tabskip 1em&
\hfil#\hfil\tabskip 1em&
\hfil#\hfil&
\hfil#\hfil&
%\hbox to 0.4in{#\leaderfil}&
\hfil#\hfil&
\hfil#\hfil\tabskip 0pt\cr
\noalign{\doubleline\vskip 2pt}
\omit&\multispan5\hfil $f_{\rm NL}$\hfil\cr
\omit&\multispan5\hrulefill\cr
Shape\hfill&KSW&Binned&Modal&******Binned $-$ KSW&Modal $-$ KSW\cr
\noalign{\vskip 4pt\hrule\vskip 6pt}
\omit\hfil{\bf Independent}\hfil&\cr
Local &!13.8 $\pm$ *5.2 & !14.1 $\pm$ *5.2 & !14.1 $\pm$ *5.3 & ******!0.3 $\pm$ *2.1 & !0.4 $\pm$ *2.6 \cr
Equilateral &!63^* $\pm$ 57^* & !62^* $\pm$ 58^* & !64^* $\pm$ 57^* & ******$-$0.9 $\pm$ 20^* & !1.0 $\pm$ 18^* \cr
Orthogonal &$-$52^* $\pm$ 37^* &$-$58^* $\pm$ 40^* &$-$54^* $\pm$ 37^* & ******$-$6.0 $\pm$ 13^* &$-$2.2 $\pm$ 12^* \cr
\noalign{\vskip 5pt}
%\hline
\noalign{\vskip 5pt}
\omit\hfil {\bf Joint}\hfil&\cr
Local & !11.7 $\pm$ *6.2 & !12.0 $\pm$ *6.6 & !12.0 $\pm$ *6.4 & ******!0.2 $\pm$ *2.7 & !0.2 $\pm$ *3.2 \cr
Equilateral & !31^* $\pm$ 59^* & !29^* $\pm$ 61^* & !31^* $\pm$ 59^* & ******$-$1.8 $\pm$ 21^* &$-$0.2 $\pm$ 19^* \cr
Orthogonal &$-$20^* $\pm$ 43^* &$-$22^* $\pm$ 47^* &$-$21^* $\pm$ 42^* & ******$-$2.1 $\pm$ 16^* &$-$0.6 $\pm$ 15^* \cr
\noalign{\vskip 3pt\hrule\vskip 4pt}}}
%\endPlancktable                    % ends one-column \halign
\endPlancktablewide                 % ends two-column \halign
\endgroup
\end{table*}                        % table* is a two-column table.  Drop the * for one column.
%%%%%%%%%%%%%%%%%%%%%%%%%%%%%%%%%%%%%%%%%%%%%%%%%%%%

%%%%%%%%%%%%%%%%%%%%%%%%%%%%%%%%%%%%%%%%%%%%%%%%%%%%
\begin{table*}[tmb]                 % table* is a two-column table.  Drop the * for one column.
\begingroup
\newdimen\tblskip \tblskip=5pt
\caption{Results from the different estimators for $f_\mathrm{NL}$ for the set 
of masked non-Gaussian simulations described in Sect.~\ref{Sec_valid_mask}.
Both the results for the estimators individually and for the differences with
KSW are given.}
\label{Tab_valid_mask}
\nointerlineskip
\vskip -3mm
\footnotesize
\setbox\tablebox=\vbox{
   \newdimen\digitwidth 
   \setbox0=\hbox{\rm 0} 
   \digitwidth=\wd0 
   \catcode`*=\active 
   \def*{\kern\digitwidth}
   \newdimen\signwidth 
   \setbox0=\hbox{+} 
   \signwidth=\wd0 
   \catcode`!=\active 
   \def!{\kern\signwidth}
    \newdimen\dotwidth 
   \setbox0=\hbox{.} 
   \dotwidth=\wd0 
   \catcode`^=\active 
   \def^{\kern\dotwidth}
\halign{\hbox to 1.5in{#\leaderfil}\tabskip 1em&
\hfil#\hfil\tabskip 1em&
\hfil#\hfil&
\hfil#\hfil&
%\hbox to 0.4in{#\leaderfil}&
\hfil#\hfil&
\hfil#\hfil\tabskip 0pt\cr
\noalign{\doubleline\vskip 2pt}
\omit&\multispan5\hfil $f_{\rm NL}$\hfil\cr
\omit&\multispan5\hrulefill\cr
Shape\hfill&KSW&Binned&Modal&******Binned $-$ KSW&Modal $-$ KSW\cr
\noalign{\vskip 4pt\hrule\vskip 6pt}
\omit\hfil {\bf Independent}\hfil&\cr
Local & !13.5 $\pm$ *7.1 & !13.1 $\pm$ *6.5 & !14.0 $\pm$ *6.8 & ******$-$0.3 $\pm$ *3.5 & !0.5 $\pm$ *4.6 \cr
Equilateral & !55^* $\pm$ 64^* & !50^* $\pm$ 59^* & !58^* $\pm$ 63^* & ******$-$4.4 $\pm$ 24^* & !3.3 $\pm$ 20^* \cr
Orthogonal &$-$50^* $\pm$ 45^* &$-$53^* $\pm$ 46^* &$-$52^* $\pm$ 45^* & ******$-$3.5 $\pm$ 16^* &$-$1.9 $\pm$ 15^* \cr
\noalign{\vskip 5pt}
%\hline
\noalign{\vskip 5pt}
\omit\hfil {\bf Joint}\hfil&\cr
Local & !11.7 $\pm$ *8.3 & !11.4 $\pm$ *7.9 & !12.2 $\pm$ *8.4 & ******$-$0.3 $\pm$ *4.3 & !0.4 $\pm$ *5.7 \cr
Equilateral & !23^* $\pm$ 66^* & !19^* $\pm$ 59^* & !24^* $\pm$ 64^* & ******$-$3.8 $\pm$ 28^* & !1.7 $\pm$ 25^* \cr
Orthogonal &$-$18^* $\pm$ 51^* &$-$20^* $\pm$ 54^* &$-$18^* $\pm$ 55^* & ******$-$1.3 $\pm$ 20^* & !0.3 $\pm$ 20^* \cr
\noalign{\vskip 3pt\hrule\vskip 4pt}}}
%\endPlancktable                    % ends one-column \halign
\endPlancktablewide                 % ends two-column \halign
\endgroup
\end{table*}                        % table* is a two-column table.  Drop the * for one column.
%%%%%%%%%%%%%%%%%%%%%%%%%%%%%%%%%%%%%%%%%%%%%%%%%%%%

A set of 96 full-sky non-Gaussian CMB simulations was created
according to the process described by \cite{2010PhRvD..82b3502F}, 
with local
$f_\mathrm{NL}^\mathrm{local}=12$, equilateral $f_\mathrm{NL}^\mathrm{equil}=35$,
and orthogonal $f_\mathrm{NL}^\mathrm{ortho}=-22$. The effect of a 5 arcmin beam 
was added, as well as realistic coloured and anisotropic noise according to the specifications
of the \SMICA\ cleaned map. The independent Fisher matrix error 
bars in that case are $5.3$ for local, $63$ for
equilateral, and $33$ for orthogonal NG, while the joint ones are respectively
$6.0$, $64$, and $37$.

The results averaged over the whole set are given in 
Table~\ref{Tab_valid_nomask} for the various estimators individually, 
as well as for the differences with respect to KSW. 
Compared to the previous case we now deviate from the exact
theoretical expectation for two reasons: we include a realistic
correlated noise component; and we have NG in the
maps. The presence of NG in the input maps will lower the agreement
between estimators with respect to the Gaussian case if the
correlation between weights is not exactly 100\%. This is even more
true in this specific case, where NG of three different kinds is
present in the input maps and also cross-correlation terms between
different expanded shapes are involved (and propagated over in the
joint analysis). Moreover, when noise is included the specific modal
expansion used for this test is 95\% correlated to the separable KSW
local shape (so there is a 3\%  reduction of the correlation compared to
the ideal case for the modal local shape); we thus expect a further
degradation of the level of agreement for this specific case. Finally,
in order to correct for noise effects, a linear term has to be added
to the estimators. Since the linear term is obtained by MC
averaging over just 80 or 96 simulations in this test (depending on the
estimator), MC errors are also adding to the measured differences. 
Of course the MC error can be reduced by increasing the number 
of simulations in the linear term sample. We do this for the analysis of 
the real data and in Sect.~\ref{Sec_valid_compsep}, but it was computationally too expensive for this set 
 of preliminary validation tests, so we decided here to just account for it in the final 
interpretation of the results.

As a consequence of the above, we can no
longer expect the map-by-map $f_\mathrm{NL}$ differences to  follow perfectly
the theoretical expectation, obtained in the previous Section in
idealized conditions (full-sky, no noise, and Gaussianity). With these
caveats in mind, the agreement between different pipelines remains very
good, being about $0.3\,\sigma$ in most cases and about $0.5\,\sigma$ for the
modal-KSW difference in the local case, which can be easily explained
by the fact that this is the set of weights with the lowest
correlation (95\%, as mentioned above). 
All estimators are unbiased and recover the correct input values.
%

%%%%%%%%%%%%%%%%%%%%%%%%%%%%%%%%%%%%%%%%%%%%%%%%%%%%
\begin{table*}[tmb] % table* is a two-column table. 
\begingroup 
\newdimen\tblskip \tblskip=5pt 
\caption{Results from the different independent estimators for $f_\mathrm{NL}$ for 99 maps 
from a set of realistic lensed 
simulations passed through the \SMICA\ pipeline, described in 
Sect.~\ref{Sec_valid_compsep}.
Both the results for the estimators individually and for the differences with
KSW are given.}
\label{Tab_valid_lensed} 
\nointerlineskip 
\vskip -6mm 
\footnotesize 
\setbox\tablebox=\vbox{ 
   \newdimen\digitwidth 
   \setbox0=\hbox{\rm 0} 
   \digitwidth=\wd0 
   \catcode`*=\active 
   \def*{\kern\digitwidth} 
   \newdimen\signwidth 
   \setbox0=\hbox{+} 
   \signwidth=\wd0 
   \catcode`!=\active 
   \def!{\kern\signwidth} 
 \newdimen\dotwidth 
   \setbox0=\hbox{.} 
   \dotwidth=\wd0 
   \catcode`^=\active 
   \def^{\kern\dotwidth}
\halign{\hbox to 0.8in{#\leaderfil}\tabskip 1em& 
\hfil#\hfil\tabskip 1em& 
\hfil#\hfil& 
\hfil#\hfil& 
\hfil#\hfil& 
%\hbox to 0.4in{#\leaderfil}&
\hfil#\hfil& 
\hfil#\hfil& 
\hfil#\hfil\tabskip 0pt\cr 
\noalign{\doubleline\vskip 2pt} 
\omit&\multispan7\hfil $f_{\rm NL}$\hfil\cr
\omit&\multispan7\hrulefill\cr
\omit&KSW&Binned&Modal&Wavelet&****Binned $-$ KSW&Modal $-$ KSW&Wavelet $-$ KSW\cr 
Parameter \hfill&&&\cr 
\noalign{\vskip 4pt\hrule\vskip 6pt} 
Local & !*7.6 $\pm$ *6.0 & !*6.8 $\pm$ *5.8 & !*7.7 $\pm$ *5.9 & *!8.1 $\pm$ *8.4 &
****$-$0.8 $\pm$ *1.2 & !0.1 $\pm$ *1.4 & !0.5 $\pm$ *6.4 \cr 
Equilateral & !*4^* $\pm$ 76^* &*$-$1^* $\pm$ 72^* & !*2^* $\pm$ 76^* &*$-$3^* $\pm$ 76^* &
****$-$5^* $\pm$ 20^* & $-$2^* $\pm$ 13^* &$-$7^* $\pm$ 91^* \cr 
Orthogonal &$-$21^* $\pm$ 42^* &$-$20^* $\pm$ 41^* &$-$21^* $\pm$ 42^* &$-$15^* $\pm$ 53^* & 
****!1.6 $\pm$ 11^* &$-$0.1 $\pm$ *8^* & !6.4 $\pm$ 48^* \cr 
\noalign{\vskip 3pt\hrule\vskip 4pt}}} 
%\endPlancktable                    % ends one-column \halign 
\endPlancktablewide                 % ends two-column \halign 
\endgroup 
\end{table*} % table* is a two-column table. 
%%%%%%%%%%%%%%%%%%%%%%%%%%%%%%%%%%%%%%%%%%%%%%%%%%%%

%%%%%%%%%%%%%%%%%%%%%%%%%%%%%%%%%%%%%%%%%%%%%%%%%%%%

\subsubsection{Impact of the mask}
\label{Sec_valid_mask}

To the simulations of Sect.~\ref{Sec_valid_nomask} we now apply the
{\it Planck} union mask - denoted U73 - masking both the Galaxy and the brightest 
point sources and leaving 73\% of the sky unmasked~\citep{planck2013-p06}.  
This is 
the same mask used to analyse \Planck\ data in Sect.~\ref{sec:Results}.
  The independent 
Fisher matrix error bars in that case 
(taking into account the $f_\mathrm{sky}$ correction)
are $6.2$ for local NG, $74$ for equilateral, and $39$ for orthogonal, while 
the joint ones are respectively $7.1$, $76$, and $44$.

All masked areas of the sky (both Galactic and point sources) are filled 
in with a simple iterative method. In this simple inpainting method each 
pixel in the mask is filled with the average of all eight surrounding pixels, 
and this is repeated 2000 times over all masked pixels. 
The filling-in helps to avoid propagating the effect of a sharp edge and 
the lack of large-scale power inside the mask to the unmasked regions during 
harmonic transforms. This inpainting method is the one that was used
to produce all NG results in this paper for methods that need it (KSW,
binned and modal). 

\begin{figure}[!t]
\includegraphics[width=\hsize]{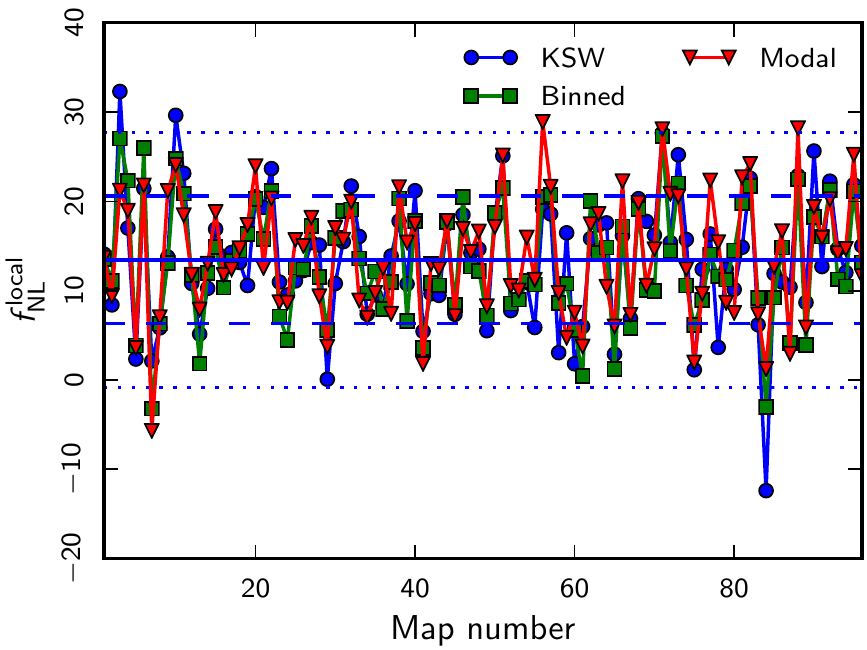}
\includegraphics[width=\hsize]{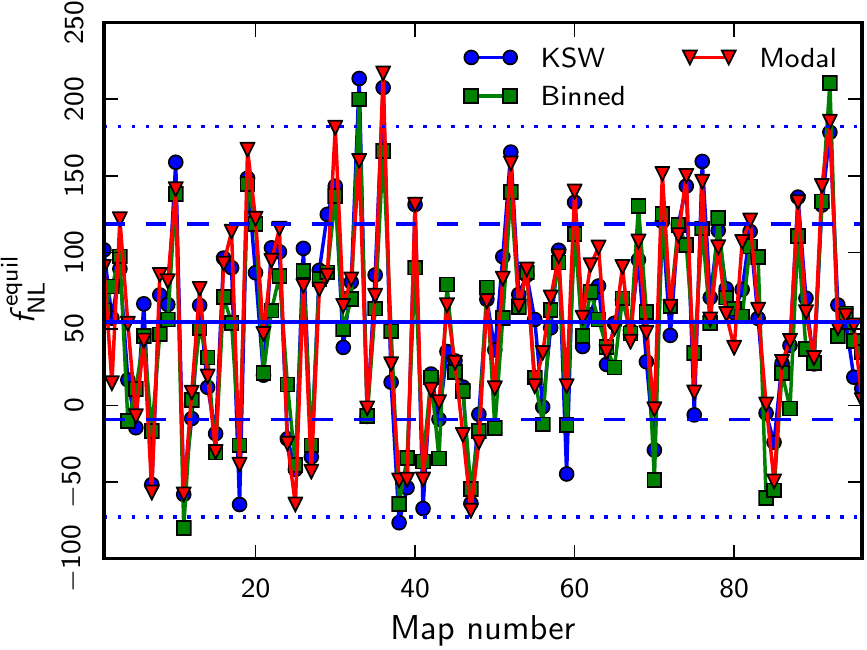}
\includegraphics[width=\hsize]{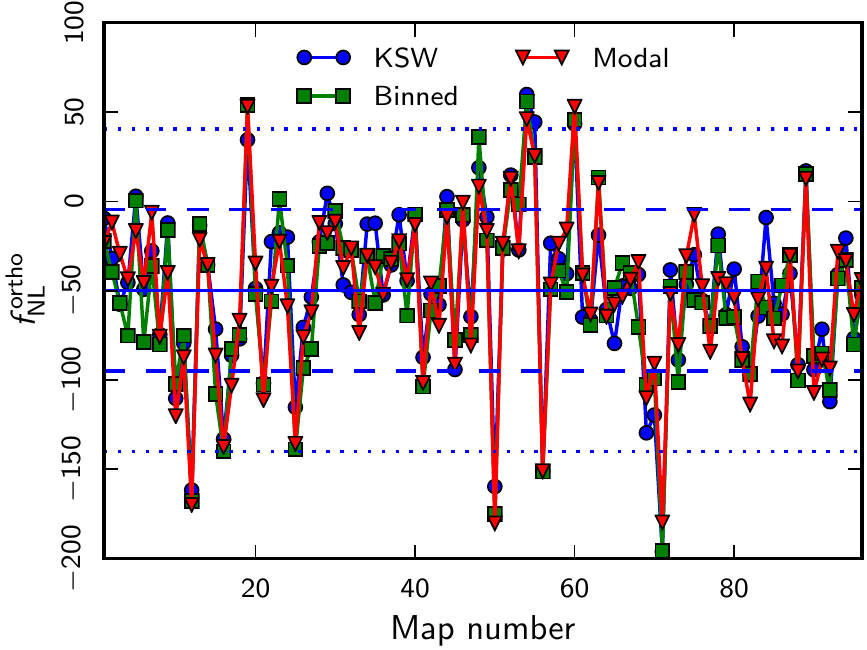}
\caption{Map-by-map comparison of the results from the different estimators 
for local (top), equilateral
(centre), and orthogonal (bottom) $f_\mathrm{NL}$ for the set of masked 
non-Gaussian simulations described in Sect.~\ref{Sec_valid_mask}, 
assuming the shapes to be independent.
The horizontal solid line is the average value of all maps for KSW, and the
dashed and dotted horizontal lines correspond to $1\sigma$ and $2\sigma$
deviations, respectively.}
\label{Fig_valid_mask}
\end{figure}

The results averaged over the whole set of simulations are given in 
Table~\ref{Tab_valid_mask} for the various estimators individually, 
as well as for the differences with respect to KSW. The
map-by-map results are shown in Fig.~\ref{Fig_valid_mask}.

This is the most realistic case we consider in this set of tests. Besides
noise, we also include a sky cut and our usual mask inpainting
procedure. All the caveats mentioned for the previous case
are still valid, and possibly emphasized by the
inclusion of mask and inpainting. In the light of this, the agreement is
still very good, worsening a bit with respect to the ``full-sky~+~noise"
case only for the local measurement, where the mask is indeed
expected to have the biggest impact.
In the joint analysis all estimators recover the correct input values
for the local and orthogonal cases, but all estimators find a value for equilateral NG
that is somewhat too low. It is unclear whether this is an effect of masking and inpainting
 on the equilateral measurement or just a statistical fluctuation for this set of simulations.  
 In any case, this potential bias is small compared to 
the statistical uncertainty, so that it would not have a significant impact on the final 
results.

To summarize the results of this Sect.~\ref{sec:Sec_valid_est}, 
we performed an extensive set of validation tests
between different $f_\mathrm{NL}$ estimators using strongly, but not
perfectly, correlated primordial NG templates in their weights. The
test consisted in comparing the $f_\mathrm{NL}$ measured by the
different estimators for different sets of simulations, on a map-by-map
basis. We started from ideal conditions: full-sky Gaussian noiseless
maps. In this case we computed a theoretical formula providing the
expected standard deviation of the $f_\mathrm{NL}$ differences, as a
function of the correlations between the input NG templates in the
different estimators. Our results match this formula very well. In the
other two simulation sets we added realistic features (noise, mask and
inpainting) and we included a linear combination of local, equilateral
and orthogonal NG. First of all we verified that all the pipelines correctly 
 recover the three $\fnl$ input values, hence they are unbiased. Moreover, we 
  observed that adding such features produces an expected slight degradation of the
level of agreement between different pipelines, that nevertheless
remains very good: about $0.3$--$0.4\,\sigma$ for equilateral and orthogonal NG,
and about $0.5$--$0.6\,\sigma$ for local NG, which is the shape most affected by
mask and noise contamination.

%%%%%%%%%%%%%%%%%%%%%%%%%%%%%%%%%%%%%%%%%%%%%%%%%%%%

\subsection{Validation of estimators on realistic {\it Planck} simulations}
\label{Sec_valid_compsep}

%%%%%%%%%%%%%%%%%%%%%%%%%%%%%%%%%%%%%%%%%%%%%%%%%%%%
%%%%%%%%%%%%%%%%%%%%%%%%%%%%%%%%%%%%%%%%%%%%%%%%%%%%

%%%%%%%%%%%%%%%%%%%%%%%%%%%%%%%%%%%%%%%%%%%%%%%%%%%%
\begin{figure}[!t] 
\includegraphics[width=\hsize]{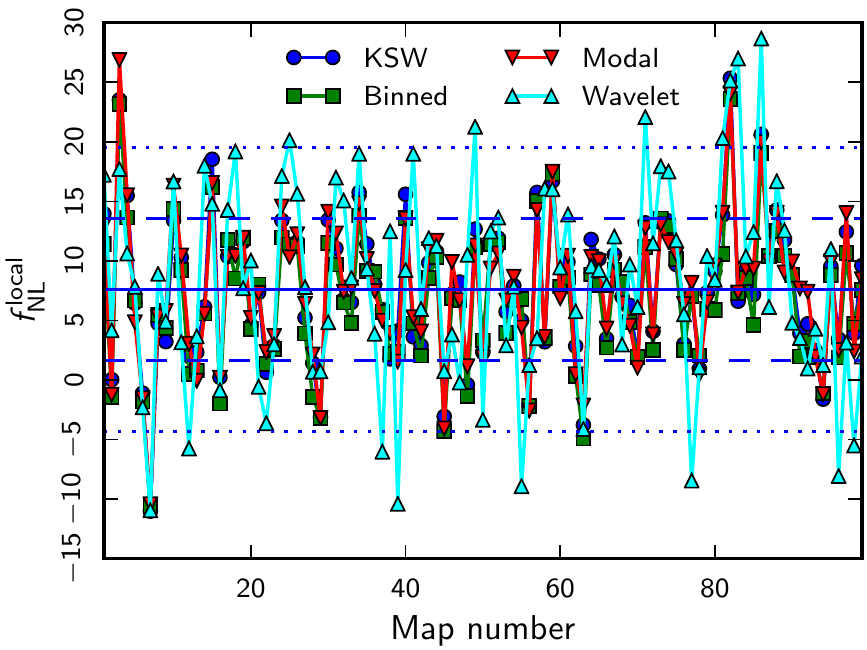} 
\includegraphics[width=\hsize]{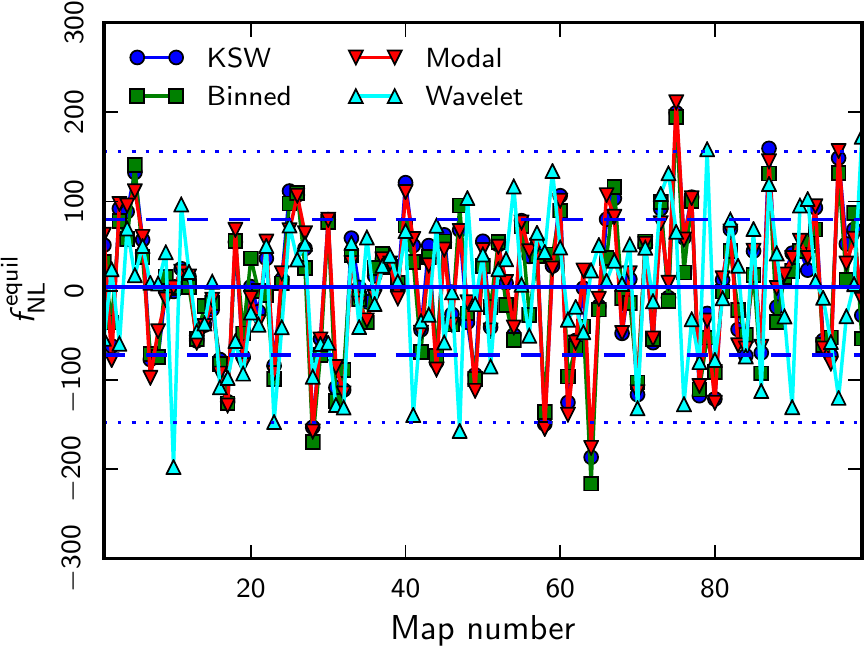} 
\includegraphics[width=\hsize]{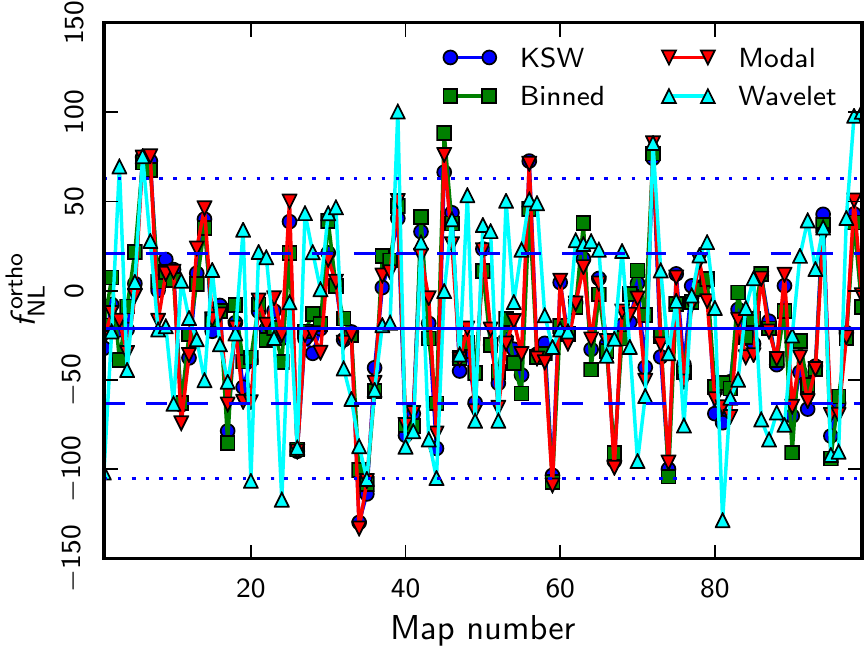} 
\caption{Map-by-map comparison of the results from the different estimators
for local (top), equilateral 
(centre), and orthogonal (bottom) $f_\mathrm{NL}$ for 99 maps from a set of 
realistic lensed simulations passed through the \SMICA\ pipeline, described 
in Sect.~\ref{Sec_valid_compsep}, assuming the shapes to be independent. 
The horizontal solid line is the average value of the maps for KSW, and the 
dashed and dotted horizontal lines correspond to $1\sigma$ and $2\sigma$ 
deviations, respectively. }
\label{Fig_valid_lensed} 
\end{figure} 
%%%%%%%%%%%%%%%%%%%%%%%%%%%%%%%%%%%%%%%%%%%%%%%%%%%%

In the tests of the previous Subsection we checked the bias of the estimators and studied 
their level of agreement, given the correlation between their 
weights, in the presence of noise and a sky cut. To speed 
 up the computation while still retaining enough accuracy for the purposes of that analysis, we used a relatively small number of maps for linear term calibrations (80--100) and used a smaller number of modes than usual in the modal estimator.
In the present Subsection we instead try to simulate as accurately as possible real data analysis conditions. Our goal is to obtain an accurate 
 MC-based expectation of the scatter between different $f_\mathrm{NL}$ measurements when the pipelines are run on actual {\it Planck} 
 maps. 
 
 To this aim we use FFP6 simulation maps described in Appendix~\ref{sec:FFP6}. 
 The original FFP6 maps were lensed using the {\tt Lenspix} algorithm, and processed through the \SMICA\ component separation pipeline. They were then multiplied by the Galactic and point source mask U73 as in the actual $\fnl$ analysis, and inpainted as usual. Since our final results show full consistency with Gaussianity for local, equilateral and orthogonal shapes, we do not include any primordial $\fnl$ in these 
  maps. We note that although the simulations were passed through \SMICA\ in order to provide a realistic filtering of the data, they did not include any foreground components. The impact of foreground residuals will be studied separately in Sect.~\ref{sec:Valid_FGresid}.

The configuration of all bispectrum pipelines was the same as used for the final data analysis, which implies a correlation of $99\%$ or better between the weights of the KSW, binned and modal estimators. Linear terms were calibrated using $200$ simulations, after verifying that this number allows accurate convergence for all the shapes. For this test we also included the wavelet bispectrum pipeline described in Sect.~\ref{sec:se}. Although this last estimator turns out to be about $30\%$ suboptimal and, in its current implementation, less correlated with the primordial templates than the other estimators, it does provide an additional interesting cross-check of our results by introducing another decomposition basis. We thus used it to analyse \SMICA\ data in Sect.~\ref{sec:Results}, while the other three pipelines were used on all maps. 

A comparison of the measured $\fnl$ map-by-map for all shapes and estimators is shown in Fig.~\ref{Fig_valid_lensed}. As an overall figure of merit of the level of agreement achieved by different pipelines we take as usual the standard deviation of the map-by-map $\fnl$ differences, $\sigma_{\delta_{\fnl}}$. Table~\ref{Tab_valid_lensed} shows that the final agreement between the three optimal pipelines (KSW, binned, and modal) is close to saturating the ideal bound in Eq.~(\ref{eqn:expectation}) determined by the imperfect correlation of the weights, i.e., it varies from about once to twice $\sigma_{\delta_{\fnl}} \simeq 0.15 \,\Delta_{\fnl}$ for an $r=0.99$ correlation. This is very consistent with the level of agreement that we find between estimators for the final results from the data, providing a good indication that no spurious NG features are present in the actual data set when compared to our simulations. 
It should be noted that we found a similarly good level of agreement between
estimators for the non-primordial shapes of point sources and ISW-lensing,
although we chose not to present those results here in order to focus on
the primordial shapes.
Finally, regarding the wavelet pipeline, the lower weight correlation and suboptimal error bars produce an expected larger scatter when compared to the other estimators. Nonetheless, the level of agreement is still of order $1\,\sigma$, which is quite acceptable for consistency checks of the optimal results. Again, this MC expectation agrees with what we see in our results on the real data.

%%%%%%%%%%%%%%% Section 7 %%%%%%%%%%%%%%%
\section{Results}
\label{sec:Results}

For our analysis of {\it Planck} data we considered foreground-cleaned maps obtained with the four component separation methods
\SMICA, \NILC, \SEVEM, and \CR. For each map, $f_{\rm NL}$ amplitudes for the
local, equilateral, and orthogonal primordial shapes have been measured using three (four for \SMICA) bispectrum estimators described in Sect.~\ref{sec:se}. The results can be found in Sect.~\ref{fnl_loc_eq_ort_results}. These estimators, as explained earlier, basically use an expansion of the theoretical bispectrum templates in different domains, and truncate the expansion when a high level of correlation with the primordial templates is achieved.
These accurate decompositions, which are highly correlated with each other, are then matched to the data in order to extract $f_{\rm NL}$.
The different expansions are all different implementations of the maximum-likelihood estimator given in Eq.~(\ref{eq:optimalestimator}). So the final estimates are all expected to be optimal, and measure $\fnl$ from nearly identical fitting templates. As discussed and tested in detail on simulations in Sect.~\ref{sec:Validation}, central $f_{\rm NL}$ values from different methods are expected to be consistent with each other within about $0.3 \sigma_{f_{\rm NL}}$.  It is then clear that comparing outputs from both different estimators and different component separation methods, as we do, allows for stringent internal consistency checks and improved robustness of the final $f_{\rm NL}$ results.

In addition, the binned and modal techniques produce shape-independent full bispectrum reconstructions in their own different domains. These reconstructions, discussed in Sect.~\ref{Sec_bisprec}, complement the standard $f_{\rm NL}$ measurements in an important way, since they allow detection of possible NG features in the three-point function of the data that do not correlate significantly with the standard primordial shapes. This advantage is shared by the skew-$C_\ell$ method, also applied to the data. A detection of such features would either produce a warning that some residual spurious NG effects are still present in the data or provide an interesting hint of  ``non-standard'' primordial NG that is not captured by the local, equilateral and orthogonal shapes. Additional constraints for a broad range of specific models are provided in Sect.~\ref{STS} (see
also Sect.~\ref{New}). In Sect.~\ref{MF} we present the constraints on local NG obtained with  Minkowski Functionals. Finally, in Sect.~\ref{trispectrum} we present our CMB trispectrum results.

%%%%%%%%%%%%%%%%%%%%%%%%%%%%%%%%%%%%%%%%%%%%%%%%%%%%

\subsection{Constraints on local, equilateral and orthogonal $f_{\rm NL}$}
\label{fnl_loc_eq_ort_results}
%

%%%%%%%%%%%%%%%%%%%%%%%%%%%%%%%%%%%%%%%%%%%%%%%%%%%%
\begin{table}[tmb]                 % table* is a two-column table.  Drop the * for one column.
\begingroup
\newdimen\tblskip \tblskip=5pt
\caption{Results for the $f_{\rm NL}$ parameters of the primordial local, equilateral, and
orthogonal shapes, determined by the KSW estimator from the
\SMICA\ foreground-cleaned map. Both independent
single-shape results and results marginalized over the point source
bispectrum and with the ISW-lensing bias subtracted are reported; error bars
are $68\%$ CL\,.  The final reported results of the paper are shown in bold face.}
\label{Tab_KSW+SMICA}
\nointerlineskip
\vskip -6mm
\footnotesize
\setbox\tablebox=\vbox{
   \newdimen\digitwidth
   \setbox0=\hbox{\rm 0}
   \digitwidth=\wd0
   \catcode`*=\active
   \def*{\kern\digitwidth}
   \newdimen\signwidth
   \setbox0=\hbox{+}
   \signwidth=\wd0
   \catcode`!=\active
   \def!{\kern\signwidth}
\newdimen\dotwidth
\setbox0=\hbox{.}
\dotwidth=\wd0
\catcode`^=\active
\def^{\kern\dotwidth}
\halign{\hbox to 1in{#\leaderfil}\tabskip 1em&
\hfil#\hfil\tabskip 1em&
\hfil#\hfil\tabskip 0pt\cr
\noalign{\vskip 10pt\doubleline\vskip 2pt}
\omit&\multispan2\hfil $f_{\rm NL}$(KSW)\hfil\cr
\omit&\multispan2\hrulefill\cr
Shape \& Method\hfill&\hfil Independent\hfil&
\hfil ISW-lensing subtracted\hfil\cr
\noalign{\vskip 2pt}
\noalign{\vskip 4pt\hrule\vskip 6pt}
\omit\hfil \SMICA\hfil&\cr
Local & *!9.8 $\pm$ *5.8 &
{\bf *!2.7 $\pm$ *5.8}\cr
Equilateral & $-$37^* $\pm$ 75^* &
{\bf $-$42^* $\pm$ 75^*}\cr
Orthogonal & $-$46^* $\pm$ 39^* &
{\bf $-$25^* $\pm$ 39^*}\cr
\noalign{\vskip 3pt\hrule\vskip 4pt}}}
\endPlancktable                    % ends one-column \halign
%\endPlancktablewide                 % ends two-column \halign
\endgroup
\end{table}                        % table* is a two-column table.  Drop the * for one column.
%%%%%%%%%%%%%%%%%%%%%%%%%%%%%%%%%%%%%%%%%%%%%%%%%%%%

%%%%%%%%%%%%%%%%%%%%%%%%%%%%%%%%%%%%%%%%%%%%%%%%%%%%
\begin{table*}[tmb]                 % table* is a two-column table.  Drop the * for one column.
\begingroup
\newdimen\tblskip \tblskip=5pt
\caption{Results for the $f_{\rm NL}$ parameters of the primordial local, equilateral, and
orthogonal shapes, determined by the KSW, binned and modal estimators from the
\SMICA, \NILC, \SEVEM, and \CR\ foreground-cleaned maps. Both independent
single-shape results and results marginalized over the point source
bispectrum and with the ISW-lensing bias subtracted are reported; error bars
are $68\%$ CL\,.  Final reported results of the paper are shown in bold.}
\label{tab:fNLsmicah}
\nointerlineskip
\vskip -3mm
\footnotesize
\setbox\tablebox=\vbox{
   \newdimen\digitwidth
   \setbox0=\hbox{\rm 0}
   \digitwidth=\wd0
   \catcode`*=\active
   \def*{\kern\digitwidth}
   \newdimen\signwidth
   \setbox0=\hbox{+}
   \signwidth=\wd0
   \catcode`!=\active
   \def!{\kern\signwidth}
\newdimen\dotwidth
\setbox0=\hbox{.}
\dotwidth=\wd0
\catcode`^=\active
\def^{\kern\dotwidth}
\halign{\hbox to 1.5in{#\leaderfil}\tabskip 1.5em&
\hfil#\hfil\tabskip 1.5em&
\hfil#\hfil&
\hfil#\hfil&
%\hbox to 0.4in{#\leaderfil}&
\hfil#\hfil&
\hfil#\hfil&
\hfil#\hfil\tabskip 0pt\cr
\noalign{\doubleline\vskip 2pt}
\omit&\multispan6\hfil $f_{\rm NL}$\hfil\cr
\omit&\multispan6\hrulefill\cr
\omit&\multicolumn{3}{c}{\hfil Independent\hfil}&
\multicolumn{3}{c}{\hfil ******ISW-lensing subtracted\hfil}\cr
\noalign{\vskip 2pt}
Shape\hfill&KSW&Binned&Modal&******KSW&Binned&Modal\cr
\noalign{\vskip 4pt\hrule\vskip 6pt}
\omit\hfil \SMICA\hfil&&\cr
Local & !*9.8 $\pm$ *5.8 & !*9.2 $\pm$ *5.9 & !*8.3 $\pm$ *5.9 &
******{\bf !*2.7 $\pm$ *5.8} & !*2.2 $\pm$ *5.9 & !*1.6 $\pm$ *6.0 \cr
Equilateral & $-$37^* $\pm$ 75^* & $-$20^* $\pm$ 73^* & $-$20^* $\pm$ 77^* &
******{\bf $-$42^* $\pm$ 75^*} & $-$25^* $\pm$ 73^* & $-$20^* $\pm$ 77^* \cr
Orthogonal & $-$46^* $\pm$ 39^* & $-$39^* $\pm$ 41^* & $-$36^* $\pm$ 41^* &
******{\bf $-$25^* $\pm$ 39^*} & $-$17^* $\pm$ 41^* & $-$14^* $\pm$ 42^* \cr
%\noalign{\vskip 5pt}
%\hline
\noalign{\vskip 5pt}
\omit\hfil \NILC\hfil&&\cr
Local & !11.6 $\pm$ *5.8 & !10.5 $\pm$ *5.8 & !*9.4 $\pm$ *5.9 &
******!*4.5 $\pm$ *5.8 & !*3.6 $\pm$ *5.8 & !*2.7 $\pm$ *6.0\cr
Equilateral & $-$41^* $\pm$ 76^* & $-$31^* $\pm$ 73^* & $-$20^* $\pm$ 76^* &
******$-$48^* $\pm$ 76^* & $-$38^* $\pm$ 73^* & $-$20^* $\pm$ 78^*\cr
Orthogonal & $-$74^* $\pm$ 40^* & $-$62^* $\pm$ 41^* & $-$60^* $\pm$ 40^* &
******$-$53^* $\pm$ 40^* & $-$41^* $\pm$ 41^* & $-$37^* $\pm$ 43^*\cr
%\noalign{\vskip 5pt}
%\hline
\noalign{\vskip 5pt}
\omit\hfil \SEVEM\hfil&&\cr
Local & !10.5 $\pm$ *5.9 & !10.1 $\pm$ *6.2 & !*9.4 $\pm$ *6.0 &
******!*3.4 $\pm$ *5.9 & !*3.2 $\pm$ *6.2 & !*2.6 $\pm$ *6.0\cr
Equilateral & $-$32^* $\pm$ 76^* & $-$21*^ $\pm$ 73^* & $-$13^* $\pm$ 77^* &
******$-$36^* $\pm$ 76^* & $-$25^* $\pm$ 73^* & $-$13^* $\pm$ 78^*\cr
Orthogonal & $-$34^* $\pm$ 40^* & $-$30*^ $\pm$ 42^* & $-$24^* $\pm$ 42^* &
******$-$14^* $\pm$ 40^* & *$-$9^* $\pm$ 42^* & *$-$2^* $\pm$ 42^*\cr
%\noalign{\vskip 5pt}
%\hline
\noalign{\vskip 5pt}
\omit\hfil \CR\hfil&&\cr
Local & !12.4 $\pm$ *6.0 & !11.3 $\pm$ *5.9 & !10.9 $\pm$ *5.9 &
******!*6.4 $\pm$ *6.0 & !*5.5 $\pm$ *5.9 & !*5.1 $\pm$ *5.9\cr
Equilateral & $-$60^* $\pm$ 79^* & $-$52^* $\pm$ 74^* & $-$33^* $\pm$ 78^* &
******$-$62^* $\pm$ 79^* & $-$55^* $\pm$ 74^* & $-$32^* $\pm$ 78^*\cr
Orthogonal & $-$76^* $\pm$ 42^* & $-$60^* $\pm$ 42^* & $-$63^* $\pm$ 42^* &
******$-$57^* $\pm$ 42^* & $-$41^* $\pm$ 42^* & $-$42^* $\pm$ 42^*\cr
\noalign{\vskip 3pt\hrule\vskip 4pt}}}
%\endPlancktable                    % ends one-column \halign
\endPlancktablewide                 % ends two-column \halign
\endgroup
\end{table*}                        % table* is a two-column table.  Drop the * for one column.
%%%%%%%%%%%%%%%%%%%%%%%%%%%%%%%%%%%%%%%%%%%%%%%%%%%%

%%%%%%%%%%%%%%%%%%%%%%%%%%%%%%%%%%%%%%%%%%%%%%%%%%%%
\begin{table}[tmb]                 % table* is a two-column table.  Drop the * for one column.
\begingroup
\newdimen\tblskip \tblskip=5pt
\caption{Results for the $f_{\rm NL}$ parameters of the primordial local, equilateral, and
orthogonal shapes, determined by the suboptimal wavelet estimator from the
\SMICA\ foreground-cleaned map. Both independent
single-shape results and results marginalized over the point source
bispectrum and with the ISW-lensing bias subtracted are reported; error bars
are $68\%$ CL. As explained in the text, our current wavelets pipeline
performs slightly worse in terms of error bars and correlation to primordial
templates than the other bispectrum estimators,
but it still provides a useful independent cross-check of other techniques.}
\label{Tab_wavelets}
\nointerlineskip
\vskip -6mm
\footnotesize
\setbox\tablebox=\vbox{
   \newdimen\digitwidth
   \setbox0=\hbox{\rm 0}
   \digitwidth=\wd0
   \catcode`*=\active
   \def*{\kern\digitwidth}
   \newdimen\signwidth
   \setbox0=\hbox{+}
   \signwidth=\wd0
   \catcode`!=\active
   \def!{\kern\signwidth}
\newdimen\dotwidth
\setbox0=\hbox{.}
\dotwidth=\wd0
\catcode`^=\active
\def^{\kern\dotwidth}
\halign{\hbox to 1in{#\leaderfil}\tabskip 1em&
\hfil#\hfil\tabskip 1em&
\hfil#\hfil\tabskip 0pt\cr
\noalign{\vskip 10pt\doubleline\vskip 2pt}
\omit&\multispan2\hfil $f_{\rm NL}$(wavelets)\hfil\cr
\omit&\multispan2\hrulefill\cr
Shape\hfill&\hfil Independent\hfil&
\hfil ISW-lensing subtracted\hfil\cr
\noalign{\vskip 4pt\hrule\vskip 6pt}
\omit\hfil \SMICA\hfil&\cr
Local & !10 $\pm$ *8.5 & !*0.9 $\pm$ *8.5\cr
Equilateral & !89 $\pm$ 84^* & !90^* $\pm$ 84^*\cr
Orthogonal & $-$73 $\pm$ 52^* & $-$45^* $\pm$ 52^*\cr
\noalign{\vskip 3pt\hrule\vskip 4pt}}}
\endPlancktable                    % ends one-column \halign
%\endPlancktablewide                 % ends two-column \halign
\endgroup
\end{table}                        % table* is a two-column table.  Drop the * for one column.
%%%%%%%%%%%%%%%%%%%%%%%%%%%%%%%%%%%%%%%%%%%%%%%%%%%%

Our goal here is to investigate the standard separable local, equilateral and orthogonal templates used e.g., in previous {\it WMAP} analyses (see e.g., \citealt{2012arXiv1212.5225B}).
When using the modal, binned, or wavelet estimator, these theoretical templates are expanded approximately (albeit very accurately) using the relevant basis functions or bins.  On the other hand, the KSW estimator by construction works with the {\em exact} templates  and, for this reason, it is chosen as the baseline to provide the final $\fnl$ results for the standard shapes (local, equilateral, orthogonal), see Table~\ref{Tab_KSW+SMICA}.   However, both the binned and modal estimators achieve optimal performance and an extremely high correlation for the standard templates ($\sim 99 \%$), so they are statistically equivalent to KSW, as demonstrated in the previous section.   This means that we can achieve a remarkable level of cross-validation for our  \Planck\ NG results.   We will be able to present consistent constraints for the local, equilateral and orthogonal models for all four \Planck\ foreground-cleaned maps, using three independent optimal estimators (refer to Table~\ref{tab:fNLsmicah}).   Regarding component separation methods, we
adopt the \SMICA\ map as the default for the final KSW results given its preferred status among foreground-separation techniques in \cite{planck2013-p06}. The other
component separation maps will be used for important cross-validation
of our results and to evaluate potential sensitivity to foreground
residuals.

All the results presented in this Section were obtained using the union mask U73, which leaves 73\% of the sky unmasked. The mask is the union
of the confidence masks of the four different
component separation methods, where each confidence mask defines the region where the corresponding
CMB cleaning is trusted (see \citealt{planck2013-p06}).
As will be shown in Sect.~\ref{sec:maskdep}, results are robust
to changes that make the mask larger, but choosing a significantly smaller mask
would leave some NG foreground contamination.
For the linear term CMB and noise calibration, and error bar determination, we used sets of realistic FFP6 maps that include all steps of data processing, and have realistic noise and beam properties (see Appendix~\ref{sec:FFP6}). The simulations were also lensed using the {\tt Lenspix} algorithm and filtered  through the component separation pipelines.

In Table~\ref{Tab_KSW+SMICA} we show results for the combination of the KSW estimator and the \SMICA\ map, at a resolution of
$\ell_{\rm max} = 2500$.
We present both ``independent'' single-shape results and ``ISW-lensing
subtracted'' ones. The former are obtained by directly fitting primordial templates to the data.
For the latter, two additional operations have been performed. In the first
place, as the name indicates, they have been corrected by subtracting the bias
due to the correlation of the primordial bispectra to the
late-time ISW-lensing contribution
(\citealt{2009PhRvD..80l3007M,2012arXiv1204.3789J,2009PhRvD..80h3004H},
see Sect.~\ref{sec:lensingISW}).
In addition, a joint fit of the primordial shape with the (Poissonian) point-source bispectrum amplitude extracted from the data has been performed on the
results marked ``ISW-lensing subtracted''.\footnote{More precisely, in the subtracted ISW-lensing results the equilateral and orthogonal primordial shapes are also fitted jointly, although this has a nearly negligible impact on the final result because the two shapes are by construction nearly perfectly uncorrelated.} Since the ISW-lensing bispectrum is peaked on squeezed configurations, its impact is well known to be largest for the local shape. The ISW-lensing bias is also important for orthogonal measurements (there is a correlation coefficient $r \sim -0.5$ between the local and orthogonal CMB templates), while it is very small in the equilateral limit. The values of the ISW-lensing bias we subtract,
summarized in Table~\ref{t:dx9pred}, are calculated assuming the {\it Planck} best-fit cosmological model as our fiducial model. The same fiducial parameters were of course consistently used to compute the theoretical bispectrum templates and the estimator normalization.  Regarding the point source contamination, we detect a Poissonian bispectrum at high significance in the \SMICA\ map, see Sect.~\ref{sec:psbisp}. However, marginalizing over point sources has negligible impact on the final primordial $\fnl$ results, because the Poisson bispectrum template has very small correlations with all the other shapes.

In light of the discussion at the beginning of this section, we take the numbers from the KSW \SMICA\ analysis in Table~\ref{Tab_KSW+SMICA} as {\em the} final local, equilateral and orthogonal $\fnl$ constraints for the current {\it Planck} data release. These results clearly show that no evidence of NG of the local, equilateral or orthogonal type is found in the data. After ISW-lensing subtraction, all $\fnl$ for the three primordial shapes are consistent with $0$ at $68 \%$ CL.  Note that these numbers have been cross-checked using two completely independent KSW pipelines, one of which is an extension to {\it Planck} resolution of the pipeline used for the {\it WMAP} analysis \citep{2012arXiv1212.5225B}.

\begin{figure}
\includegraphics[width=\hsize]{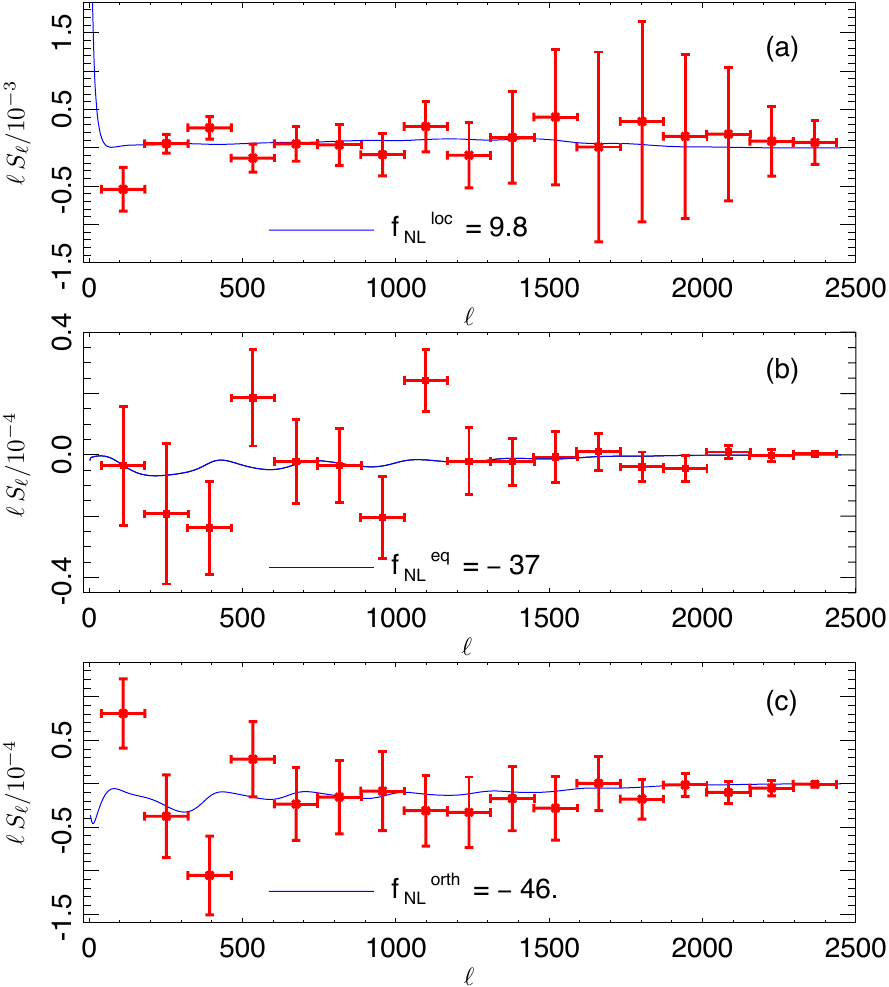}
\caption{Binned skew-$C_\ell$ statistics from the \SMICA\ map for
(a) local, (b) equilateral, and (c) orthogonal.
Theoretical curves are not fitted to the data shown, but are plotted with
the amplitude (the only free parameter) determined from the KSW technique.
There is no evidence for detection of primordial NG. Error bars are derived from the covariance of estimates from 1000 simulations.  There are mild correlations between data points in all figures, but very strong correlations at high $\ell$ in the local case, reaching correlation coefficients $r>0.99$ for $\ell>1750$.} 
\label{fig:skewcllocal}
\end{figure}

Unlike other methods, the KSW technique is not designed to provide a reconstruction of the full bispectrum of the data. However, the related skew-$C_\ell$ statistic described in Sect.~\ref{skewCl} allows, for each given shape, visualization and study of the  contribution to the measured $f_{\rm NL}$ from separate $\ell$-bins. This is a useful tool to study potential spurious NG contamination in the data.
We show for the \SMICA\ map in Fig.~\ref{fig:skewcllocal} the
measured skew-$C_\ell$ spectrum for optimal detection of primordial local,
equilateral and orthogonal NG, along with the best-fitting estimates of
$f_{\rm NL}$ from the KSW method.
Contrary to the case of the point source and ISW-lensing foregrounds (see
Sect.~\ref{sec:npNG}), the skew-$C_\ell$ statistics do not show
convincing evidence for detection of the primordial shapes, suggesting that these primordial effects are not significant sources of the NG in the map.  Again, point sources contribute very little to this
statistic; ISW-lensing contributes, but only a small fraction of
the amplitude.
%%%%%%%%%%%%%%%%%%%%%%%%%%%%%%%%%%%%%%%%%%%%%%%%%%%%
%%%%%%%%%%%%%%%%%%%%%%%%%%%%%%%%%%%%%%%%%%%%%%%%%%%%

As mentioned before, our analysis went beyond the simple application of the KSW estimator to the \SMICA\ map. All $\fnl$ pipelines developed for {\it Planck} analysis were applied to all component-separated maps by \SMICA, \NILC, \SEVEM, and \CR. We found from simulations in the previous Sections that the KSW, binned, and modal pipelines saturate the Cram\' er-Rao bound, while the wavelet estimator in its current implementation provides slightly suboptimal results. Wavelets remain however a useful cross-check of the other methods, also given some technical complementarities, e.g., they are the only approach that does not require inpainting, as explained in Sect.~\ref{sec:se}. Hence we include wavelet results, but only for \SMICA. The $\fnl$ results for the optimal KSW, binned and modal bispectrum estimators, for the four component separation methods, are summarized in Table~\ref{tab:fNLsmicah}, one of the main products of our analysis of {\it Planck} data. The wavelet bispectrum analysis of \SMICA\ is reported in Table~\ref{Tab_wavelets}. In the analysis, the KSW and binned bispectrum estimators considered multipoles up to $\ell_\mathrm{max}=2500$, while the modal estimator went to $\ell_\mathrm{max}=2000$.
As shown in Sect.~\ref{Sec_deplmax} and Table~\ref{Tab_lmaxdep}, error bars and central values for the three primordial shapes have converged at $\ell_{\rm max} = 2000$, so the final primordial $\fnl$ estimates from the three pipelines are directly comparable.\footnote{The lower $\ell_{\rm max}$ for the modal pipeline is also a conservative choice in view of the large survey of ``non-standard" models that will be presented in Sect.~\ref{STS}}

The binned bispectrum estimator used 51 bins, which were determined by optimizing the expected variance of the different $f_\mathrm{NL}$ parameters, focusing in particular on the primordial shapes.\footnote{The boundary values of the bins are: 2, 4, 10, 18, 27, 39, 55, 75, 99, 130, 170, 224, 264, 321, 335, 390, 420, 450, 518, 560, 615, 644, 670, 700, 742, 800, 850, 909, 950, 979, 1005, 1050, 1110, 1150, 1200, 1230, 1260, 1303, 1346, 1400, 1460, 1510, 1550, 1610, 1665, 1725, 1795, 1871, 1955, 2091, 2240, and 2500 (i.e., the first bin is [2,3], the second [4,9], etc., while the last one is [2240,2500]).} The modal estimator employed a polynomial basis ($n_{\rm max}=600$) previously described in \citet{2010PhRvD..82b3502F}, but augmented with a local shape mode (approximating the SW large-angle local solution) to improve convergence in the squeezed limit. The above choices for the binned and modal methods produce a very high correlation (generally $99\%$ or better) of the expanded/binned templates with the exact ones used by the KSW estimator.
The wavelet estimator is based on third-order statistics generated by the different possible combinations of the wavelet coefficient maps
of the SMHW evaluated at certain angular scales. See for example \citet{antoine:1998} and \citet{martinez:2002} for detailed information about
this wavelet. We considered a set of 15 scales logarithmically spaced between 1.3 and 956.3 arcmin and we also included the unconvolved
map. The wavelet map $w(R_i; \vec{b})$ (Eq.~(\ref{waveneedcomp})) for each angular scale $R_i$ has an associated extended mask generated from the mask U73 following the procedure described
and extensively used in~\citealt{curto:2009a,curto:2009b,curto:2011a,curto:2011b,curto:2012,donzelli:2012,regan:2013}. The wavelet coefficient maps are later combined into the third-order
moments $q_{ijk}$ (Eq.~(\ref{waveneedbisp})), for a total 816 different statistics, and these statistics are used to constrain $f_{\rm NL}$ through a $\chi^2$ test.

The high level of agreement between results from the KSW, binned and modal $f_{\rm NL}$ estimators, and from all the component separation pipelines, is representative of the robustness of our results with respect to residual foreground contamination, and is fully consistent with our
preliminary MC analysis shown in Sect.~\ref{sec:Validation}. The scatter with wavelets is a bit larger, but this was expected due to the suboptimality of the wavelet estimator and is also in agreement with our MC expectations from the tests. Therefore wavelets do provide another successful cross-check.

%%%%%%%%%%%%%%%%%%%%%%%%%%%%%%%%%%%%%%%%%%%%%%%%%%%%

\subsection{Bispectrum reconstruction}
\label{Sec_bisprec}

As previously explained (see Sect.~\ref{sec:se}),
in addition to looking in specific bispectrum-space directions and
extracting the single number $\fnl$ for given shapes,
the binned and modal pipelines have the capability to generate a smoothed (i.e., either coarse-grained in $\ell$-space, or truncated at a given expansion eigenmode) reconstruction of the full bispectrum of the data.
See also \cite{planck2013-p09}.

\subsubsection{Modal bispectrum reconstruction}

%%%%%%%%%%%%%%%%%%%%%%%%%%%%%%%%%%%%%%%%%%%%%%%%%%%%
\begin{figure*}
\sidecaption
\includegraphics[width=18cm]{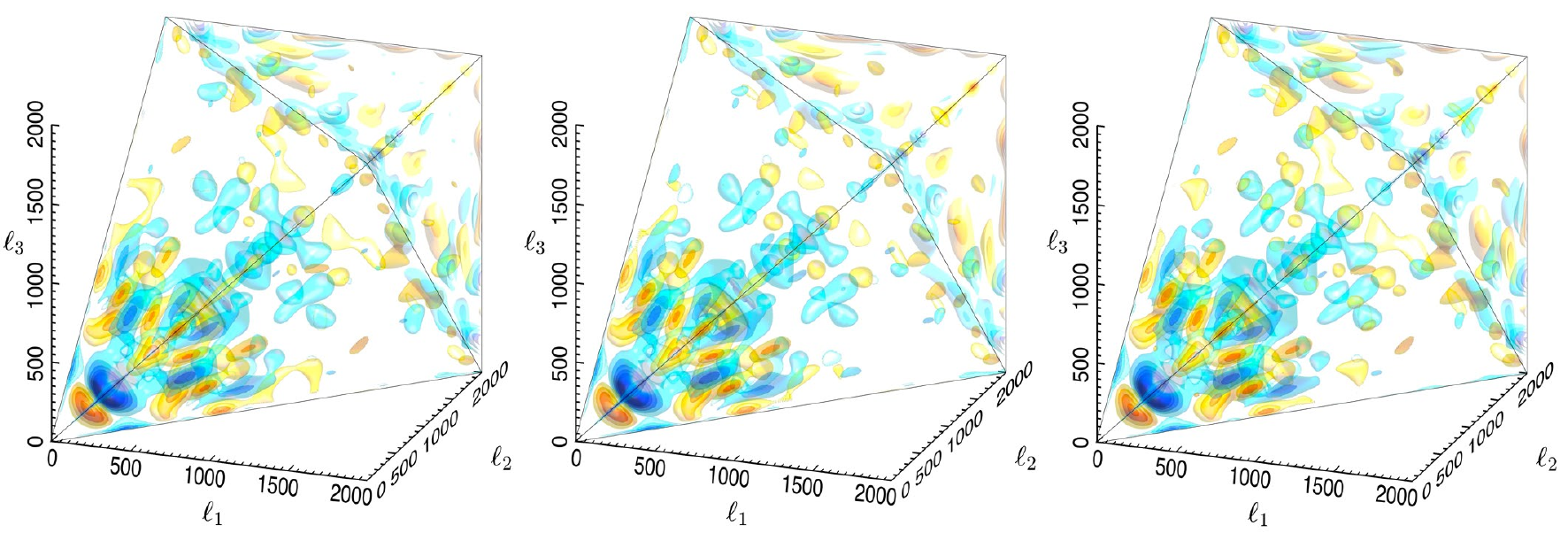}
\caption[Modal reconstruction of the Planck CMB bispectrum]{Full 3D
  CMB bispectrum recovered from the {\it Planck} foreground-cleaned maps, including \SMICA\ (left), \NILC\ (centre) and \SEVEM\ (right), using the hybrid Fourier mode coefficients illustrated in Fig.~\ref{fig:DDX9_modal_beta},    These are plotted in three-dimensions with multipole coordinates $\{\ell_1,\,\ell_2,\,\ell_3\}$ on the tetrahedral domain shown in Fig.~\ref{fig:tetrapyd} out to  $\ell_{\rm max}=2000$.   Several density contours are plotted  with red positive and blue negative. The bispectra extracted from the different foreground-separated maps appear to be almost indistinguishable.}
\label{fig:DDX9_smica_3D}\label{fig:recon}
\end{figure*}
%%%%%%%%%%%%%%%%%%%%%%%%%%%%%%%%%%%%%%%%%%%%%%%%%%%%

%%%%%%%%%%%%%%%%%%%%%%%%%%%%%%%%%%%%%%%%%%%%%%%%%%%%
\begin{figure*}
\includegraphics[width=18cm]{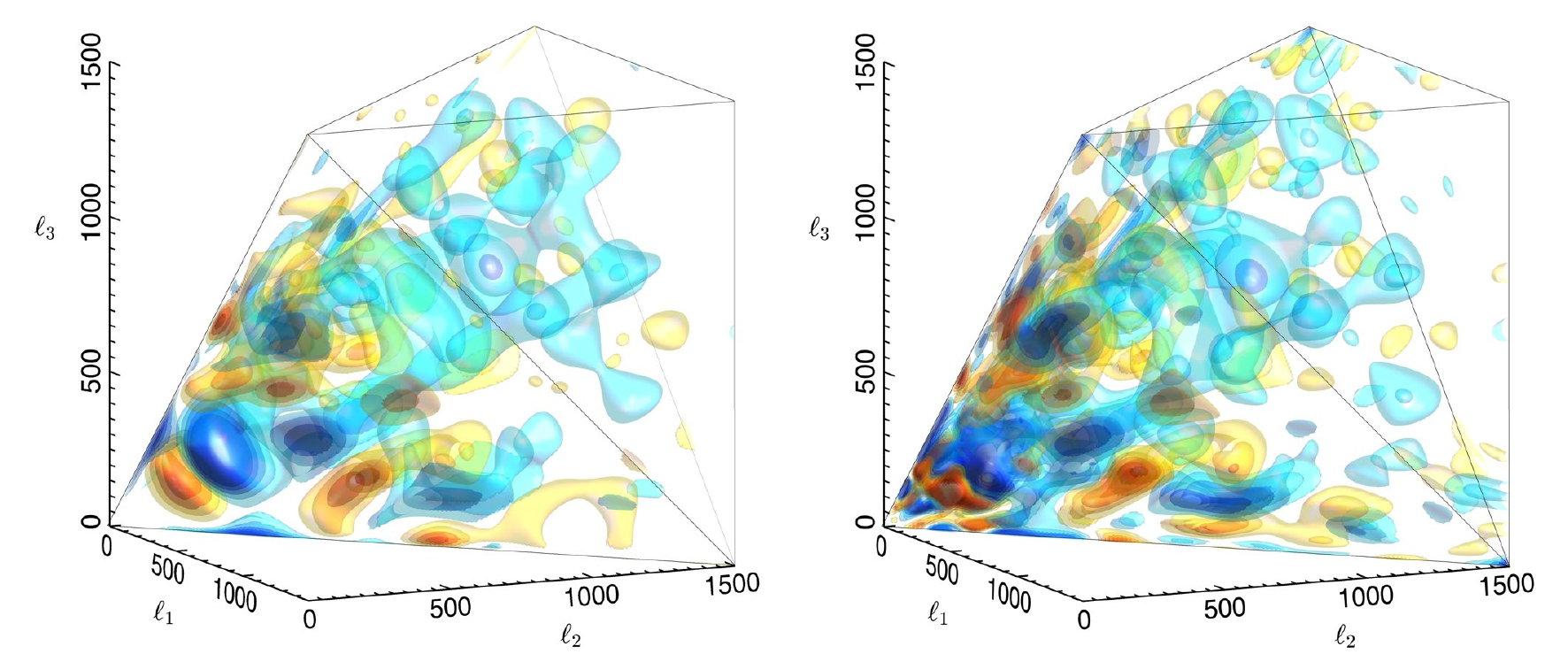}
\caption{\Planck\ CMB bispectrum detail in the signal-dominated regime showing a comparison between full 3D reconstruction using hybrid Fourier modes (left) and hybrid polynomials (right).  Note the consistency of the main bispectrum properties which include an apparently `oscillatory' central feature for low-$\ell$ together with a flattened signal beyond to $\ell\lesssim 1400$.   Note also the periodic CMB ISW-lensing signal in the squeezed limit along the edges of the tetrapyd.}
\label{fig:recondetail}
\end{figure*}
%%%%%%%%%%%%%%%%%%%%%%%%%%%%%%%%%%%%%%%%%%%%%%%%%%%%
%%%%%%%%%%%%%%%%%%%%%%%%%%%%%%%%%%%%%%%%%%%%%%%%%%%%
\begin{figure}
\includegraphics[width=\hsize]{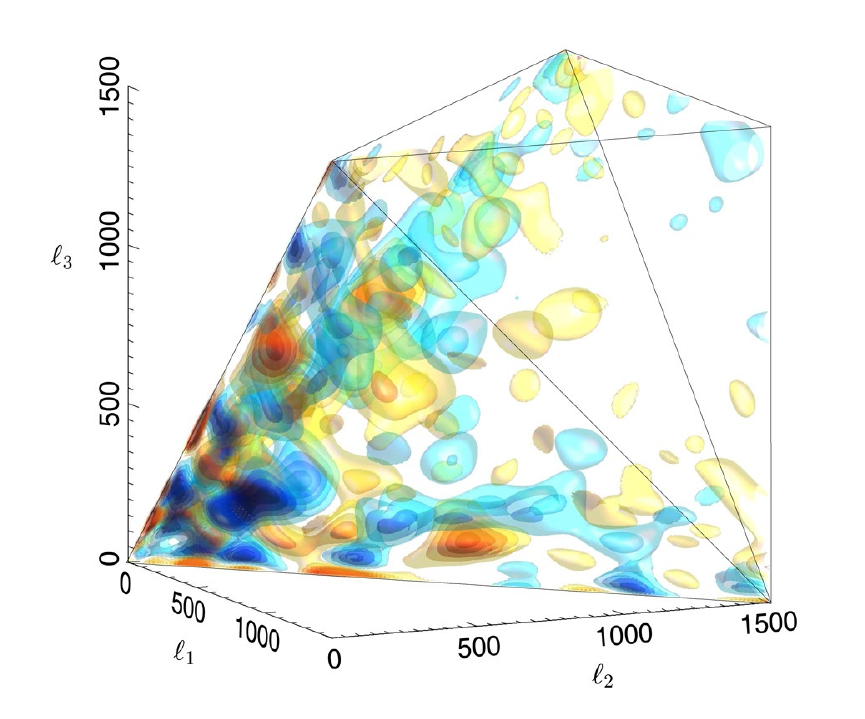}
\caption{CMB bispectrum from a Gaussian simulation including gravitational lensing. This reconstruction uses the same hybrid polynomials as for the \Planck\ bispectrum in Fig.~\ref{fig:recondetail} (right), with which it can be compared. Note that an indication of a ISW-lensing bispectrum signal can be seen along the edges of the tetrahedron.}
\label{fig:reconlensed}
\end{figure}
%%%%%%%%%%%%%%%%%%%%%%%%%%%%%%%%%%%%%%%%%%%%%%%%%%%%
\begin{figure}
\includegraphics[width=8.8cm]{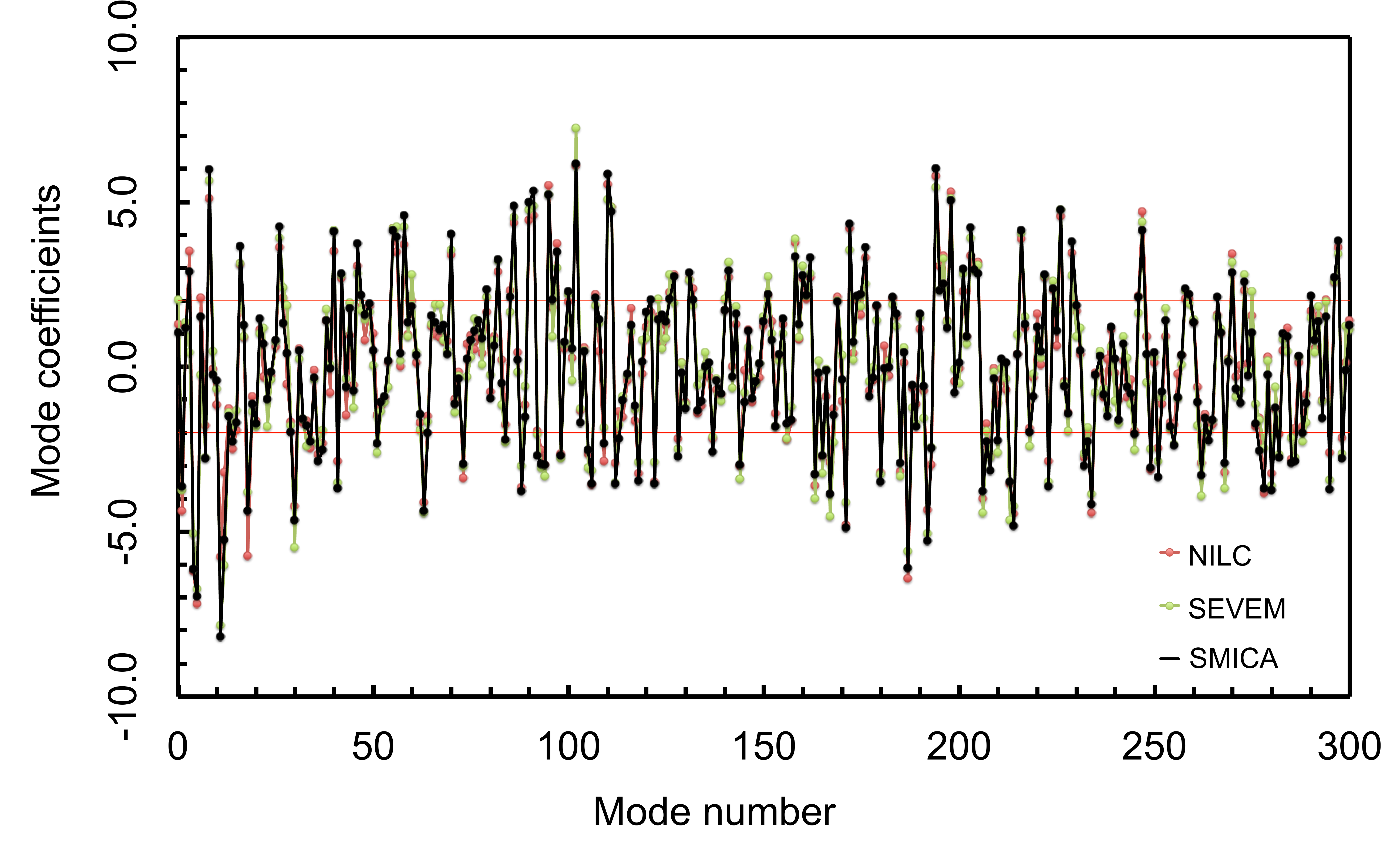}
\caption{Modal bispectrum coefficients $\beta^R_n$ for the mode expansion (Eq.~ \eqref{eq:cmborthomodes}) obtained from \Planck\ foreground-cleaned maps using hybrid Fourier modes.   The different component separation methods, \SMICA, \NILC\ and \SEVEM\ exhibit remarkable agreement.  The variance from 200 simulated noise maps was nearly constant for each of the 300 modes, with the average $\pm$1$\sigma$ variation shown in red.\bigskip\bigskip}
\label{fig:DDX9_modal_beta}
\end{figure}
%%%%%%%%%%%%%%%%%%%%%%%%%%%%%%%%%%%%%%%%%%%%%%%%%%%%

%%%%%%%%%%%%%%%%%%%%%%%%%%%%%%%%%%%%%%%%%%%%%%%%%%%%
\begin{figure}
\includegraphics[width=8.8cm]{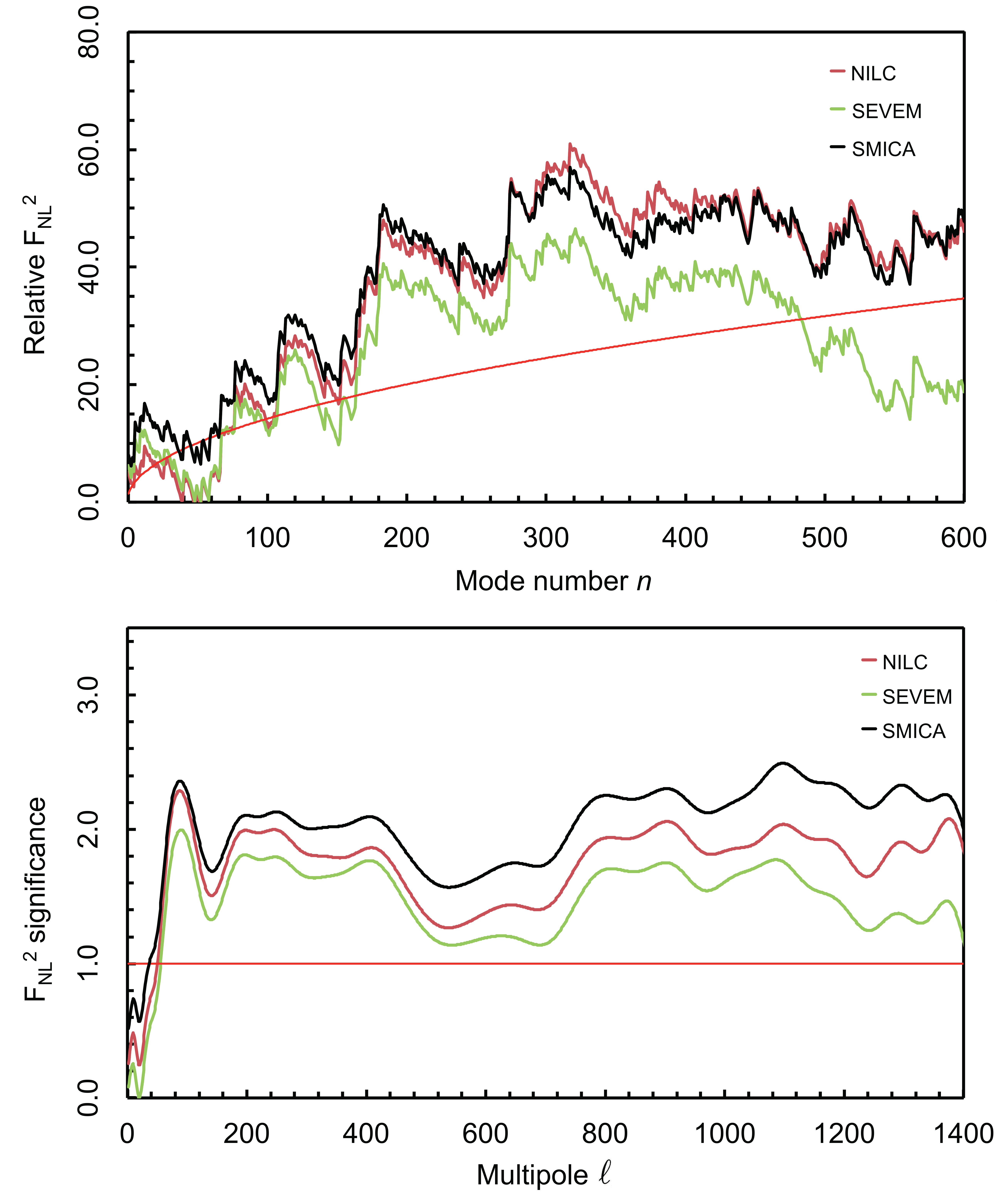}
\caption{
The total integrated bispectrum $F_{\rm NL}^2$ defined in 
Eq.~\eqref{eq:totalbispectrum} as a cumulative sum over orthonormal modal coefficients 
$\beta^R_n{}^2$ (upper panel) and 
over multipoles up to a given $\ell$ (lower panel).   Above, the relative quantity $F_{\rm NL}^2 
\equiv \bar F_{\rm NL}^2 -  {F^{\rm G}_{\rm NL}} {}^2$ is plotted, where $ {F^{\rm G}_{\rm NL} 
{}^2}$ is the mean 
obtained from 200 CMB Gaussian maps with the standard deviation shown as the red line. 
Below the square of the bispectrum is integrated over the tetrapyd out to  $\ell$ and its 
significance plotted relative to the Gaussian standard deviation (1$\sigma$ red line).   A hybrid 
polynomial basis $n_{\rm max}=600$ is employed in the signal-dominated region 
$\ell\le1500$. 
}
\label{fig:DDX9_modal_Fnl}
\end{figure}
%%%%%%%%%%%%%%%%%%%%%%%%%%%%%%%%%%%%%%%%%%%%%%%%%%%%

The modal pipeline was applied to the {\it Planck} temperature maps for the foreground-separation techniques \SMICA, \NILC, and \SEVEM\ \citep{2010PhRvD..82b3502F}.  For this analysis we used two alternative sets of hybrid basis functions in order to cross-check results and identify particular signals.  First, we employed trigonometric functions ($n_{\rm max}=300$) augmented with the SW local mode, together with the three separable modes contributing to the CMB ISW-lensing signal.  Secondly,  we employed the same polynomial basis ($n_{\rm max}=600$) with local SW mode as was used for $\fnl$ estimation.

The modal coefficients $\beta^{\rm R}_n$ extracted from the {\it Planck} \SMICA, \NILC, and \SEVEM\ maps are shown in Fig.~\ref{fig:DDX9_modal_beta}.  Here we have used the hybrid Fourier modes with local and ISW-lensing modes. These amplitudes show remarkable consistency between the different maps, demonstrating that the alternative foreground separation techniques do not appear to be introducing spurious NG.   Note that here the $\beta^{\rm R}_n$ coefficients are for the orthonormalized modes $R_n$  (Eq.~\eqref{eq:cmborthomodes}) and they have a roughly constant variance, so anomalously large modes can be easily identified.  It is evident, for example, that  among the low modes there are large signals, which include the ISW-lensing signal and point source contributions.

Using the modal expansion of Eq.~(\ref{eq:cmbestmodes}) with
Eq.~(\ref{eq:cmborthomodes}), we have reconstructed the full 3D \textit{Planck} bispectrum.
This is illustrated in Fig.~\ref{fig:DDX9_smica_3D}, where we show ``tetrapyd'' comparisons between different foreground
cleaned maps. The tetrapyd (see Fig.~\ref{fig:tetrapyd}) is the region defined
by the multipoles that obey the
triangle condition, with $\ell \leq \ell_{\rm max}$. The 3D plots show
the reduced bispectrum of the map, divided by a Sachs-Wolfe CMB bispectrum
solution for a constant
primordial shape, $S(k_1,k_2,k_3)=1$. This constant primordial bispectrum template
normalizaton is carried out
in order to remove an $\sim \ell^4$ scaling from the starting bispectrum (it
is analogous to multiplication of the power spectrum by $\ell(\ell+1)$).
To facilitate the interpretation of 3D bispectrum figures, note that squeezed
configurations lie on the edges of the tetrapyd, flattened on the faces and
equilateral in the interior, with $b_{\ell \ell \ell}$ on the diagonal. The
colour levels are equally spaced with
red denoting positive values, and blue denoting negative.  Given the correspondence of the $\beta^{\rm R}_n$ coefficients for \SMICA, \NILC, and \SEVEM, the reconstructed 3D signals also appear remarkably consistent, showing similar contours out to $\ell \lesssim 1500$.      At large multipoles $\ell$ approaching $\ell_{\rm max}=2000$, there is increased randomness in the reconstruction due to the rise in experimental noise and some evidence for a residual point source contribution.

There are some striking features evident in the 3D bispectrum reconstruction which appear in both Fourier and polynomial representations, as shown in more detail in Fig.~\ref{fig:recondetail}.   There is an apparent oscillation at low $\ell \lesssim 500$ already seen in {\it WMAP}-7 \citep{2010arXiv1006.1642F}.    Beyond out to $\ell \sim 1200$ there are further distinct features (mostly ``flattened'' on the walls of the tetrapyd), and an oscillating ISW-lensing contribution can be discerned in the squeezed limit. For comparison, the bispectrum reconstruction for one of the lensed Gaussian simulations 
described in sec. 6.2 is shown in Fig.~\ref{fig:reconlensed}. ISW-lensing oscillating signatures in the squeezed 
limit are visible also in this case. When comparing results from a single simulation to the data, 
it is however always useful to keep in mind that the observed tetrapyd pattern  is quite 
realization dependent, since full bispectrum reconstructions are noisy. Whatever its origin, Gaussian or otherwise, Fig.~\ref{fig:recondetail} reveals the CMB bispectrum of our Universe as observed by \Planck.

The cumulative sum $\Fnl^2$ over the squared orthonormal coefficients $\beta_n^{\rm R\,}{}^2$ from Eq.~\eqref{eq:totalbispectrum} for the {\it Planck} data is illustrated in Fig.~\ref{fig:DDX9_modal_Fnl} (upper panel).     The {\it Planck} bispectrum contribution can be directly compared with Gaussian expectations averaged from 200 lensed Gaussian maps with simulated residual foregrounds. It is interesting to note that the integrated bispectrum signal fairly consistently exceeds the Gaussian mean by around $2\sigma$ over much of the domain.  This includes the ISW and PS contributions for which subtraction only has a modest effect.   Also shown (lower panel) is the corresponding cumulative $\Fnl^2$ quantity as a function of multipole $\ell$, for which features have visible counterparts at comparable $\ell$ in Fig.~\ref{fig:recondetail}.   Despite the high bispectrum signal, this $\chi^2$-test over the orthonormal mode coefficients  $\beta_n^{\rm R}$ is cumulatively consistent with Gaussianity
for each of the three component separation 
methods considered. It is however important to stress that 
this is a model independent integrated measurement of NG. Fitting the data with specific 
bispectrum templates can of course enhance the signal-to-noise ratio for given models, 
especially in light of  the $1$ to $2$ $\sigma$ excess 
with respect to the Gaussian expectation, shown in the lower panel. This measurement is thus 
not in disagreement with detection of a residual point source (Poisson) 
bispectrum in the same maps 
(see Table~\ref{tab:PB}), or with the results shown in our feature models survey 
of Section~\ref{featureres}. 
We also note some differences in the high mode region between \SEVEM\ and the other two 
methods, with the {\SEVEM} results being closer to the Gaussian expectation. 
This is consistent with tests on 
simulations and with {\SEVEM} measuring a slightly lower ISW-lensing amplitude than the other 
methods (see Table ~\ref{tab:ISWL}). On the other hand the discrepancies are well within statistical 
bounds, as it can be seen by comparing them to the Gaussian standard deviation from simulations (red 
line). We thus conclude that, even accounting for these high modes deviations, the different component 
separation methods display a good level of internal consistency. 
It is also important to notice that this already good level of agreement becomes even stronger in ``non-blind'' 
$\fnl$ measurements, when 
primordial bispectrum templates are fit to the data, and specific theoretically relevant 
regions of the tetrapyd are selected. This can be clearly seen from Table~\ref{tab:fNLsmicah}.

\subsubsection{Binned bispectrum reconstruction}

\begin{figure*}[!t]
\centering
\includegraphics{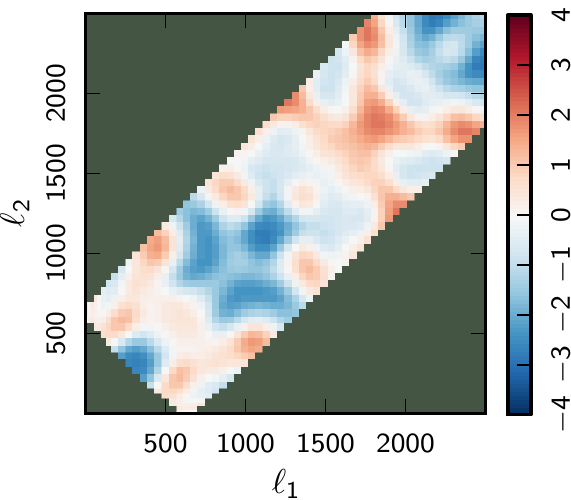}
\includegraphics{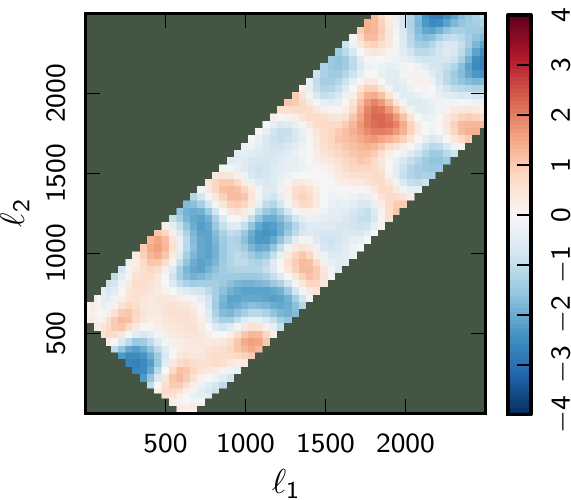}
\includegraphics{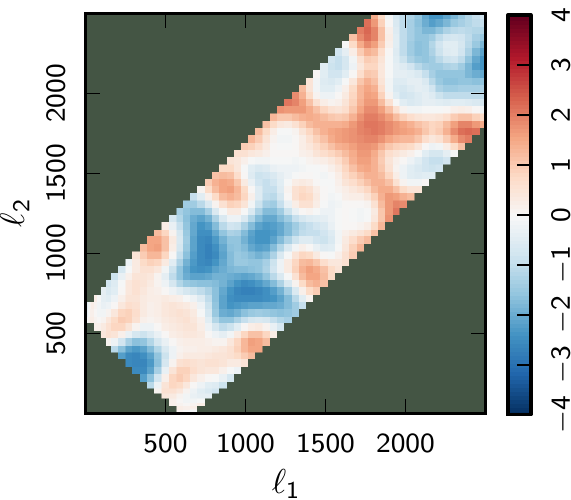}

\includegraphics{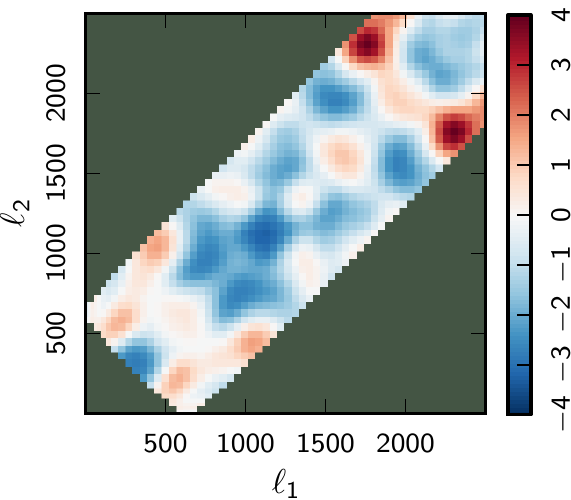}
\includegraphics{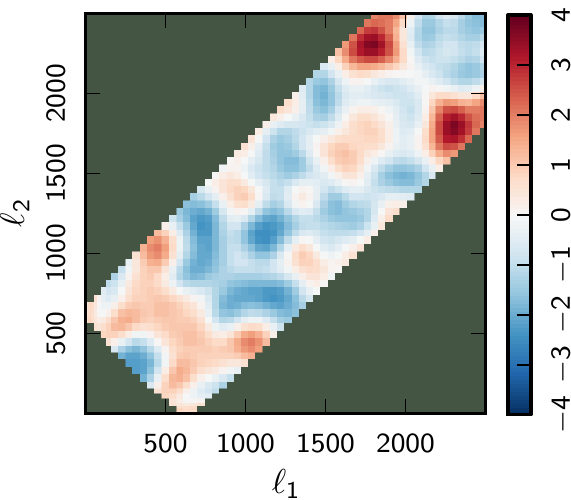}
\includegraphics{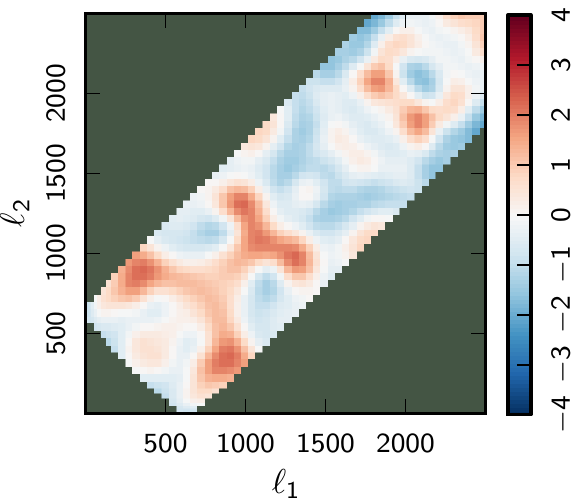}
\caption{Smoothed observed bispectrum as determined with the
binned estimator divided by its expected standard
deviation, as a function of $\ell_1$ and $\ell_2$, with $\ell_3$ in the bin
[610,654]. From left to right on the top row are shown: \SMICA, \NILC,
and \SEVEM; and on the bottom row: \CR\ and the raw 143~GHz channel.
For comparison purposes the last figure on the bottom row shows the
same quantity for one of the lensed Gaussian simulations described in 
Sect.~\ref{Sec_valid_compsep}.}
\label{Fig_slice1}
\end{figure*}

\begin{figure*}[!t]
\centering
\includegraphics{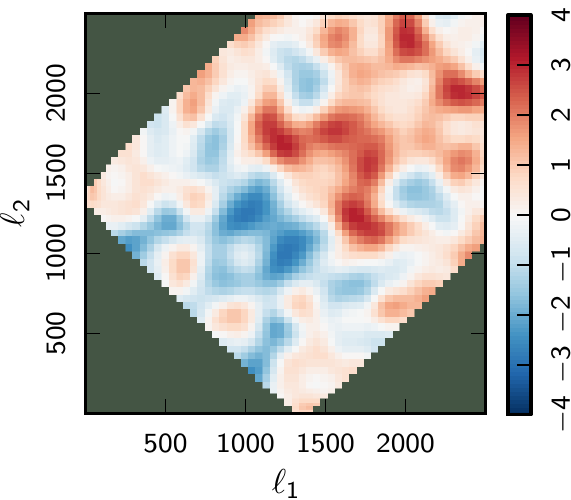}
\includegraphics{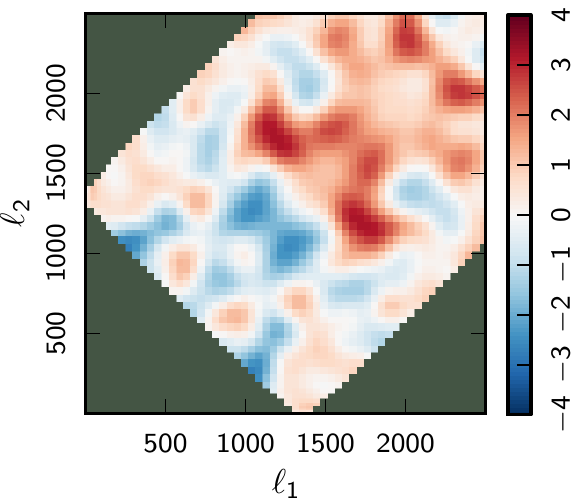}
\includegraphics{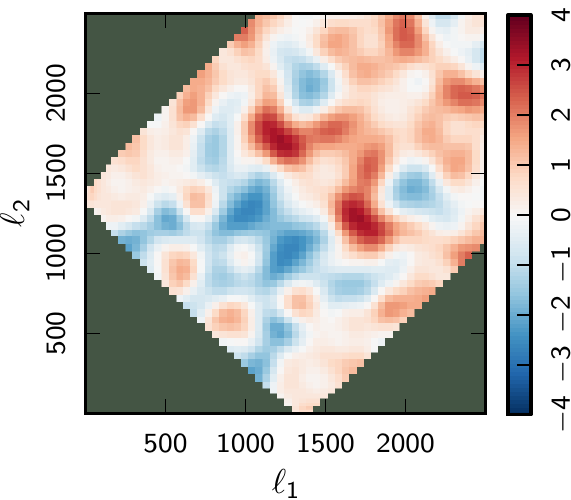}

\includegraphics{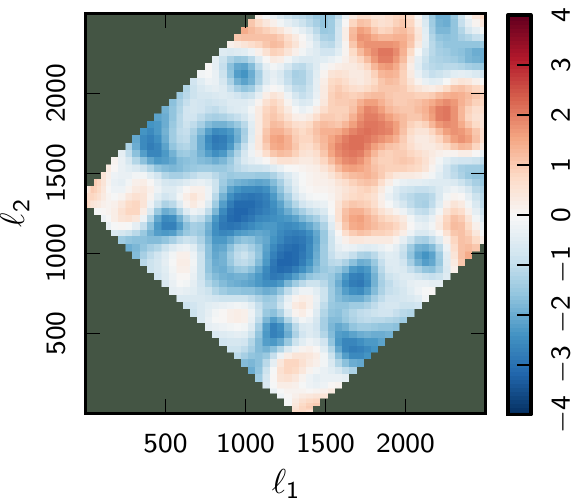}
\includegraphics{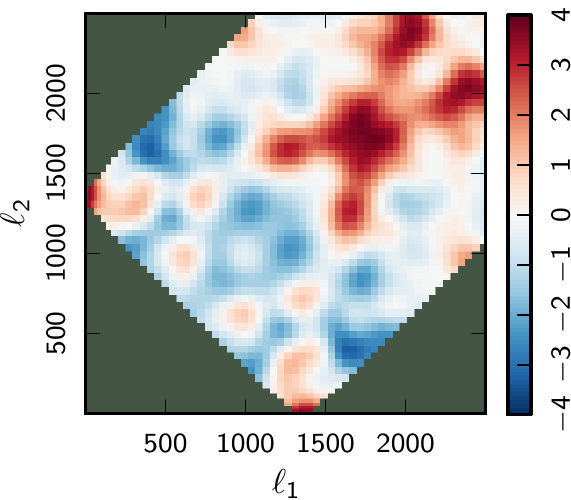}
\includegraphics{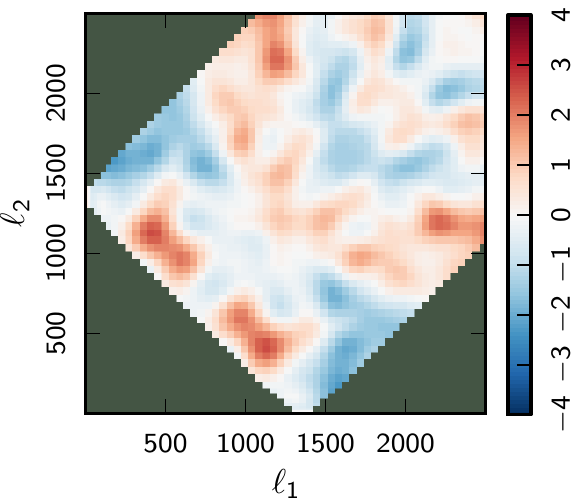}
\caption{Similar to Fig.~\ref{Fig_slice1}, but with $\ell_3$ in the
bin [1330,1374].}
\label{Fig_slice2}
\end{figure*}

As explained in Sect.~\ref{binnedrec}, it is interesting to study the
smoothed observed bispectrum divided by its expected standard deviation, since
this will indicate if there is a significant deviation from
Gaussianity for certain regions of $\ell$-space. This quantity is shown in
Figs.~\ref{Fig_slice1} and~\ref{Fig_slice2} as a function of $\ell_1$
and $\ell_2$, for two different values (or rather, bins) of $\ell_3$: the
intermediate value [610,654] in Fig.~ \ref{Fig_slice1} and the high value
[1330,1374] in Fig.~\ref{Fig_slice2}.
Each figure shows the results for the \SMICA, \NILC, \SEVEM, and \CR\ cleaned
maps as well as for the raw 143~GHz channel map. 
For comparison, the result for one of the lensed Gaussian simulations
described in Sect.~\ref{Sec_valid_compsep} is also shown.
The bispectra were obtained
with the binned bispectrum estimator and smoothed with a Gaussian kernel
as explained in Sect.~\ref{binnedrec}.
Very blue or red regions indicate
significant NG, regions that are less red or blue just
represent expected fluctuations of a Gaussian distribution.

From Fig.~\ref{Fig_slice1} at an intermediate value of $\ell_3$ we can
conclude that there is a very good agreement between \SMICA, \NILC, and
\SEVEM\ for all values of $\ell_1$ and $\ell_2$, and with \CR\ up to about
$\ell_1, \ell_2 \sim 1500$. In fact, up to $1500$ there is also a good
agreement with the raw 143~GHz channel.
We also see no significant non-Gaussian features in this figure (except
maybe in the \CR\ and raw maps at $\ell_1 , \ell_2 > 2000$).
The lensed Gaussian simulation in the last panel looks quite Gaussian 
as well (as it should), but very different from the others. This is not 
surprising, as all we are seeing here are the small random fluctuations 
that exist in any Gaussian realization, and the simulation represents of 
course another realization than the real data. (Note that the ISW-lensing NG,
while present in the simulation, is not really visible in the particular 
slices shown here due to the linear scale of the axes and the fact 
that the ISW-lensing NG peaks in the squeezed configuration.)

Figure~\ref{Fig_slice2} at a high value of $\ell_3$, on the
contrary, shows significant non-Gaussian features in the raw map, but much
less NG in the cleaned maps. In particular one can see the
point-source bispectral signal at high-$\ell$ approximately at equilateral
configurations in the data. This is absent in the the lensed Gaussian simulation, which has no point sources.
There is still an excellent agreement between \SMICA, \NILC, and \SEVEM.
The \CR\ map shows less NG than the other three cleaned maps, which
is consistent with the absence of a detection of the Poisson point source
bispectrum for \CR, see Table~\ref{tab:PB}.

%%%%%%%%%%%%%%%%%%%%%%%%%%%%%%%%%%%%%%%%%%%%%%%%%%%%
%%%%%%%%%%%%%%%%%%%%%%%%%%%%%%%%%%%%%%%%%%%%%%%%%%%%

%%%%%%%%%%%%%%%%%%%%%%%%%%%%%%%%%%%%%%%%%%%%%%%%%%%%
%%%%%%%%%%%%%%%%%%%%%%%%%%%%%%%%%%%%%%%%%%%%%%%%%%%%

%%%%%%%%%%%%%%%%%%%%%%%%%%%%%%%%%%%%%%%%%%%%%%%%%%%%

\subsection{Constraints on specific targeted shapes}
\label{STS}

We have deployed the modal estimator to investigate a  wide range of the inflationary models described in Sect.~\ref{SectionII}.   This is the same validated estimator for which the standard $\fnl$ results have been reported in the Sect.~\ref{sec:Results}, but it is augmented with the primordial modal decomposition and projection described in Sect.~\ref{Sec_modalest}.   The resulting modal-projected local, equilateral and orthogonal shapes are $\sim$99\% correlated with those found using direct integration  of Eq.~\eqref{eq:presutti3} (as for the analysis above).   Modal correlations for the other models investigated were determined for both the primordial shapes and the late-time projected decompositions and were all above 90\%, unless stated otherwise.    This primordial modal estimator pipeline has been applied already extensively to the {\it WMAP}-7 data (\citealt{2010arXiv1006.1642F}).

\subsubsection{Nonseparable single-field bispectrum shape results}
\label{nonsep}
Having characterised single-field inflation bispectra using combinations of the separable equilateral and orthogonal  ans\"atze, we note that the actual leading-order non-separable contributions  (Eqs.~(\ref{EFT2Bis},~\ref{dbiBis})) exhibit significant differences in the collinear (flattened) limit.  For this reason we provide constraints on DBI inflation (Eq.~\eqref{dbiBis})  and the two effective field theory shapes (Eqs.~(\ref{EFT1Bis},~\ref{EFT2Bis})), as well as the ghost inflation bispectrum, which is an exemplar of higher-order derivative theories (specifically Eq.~(3.8) in \citealt{2004JCAP...04..001A}).  Using the primordial modal estimator, with the \SMICA\ foreground-cleaned data, we find:
\eq
\label{fnlequilfamily}
\fnl^{\rm DBI} & = & \,\,\,\, 11 \pm 69\quad\quad  (\Fnl^{\rm DBI-eq} \,\,= \,\,\,\,10 \pm 77)\,,\nn\\
\fnl^{\rm EFT1}&=& \quad\, 8 \pm 73\quad\quad (\Fnl^{\rm EFT1-eq}= \quad\,8 \pm 77)\,,\nn\\
\fnl^{\rm EFT2}&=& \,\,\,\,19 \pm 57\quad\quad (\Fnl^{\rm EFT2-eq}= \,\,\,\,27 \pm 79)\,,\nn\\
\fnl^{\rm Ghost}&=& -23 \pm 88\quad\quad (\Fnl^{\rm Ghost-eq}= -20 \pm 75)\,.
\qe
where we have normalized with the usual primordial $\fnl$ convention which is shape-dependent (i.e., the central value of the shape function is taken such that $S(k,k,k)=1$).  In parentheses we also give a reweighted $\Fnl^{\rm equil}$ constraint for easier comparison with the equilateral constraint from the same modal estimator, i.e., we have rescaled using the Fisher variance for the closely-related equilateral shape.
Given the strong cross-correlation (above 95\%) between all these models, the equilateral family results of~\eqref{fnlequilfamily} reveal larger differences around $\sigma/3$  than might be expected (and somewhat larger than observed previously in the {\it WMAP} data (\citealt{2010arXiv1006.1642F})). The reason for this variation between the equilateral shapes in {\it Planck} appears to be the additional signal observed in the flattened limit in the bispectrum reconstruction beyond the {\it WMAP} signal-dominated range (see Fig.~\ref{fig:recon}).  There is also a contribution from the small correlation difference between equilateral models from primordial modal and KSW methods.   The results for these models for all the \SMICA, \NILC\ and \SEVEM\ foreground-separated maps are given in Appendix~\ref{sec:AB} (Table~\ref{tab:modalmapmethods}).
%

%%%%%%%%%%%%%%%%%%%%%%%%%%%%%%%%%%%%%%%%%%%%%%%%%%%%
\begin{table*}[t]
\begingroup
\newdimen\tblskip \tblskip=5pt
\caption{Constraints on flattened or collinear bispectrum models (and related models) using the \SMICA\ foreground-cleaned {\it Planck} map.   These bispectrum shapes, with equation numbers given, are described in detail in the text.}
\label{tab:fnlnonstandard}
\medskip
\nointerlineskip
\vskip -4mm
\footnotesize
\setbox\tablebox=\vbox{
   \newdimen\digitwidth
   \setbox0=\hbox{\rm 0}
   \digitwidth=\wd0
   \catcode`*=\active
   \def*{\kern\digitwidth}

   \newdimen\signwidth
   \setbox0=\hbox{+}
   \signwidth=\wd0
   \catcode`!=\active
   \def!{\kern\signwidth}
\newdimen\dotwidth
\setbox0=\hbox{.}
\dotwidth=\wd0
\catcode`^=\active
\def^{\kern\dotwidth}
    \halign{\hbox to 2.15in{#\leaderfil}\tabskip 1.0em&
            \hfil#\hfil&
            \hfil#\hfil&
            \hfil#\hfil&
            \hfil#\hfil&
            \hfil#\hfil\tabskip 0pt\cr
    \noalign{\doubleline\vskip 2pt}
 \omit   Flattened model (Eq. number) \hfil&Raw $\fnl$ & Clean $\fnl$ & $\Delta \fnl$ & $\sigma$  & Clean $\sigma$  \cr
     \noalign{\vskip 4pt\hrule\vskip 6pt}
Flat model \eqref{flatBis}  & $!*70^*$ & $!*37^*$ & $*77^*$ & $!0.9$ & $!0.5$\cr
Non-Bunch-Davies (NBD)  & $!178^*$ & $!155^*$ & $*78^*$  & $!2.2$ & $!2.0$\cr
Single-field NBD1 flattened \eqref{NBD2Bis} & $!*31^*$ & $!*19^*$ & $*13^*$ &  $!2.4$ & $!1.4$\cr
Single-field NBD2 squeezed  \eqref{NBD2Bis} & $!**0.8$ & $!**0.2$ & $**0.4$ &  $!1.8$ & $!0.5$\cr
Non-canonical NBD3 \eqref{NBD3Bis} & $!*13^*$ & $!**9.6$ & $**9.7$ &  $!1.3$ & $!1.0$\cr
Vector model $L=1$ \eqref{vectorBis}  & $*-18^*$ & $**-4.6$ & $*47^*$ &  $-0.4$ & $-0.1$\cr
Vector model $L=2$ \eqref{vectorBis} & $!**2.8$ & $**-0.4$ & $**2.9$ & $!1.0$ & $-0.1$\cr
 \noalign{\vskip 3pt\hrule\vskip 4pt}
}}
\endPlancktablewide

\endgroup
\end{table*}
%%%%%%%%%%%%%%%%%%%%%%%%%%%%%%%%%%%%%%%%%%%%%%%%%%%%

%%%%%%%%%%%%%%%%%%%%%%%%%%%%%%%%%%%%%%%%%%%%%%%%%%%%

\subsubsection{Non-Bunch-Davies vacuum results}

We have investigated the non-separable shapes arising from excited initial states (non-Bunch-Davies vacuum models) which usually peak in the flattened or collinear limit.  In particular, we have searched for the four non-separable bispectra described in Eqs.~\eqref{NBD2Bis} and \eqref{NBD3Bis}, as well as the original flattened shape $B^{\rm NBD}_\Phi$ (Eq.~(6.2-3) in \citealt{2007JCAP...01..002C}). This entails choosing suitable cut-offs $k_{\rm c}$ to ensure that the signal is strongly flattened (i.e., distinct from flat in Eq.~\eqref{flatBis}), while also accurately represented by the modal expansion at both early and late  times (Eqs.~(\ref{primmodes},~\ref{eq:transform})).  For $B^{\rm NBD}_\Phi$, we adopted the same edge truncation and mild Gaussian filter described in \citet{2010arXiv1006.1642F}, while for $B^{\rm NBD1}_\Phi$ and $B^{\rm NBD2}_\Phi$, which are
described by Eq.~\eqref{NBD2Bis}, we chose $k_{\rm c}=0.001$, and in Eq.~\eqref{NBD3Bis} we take $k_{\rm c}=0.01$.  The shape correlations for most non-Bunch-Davies vacua were good (above 90\%), except for the strongly squeezed model with oscillations of Eq.~\eqref{NBD2Bis} which was relatively poor (60\%).   Together with the orthogonal (Eq.~\eqref{orthogonalBis}), flat (Eq.~\eqref{flatBis}) and vector (Eq.~\eqref{vectorBis}) shapes, these non-Bunch-Davies models explore a broad range of flattened models, with a variety of different widths for picking out signals around the faces of the tetrapyd (see Fig.~ \ref{fig:tetrapyd}).

The $\fnl$ results obtained for the non-Bunch-Davies models from the different foreground-cleaned map bispectra were consistent and the constraints from \SMICA\ (for brevity) are given in Table~\ref{tab:fnlnonstandard}.  More comprehensive results from \SMICA, \NILC\ and \SEVEM\ can be found in Table~\ref{tab:modalmapmethods} in Appendix~\ref{sec:AB}.  Both $B^{\rm NBD}_\Phi$ and $B^{\rm NBD2}_\Phi$  (Eq.~\eqref{NBD2Bis}) produced raw results above $2\sigma$,  in part picking out the flattened signal observed in the bispectrum reconstruction in Fig.~\ref{fig:recon}.   However, these flattened squeezed signals are also correlated with CMB ISW-lensing and so, after subtracting the predicted ISW bias (as well as the measured point source signal), most NBD $\fnl$ results were reduced to $1\sigma$ or less (see ``Clean $\fnl$'' column in Table~\ref{tab:fnlnonstandard}).  The exception was the most flattened model $B^{\rm NBD}_\Phi$ which remained higher
$\fnl^{\rm NDB}=178 \pm 78$, i.e., with signals at $2.0\sigma$, $1.8\sigma$ and $2.1\sigma$ for \SMICA, \NILC\ and \SEVEM\ respectively.

We emphasise that this has to be considered just as preliminary study of flattened NG in the {\it Planck} data using four exemplar models. In order to reach a complete statistical assessment of constraints regarding flattened models in forthcoming analyses,  we will have to undertake a systematic search for best-fit {\it Planck} NBD models using the parameter freedom available.

\subsubsection{Scale-dependent feature and resonant model results}
\label{featureres}

We have investigated whether the {\it Planck} bispectrum reconstructions include oscillations expected in feature or resonant  models (Eqs.~(\ref{featureBis},~\ref{resonantBis})).    Although poorly correlated with scale-invariant shapes, the feature and resonant models have (at least) two free parameters - the period $k_{\rm c}$ and the phase $\phi$ - forming a model space which must be scanned to determine if there is any significant correlation (in the absence of any physical motivation for restricting attention to specific periodicities).   We have undertaken an initial survey of these models with the wavelength range defined by the native resolution of the present modal estimator (hybrid local polynomials with 600 modes), similar to the feature model search in {\it WMAP} data in \cite{2010arXiv1006.1642F}.   For feature models  of Eq.~\eqref{featureBis} we can obtain high correlations (above 95\%) for the predicted CMB bispectrum if we take $k_{\rm c}> 0.01$, that is, for an effective multipole periodicity $\ell_{\rm c} > 140$ feature models are accurately represented.
%

%%%%%%%%%%%%%%%%%%%%%%%%%%%%%%%%%%%%%%%%%%%%%%%%%%%%
\begin{table*}[t]
\begingroup
\newdimen\tblskip \tblskip=5pt
\caption{\Planck\ bispectrum estimation results for feature models compared to the \SMICA\ foreground-cleaned maps.   This preliminary survey on a coarse grid in the range
$0.01\le k_{\rm c} \le 0.025$ and $0\le \phi<\pi$ finds specific models with significance up to 99.7\%.
}
\label{tab:fnlfeature}
\medskip
\nointerlineskip
\vskip -6mm
\footnotesize
\setbox\tablebox=\vbox{
   \newdimen\digitwidth
   \setbox0=\hbox{\rm 0}
   \digitwidth=\wd0
   \catcode`*=\active
   \def*{\kern\digitwidth}
   \newdimen\signwidth
   \setbox0=\hbox{+}
   \signwidth=\wd0
   \catcode`!=\active
   \def!{\kern\signwidth}
\newdimen\dotwidth
\setbox0=\hbox{.}
\dotwidth=\wd0
\catcode`^=\active
\def^{\kern\dotwidth}
    \halign{\hbox to 1.2in{#\leaderfil}\tabskip 1.0em&
            \hfil#&
            \hfil#&
            \hfil#&
            \hfil#\tabskip 0pt\cr
    \noalign{\doubleline\vskip 2pt}
    \omit&\multispan4\hfil $\fnl\pm\Delta\fnl ~~(\sigma)$\hfil\cr
\omit&\multispan4\hrulefill\cr
 Wavenumber $k_{\rm c}$\hfill& $\phi=0$ \hfil &$\phi=\pi/4 $ \hfil & $\phi=\pi/2$ \hfil & $\phi=3\pi/4$ \hfil \cr
 %\omit   Wavenumber  \hfil& $\fnl\pm\Delta\fnl ~~(\sigma)$ &$\fnl\pm\Delta\fnl ~~(\sigma)$ & $\fnl\pm\Delta\fnl ~ ~(\sigma)$ & $\fnl\pm\Delta\fnl  ~~(\sigma)$ \cr
     \noalign{\vskip 4pt\hrule\vskip 6pt}
$0.01000$ & $-110 \pm 159 ~~(-0.7)$ & $*-98 \pm 167 ~~(-0.6)$ & $*-17 \pm 147 ~~(-0.1)$ & $!*56 \pm 142 ~~(!0.4)$ \cr
$0.01125$ & $!434 \pm 170 ~~(!2.6)$ & $!363 \pm 185 ~~(!2.0)$ & $!*57 \pm 183 ~~(!0.3)$ & $-262 \pm 168 ~~(-1.6)$ \cr
$0.01250$ & $*-70 \pm 158 ~~(-0.4)$ & $!130 \pm 166 ~~(!0.8)$ & $!261 \pm 167 ~~(!1.6)$ & $!233 \pm 159 ~~(!1.5)$ \cr
$0.01375$ & $!*35 \pm 162 ~~(!0.2)$ & $!291 \pm 145 ~~(!2.0)$ & $!345 \pm 147 ~~(!2.3)$ & $!235 \pm 162 ~~(!1.5)$ \cr
$0.01500$ & $-313 \pm 144 ~~(-2.2)$ & $-270 \pm 137 ~~(-2.0)$ & $*-95 \pm 145 ~~(-0.7)$ & $!179 \pm 154 ~~(!1.2)$ \cr
$0.01625$ & $!*81 \pm 126 ~~(!0.6)$ & $!177 \pm 141 ~~(!1.2)$ & $!165 \pm 144 ~~(!1.1)$ & $!*51 \pm 129 ~~(!0.4)$ \cr
$0.01750$ & $-335 \pm 137 ~~(-2.4)$ & $-104 \pm 128 ~~(-0.8)$ & $!181 \pm 117 ~~(!1.5)$ & $!366 \pm 126 ~~(!2.9)$ \cr
$0.01875$ & $-348 \pm 118 ~~(-3.0)$ & $-323 \pm 120 ~~(-2.7)$ & $-126 \pm 119 ~~(-1.1)$ & $!137 \pm 117 ~~(!1.2)$ \cr
$0.02000$ & $-155 \pm 110 ~~(-1.4)$ & $-298 \pm 119 ~~(-2.5)$ & $-241 \pm 113 ~~(-2.1)$ & $*-44 \pm 105 ~~(-0.4)$ \cr
$0.02125$ & $*-43 \pm *96 ~~(-0.4)$ & $-186 \pm 107 ~~(-1.7)$ & $-229 \pm 115 ~~(-2.0)$ & $-125 \pm 104 ~~(-1.2)$ \cr
$0.02250$ & $!*22 \pm *95 ~~(!0.2)$ & $-115 \pm *92 ~~(-1.2)$ & $-194 \pm 105 ~~(-1.8)$ & $-148 \pm 107 ~~(-1.4)$ \cr
$0.02375$ & $!*70 \pm 100 ~~(!0.7)$ & $*-56 \pm *94 ~~(-0.6)$ & $-159 \pm *93 ~~(-1.7)$ & $-164 \pm 101 ~~(-1.6)$ \cr
$0.02500$ & $!106 \pm *93 ~~(!1.1)$ & $!**6 \pm *97 ~~(!0.1)$ & $-103 \pm *98 ~~(-1.1)$ & $-153 \pm *94 ~~(-1.6)$ \cr
\noalign{\vskip 3pt\hrule\vskip 4pt}
}}
\endPlancktablewide
\endgroup
\end{table*}
%%%%%%%%%%%%%%%%%%%%%%%%%%%%%%%%%%%%%%%%%%%%%%%%%%%%

The results of a first survey of feature models in the {\it Planck} data is shown in Table~\ref{tab:fnlfeature} for $0.01\le k_{\rm c}\le 0.1$ and phases $\phi = 0,\pi/4,\pi/2,3\pi/4$ (for $\phi\ge\pi$ we will identify a correlation with the opposite sign).   Again, there was good consistency between the different foreground-separation methods \SMICA, \NILC\ and \SEVEM\, showing that the results are robust to potential residual foreground contamination in the data. For brevity we only give \SMICA\ results here, while providing measurements from other component separation methods in Appendix~\ref{sec:AB}. Feature signals are typically largely uncorrelated with the ISW-lensing or point sources, but nevertheless we subtract these signals and give results for the cleaned $\fnl$.    The Table~\ref{tab:fnlfeature} results show that there is a parameter region around $0.01 \le k_{\rm c} \le 0.025$ for which signals well in excess of $2\sigma$ are possible (we undertook a broader search with $0.01 \le k_{\rm c} \le 0.1$ but found only a low signal beyond $k>0.3$).   It appears that some feature models are able to match the low-$\ell$ `plus-minus' and other features in the {\it Planck} bispectrum reconstruction (see Fig.~\ref{fig:recon}).   The best fit model has $k_{\rm c}=0.0185$ ($\ell_{\rm c}\approx 260$) and phase $\phi = 0$ with a signal $-3\sigma$. As a further validation step of our results, we also re-analysed the models with $>2.5 \sigma$ significance using a different modal decomposition, namely an oscillating Fourier basis ($n_{\rm max} = 300$)  augmented with a local SW mode (the same used for the reconstruction plots in Sect.~\ref{sec:Results}). The results from this basis are shown in Appendix~\ref{sec:AB} and they are fully consistent with the polynomial measurements presented here. The previous best-fit {\it WMAP} feature model, $k_{\rm c}=0.014$ ($\ell_{\rm c}\approx200$) and phase $\phi=3\pi/4$, attained a 2.15$\sigma$ signal  with $\ell<500$ (\citealt{2010arXiv1006.1642F}), but it only remains at this level for {\it Planck}.

We note however that the apparently high statistical significance of these results is much lower if we consider this to be a blind survey of feature models, because we are seeking several  uncorrelated models simultaneously.  Following what we did for our study of impact of foregrounds in Sect.~\ref{Sec_valid_data},  we considered a set of $200$ realistic lensed FFP6 simulations, processed through the \SMICA\ pipeline, and including realistic foreground residuals.
If we use this accurate MC sample to search for the same grid of 52 feature models as in Table~\ref{tab:fnlfeature},  we find a  typical maximum signal of $2.23 (\pm 0.56)\,\sigma$.   Searching across all feature models (see below) studied here yields an expected maximum $2.37 (\pm 0.53)\,\sigma$ (whereas the survey for all 511 models from all paradigms investigated yielded $2.55(\pm0.52)\,\sigma$). This means that our best-fit model from data has a statistical significance below $1.5\,\sigma$ above the maximum signal expectation from simulations, so we conclude that we have
 {\em no significant detection of feature models} from {\it Planck} data.

%%%%%%%%%%%%%%%%%%%%%%%%%%%%%%%%%%%%%%%%%%%%%%%%%%%%
\begin{figure}[h]
\centering
\includegraphics[width=\hsize]{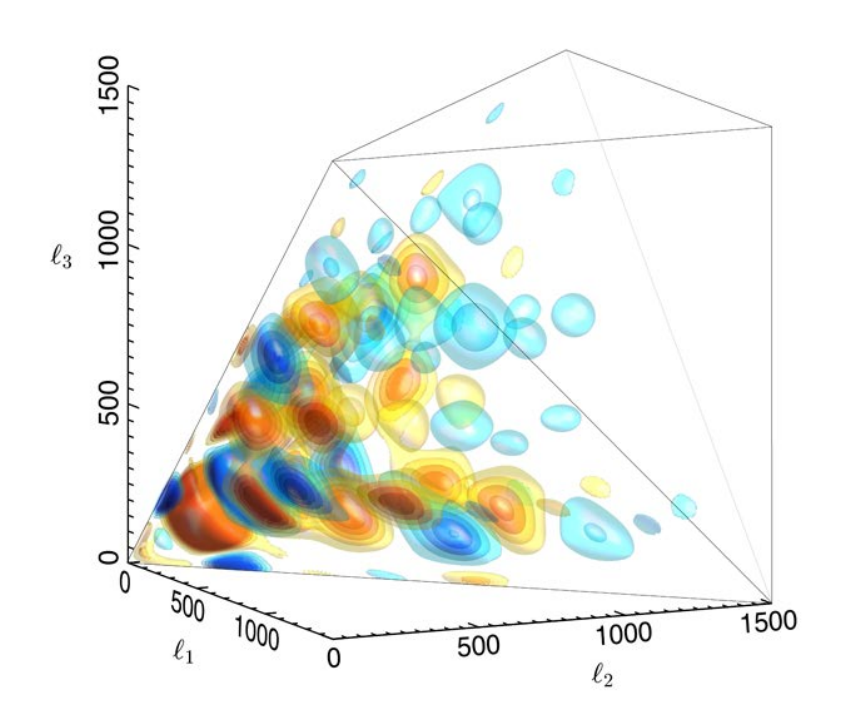}
\caption[]{CMB bispectrum shown for the best-fit feature model with an envelope with parameters $k=0.01875$, phase  $\phi=0$ and $\Delta k=0.045$ (see Table~\ref{tab:fnlfeatureenv}).  Compare with the Planck bispectrum reconstruction, Fig.~\ref{fig:recondetail}.}
\label{fig:featureenv}
\end{figure}
%%%%%%%%%%%%%%%%%%%%%%%%%%%%%%%%%%%%%%%%%%%%%%%%%%%%

%%%%%%%%%%%%%%%%%%%%%%%%%%%%%%%%%%%%%%%%%%%%%%%%%%%%
\begin{table*}[t]
\begingroup
\newdimen\tblskip \tblskip=5pt
\caption{Feature model results with an envelope decay function of width $\Delta k$. Results are only presented for feature models with better than 95\% CL result on the full domain (see Table~\ref{tab:fnlfeature}).}
\label{tab:fnlfeatureenv}
\medskip
\nointerlineskip
\vskip -6mm
\footnotesize
\setbox\tablebox=\vbox{
   \newdimen\digitwidth
   \setbox0=\hbox{\rm 0}
   \digitwidth=\wd0
   \catcode`*=\active
   \def*{\kern\digitwidth}
   \newdimen\signwidth
   \setbox0=\hbox{+}
   \signwidth=\wd0
   \catcode`!=\active
   \def!{\kern\signwidth}
\newdimen\dotwidth
\setbox0=\hbox{.}
\dotwidth=\wd0
\catcode`^=\active
\def^{\kern\dotwidth}
    \halign{\hbox to 1.2in{#\leaderfil}\tabskip 1.0em&
            \hfil#&
            \hfil#&
            \hfil#&
            \hfil#\tabskip 0pt\cr
    \noalign{\doubleline\vskip 2pt}
        \omit&\multispan4\hfil $\fnl\pm\Delta\fnl ~~(\sigma)$\hfil\cr
\omit&\multispan4\hrulefill\cr
\omit    Wavenumber $k_{\rm c}$; phase\hfil& $\Delta k=0.015$ \hfil &$\Delta k=0.03$ \hfil & $\Delta k=0.045$ \hfil & Full \hfil \cr
     \noalign{\vskip 4pt\hrule\vskip 6pt}
$0.01125;\,\phi=0$ & $!765 \pm 275 ~~(!2.8)$ & $!703 \pm 241 ~~(!2.9)$ & $!648 \pm 218 ~~(!3.0)$ & $!434 \pm 170 ~~(!2.6)$ \cr
$0.01750;\,\phi=0$ & $-661 \pm 234 ~~(-2.8)$ & $-494 \pm 192 ~~(-2.6)$ & $-425 \pm 171 ~~(-2.5)$ & $-335 \pm 137 ~~(-2.4)$ \cr
$0.01750;\,\phi=3\pi/4$ & $!399 \pm 207 ~~(!1.9)$ & $!438 \pm 183 ~~(!2.4)$ & $!442 \pm 165 ~~(!2.7)$ & $!366 \pm 126 ~~(!2.9)$ \cr
$0.01875;\,\phi=0$ & $-562 \pm 211 ~~(-2.7)$ & $-559 \pm 180 ~~(-3.1)$ & $-515 \pm 159 ~~(-3.2)$ & $-348 \pm 118 ~~(-3.0)$ \cr
$0.01875;\,\phi=\pi/4$ & $-646 \pm 240 ~~(-2.7)$ & $-525 \pm 189 ~~(-2.8)$ & $-468 \pm 164 ~~(-2.9)$ & $-323 \pm 120 ~~(-2.7)$ \cr
$0.02000;\,\phi=\pi/4$ & $-665 \pm 229 ~~(-2.9)$ & $-593 \pm 185 ~~(-3.2)$ & $-500 \pm 160 ~~(-3.1)$ & $-298 \pm 119 ~~(-2.5)$ \cr
\noalign{\vskip 3pt\hrule\vskip 4pt}
}}
\endPlancktablewide
\endgroup
\end{table*}
%%%%%%%%%%%%%%%%%%%%%%%%%%%%%%%%%%%%%%%%%%%%%%%%%%%%

Feature models typically have a damping envelope representing the decay of the oscillations as the inflaton returns to its background slow-roll evolution.   Indeed, the feature envelope is a characteristic of the primordial mechanism producing the fluctuations, decaying as $k$ increases for inflation while rising for contracting models like the ekpyrotic case (\citealt{Chen:2011tu}).  We have made an initial survey to determine whether a decaying envelope improves the significance of any feature models.  The envelope employed was a Gaussian centred at $k_{\rm c}=0.045$ with a falloff $\Delta k = 0.015, 0.02, 0.025, 0.03, 0.035, 0.04, 0.045$ and results for specific parameters are given in Table~\ref{tab:fnlfeatureenv}.  The best fit model remains $k=0.01875$ ($\ell_{\rm c} = 265$) with phase  $\phi=0$ and the significance rises to $3.23\sigma$, together with a second model $k=0.02$ ($\ell_{\rm c} = 285$) $\phi=\pi/4$. However the caveats about blind survey statistics previously noted also do not allow a claim of any detection in this case.   A plot of the best-fit feature model with a decay envelope is shown in Fig.~\ref{fig:featureenv}, for which the main features should be compared with those in Fig.~\ref{fig:recondetail}.     Non-Gaussian bispectrum signals from feature models typically produce counterparts in the power spectrum as will be described in Sect.~\ref{sec:Implications}. An improved statistical interpretation of the results presented in this Section will be possible when this additional investigation will be completed.

We have also undertaken a survey of resonant models and the non-Bunch-Davies resonant models (or enfolded resonance models).  With the modal estimator, we can achieve high accuracy for the predicted bispectrum for $k_{\rm c}> 0.001$ (note that this has a different logarithmic dependence to feature models and a varying effective $\ell_{\rm c}$).    For the resonance model shape of Eq.~\eqref{resNBDBis}, we have not undertaken an extensive survey, except selecting a likely range for a high signal with periodicity comparable to the feature model, that is, with $ 0.25 < k_{\rm c} <0.5$ and phases $\phi = 0,\; \pi/4,\;\pi/2,\;3\pi/4,\; \pi$.   However, no significant signal was found (all below 1$\sigma$), as can be verified in Table~\ref{tab:fnlresonant} in Appendix~\ref{sec:AB}.    For the enfolded resonance model shape of Eq.~\eqref{resNBDBis} , we have undertaken a preliminary search in the range $ 4 < k_{\rm c} <12$ with the same phases.  Again, no significant signal emerges from the {\it Planck} data, as shown in Table~\ref{tab:fnlresNBD} in Appendix~\ref{sec:AB}.

\subsubsection{Directional dependence motivated by gauge fields}
\medskip
\noindent We have investigated whether there is significant NG from bispectrum shapes with non-trivial directional dependence (Eq.~ \eqref{vectorBis}), which are motivated by inflationary models with vector fields.  Using the primordial modal estimator we obtained a good correlation with the $L=1$ flattened-type model, but the squeezed $L=2$ model produced a relatively poor correlation of only 60\%, given the complexity of the dominant squeezed limit.   Preliminary constraints on these models are given in the Table~\ref{tab:fnlnonstandard}, showing no evidence of a significant signal.

\subsubsection{Warm inflation}
\label{warmres}
Warm inflation produces a related shape with a sign change in the squeezed limit.   This also had a  poor correlation, until smoothing (WarmS) was applied as described in \citet{2010arXiv1006.1642F}.   The resulting bispectrum shows no evidence for significant correlation with {\it Planck} data (\SMICA),
\eq\label{fnlwarmS}
\fnl^{\rm WarmS}  =  4 \pm 33\,.
\qe
The full list of constraints for \SMICA, \NILC\ and \SEVEM\ models can be found for warm inflation and vector models in Table~\ref{tab:modalmapmethods} in Appendix~\ref{sec:AB}.

\subsubsection{Quasi-single-field inflation}
\label{qsfresults}
Finally, quasi-single-field inflation has been analysed constraining the bispectrum shape of Q (Eq.~(\ref{qsiBis})), that depends on two parameters, $\nu$ and $f_{\rm NL}^{\rm QSI}$.
In order to constrain this model we have calculated modal coefficients for $0 \leq \nu \leq1.5$ in steps of $0.01$ (so 151 models in total). These were then applied to the data and the one with the greatest significance was selected. Results are shown in Fig.~\ref{QSF}. The maximum signal occurred at $\nu = 1.5$, $f_{\rm NL}^{\rm QSI} = 4.79$  $(0.31 \sigma)$. To obtain error curves we performed a full likelihood using 2 billion simulations following the method described in~\cite{Sefusatti:2012ye}.
Such a large number of simulations was possible as they were generated from the modal $\beta$-covariance matrix which is calculated once from the 200 \textit{Planck} realistic CMB simulations, rather than repeatedly from the CMB simulations themselves. The procedure is to take the $151\times151$ correlation matrix for the models (this is just the normalized dot product of the modal coefficients). This is then diagonalised using PCA, after which only the first 5 eigenvalues are kept as the remaining eigenvalues are $<10^{-10}$. The $\beta$-covariance matrix is projected into the same sub-basis where it is also diagonalised via PCA into 5 orthonormal modes, with the two leading modes closely correlated with local and equilateral.  The procedure by which to produce a simulation is to generate five Gaussian random numbers and add the mean values obtained from the \Planck\ data, rotating them to the sub-basis where we determine the $\nu$ with the greatest significance. The result is then projected back to the original space to determine the related $f_{\rm NL}$.  The two billion results from this MC analysis are then converted into confidence curves plotted in Fig.~\ref{QSF}.  The curve shows that there is no preferred value for $\nu$ with all values allowed at $3\sigma$.  This reflects the results obtained from data previously, where we found the least preferred value of $\nu=0.86$ had only a marginally lower significance of $0.28\sigma$ \citep{Sefusatti:2012ye}. Of course, these conclusions are directly related to the null results for both local and equilateral templates.

%%%%%%%%%%%%%%%%%%%%%%%%%%%%%%%%%%%%%%%%%%%%%%%%%%%%

\subsection{Constraints on local non-Gaussianity with Minkowski Functionals}
\label{MF}

%%%%%%%%%%%%%%%%%%%%%%%%%%%%%%%%%%%%%%%%%%%%%%%%%%%%
\begin{figure}[!t]
\includegraphics[width=\hsize]{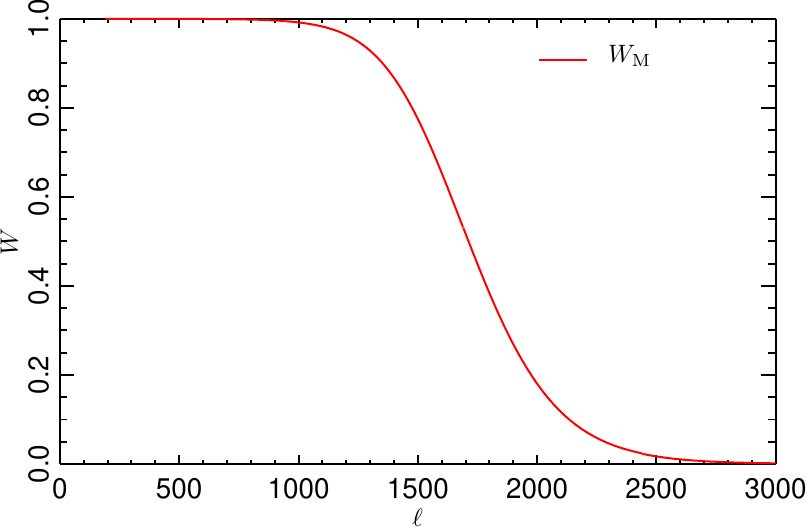}
\caption{The Wiener filter $W_{\rm M}$ used to constrain $\fnllocal$ with MFs.}
\label{fig:Wiener_filter_MFs}
\end{figure}
%%%%%%%%%%%%%%%%%%%%%%%%%%%%%%%%%%%%%%%%%%%%%%%%%%%%

%%%%%%%%%%%%%%%%%%%%%%%%%%%%%%%%%%%%%%%%%%%%%%%%%%%%
\begin{figure*}[!h]
\includegraphics[width=\hsize]{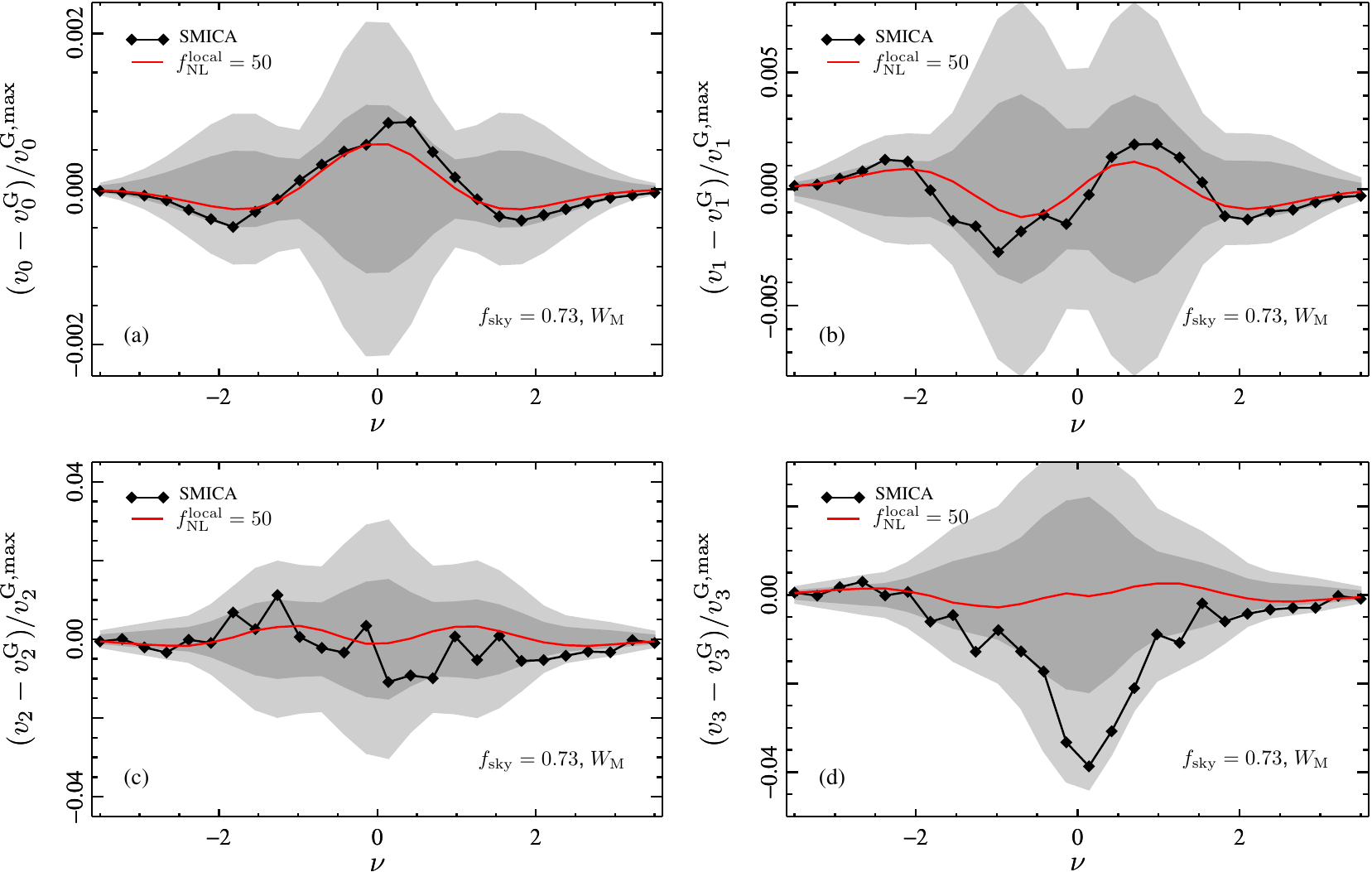}
\caption{The MFs curves for \SMICA\ at $N_{\rm side}=1024$ and $\ell_\mathrm{max}=2000$, for the four functionals $v_k$: (a) Area, (b) Perimeter, (c) Genus, and (d) $N_{\rm cluster}$. The curves are the difference of each normalized MF, measured from the data, to the average from Gaussian \textit{Planck} realistic simulations (not lensed). The difference curves are normalized by the maximum of the Gaussian curve. To compare the curves to the presence of primordial NG, the average difference curves for non-Gaussian simulations with $\fnllocal=50$ is also represented (100 simulations).}
\label{fig:curves_MFs_wd0}
\end{figure*}
%%%%%%%%%%%%%%%%%%%%%%%%%%%%%%%%%%%%%%%%%%%%%%%%%%%%
%
%%%%%%%%%%%%%%%%%%%%%%%%%%%%%%%%%%%%%%%%%%
%

%%%%%%%%%%%%%%%%%%%%%%%%%%%%%%%%%%%%%%%%%%%%%%%%%%%%
\begin{table}[tmb]  % use table for a one-column table
%\begin{table*}[tmb]	% use table* for a two-column table
\begingroup	% this + \endgroup at the end keep table things local
\newdimen\tblskip \tblskip=5pt
\caption{Validation tests with MFs: results for $\fnllocal$ obtained using the filter $W_{\rm M}$, for $\ell_\mathrm{max}=2000$ and $N_{\rm side}=2048$.}
\label{tab:validation_mfs}
\nointerlineskip
\vskip -6mm
\footnotesize  % good font size for a table, but can be changed
\setbox\tablebox=\vbox{
\newdimen\digitwidth  % see \S\,18.12 for the purpose of the next 10 lines
\setbox0=\hbox{\rm 0}
\digitwidth=\wd0
\catcode`*=\active
\def*{\kern\digitwidth} %
\newdimen\signwidth
\setbox0=\hbox{+}
\signwidth=\wd0
\catcode`!=\active
\def!{\kern\signwidth} %
\newdimen\dotwidth
\setbox0=\hbox{.}
\dotwidth=\wd0
\catcode`^=\active
\def^{\kern\dotwidth}
\halign{ #\hfil \tabskip=1em	& \hfil#\hfil  \tabskip=1em \cr			% template goes here.
\noalign{\doubleline}
Setup \hfill&$\fnllocal$(MFs)\cr
\noalign{\vskip 4pt\hrule\vskip 6pt}
Gaussian ideal                                   & *0.2 $\pm$ 10.7    \cr
Realistic noise, $\fnllocal=12$         & 12^* $\pm$ 13^*   \cr
Mask ($f_{\rm sky}=0.73$), realistic noise $\fnllocal=12$ & 12^* $\pm$ 18^*   \cr
%Gaussian Ideal                                   & *0.21 $\pm$ 10.71     \cr
%Realistic Noise, $\fnllocal=12$         & 12.14 $\pm$ 13.12   \cr
%Mask ($f_{\rm sky}=0.73$), realistic noise $\fnllocal=12$ & 12.18 $\pm$ 18.13    \cr
\noalign{\vskip 5pt\hrule\vskip 3pt}}}
\endPlancktable	 	% for a one-column table; defined in Planck.tex.
%\endPlancktablewide	% for a two-column table; defined in Planck.tex.
\endgroup
\end{table}	%or \end{table*}
%%%%%%%%%%%%%%%%%%%%%%%%%%%%%%%%%%%%%%%%%%%%%%%%%%%%

%
%
%%%%%%%%%%%%%%%%%%%%%%%%%%%%%%%%%%%%%%%%%%%%%%%%%%%%
\begin{table*}[tmb]	% use table* for a two-column table
\begingroup	% this + \endgroup at the end keep table things local
\newdimen\tblskip \tblskip=5pt
\caption{Estimates of $\fnllocal$ obtained with MFs on {\it Planck} data. Foreground and secondary effects are evaluated in terms of $\fnllocal$. Results are for \SMICA\ at $N_{\rm side}=1024$ and $\ell_\mathrm{max}=2000$.}
\label{tab:results_mfs}
\nointerlineskip
\vskip -3mm
\footnotesize  % good font size for a table, but can be changed
\setbox\tablebox=\vbox{
\newdimen\digitwidth  % see \S\,18.12 for the purpose of the next 10 lines
\setbox0=\hbox{\rm 0}
\digitwidth=\wd0
\catcode`*=\active
\def*{\kern\digitwidth} %
\newdimen\signwidth
\setbox0=\hbox{+}
\signwidth=\wd0
\catcode`!=\active
\def!{\kern\signwidth} %
\halign{ \hbox to 1.2in{#\leaderfil}\tabskip 1.0em&	& \hfil#\hfil \tabskip=1em & \hfil#\hfil  \tabskip=1em & \hfil#\hfil \cr
%\halign{ #\hfil\tabskip=2em	& \hfil#\hfil \tabskip=1em & \hfil#\hfil  \tabskip=1em & \hfil#\hfil \cr			% template goes here.
\noalign{\doubleline}
Map\hfill                &  $\fnllocal$  & Source & Corresponding $\Delta \fnllocal$ \cr % heading goes here.
\noalign{\vskip 3pt\hrule\vskip 5pt}
Raw map                   &  	19.1 $\pm$ 19.3 &  &    \cr
%\noalign{\vskip 3pt\hrule\vskip 5pt}
 Lensing subtracted  &  	*8.5 $\pm$ 20.5 & Lensing  & $+$10.6 \cr
 %\noalign{\vskip 3pt\hrule\vskip 5pt}
  Lensing+PS subtracted  &  *7.7 $\pm$ 20.3	& Point sources   & *$+$0.8 \cr
  Lensing+CIB subtracted &  *7.5 $\pm$ 20.5	& CIB   &  *$+$1.0 \cr
  Lensing+SZ subtracted  &  *6.0 $\pm$ 20.4 & SZ  & *$+$2.5 \cr
  %\noalign{\vskip 3pt\hrule\vskip 5pt}
   All subtracted         &  *4.2 $\pm$ 20.5	& All   & $+$14.9  \cr  % table lines go here.
\noalign{\vskip 5pt\hrule\vskip 3pt}}} %\endPlancktable	 or	% for a one-column table; defined in Planck.tex.
\endPlancktablewide	% for a two-column table; defined in Planck.tex.
\endgroup %\end{table}	or
\end{table*}
%%%%%%%%%%%%%%%%%%%%%%%%%%%%%%%%%%%%%%%%%%%%%%%%%%%%

In this Subsection, we present constraints on local NG  obtained with Minkowski Functionals (MFs). MFs describe the morphological properties of the CMB field and can be used as generic estimators of NG
\citep{2003ApJS..148..119K,2004ApJ...612...64E,2007ApJ...670L..73D,hikage2008,2008A&A...486..383C,2010MNRAS.408.1658N,hikage2012,2013MNRAS.428..551M}. As they are sensitive to every order of NG, they can be used to constrain different bispectrum and trispectrum shapes
\citep{hikage2006,hikage2008,hikage2012}. They are therefore  complementary to, and a useful validation of, optimal estimators.
Their precise definition and analytic formulations are presented in \cite{planck2013-p09}. The MF technique is also used in the companion paper \cite{planck2013-p20}.

We review here the properties of MFs, as a complementary tool to poly-spectrum based estimators.

First, they are defined in real space, which makes MFs robust to masking effects and no  linear term is needed to take into account the anisotropy of  the data model.
Second, as MFs are sensitive to every non-Gaussian feature in the maps, they can be a useful probe of every potential bias in the bispectrum measurement, in particular the different astrophysical contaminations (foregrounds and secondaries).

There is a limitation to MF studies: they can be expressed in terms of weighted sums of the bispectrum (and trispectrum) in harmonic space \citep{matsubara2010}, hence the angle-dependence of the bispectrum is partially lost. This makes MFs suboptimal in two ways: increasing error bars for constraints on specific shapes and reducing the distinguishability of different bispectrum shapes. This lack of specificity can introduce biases, as MFs will partially confuse  primordial  and non-primordial sources of NG and  can introduce degeneracies between different primordial shapes. Constraints on orthogonal and equilateral shapes are quite degenerate with MFs, we therefore chose here to focus on the local bispectrum shape. We also leave trispectrum analyses for future studies.

An attractive feature of MFs is their linearity for weak NG ($\fnl$) and weak signals (such as point sources, and Galactic residuals after masking and component separation) \citep{ducout2012}. This property can be used to estimate different known non-primordial contributions.

\subsubsection{Method}
We constrain  $\fnllocal$ using the optimized procedure described in \cite{ducout2012}.  To obtain  constraints on $\fnllocal$, we apply a specific Wiener filter on the map ($W_{\rm M}$), shown in Fig.~\ref{fig:Wiener_filter_MFs}. We do not use here the filter designed to enhance the information from the gradients of the map ($W_{\rm D1}=\sqrt{\ell (\ell +1) }W_{\rm M}$), because this filter is very sensitive to small-scale effects and may be biased by foreground residuals.

We use maps at HEALPix resolution $N_{\rm side}=1024$~\citep{gorski2005} and $\ell_{\rm max}=2000$.
Our results are based on the four normalized\footnote{Raw MFs $V_k$ depend on the Gaussian part of fields through a normalization factor $A_k$ that is a function only of the power spectrum shape. We therefore normalize functionals $v_k=V_k/A_k$ to focus on NG; see \cite{planck2013-p09} and references therein.} functionals $v_k\; (k=0,3)$ (respectively Area, Perimeter, Genus and $N_{\rm cluster}$), computed on $n_{\rm th}=26$ thresholds $\nu$, between $\nu_{\rm min}=-3.5$ and $\nu_{\rm max}=+3.5$ in units of the standard deviation of the map.

We combine all functionals into one vector $y$ (of size $n=104$). We then analyse this vector in a Bayesian way to obtain a posterior for the $\fnllocal$ , and hence an estimate of this parameter.
The principle is to compare the vector measured on the data $\hat{y}$ to the ones measured on non-Gaussian simulations with the same systematic effects (realistic noise, effective beam) and data processing (Wiener filtering) as the data, $\bar{y}( \fnllocal ) $. Modelling the MFs as multivariate Gaussians we obtain the posterior distribution for $\fnllocal$ with a $\chi^{2}$ test :
\begin{equation}
P(\fnllocal|\,\hat{y}) \propto \exp \left[ -\dfrac{\chi^{2}(\hat{y},\fnllocal)}{2} \right]
\label{eq:postsimp}
\end{equation}
with
\begin{equation}
\chi^{2}(\hat{y}, \fnllocal)  \equiv \left[ \hat{y}- \bar{y}( \fnllocal )   \right]^{T}C^{-1} \left[ \hat{y}- \bar{y}( \fnllocal )  \right],
\label{eq:mychi2}
\end{equation}

Since NG is weak, the covariance matrix $C$ is computed with $10^4$ Gaussian simulations, again reproducing effective beam, realistic noise and filtering of the data. The dependence of the MFs on $\fnllocal$, $\bar{y}( \fnllocal )$, is  obtained as an average of $\hat{y}$ measured on 100 simulations. The simulations used here are based on the {\it WMAP}-7  best-fit power spectrum \citep{2011ApJS..192...18K}, using the procedure described in \cite{2009ApJS..184..264E}.

\subsubsection{Validation tests}
We report here validation of the MFs estimator on $\fnllocal$ in thoroughly realistic \textit{Planck} simulations. This validation subsection is analogous to Sect.~\ref{sec:Sec_valid_est} concerning bispectrum-based estimators. The same tests (ideal Gaussian maps, full-sky non-Gaussian maps with noise and non-Gaussian maps with noise and mask) are performed, but different non-Gaussian simulations are used. Non-Gaussian CMB simulations as processed in \cite{2010PhRvD..82b3502F} only guarantee the correctness of the 3-point correlations. Since the MFs are sensitive to higher-order $n$-point functions,  they were validated with physical simulations  \citep{2009ApJS..184..264E}.

The first test consists of 100 simulations of a full-sky Gaussian CMB,
with a Gaussian beam with FWHM of 5 arcmin and without any noise, cut off at
$\ell_\mathrm{max}=2000$, with $N_{\rm side}=2048$. Here validation tests were made at $N_{\rm side}=2048$, but results (estimate and error bars) remain the same at $N_{\rm side}=1024$ as we keep the same $\ell_\mathrm{max}$. The second test includes non-Gaussian simulations with $\fnllocal=12$ and realistic coloured and anisotropic noise, processed through the \textit{Planck} simulation pipeline and the component-separation method \SMICA. Finally, in the third test we add the union mask U73 to the previous simulations,
masking both the Galaxy and the brightest point sources, and leaving $73\%$
of the sky unmasked. Only the inpainting of the smallest holes in the point sources part of the mask was performed. For these three tests, the results are presented in Table~\ref{tab:validation_mfs}. We give here the average estimate and error bar obtained on the 100 simulations, when we use the four functionals. The results show that the MF estimator is unbiased, robust, and a competitive alternative to bispectrum-based estimators.
Moreover a map-by-map comparison of the results obtained on $\fnllocal$ with KSW and MFs estimators showed a fair agreement between the two methods.

\subsubsection{Results}

For our analysis we considered a foreground-cleaned map obtained with the component separation method \SMICA\ at $N_{\rm side}~=~1024$ and $\ell_\mathrm{max}=2000$.  As for the previous results in this Section, we used the union mask U73, which leaves $f_{\rm sky}=73$\% of the sky after masking Galaxy and point sources. To take into account some instrumental effects (asymmetry of beams, component separation processing) and known non-Gaussian contributions (lensing), we used realistic unlensed and lensed simulations ($10^3$) of \textit{Planck} data (FFP6 simulations, see Appendix~\ref{sec:FFP6}). First, MFs were applied to the unlensed simulations and the resulting curves served to calibrate the estimator, as the Gaussian part of the NG curves $\bar{y}( \fnllocal )$\footnote{The overall effect of data processing on the $\fnllocal$ constraint from MFs was evaluated as $\fnllocal (\rm process.)\sim 3$.}. This estimate is referred to as the ``raw map''. Secondly, MFs were applied to the lensed simulations, and the same procedure was applied, the result being referred as ``lensing-subtracted''. We summarize the procedure in the following equation:
\begin{equation}
\bar{y}( \fnllocal )=\bar{y}^{\rm G}_{\mathit{Planck}\: \rm simulations, \: lensed }+\Delta \bar{y}_{\fnl ,\: \rm NG\: simulations}^{\rm NG}
\end{equation}
Here we assume that the MF respond linearly to  lensing at first order and that primordial NG and lensing contributions are therefore additive.

Additionally, we tried to characterize other non-primordial contributions that one could expect in masked  \SMICA-cleaned maps covering  $73$\% of the sky. To this end, we used simulations of extragalactic foregrounds and secondary anisotropies: uncorrelated (Poissonian) point sources; clustered CIB; and SZ clusters. These component simulations reproduce accurately the whole \textit{Planck} data processing pipeline (beam asymmetry, component separation method). Using the linearity of MFs \citep{ducout2012}, we could introduce these effects as a simple additive bias on the curves following
\begin{equation}
\hat{y}=\hat{y}^{\rm FG subtracted}+\Delta \bar{y}^{\rm PS}+\Delta \bar{y}^{\rm CIB}+\Delta \bar{y}^{\rm SZ}
\end{equation}
where $\Delta \bar{y}^{\rm PS, ...}$ is the average bias measured on 100 simulations. Note that the SZ simulation does not take into account the SZ-lensing correlation, which  is expected to be negligible given the error bars.

Results are summarized in Table~\ref{tab:results_mfs} and MFs curves are shown in Fig.~\ref{fig:curves_MFs_wd0}, without including the lensing subtraction (``raw curves''). Considering the larger error bars of MF estimators, the constraints obtained are consistent with those from the bispectrum-based estimators, even without subtracting the expected non-primordial  contributions.
Moreover, results are quite robust to Galactic residuals: constraints obtained with other component separation methods (\NILC\ and \SEVEM), with different sky coverage,  differ from the \SMICA\ results presented here by less than  $\Delta \fnllocal = 1$.

%%%%%%%%%%%%%%%%%%%%%%%%%%%%%%%%%%%%%%%%%%%%%%%%%%%%
\begin{figure}[!h]
\centering
\includegraphics[width=0.95\hsize]{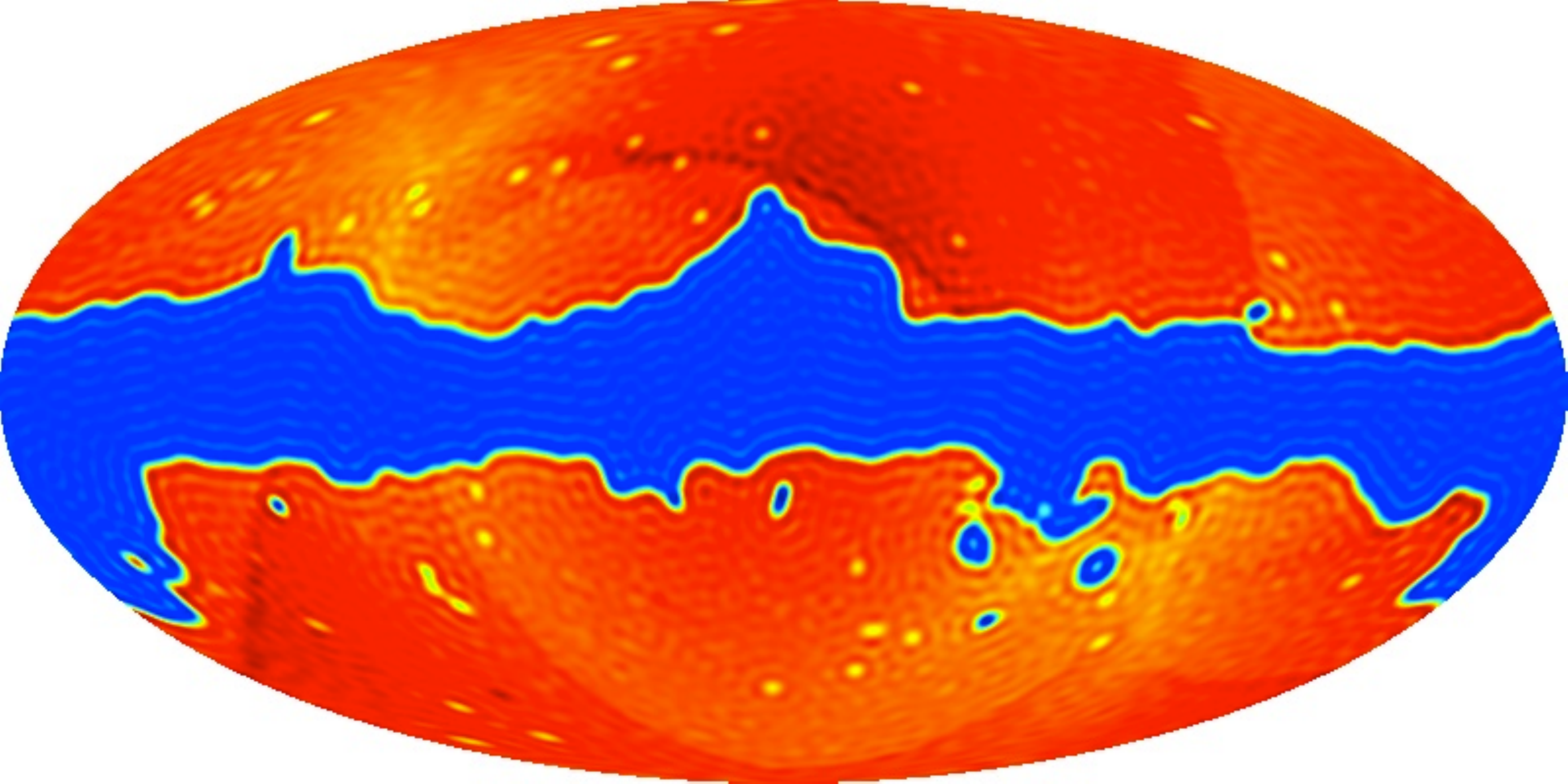}
\includegraphics[width=0.95\hsize]{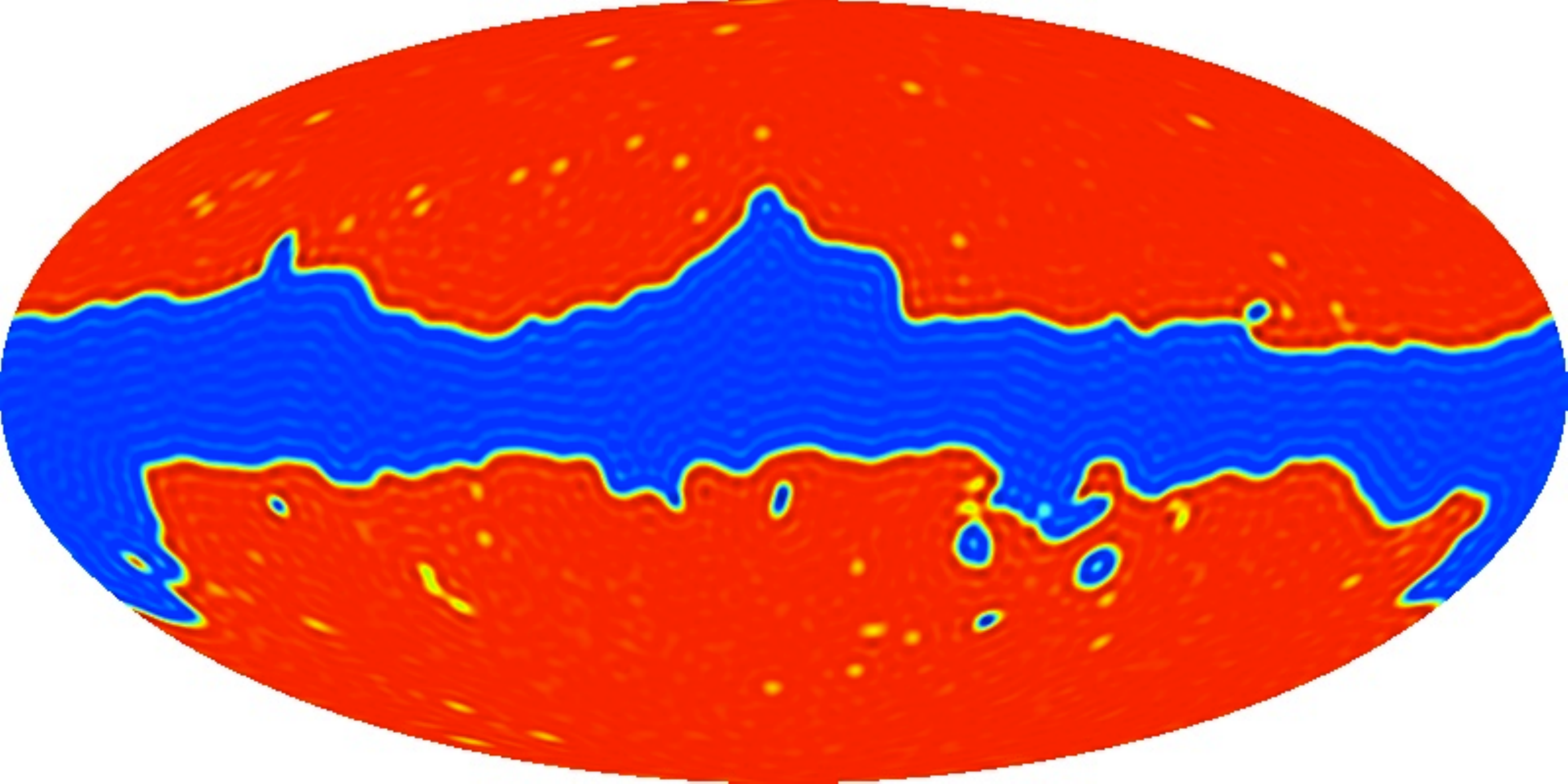}
\includegraphics[width=\hsize]{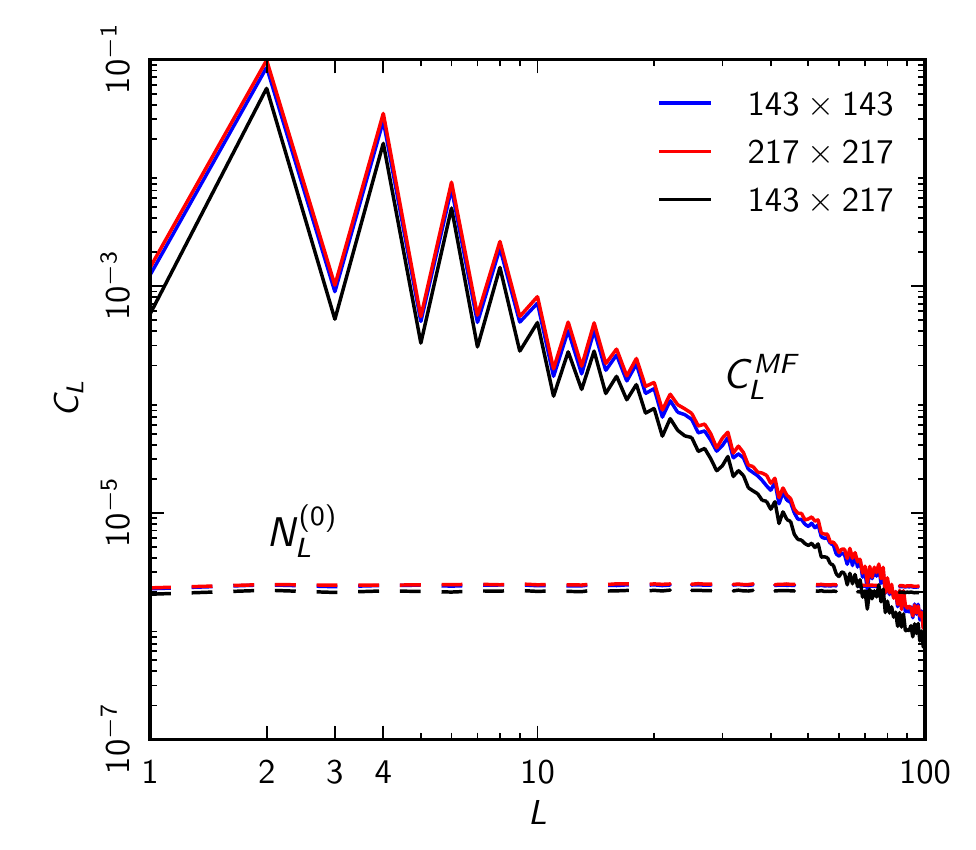}
\caption{
The two upper maps show the modulation reconstruction mean field $\mfmodulation(\hat{\vn})$ at $L\le 100$ , which is essentially a map of the expected total small-scale power on the masked map as a function of position (assuming there is no primordial power modulation). The top mean field map from the $143$ GHz auto estimator has a large signal from both the cut (which can be calculated accurately), and from the noise anisotropy (aligned roughly with the ecliptic, which cannot be estimated very accurately from simulations). The lower mean field is the $143\times 217$ GHz cross-estimator map, and does not have the contribution from the noise anisotropy (note the colour scale is adjusted). The lower plot shows the corresponding mean field power spectra compared to the reconstruction noise $N_L^{(0)}$ (connected part of the trispectrum); the reconstruction noise is much smaller than both the detector noise and mask contributions to the mean field. Since any $\taunl$ signal is all on large scales we do not reconstruct modes above $L_{\rm max} =100$.
\label{modulation_mean_fields}
}
\end{figure}
%%%%%%%%%%%%%%%%%%%%%%%%%%%%%%%%%%%%%%%%%%%%%%%%%%%%

%%%%%%%%%%%%%%%%%%%%%%%%%%%%%%%%%%%%%%%%%%%%%%%%%%%%
\begin{figure}[t]
\includegraphics[width=\hsize]{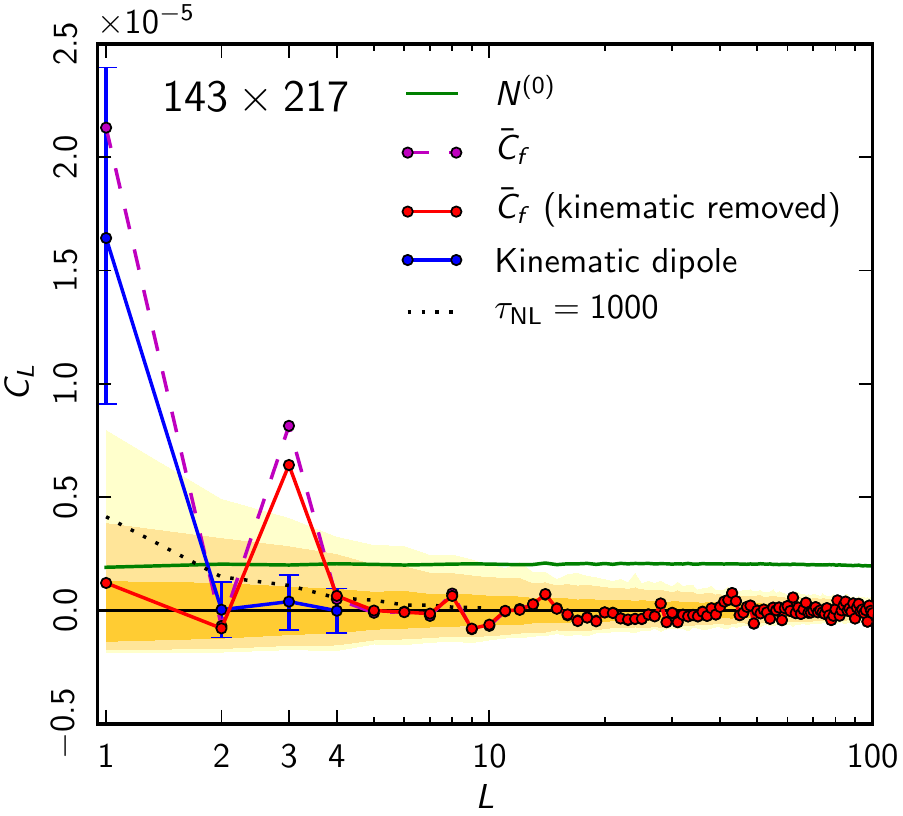}
\caption{Power spectrum of the power modulation reconstructed from $143\times 217$ GHz maps. Shading shows the $68\%$, $95\%$ and $99\%$ CL intervals from simulations with no modulation or kinematic signal. The dashed lines are when the mean field simulations include no kinematic effects, showing a clear detection of a modulation dipole. The blue points show the expected kinematic modulation dipole signal from simulations, along with $1\sigma$ error bars (only first four points shown for clarity).
The solid line subtracts the dipolar kinematic signal in the mean fields from simulations including the expected signal, and represents out best estimate of the non-kinematic signal (note this is not just a subtraction of the power spectra since the mean field takes out the fixed dipole anisotropy in real space before calculating the remaining modulation power).
The dotted line shows the expected signal for $\taunl=1000$.
}
\label{modulation_cl_results}
\end{figure}
%%%%%%%%%%%%%%%%%%%%%%%%%%%%%%%%%%%%%%%%%%%%%%%%%%%%

%%%%%%%%%%%%%%%%%%%%%%%%%%%%%%%%%%%%%%%%%%%%%%%%%%%%
\begin{figure}[t]
%\center
\includegraphics[width=\hsize]{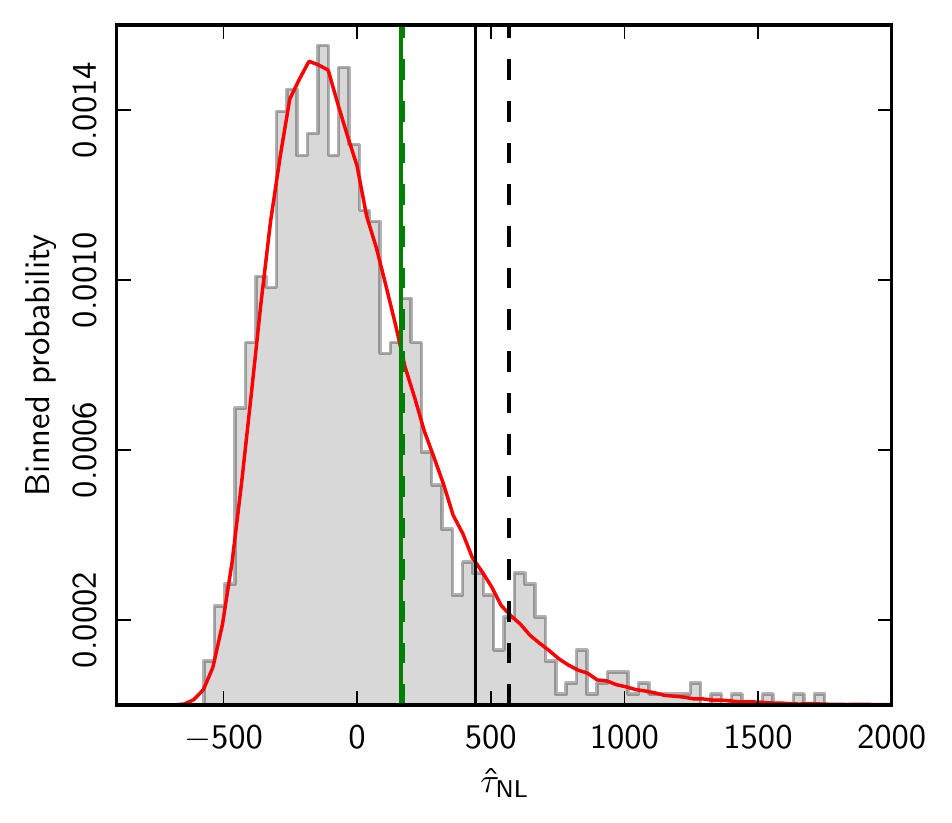}
%~~
\caption{Distribution of $\htaunl$ estimators from Gaussian simulations ($\Lmax=10$) compared to data estimates (vertical lines).
The distribution is rather skewed because the main contributions are from $L\alt 4$ where the modulation power spectra have skewed $\chi^2$ distributions with low degrees of freedom. The red line shows the predicted distribution for a weighted sum of $\htaunl(L)$ estimators assuming the reconstructed modulation modes are Gaussian with $2L+1$ modes measured per $L$, which fits the full simulations well.
The black vertical lines show the data estimates from $\Lmax=10$, and should be compared to the green which have $\Lmax=2$ and hence are insensitive to the anomalous octopole signal. The dashed lines are $\tau_{{\rm NL},1}$, the slightly more optimal variant of the estimator.
}
\label{taunl_hist}
\end{figure}
%%%%%%%%%%%%%%%%%%%%%%%%%%%%%%%%%%%%%%%%%%%%%%%%%%%%

%%%%%%%%%%%%%%%%%%%%%%%%%%%%%%%%%%%%%%%%%%%%%%%%%%%%

\subsection{CMB trispectrum results}
\label{trispectrum}

As shown in Fig.~\ref{modulation_mean_fields} the modulation reconstruction mean field has two large contributions, one from the mask and one from anisotropic noise, reflecting the fact that they both look like a large spatially-varying modulation of the fluctuation power.
The noise we use to estimate the mean field is taken from FFP6 simulations, adjusted with an additional $10\,\mu {\rm K}\,{\rm arcmin}$ white noise component to match the power spectrum in the observed maps. However this is still only an approximate description of the instrumental noise present in the data.
The mean field from non-independent maps (e.g., $143\times 143$ GHz maps) shows a large noise anisotropy that is primarily quadrupolar before masking, and any mismatch between the simulated noise and reality would lead to a large error in the mean field subtraction.
By instead using the modulation estimator for $143\times 217$ GHz maps errors due to misestimation of the noise are avoided, and the mean field is then dominated by the shape of the Galactic cut, which is well known, and a smaller uncertainty from assumed simulation power spectrum and beam errors (see Fig.~\ref{modulation_mean_fields}). For this reason for our main result we work with modulation reconstructions generated from  $143\times 217$ GHz maps with independent noise, which removes the leading error due to noise mean field misestimation.

Figure~\ref{modulation_cl_results} shows the reconstructed modulation power from $143\times 217$ GHz maps that we use for our analysis. We show two results: one where we do not include the expected kinematic dipole signal in the mean field subtraction (see Sect.~\ref{taunl_nonprimordial}), and one were we do so that the reconstruction should then be dominated by the primordial and any unmodelled systematic effects. In the first case the $143\times 217$ result gives a clear first detection of the dipolar kinematic modulation signal of roughly the expected magnitude (see~\cite{planck2013-pipaberration} for a more detailed discussion of this signal).
Including the expected kinematic signal in the simulations (and hence the mean field) removes this signal, giving a cosmological modulation reconstruction that is broadly consistent with no modulation (statistical isotropy) except for the anomalous very significant signal in the modulation octopole.

Note that only the two-point reconstruction is free from noise bias, the four-point is still sensitive to noise modelling at the level of the subtraction of the $N^{(0)}_L$ reconstruction noise power spectrum (Eq.~(\ref{N_L0})). However as shown in the Fig.~\ref{modulation_cl_results}, $N^{(0)}_L$ is not that much larger than the reconstruction scatter at low multipoles where the $\taunl$ signal peaks, so the sensitivity to noise misestimation is much less than in the mean field subtraction (where the large-scale noise anisotropy gives a  large-scale mean field in the auto-estimators orders of magnitude larger, Fig.~\ref{modulation_mean_fields}).

%%%%%%%%%%%%%%%%%%%%%%%%%%%%%%%%%%%%%%%%%%%%%%%%%%%%
\begin{figure}[h]
\center
\includegraphics[width=\hsize]{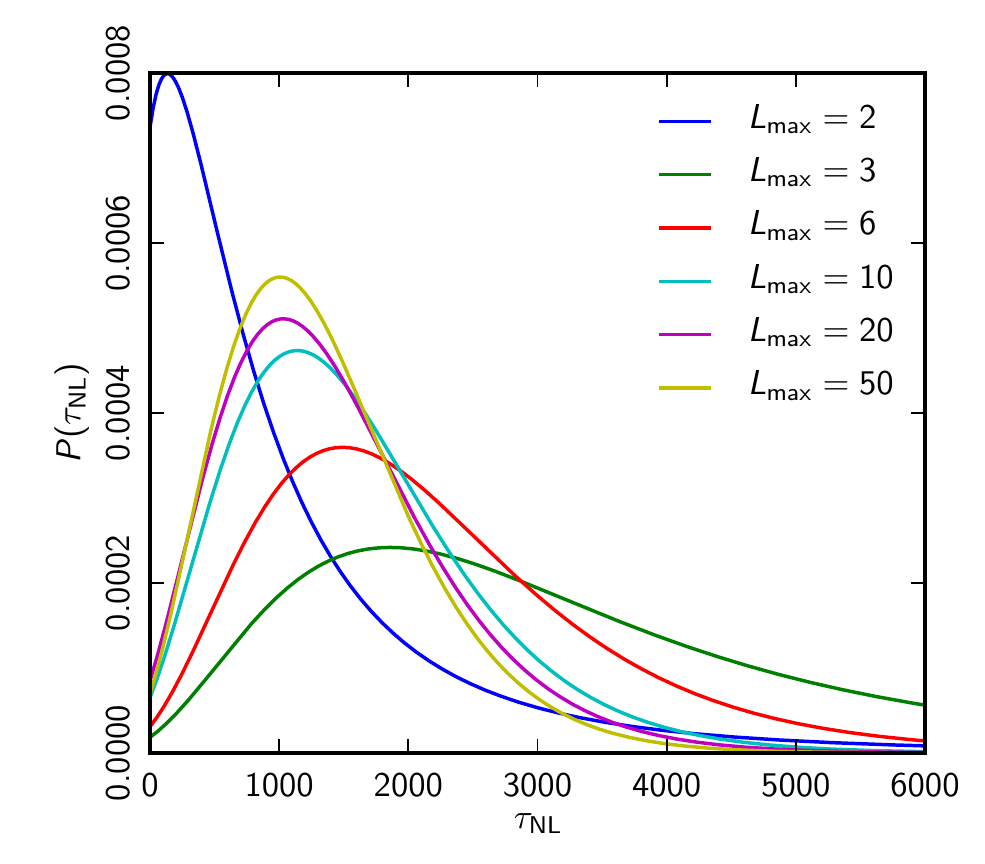}
%~~
\caption{Approximate posterior distributions $P(\taunl|\htaunl(L))$ for a range of $L_{\rm max}$. The distributions have broad tails to high values because of the small number of large-scale modulation modes that are measured, and hence large cosmic variance. For $\Lmax\ge 3$ the distributions are pulled away from zero by the significant octopole modulation signal observed, and are gradually move back towards zero as we include more modulation modes that are inconsistent with large $\taunl$ values. As shown in Fig.~\ref{modulation_freq} the octopole has significant frequency dependence and is therefore unlikely to be physical.
}
\label{taunlPosteriors}
\end{figure}
%%%%%%%%%%%%%%%%%%%%%%%%%%%%%%%%%%%%%%%%%%%%%%%%%%%%

%%%%%%%%%%%%%%%%%%%%%%%%%%%%%%%%%%%%%%%%%%%%%%%%%%%%
\begin{figure}[h]
%\center
\includegraphics[width=\hsize]{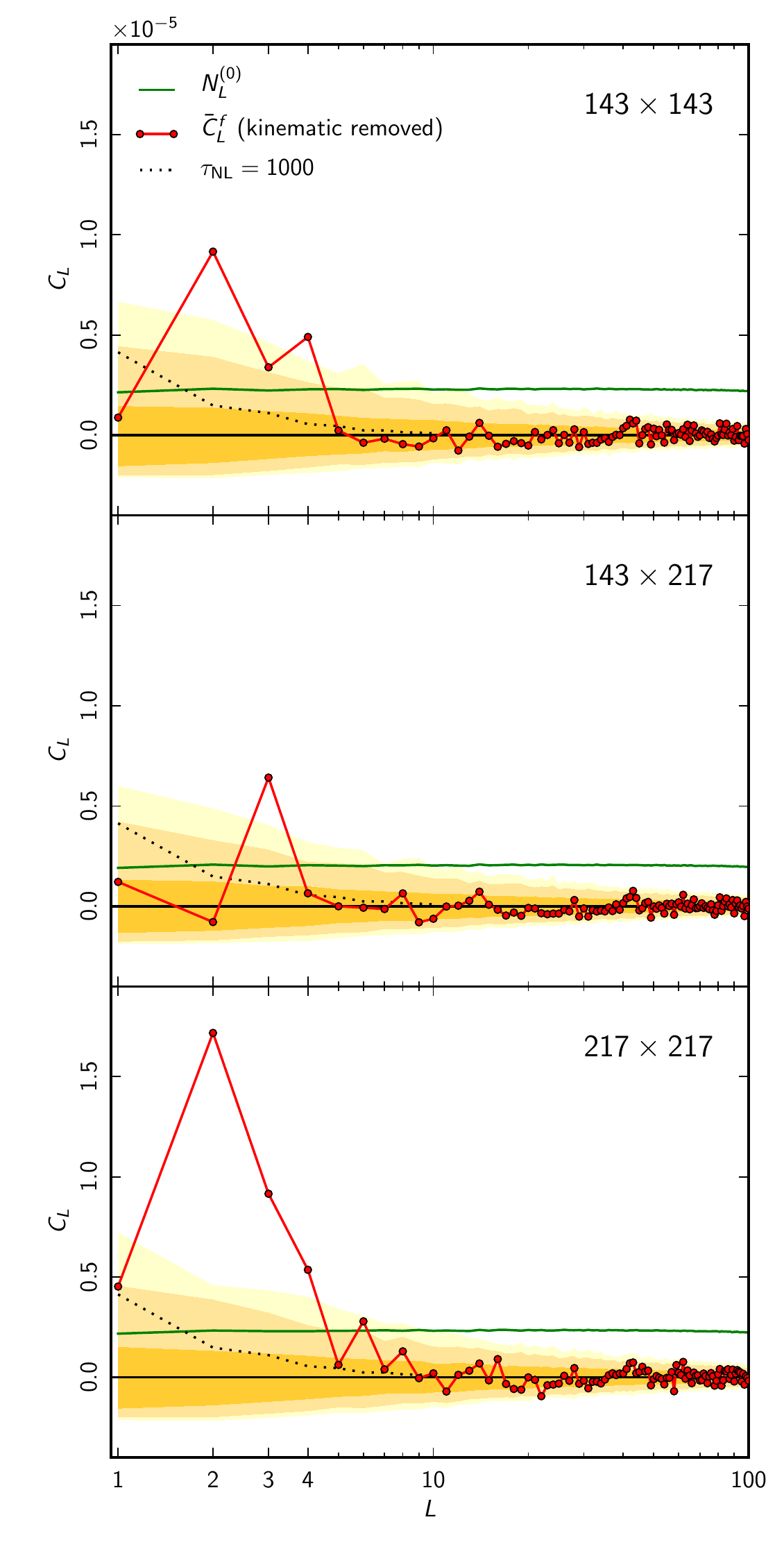}
%~~
\caption{Comparison of the un-normalized modulation power $\bar{C}^f_L$ with various combinations of frequencies. The middle plot shows the results used for our main $\taunl$ result, since it removes the significant largely-quadrupolar signal from anisotropic noise misestimation seen in the two other plots. The noticeable difference in the odd octopole signal between channels indicates that the residual signal in $143\times 217$ GHz is unlikely to be physical, but we cannot currently identify its origin.
}
\label{modulation_freq}
\end{figure}
%%%%%%%%%%%%%%%%%%%%%%%%%%%%%%%%%%%%%%%%%%%%%%%%%%%%

The $\taunl$ estimator from the $143\times 217$ GHz modulation reconstruction gives $\htaunl = 442$, compared to a null hypothesis distribution $-452<\htaunl<835$ at $95\%$ CL ($\sigma_{\taunl}\approx 335$). Our quoted error bar is assuming zero signal so that there is no signal cosmic variance contribution, and the bulk of the apparent signal is coming from the high octopole seen in Fig.~\ref{modulation_cl_results}. The alternative estimator $\hat{\tau}_{{\rm NL},1}$ gives a slightly different weighting to the octopole, giving $\hat{\tau}_{{\rm NL},1} = 569$ with an expected null-hypothesis $\sigma_{\taunl} \approx 332$. The surprisingly large difference between the estimators
can be explained as due to the large octopole signal, which has $\htaunl(L=3)\approx 6000$. However the shape of the total signal would not be expected from a genuine $\taunl$ signal, since as shown in Fig.~\ref{modulation_cl_results} on average the latter is expected to fall off approximately proportional to $1/L^2$ (i.e., a large primordial $\taunl$ would be expected in most realisations to give large dipole and quadrupole signals that we do not see). If we estimate $\taunl$ using $\Lmax=2$ we obtain $\htaunl=165$ with only a slightly larger null-hypothesis error $\sigma_{\taunl}=349$, where in this case $\hat{\tau}_{{\rm NL},1}=172$.

Note that the distribution of $\htaunl$ is quite skewed because the signal is dominated by the very-low multipole modulation power spectra which have skewed $\chi^2$-like distributions due to the large cosmic variance there (see Fig.~\ref{taunl_hist}; ~\citealt{Hanson:2009gu,Smith:2012ty}). The reconstruction noise acts like nearly-uncorrelated Gaussian white noise, so each of the $\bar{C}_L^f$ comes from a sum of squares of $\sim 2L+1$ modulation reconstruction modes; the shape of the $\htaunl$ distribution is consistent with what would come from calculating a weighted sum of $\chi^2$-distributed random variables. If we assume that any primordial modulation modes giving rise to a physical $\taunl$ signal are also Gaussian, for any given physical $\taunl$ the $\htaunl(L)$ estimators would also have $\chi^2$ distributions. This allows us to evaluate the posterior distribution of $\taunl$ given the observed $\htaunl$, in exactly the same way that one can do for the CMB temperature power spectrum. For each $L$ the posterior distribution $P(\taunl(L)|\htaunl(L))$ on the full sky would have an inverse-gamma distribution. We follow~\cite{Hamimeche:2008ai} by generalizing this to a cut-sky approximation for a  range of multipoles:
\begin{multline}
-2\ln P(\{\taunl(L)\}|\{\htaunl(L)\}) \approx  \\
 \sum_{LL'} \left[g(x(L)) N^{(0)}_{\taunl}(L)\right] \left[M^{-1}\right]_{LL'}\left[N^{(0)}_{\taunl}(L') g(x(L'))\right]
\end{multline}
where $M_{LL'}$ is the covariance of the estimators calculated from null hypothesis simulations,  $N^{(0)}_{\taunl}(L) =k_L N^{(0)}_L/C_L^{\zeta_*}$ is the estimator reconstruction noise,
\be
x(L) \equiv \frac{\htaunl(L) + N^{(0)}_{\taunl}(L)}{\taunl(L) + N^{(0)}_{\taunl}(L)},
\ee
and
\be
g(x) \equiv {\rm sign}(x-1)\sqrt{2(x-\ln(x)-1)}.
\ee
For uncorrelated $\chi^2$-distributed estimators this distribution reduces to the exact distribution, a product of inverse-gamma distributions. This approximation to the posterior can be used to evaluate the probability of any scale dependent $\taunl(L)$, and does not rely on compression into a single $\htaunl$ estimator; it can therefore fully account for the observed distribution of modulation power between $L$.
Here we focus on the main case of interest where $\taunl(L)$ is nearly-scale-invariant so that for all $L$ we have $\taunl(L)=\taunl$.

The resulting posterior distributions of $\taunl$ are shown in Fig.~\ref{taunlPosteriors} for a range of $L_{\rm max}$. These are strongly skewed, in the same way as the posterior from the low quadrupole in the CMB temperature data.
The high octopole is pulling the distributions up to higher $\taunl$ values, but increasing $L_{\rm max}$ can reduce the high-$L$ tail because very large $\taunl$ values are inconsistent with the low modulation power seen at $L\ne 3$. With $L_{\rm max}=2$ the posterior peaks near zero, but the distribution is then very broad because there are only about 8 modes, which therefore have large cosmic variance. For $L_{\rm max}=50$ we find $\taunl < 2800$ at $95\%$ CL, which we take as our upper limit.

Figure~\ref{modulation_freq} shows the modulation reconstructions for the 143 GHz and 217 GHz maps separately compared to the cross estimator.
 The picture is complicated here by the large signals from noise misestimation seen in the $143\times143$ and $217\times217$ estimators, however the fact that the octopole in $143\times143$  is \emph{lower} than in the cross-estimator indicates that the octopole signal is very unlikely to be mainly physical.
Our measured $\taunl$ limit in practice represents a strong upper limit on the level of primordial $\taunl$ that could be present, since unmodelled varying small-scale foreground or non-constant gain/calibration would also only serve to increase the measured estimate compared to primordial on average. The octopole signal does vary slightly with Galactic mask, though at present we cannot clearly isolate its origin. If more extensive analysis (for example using cross-map estimators at the same frequency) can identify a non-physical origin and remove it, the quoted upper limit on $\taunl$ would become significantly tighter. For a more extensive discussion of possible foreground and systematic effect issues that can affect 4-point estimators see~\cite{planck2013-pipaberration,planck2013-p12}.

%%%%%%%%%%%%%%%%%%%%%%%%%%%%%%%%%%%%%%%%%%%%%%%%%%%%
\begin{figure}[h]
\center
\includegraphics[width=\hsize]{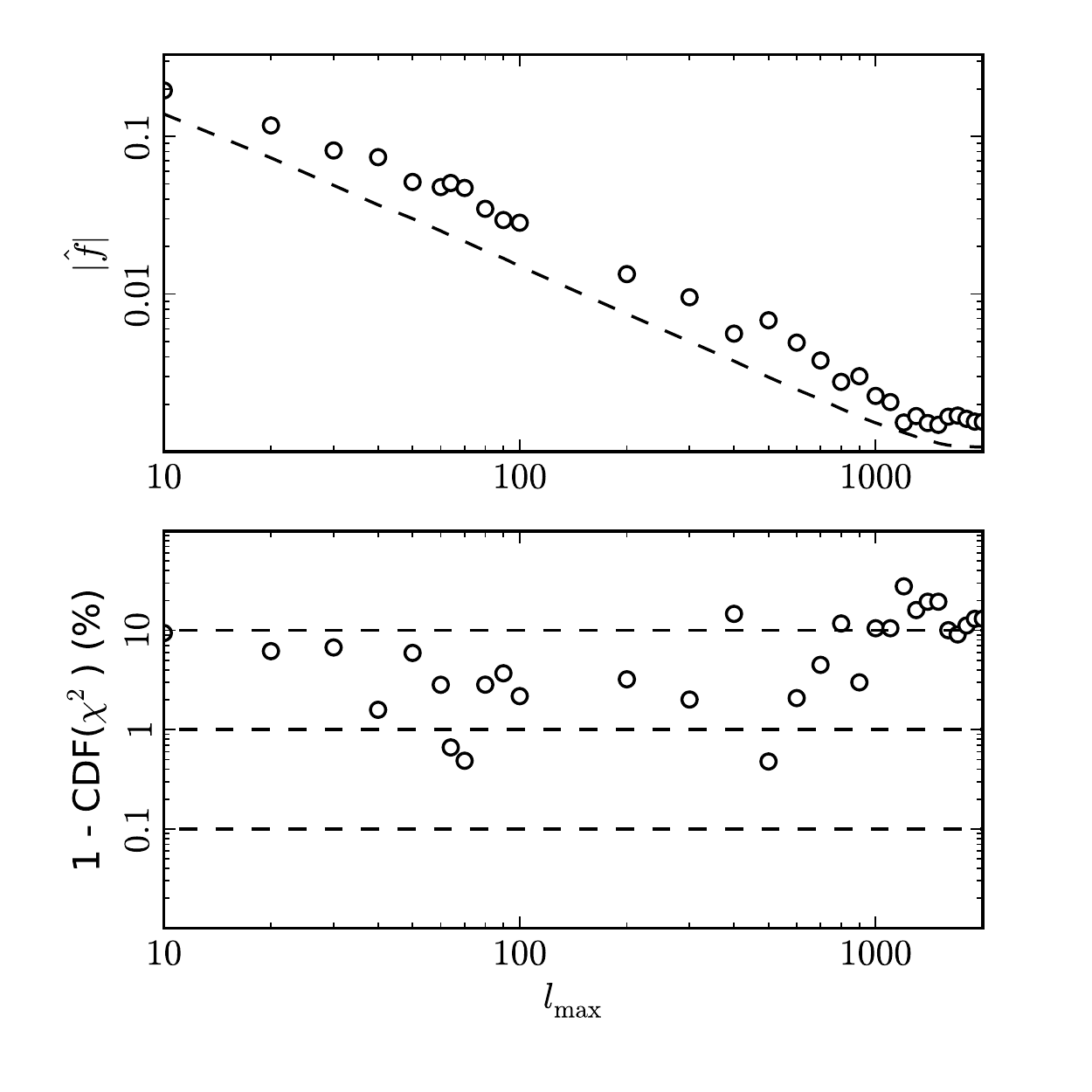}
%~~
\caption{
 The dependence of the dipole modulation amplitude as a function of $\lmax$.
\emph{Upper panel}: the amplitude $|\hat{f}|$ of the dipole of the reconstructed power modulation from $143\times 217~{\rm GHz}$ maps with the kinematic dipole subtracted; the dashed line shows that the observed signal decreases with $\lmax$ in the same way as the r.m.s. value expected from simulations.
\emph{Lower panel}: corresponding confidence values for the observed dipole of the modulation power spectrum $\bar{C}_f$ assuming it followed a chi-squared distribution; this shows the anomalous-looking results for $\lmax\sim 60$ and $\lmax\sim 600$ consistent with~\cite{planck2013-pipaberration}, but that on smaller scales the observed signal is consistent with isotropy as expected from the scale-invariant $\bar{C}_f(L=1)$ constraint using $\lmax=2000$ shown in Fig.~\ref{modulation_cl_results}.
\label{taunl_lmax}
}
\end{figure}
%%%%%%%%%%%%%%%%%%%%%%%%%%%%%%%%%%%%%%%%%%%%%%%%%%%%

We have focussed on nearly scale invariant modulations here. As discussed in~\cite{planck2013-p09} there are some potentially interesting ``anomalies'' in the distribution of power in the \planck\ data, especially the hemispherical power asymmetry 
at $\lmax \le 600$.
If these power asymmetries are physical rather than statistical flukes, they must be strongly scale-dependent, and would correspond to a scale-dependent $\taunl$-like trispectrum.
As we have shown here the dipole power asymmetry does not persist to small scales, except for the signal aligned with the CMB dipole expected from kinematic effects: from our analysis using $\lmax=2000$ we find a primordial $\htaunl(L=1)$ consistent with Gaussian simulations, corresponding to a dipole modulation amplitude $|f| \sim 0.2\%$. To be consistent with a $\sim 7\%$ modulation at $\lmax \sim 60$ and $\sim 1\%$ modulation at $\lmax~ 600$, as seen in the lower-$\ell$ anisotropies of both {\it WMAP} and \Planck\, the primordial trispectrum would have to be strongly scale-dependent on small-scale modes, so that larger small-scale modes are modulated more than the smallest ones  (rather than just $\taunl=\taunl(L)$). Fig.~\ref{taunl_lmax} shows how the allowed dipole modulation amplitude drops as $\lmax$ increases (similar to Fig.~3 of~\cite{Hanson:2009gu} for {\it WMAP}, but now extending to the higher $\lmax$ available from \planck). This result is consistent with~\cite{planck2013-p09}, where different estimators and analysis cuts also find no evidence of primordial dipole-like power asymmetry on small scales, but confirm the large-scale distribution of power seen with {\it WMAP} and also a marginally-significant smaller-amplitude ($f\sim 1\%$) dipole-like modulation at $\lmax \sim 600$.

%%%%%%%%%%%%%%% Section 8 %%%%%%%%%%%%%%%
\section{Validation of \textit{Planck} results}
\label{Sec_valid_data}

%%%%%%%%%%%%%%%%%%%%%%%%%%%%%%%%%%%%%%%%%%%%%%%%%%%%
\begin{figure*}[!t]
\centering
\includegraphics{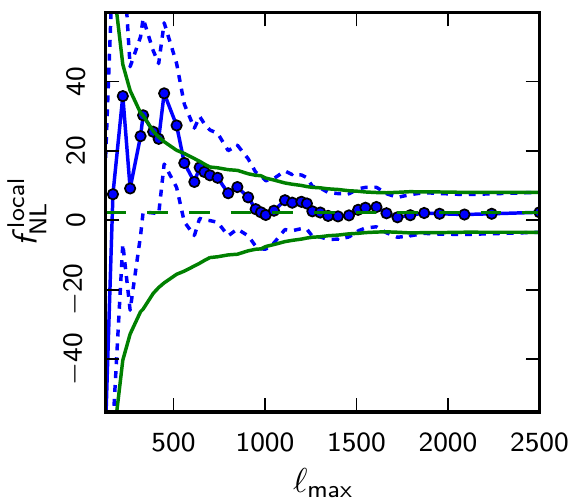}
\includegraphics{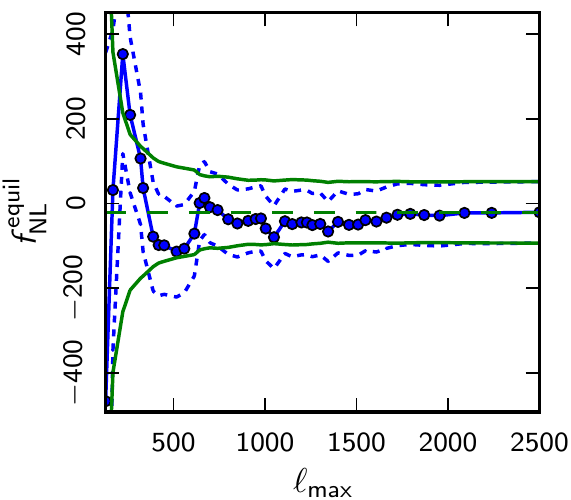}
\includegraphics{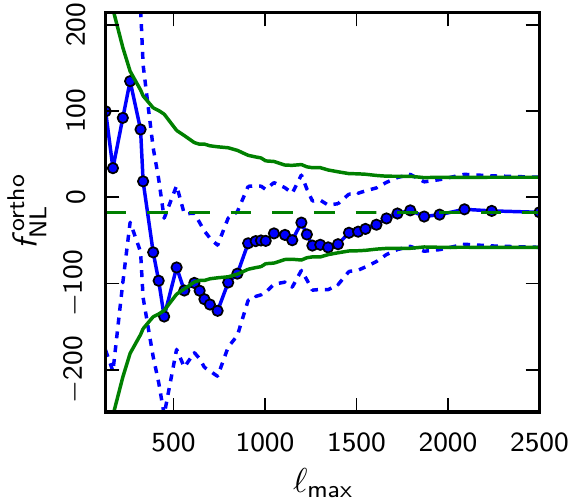}

\includegraphics{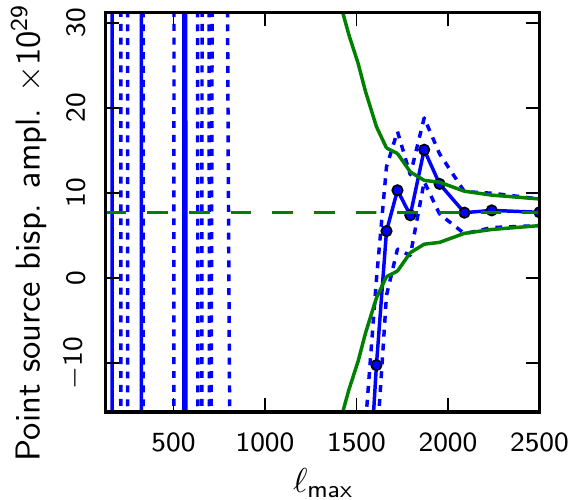}
\includegraphics{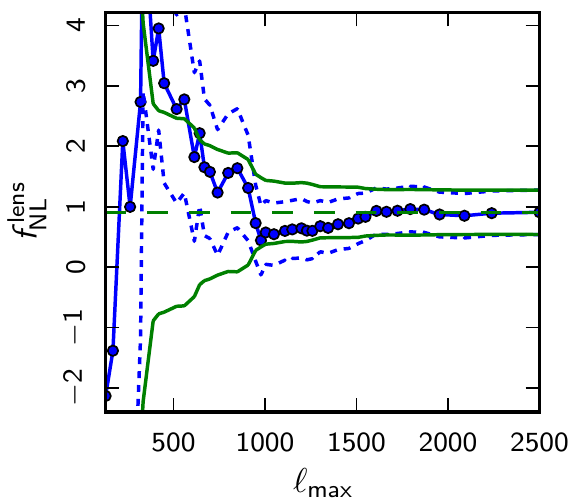}
\includegraphics{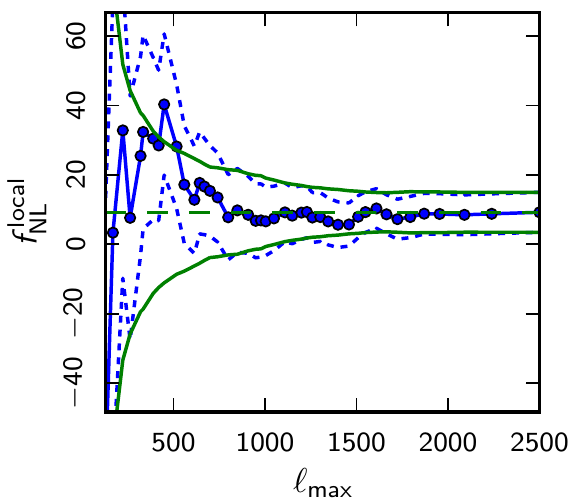}
\caption{Evolution of the $f_\mathrm{NL}$ parameters 
(solid blue line with data points) and their uncertainties (dashed lines) 
for the five bispectrum templates as a function of the 
maximum multipole number $\ell_\mathrm{max}$ used in the analysis. From 
left to right and top to bottom the figures show respectively local, 
equilateral, orthogonal, diffuse point sources (all four with the ISW-lensing
bias subtracted), ISW-lensing and local again (the last two without
subtracting the ISW-lensing bias). To show better
the evolution of the uncertainties, they are also plotted around
the final value of $f_\mathrm{NL}$ (solid green lines without data points).
The results are for \SMICA, assume all shapes to be independent, and have
been determined with the binned bispectrum estimator.}
\label{Fig_lmaxdep}
\end{figure*}
%%%%%%%%%%%%%%%%%%%%%%%%%%%%%%%%%%%%%%%%%%%%%%%%%%%%
%

%%%%%%%%%%%%%%%%%%%%%%%%%%%%%%%%%%%%%%%%%%%%%%%%%%%%
\begin{table*}[tmb]                 % table* is a two-column table.  Drop the * for one column.
\begingroup
\newdimen\tblskip \tblskip=5pt
\caption{Results for $f_\mathrm{NL}$ (assumed independent, without any correction
for the ISW-lensing bias) of the \SMICA\ 
cleaned map using different values of $\ell_\mathrm{max}$, for the KSW and 
binned estimators.} 
\label{Tab_lmaxdep}
%\nointerlineskip
\vskip -6mm
\footnotesize
\setbox\tablebox=\vbox{
   \newdimen\digitwidth 
   \setbox0=\hbox{\rm 0} 
   \digitwidth=\wd0 
   \catcode`*=\active 
   \def*{\kern\digitwidth}
   \newdimen\signwidth 
   \setbox0=\hbox{+} 
   \signwidth=\wd0 
   \catcode`!=\active 
   \def!{\kern\signwidth}
\newdimen\dotwidth
\setbox0=\hbox{.}
\dotwidth=\wd0
\catcode`;=\active
\def;{\kern\dotwidth}
\halign{\hbox to 1in{#\leaderfil}\tabskip 1em&
\hfil#\hfil\tabskip 0.8em&
\hfil#\hfil&
\hfil#\hfil&
\hfil#\hfil&
\hfil#\hfil\tabskip 0pt\cr
\noalign{\doubleline\vskip 2pt}
\omit&\multispan5\hfil $f_{\rm NL}$\hfil\cr
\omit&\multispan5\hrulefill\cr
Shape\hfill&$\ell_\mathrm{max}=500$&$\ell_\mathrm{max}=1000$&$\ell_\mathrm{max}=1500$&$\ell_\mathrm{max}=2000$&$\ell_\mathrm{max}=2500$ \cr
\noalign{\vskip 4pt\hrule\vskip 6pt}
\omit\hfil {\bf KSW}\hfil&&&\cr
Local & *!38;* $\pm$ *18;* & !*6.4* $\pm$ *9.7* & !*6.9* $\pm$ *6.2* & !*9.1* $\pm$ *5.8* & !*9.8* $\pm$ *5.8*\cr
Equilateral & $-$119;* $\pm$ 121;* & $-$45;** $\pm$ 88;** & $-$41;** $\pm$ 75;** & $-$40;** $\pm$ 75;** & $-$37;** $\pm$ 75;**\cr
Orthogonal & $-$163;* $\pm$ 109;* & $-$89;** $\pm$ 52;** & $-$57;** $\pm$ 45;** & $-$45;** $\pm$ 40;** & $-$46;** $\pm$ 39;**\cr
Diff.ps $/10^{-29}$ & ($-$1.5 $\pm$ 1.3)$\times 10^4$ & ($-$7.9 $\pm$ 3.1)$\times 10^2$ & $-$39;** $\pm$ 18;** & !10.0* $\pm$ *3.1* & !*7.7* $\pm$ *1.5* \cr
ISW-lensing &  !**3.2 $\pm$ **1.2 & !*1.00 $\pm$ *0.43 & !*1.00 $\pm$ *0.35 & !*0.83 $\pm$ *0.31 & !*0.81 $\pm$ *0.31\cr
\noalign{\vskip 5pt}
%\hline
%\noalign{\vskip 5pt}
\omit\hfil {\bf Binned}\hfil&&&\cr
Local & !*33;* $\pm$ *18;* & !*6.6* $\pm$ *9.8* & !*7.1* $\pm$ *6.1* & !*8.5* $\pm$ *5.9* & !*9.2* $\pm$ *5.9*\cr
Equilateral & *$-$95;* $\pm$ 107;* & $-$55;** $\pm$ 77;** & $-$47;** $\pm$ 72;** & $-$22;** $\pm$ 73;** & $-$20;** $\pm$ 73;**\cr 
Orthogonal & $-$102;* $\pm$ *94;* & $-$69;** $\pm$ 58;** & $-$60;** $\pm$ 44;** & $-$35;** $\pm$ 40;** & $-$39;** $\pm$ 41;**\cr
Diff.ps $/10^{-29}$ & ($-$1.4 $\pm$ 1.2)$\times 10^4$ & ($-$8.2 $\pm$ 2.9)$\times 10^2$ & $-$42;** $\pm$ 17;** & !*9.9* $\pm$ *2.9* & !*7.7* $\pm$ *1.6*\cr
ISW-lensing & !**2.6 $\pm$ **1.6 & !*0.57 $\pm$ *0.52 & !*0.80 $\pm$ *0.42 & !*0.85 $\pm$ *0.38 & !*0.91 $\pm$ *0.37\cr
\noalign{\vskip 3pt\hrule\vskip 4pt}}}	
                      % Template goes here. 
%\noalign{\doubleline}
                                    % Table headings go here.
%\noalign{\vskip 3pt\hrule\vskip 5pt}
                                    % Body of table goes here.
%\noalign{\vskip 5pt\hrule\vskip 3pt}}}
%\endPlancktable                    % ends one-column \halign
\endPlancktablewide                 % ends two-column \halign
%\tablenote a Footnote a.\par
%\tablenote b Footnote b.\par
\endgroup
\end{table*}                        % table* is a two-column table.  Drop the * for one column.
%%%%%%%%%%%%%%%%%%%%%%%%%%%%%%%%%%%%%%%%%%%%%%%%%%%%

Here we perform a set of tests to check the robustness 
and stability of our $\fnl$ measurements. As these are validation tests of {\it Planck} results, 
and not internal comparisons of bispectrum pipelines (already shown to be in tight agreement 
in Sect.~\ref{sec:Validation} and \ref{sec:Results})
we will not employ all the bispectrum estimators on each test. In general 
we choose to use two estimators on each test, in order to have a cross-check 
of the outcomes without excessive redundancy. 

%

%%%%%%%%%%%%%%%%%%%%%%%%%%%%%%%%%%%%%%%%%%%%%%%%%%%%
\begin{table*}[tmb]                 % table* is a two-column table.  Drop the * for one column.
\begingroup
\newdimen\tblskip \tblskip=5pt
\caption{Results for $f_\mathrm{NL}$ (assumed independent) of the \SMICA\
cleaned map using different masks as described in the main text (Sect.~\ref{sec:maskdep}). 
Results are given for the binned and modal estimators. Uncertainties
for the binned estimator in this table and the next are Fisher error bars. The
modal estimator uses a faster and slightly less correlated expansion of 
the primordial templates for this test and the next than for other analyses 
(see Sect.~\ref{sec:maskdep} for more explanations). These caveats explain why the 
results shown in this table for the $f_{\rm sky} = 0.73$ mask
display small differences with respect to the corresponding numbers in the 
main results tables of Sect.~\ref{sec:Results}, for both estimators.
We also note that the binned estimator uses $\ell_\mathrm{max}=2500$ and the 
modal estimator $\ell_\mathrm{max}=2000$, which has an impact on the point 
source results as explained in the main text.}
\label{Tab_maskcomp}
%\nointerlineskip
\vskip -6mm
\footnotesize
\setbox\tablebox=\vbox{
   \newdimen\digitwidth 
   \setbox0=\hbox{\rm 0} 
   \digitwidth=\wd0 
   \catcode`*=\active 
   \def*{\kern\digitwidth}
   \newdimen\signwidth 
   \setbox0=\hbox{+} 
   \signwidth=\wd0 
   \catcode`!=\active 
   \def!{\kern\signwidth}
\newdimen\dotwidth
\setbox0=\hbox{.}
\dotwidth=\wd0
\catcode`;=\active
\def;{\kern\dotwidth}
\halign{\hbox to 1in{#\leaderfil}\tabskip 1em&
\hfil#\hfil\tabskip 0.8em&
\hfil#\hfil&
\hfil#\hfil&
\hfil#\hfil\tabskip 0pt\cr
\noalign{\doubleline\vskip 2pt}
\omit&\multispan4\hfil $f_{\rm NL}$\hfil\cr
\omit&\multispan4\hrulefill\cr
Shape\hfill&$f_\mathrm{sky}=0.89$&$f_\mathrm{sky}=0.73$&$  f_\mathrm{sky}=0.56 $&$ f_\mathrm{sky}=0.32$\cr
\noalign{\vskip 4pt\hrule\vskip 6pt}
\omit\hfil {\bf Binned}\hfil&&&\cr
Local & !13.0* $\pm$ *5.4* & !*9.2* $\pm$ *5.9* & !11;** $\pm$ *6.8* & !*6.1* $\pm$ **8.9* \cr
Equilateral & !35;** $\pm$ 66;** & $-$20;** $\pm$ 73;** & $-$20;** $\pm$ 83;** & !39;** $\pm$ 109;** \cr
Orthogonal & $-$18;** $\pm$ 36;** & $-$39;** $\pm$ 39;** & $-$45;** $\pm$ 45;** & *$-$5;** $\pm$ *59;** \cr
Diff.ps $/10^{-29}$ & !14.0* $\pm$ *1.3* & !*7.7* $\pm$ *1.4* & !*9.0* $\pm$ *1.7* & !10.3* $\pm$ **2.2* \cr
ISW-lensing & !*0.69 $\pm$ *0.26 & !*0.91 $\pm$ *0.29 & !*0.84 $\pm$ *0.33 & !*0.81 $\pm$ **0.43 \cr
\noalign{\vskip 5pt}
%\hline
%\noalign{\vskip 5pt}
\omit\hfil {\bf Modal}\hfil&&&\cr
Local &  12.1* $\pm$ *5.5*   &  !*8.4* $\pm$ *6.0* &  !12.3 $\pm$ *7.1* &  !*9.2* $\pm$ **8.7*  \cr
Equilateral & 52;** $\pm$  66;**  &   $-$56;** $\pm$ 72;**  &  $-$31;* $\pm$ 84;** &  !42;** $\pm$ 104;**  \cr 
Orthogonal & *3.3* $\pm$ 35;**  &  $-$31;** $\pm$ 40;** & $-$50;* $\pm$ 47;** &  $-$27;** $\pm$ *59;** \cr
Diff.ps  $/10^{-29}$ & 20.6* $\pm$ *2.5*  &  !11.4* $\pm$ *2.8* &  !10.7 $\pm$ *3.2* & !12.7* $\pm$ **3.9* \cr
ISW-lensing & *0.42 $\pm$ *0.35 &  !*0.62 $\pm$ *0.40 &  !*1.1 $\pm$ *0.45 & !*0.80 $\pm$ **0.48  \cr
\noalign{\vskip 3pt\hrule\vskip 4pt}}}	
                      % Template goes here. 
%\noalign{\doubleline}
                                    % Table headings go here.
%\noalign{\vskip 3pt\hrule\vskip 5pt}
                                    % Body of table goes here.
%\noalign{\vskip 5pt\hrule\vskip 3pt}}}
%\endPlancktable                    % ends one-column \halign
\endPlancktablewide                 % ends two-column \halign
%\tablenote a Footnote a.\par
%\tablenote b Footnote b.\par
\endgroup
\end{table*}                        % table* is a two-column table.  Drop the * for one column.
%%%%%%%%%%%%%%%%%%%%%%%%%%%%%%%%%%%%%%%%%%%%%%%%%%%%

%%%%%%%%%%%%%%%%%%%%%%%%%%%%%%%%%%%%%%%%%%%%%%%%%%%%

\subsection{Dependence on maximum multipole number}
\label{Sec_deplmax}

The dependence on the maximum multipole number $\ell_\mathrm{max}$ of the
\SMICA\ results (assuming independent shapes) is shown in 
Fig.~\ref{Fig_lmaxdep} (for the binned estimator) and Table~\ref{Tab_lmaxdep}
(for both the KSW and binned estimators). Testing the $\ell_\mathrm{max}$
dependence is easiest for the binned estimator, where one can simply omit
the highest bins in the final sum when computing $f_\mathrm{NL}$.
It is clear that we 
have reached convergence both for the values of $f_\mathrm{NL}$ and for their 
error bars at $\ell_\mathrm{max}=2500$, with the possible exception of the 
error bars of the diffuse point-source bispectrum. The diffuse point-source 
bispectrum template is dominated by equilateral configurations at high $\ell$.
Moreover, for all the shapes except point sources, results at 
$\ell_\mathrm{max}=2000$ are very close to those at 
$\ell_\mathrm{max}=2500$, taking into account the size of the error bars. 

It is very interesting to see that at $\ell_\mathrm{max}\sim 500$ we find
a local $f_\mathrm{NL}$ result in very good agreement with the {\it WMAP}-9 
value of $39.8 \pm 19.9$~\citep{2012arXiv1212.5225B} (or $37.2 \pm 19.9$ 
after ISW-lensing bias subtraction).
At these low $\ell_\mathrm{max}$ values we also 
find negative values for orthogonal $f_\mathrm{NL}$, although not as large 
or significant as the {\it WMAP}-9 value (which is $-245 \pm 100$). One can
clearly see the importance of the higher resolution of \Planck\, both for
the values of the different $f_\mathrm{NL}$ parameters and for their error
bars.

It is also clear that the higher resolution of {\it Planck} is
absolutely crucial for the ISW-lensing bispectrum; this is simply
undetectable at {\it WMAP} resolution.  On the other hand, the high
sensitivity of {\it Planck} measurements also exposes us to a larger
number of potentially spurious effects. For example we see that the
bispectrum of point sources is also detected at high significance by
{\it Planck} at $\ell_\mathrm{max}\geq 2000$, while remaining undetectable
at lower resolutions. The presence of this bipectrum in the data could
in principle contaminate our primordial $\fnl$ measurements.  For this
reason, the presence of a large point source signal has been
accounted for in previous Sections by always including the Poisson
bispectrum in a joint fit with primordial shapes. Fortunately, it
turns out that the very low correlation between the primordial
templates and the Poisson one makes the latter a negligible
contaminant for $\fnl$, even when the residual point source amplitude
is large.

%%%%%%%%%%%%%%%%%%%%%%%%%%%%%%%%%%%%%%%%%%%%%%%%%%%%

\subsection{Dependence on mask and consistency between frequency channels}
\label{sec:maskdep}

%

%%%%%%%%%%%%%%%%%%%%%%%%%%%%%%%%%%%%%%%%%%%%%%%%%%%%
\begin{table*}[tmb]                 % table* is a two-column table.  Drop the * for one column.
\begingroup
\newdimen\tblskip \tblskip=5pt
\caption{Results for $f_\mathrm{NL}$ (assumed independent) for the raw frequency
maps at 70, 100, 143, and 217~GHz with a very large mask ($f_\mathrm{sky}=0.32$)
compared to the \SMICA\ result with the union mask U73 ($f_\mathrm{sky}=0.73$),
as determined by the binned (with $\ell_\mathrm{max}=2500$) and modal 
(with $\ell_\mathrm{max}=2000$) estimators. The same caveats as for the
previous table (Table~\ref{Tab_maskcomp}) apply here as well.
}
\label{Tab_freqcomp}
\nointerlineskip
\vskip -3mm
\footnotesize
\setbox\tablebox=\vbox{
   \newdimen\digitwidth 
   \setbox0=\hbox{\rm 0} 
   \digitwidth=\wd0 
   \catcode`*=\active 
   \def*{\kern\digitwidth}
   \newdimen\signwidth 
   \setbox0=\hbox{+} 
   \signwidth=\wd0 
   \catcode`!=\active 
   \def!{\kern\signwidth}
\newdimen\dotwidth
\setbox0=\hbox{.}
\dotwidth=\wd0
\catcode`;=\active
\def;{\kern\dotwidth}
\halign{\hbox to 1in{#\leaderfil}\tabskip 1em&
\hfil#\hfil\tabskip 0.8em&
\hfil#\hfil&
\hfil#\hfil&
\hfil#\hfil&
\hfil#\hfil\tabskip 0pt\cr
\noalign{\doubleline\vskip 2pt}
\omit&\multispan5\hfil $f_{\rm NL}$\hfil\cr
\omit&\multispan5\hrulefill\cr
Shape\hfill&\SMICA&70 GHz&100 GHz&143 GHz&217 GHz\cr
\noalign{\vskip 4pt\hrule\vskip 6pt}
\omit\hfil {\bf Binned}\hfil&&&\cr
Local & !*9.2* $\pm$ *5.9* & !*19.7 $\pm$ *26.0 & **$-$2.5* $\pm$ *13.2* & !*10.4* $\pm$ **9.8* & **$-$4.7* $\pm$ **9.6*  \cr
Equilateral & $-$20;** $\pm$ 73;** & !159;* $\pm$ 188;* & *!70;** $\pm$ 132;** & *!48;** $\pm$ 114;** & **$-$9;** $\pm$ 114;**   \cr
Orthogonal & $-$39;** $\pm$ 39;** & *$-$78;* $\pm$ 139;*  &  $-$106;** $\pm$ *81;** & $-$101;** $\pm$ *64;** & *$-$84;** $\pm$ *63;**  \cr
Diff.ps $/10^{-29}$ & !*7.7* $\pm$ *1.4* & ($-$1.4 $\pm$ 2.3)$\times 10^3$ & **$-$4.0* $\pm$ *64;** & !**8.7* $\pm$ **6.1* & !*14.2* $\pm$ **3.0*   \cr
ISW-lensing & !*0.91 $\pm$ *0.29 & !**3.5 $\pm$ **2.2 & **!0.35 $\pm$ **0.78 & !**0.89 $\pm$ **0.50 & !**0.87 $\pm$ **0.48  \cr
\noalign{\vskip 5pt}
%\hline
%\noalign{\vskip 5pt}
\omit\hfil {\bf Modal}\hfil&&&\cr
Local               &  !*8.4*  $\pm$ *6.0*   & !*36.5 $\pm$ *27.2  & **$-$6.6*  $\pm$ *13.6*  & !**6.6*    $\pm$ **9.4*    & **$-$6.5*  $\pm$ **8.9*	    \cr
Equilateral         & $-$56;** $\pm$ 72;**  & *!74;* $\pm$ 193;*  & *!49;**  $\pm$ 123;** &  *!81;** $\pm$ 111;** & !*29;**  $\pm$ 110;**    \cr
Orthogonal          & $-$31;** $\pm$ 40;**  & $-$225;* $\pm$ 119;* & *$-$75;** $\pm$ *80;**  &  $-$133;** $\pm$ *62;** & $-$112;** $\pm$ *61;**     \cr
Diff.ps $/10^{-29}$ &  !11.4* $\pm$ *2.8*   & (-2.5 $\pm$ 2.8)$\times 10^3$  & *$-$45;** $\pm$ *64;** & !**5.7*    $\pm$ **7.0*    & *!25;** $\pm$ **5.0*      \cr
ISW-lensing         &  !*0.62 $\pm$ *0.40   & !**2.6 $\pm$ **2.3 & !**0.92  $\pm$ **0.80  &  **!0.78   $\pm$ **0.60   & **!0.85   $\pm$ **0.56     \cr
\noalign{\vskip 3pt\hrule\vskip 4pt}}}
\endPlancktablewide                 % ends two-column \halign
%\tablenote a Footnote a.\par
%\tablenote b Footnote b.\par
\endgroup
\end{table*}                        % table* is a two-column table.  Drop the * for one column.
%%%%%%%%%%%%%%%%%%%%%%%%%%%%%%%%%%%%%%%%%%%%%%%%%%%%

To test the dependence on the mask, we have analysed the \SMICA\ maps applying four
different masks. Firstly the union mask U73 used for the final results in Sect.~\ref{sec:Results}, 
which leaves 73\% of the sky unmasked. Secondly we used the confidence mask CS-SMICA89
of the \SMICA\ technique, which leaves 89\% of the sky.
% after masking the region of the sky where the \SMICA\ foregrounds separation is not trusted. 
Next, a bigger mask constructed by multiplying the union mask U73 with the 
 {\it Planck} Galactic mask CG60,
leading to a mask that leaves 56\% of the sky. And
finally a very large mask, leaving only 32\% of the sky, which is the union of the
mask CL31 - used for power spectrum estimation on the raw frequency maps - with the union mask U73 (for mask details see \citealt{planck2013-p06} for U73, CS-SMICA89, and CG60;~\citealt{planck2013-p08} for CL31).
The results of this analysis are presented in Table~\ref{Tab_maskcomp}
for two different estimators: binned and modal. The $f_\mathrm{NL}$ are
assumed independent here. In order to correctly interpret our results and 
conclusions, an important point to note is that binned results 
have been obtained choosing $\ell_\mathrm{max} = 2500$, while 
modal results correspond to $\ell_\mathrm{max} = 2000$. Primordial shape 
and ISW-lensing results and error bars saturate at $\ell_\mathrm{max} = 2000$ 
(see Sect.~\ref{Sec_deplmax}), so the results from the two estimators are 
directly comparable in this case. The Poisson (point sources) bispectrum is 
however dominated by high-$\ell$ equilateral configurations and the signal 
for this specific template still changes from $\ell=2000$ to $\ell=2500$. 
The differences in central values and uncertainties between the two estimators 
for the Poisson shape are fully consistent with the different 
$\ell_\mathrm{max}$ values. Direct comparisons on data and simulations
between these two estimators and the KSW estimator showed that Poisson 
bispectrum results match each other very well when the same 
$\ell_\mathrm{max}$ is used.

Results from the modal pipeline have uncertainties determined from
MC simulations, while the results from the binned pipeline
(in Table~\ref{Tab_maskcomp} and the next only) are given with Fisher error bars. It
is very interesting to see that even with the large 
$f_\mathrm{sky}=0.32$ mask, the simple inpainting technique still
allows us to saturate the (Gaussian) Cram\' er-Rao bound, except for the
ISW-lensing shape where we have a significant detection of
NG in a squeezed configuration (so that an error estimate
assuming Gaussianity is not good enough). Finally we note that only for the tests in this and in the next 
 paragraph we adopted a faster but slightly less accurate version of the modal estimator  
  than the one used to obtain the final $\fnl$ constraints in Sect.~\ref{sec:Results}. In this faster 
 implementation we use fewer modes in order to increase computational speed, and consequently we get a 
slight degrading of the level of correlation of our expanded templates with the initial primordial shapes. 
Note that the changes  are small: we go from $99\%$ correlation for local, equilateral and orthogonal 
 shapes in the most accurate (and slower) implementation to $98 \%$ correlation for equilateral and orthogonal snapes, 
 and $95 \%$ correlation for local shape in the faster version. This of course still allows for very stringent validation 
tests for all the primordial shapes, and produces results very close to the high-accuracy pipeline, 
while at the same time increasing overall speed by almost a factor 2. Both versions of the modal pipeline were 
 separately validated on simulations (see Sect.~\ref{sec:Validation}).

Besides confirming again the good level of agreement between the two
estimators already discussed in Sects.~\ref{sec:Validation} and \ref{sec:Results},
the main conclusion we draw from this analysis is that our
measurements for all shapes are robust to changes in sky coverage, taking 
into account the error bars and significance levels, at
least starting from a certain minimal mask. The $f_{\rm sky}=0.89$
mask is probably a bit too small, likely leaving foreground
contamination around the edges of the mask, though even for this mask
the results are consistent within $1\sigma$, except for point
sources (which might suggest the presence of residual Galactic point source contamination 
for the small mask). The results from the $f_{\rm sky}=0.73$ and $f_{\rm sky}=0.56$ 
masks are highly consistent. This conclusion does not
really change when going down to $f_{\rm sky}=0.32$, although uncertainties 
of course start increasing significantly for this large mask.

We also investigate if there is consistency between frequency channels
when the largest mask with $f_\mathrm{sky}=0.32$ is used, and if these
results agree with the \SMICA\ results obtained with the common
mask. The results (assuming independent $f_\mathrm{NL}$) are given
both for the binned and the modal estimator in
Table~\ref{Tab_freqcomp}. As in the previous table, the full modal
pipeline (faster but slightly less accurate version) has been run here, 
obtaining both central values from data
and MC error bars from simulations, while the binned pipeline
(which is slower in determining full error bars than the modal pipeline)
is used to cross-check the modal measurements and has error bars given
by simple Fisher matrix estimates. As one can see here and as was also
checked explicitly in many other cases, the error bars from different
estimators are perfectly consistent with each other and saturate the
Cram\' er-Rao bound (except in the case of a significant non-Gaussian
ISW-lensing detection).

A detailed analysis of Table~\ref{Tab_freqcomp} might actually suggest that 
the agreement between the two estimators employed for this test, although still 
 clearly good, is slightly 
degraded when compared to their performance on clean maps from different 
component separation pipelines.
If we compare e.g., \SMICA\ results in Table~\ref{Tab_maskcomp} to raw data 
results in Table~\ref{Tab_freqcomp}, we see that in the former case 
the discrepancy between the two estimators is at most of order 
${\sigma_{f_{\rm NL}}/3}$, and smaller in most cases. In the latter case, however, 
we notice several measurements displaying differences of order $\sigma_{f_{\rm NL}}/2$ 
between the two pipelines, and the value of $f_\textrm{NL}^\textrm{ortho}$
at 70~GHz being ${1\sigma}$ away. 
We explain these larger differences as follows.
For \SMICA\ runs we calibrated
the estimator linear terms using FFP6 simulations,
accurately reproducing noise properties and correlations (see Appendix~\ref{sec:FFP6}). On the other hand,
for the present tests on raw frequency channels we adopted a simple noise 
model, based on generating uncorrelated noise multipoles with a power spectrum 
as extracted from the half-ring null map, and remodulating the noise 
in pixel space according to the hit-count distribution. This approximation is
expected to degrade the accuracy of the linear term calibration, and
thus to produce a slightly lower agreement of different pipelines for
shapes where the linear correction is most important. Those are
the shapes that take significant contributions from squeezed
triangles: local and ISW-lensing, and to a smaller but non-negligible 
extent orthogonal, i.e., exactly the shapes for which we find slightly 
larger differences.

We conclude that no significant fluctuations are observed when comparing 
measurements from different frequency channels (between themselves or
with the clean and co-added \SMICA\ map) or from different estimators 
on a given channel for the primordial shapes. The same is true for the
ISW-lensing shape, although it should be noted that in particular the 
70~GHz channel (like {\it WMAP}) does not have sufficient resolution to
measure either the lensing or point source contributions. The uncertainties of the point source 
contribution vary significantly between frequency channels, although
results remain consistent between channels given the error bars
(when all multipoles up to $\ell_\mathrm{max}=2500$ are taken into account,
as performed by the binned estimator). This is due to the fact that this shape 
is dominated by high-$\ell$ equilateral configurations, the signal-to-noise 
of which depends crucially on the beam and noise characteristics, which vary 
from channel to channel. In the \SMICA\ map point sources are partially removed 
by foreground cleaning, explaining the significantly lower value than for 
217~GHz. As explained before, differences between the binned and modal 
estimators regarding point sources are due to the different values of 
$\ell_\mathrm{max}$ (2500 for binned and 2000 for modal), which particularly
affects the 217~GHz channel and the \SMICA\ cleaned map. 

%%%%%%%%%%%%%%%%%%%%%%%%%%%%%%%%%%%%%%%%%%%%%%%%%%%%

\subsection{Null tests}
%

%%%%%%%%%%%%%%%%%%%%%%%%%%%%%%%%%%%%%%%%%%%%%%%%%%%%
\begin{table*}[tmb]                 % table* is a two-column table.  Drop the * for one column.
\begingroup
\newdimen\tblskip \tblskip=5pt
\caption{Results for $f_\mathrm{NL}$ (assumed independent) of the \SMICA\  half-ring null maps, determined by the KSW, binned and modal estimators.}
\label{Tab_jaccknife_SMICA}
\nointerlineskip
\vskip -3mm
\footnotesize
\setbox\tablebox=\vbox{
   \newdimen\digitwidth 
   \setbox0=\hbox{\rm 0} 
   \digitwidth=\wd0 
   \catcode`*=\active 
   \def*{\kern\digitwidth}
   \newdimen\signwidth 
   \setbox0=\hbox{+} 
   \signwidth=\wd0 
   \catcode`!=\active 
   \def!{\kern\signwidth}
   \newdimen\dotwidth 
   \setbox0=\hbox{.} 
   \dotwidth=\wd0 
   \catcode`;=\active 
   \def;{\kern\dotwidth}
   \halign{\hbox to 1in{#\leaderfil}\tabskip 1em&
\hfil#\hfil\tabskip 0.8em&
\hfil#\hfil&
\hfil#\hfil&
\hfil#\hfil\tabskip 0pt\cr
\noalign{\doubleline\vskip 2pt}
\omit&\multispan3\hfil $\fnl$\hfil\cr
\omit&\multispan3\hrulefill\cr
Shape\hfill&KSW&Binned&Modal\cr
\noalign{\vskip 4pt\hrule\vskip 6pt}
\omit\hfil \SMICA\  {\bf half-ring}\hfil&&&\cr
Local & $-$0.008 $\pm$ 0.18 & $-$0.086 $\pm$ 0.20 & $-$0.13* $\pm$ 0.35* \cr
Equilateral & $-$0.16*  $\pm$ 2.2* & !1.3** $\pm$ 2.1* & !0.66* $\pm$ 2.0** \cr
Orthogonal & $-$0.035  $\pm$ 0.57 & !0.51* $\pm$ 0.57 & !0.14* $\pm$ 0.60* \cr
Diff.ps $/10^{-29}$ & $-$0.05* $\pm$ 0.60 & !0.03* $\pm$ 0.68 & !0.05* $\pm$ 0.65* \cr
ISW-lensing & (-0.06 $\pm$ 2.0)$\times 10^{-3}$ & ($-$2.2 $\pm$ 4.7)$\times 10^{-3}$ & !0.009 $\pm$ 0.030 \cr
\noalign{\vskip 3pt\hrule\vskip 4pt}}}
\endPlancktablewide                 % ends two-column \halign
\endgroup
\end{table*}                        % table* is a two-column table.  Drop the * for one column.
%%%%%%%%%%%%%%%%%%%%%%%%%%%%%%%%%%%%%%%%%%%%%%%%%%%%

%

%%%%%%%%%%%%%%%%%%%%%%%%%%%%%%%%%%%%%%%%%%%%%%%%%%%%
\begin{table*}[tmb]                 % table* is a two-column table.  Drop the * for one column.
\begingroup
\newdimen\tblskip \tblskip=5pt
\caption{Results for $f_\mathrm{NL}$ (assumed independent) of several null
maps determined by the binned estimator. We consider half-ring
$(r1-r2)/2$, survey $(s2-s1)/2$, and detector set $(d1-d2)/2$ difference maps for \SMICA\
and the raw 143 GHz channel.}
\label{Tab_jackknives}
\nointerlineskip
\vskip -3mm
\footnotesize
\setbox\tablebox=\vbox{
   \newdimen\digitwidth 
   \setbox0=\hbox{\rm 0} 
   \digitwidth=\wd0 
   \catcode`*=\active 
   \def*{\kern\digitwidth}
   \newdimen\signwidth 
   \setbox0=\hbox{+} 
   \signwidth=\wd0 
   \catcode`!=\active 
   \def!{\kern\signwidth}
   \newdimen\dotwidth 
   \setbox0=\hbox{.} 
   \dotwidth=\wd0 
   \catcode`;=\active 
   \def;{\kern\dotwidth}
\halign{\hbox to 1in{#\leaderfil}\tabskip 1em&
\hfil#\hfil\tabskip 0.8em&
\hfil#\hfil&
\hfil#\hfil&
\hfil#\hfil\tabskip 0pt\cr
\noalign{\doubleline\vskip 2pt}
\omit&\multispan4\hfil $\fnl$\hfil\cr
\omit&\multispan4\hrulefill\cr
\omit&\SMICA&143 GHz&143 GHz&143 GHz\cr
Shape\hfill&half-ring&half-ring&survey&detector set\cr
\noalign{\vskip 4pt\hrule\vskip 6pt}
\omit\hfil {\bf Binned}\hfil&&&\cr
Local & $-$0.086 $\pm$ 0.20 & $-$0.016 $\pm$ 0.073 & !0.43 $\pm$ 0.56 & !1.9** $\pm$ 1.7**  \cr
Equilateral & !1.3** $\pm$ 2.1* & !3.2** $\pm$ 1.8** & $-$1.5* $\pm$ 4.2* & !0.9** $\pm$ 5.8** \cr
Orthogonal & !0.51* $\pm$ 0.57 & !1.2** $\pm$ 0.6** & $-$1.7* $\pm$ 1.3*  & $-$1.3** $\pm$ 1.8** \cr
Diff.ps  $/10^{-29}$ & !0.03* $\pm$ 0.68 & !0.2** $\pm$ 1.9** & !3.4* $\pm$ 3.2*  & $-$1.0** $\pm$ 4.3** \cr
ISW-lensing & ($-$2.2 $\pm$ 4.7)$\times 10^{-3}$ & ($-$0.5 $\pm$ 1.7)$\times 10^{-3}$ & ($-$0.6 $\pm$ 11)$\times 10^{-3}$ & !0.033 $\pm$ 0.026  \cr
\noalign{\vskip 3pt\hrule\vskip 4pt}}}
\endPlancktablewide                 % ends two-column \halign
\endgroup
\end{table*}                        % table* is a two-column table.  Drop the * for one column.
%%%%%%%%%%%%%%%%%%%%%%%%%%%%%%%%%%%%%%%%%%%%%%%%%%%%

%
To make sure there is no hidden NG in the instrumental
noise, we performed a set of tests on null maps. These
are noise-only maps obtained from differences between maps with 
the same sky signal. In
the first place we constructed half-ring null maps, i.e., maps
constructed by taking the difference between the first and second
halves of each pointing period, divided by 2.  Secondly, we
constructed a survey difference map (Survey 2 minus Survey 1 divided by
2). A ``survey" is defined as half a year of data, roughly the time
needed to scan the full sky once; the nominal period of {\it Planck}
data described by these papers contains two full surveys. Finally we
constructed the detector set difference map (``detset 1'' minus ``detset 2''
divided by 2). The four polarized detectors at each frequency from
100 to 353~GHz are split into two detector sets per frequency, in such a 
way that each set can measure all polarizations and the detectors in a set are
aligned in the focal plane (see \cite{planck2013-p03} and \cite{planck2013-p06} for details 
on the null maps analysed in this Section).

All these maps are analysed using the union mask U73 used for the final
data results. However, in the case of the survey and detector set difference maps 
this mask needs to be increased by the unseen pixels. That effect only concerns
a few additional pixels for the detector set null map, but is 
particularly important for the survey difference map, since a survey only
approximately covers the full sky. The final $f_\mathrm{sky}$ of the mask 
used for the survey difference map is 64\%.

The test consisted of extracting $f_{\rm NL}$  from the null maps described 
above, using only the cubic part of the bispectrum estimators (i.e., no 
linear term correction), and keeping the same weights as for the full 
``signal $+$ noise" analysis. 
This means that the weights were not optimized for noise-only maps, as our 
aim was not to study the bispectrum of the noise per se but rather 
to check whether the noise alone produces a three-point function 
detectable by our estimators when they are run in the {\em same} 
configuration as for the actual CMB data analysis. For a similar reason it 
would have been pointless to introduce a linear term in this test. 
The purpose of the linear correction is in fact that of decreasing the 
error bars by accounting for off-diagonal covariance terms introduced by 
sky cuts and noise correlations when optimal weights are used, which is not 
the case here.

Our $f_{\rm NL}$ error bars for this test are obtained by running the 
estimators' cubic part on Gaussian noise simulations including realistic 
correlation properties. In the light of the above paragraph it is clear that 
such uncertainties have nothing to do with the actual uncertainties from CMB data, 
and cannot be compared to them.

Since \SMICA\ was the main component-separation 
method for our final analysis of data,
we present in Table~\ref{Tab_jaccknife_SMICA} the results of our 
\SMICA\ half-ring study using the KSW, binned and modal estimators, 
i.e., all the three main pipelines used in this paper. For the
cleaned maps we do not have survey or detector set difference maps. Those are, 
however, available for single frequency channels. Thus we also studied all
three types of null maps for the raw 143~GHz channel in 
Table~\ref{Tab_jackknives}, using the binned estimator. 
In both tables all $f_\mathrm{NL}$ shapes are assumed to be independent.
The binned estimator is best suited for these specific tests as its cubic part is 
less sensitive to masking compared to other pipelines, especially modal. 
Therefore in this ``cubic only" test, the binned results provide the most 
stringent constraints in terms of final error bars. 

As one can see {\it Planck} passes these null tests without any problems: 
all values found for $f_\mathrm{NL}$ in these null maps are completely 
negligible compared to the final measured results on the data maps, and 
consistent with zero within the error bars.

%%%%%%%%%%%%%%%%%%%%%%%%%%%%%%%%%%%%%%%%%%%%%%%%%%%%

\subsection{Impact of foreground residuals}
\label{sec:Valid_FGresid}
%
%%%%%%%%%%%%%%%%%%%%%%%%%%%%%%%%%%%%%%%%%%%%%%%%%%%% 
\begin{table*}[tmb] 
\begingroup 
\newdimen\tblskip \tblskip=5pt 
\caption{Summary of our $\fnl$ analysis of foreground residuals. For realistic lensed FFP6
 simulations processed through the \SMICA\ and \NILC\ component separation pipelines, we report: the average $\fnl$ with and without 
foreground residuals added to the maps, the $\fnl$ standard deviation in the same two cases, and the standard deviation of the map-by-map $\fnl$ difference 
 between the ``clean'' and ``contaminated'' sample. The impact of foreground residuals is clearly subdominant when compared to statistical error bars 
for all shapes. Results reported below have been obtained using the modal estimator.} 
\label{Tab_fgresid} 
\nointerlineskip 
\vskip -3mm 
\footnotesize 
\setbox\tablebox=\vbox{ 
   \newdimen\digitwidth 
   \setbox0=\hbox{\rm 0} 
   \digitwidth=\wd0 
   \catcode`*=\active 
   \def*{\kern\digitwidth} 
   \newdimen\signwidth 
   \setbox0=\hbox{+} 
   \signwidth=\wd0 
   \catcode`!=\active 
   \def!{\kern\signwidth} 
   \newdimen\dotwidth 
   \setbox0=\hbox{.} 
   \dotwidth=\wd0 
   \catcode`;=\active 
   \def;{\kern\dotwidth}
\halign{\hbox to 1in{#\leaderfil}\tabskip 1em& 
\hfil#\hfil\tabskip 0.8em& 
\hfil#\hfil& 
\hfil#\hfil& 
\hfil#\hfil& 
\hfil#\hfil& 
\hfil#\hfil\tabskip 0pt\cr 
\noalign{\doubleline\vskip 2pt} 
\omit&\multispan6\hfil $\fnl$\hfil\cr
\omit&\multispan6\hrulefill\cr
\omit& \SMICA\ & \SMICA\ & \NILC\ & \NILC\ & \SMICA\ & \NILC\cr 
Shape\hfill& clean & residual & clean & residual & residual $-$ clean & 
residual $-$ clean\cr 
\noalign{\vskip 4pt\hrule\vskip 6pt}
\omit\hfil {\bf Modal}\hfil&&&\cr  
Local & !*7.7* $\pm$ *5.9* & !*7.8* $\pm$ *5.9* & !*7.7* $\pm$ *5.8* & !*7.4* $\pm$ *6.0* & 
!0.04* $\pm$ 1.0** & $-$0.27* $\pm$ 1.1** \cr 
Equilateral & *$-$0.5* $\pm$ 77;** & *$-$8.7* $\pm$ 79;** & *$-$0.6* $\pm$ 78;** & *$-$9.0* $\pm$ 
79;** & $-$8.3** $\pm$ 8.2** & $-$8.4** $\pm$ 8.3** \cr 
Orthogonal & $-$23;** $\pm$ 41;** & $-$25;** $\pm$ 41;** & $-$24;** $\pm$ 40;** & $-$26;** $\pm$ 41;** & 
$-$2.0** $\pm$ 4.7** & $-$2.4** $\pm$ 4.8** \cr 
ISW-lensing & *!1.00 $\pm$ *0.38 & !*1.01 $\pm$ *0.38 & !*1.01 $\pm$ *0.38 & !*1.02 
$\pm$ *0.38 & !0.006 $\pm$ 0.052 & !0.013 $\pm$ 0.052 \cr 
\noalign{\vskip 3pt\hrule\vskip 4pt}}} 
\endPlancktablewide                 % ends two-column \halign 
\endgroup 
\end{table*} 
%%%%%%%%%%%%%%%%%%%%%%%%%%%%%%%%%%%%%%%%%%%%%%%%%%%%

%%%%%%%%%%%%%%%%%%%%%%%%%%%%%%%%%%%%%%%%%%%%%%%%%%%%  
\begin{figure}[!t]
\includegraphics{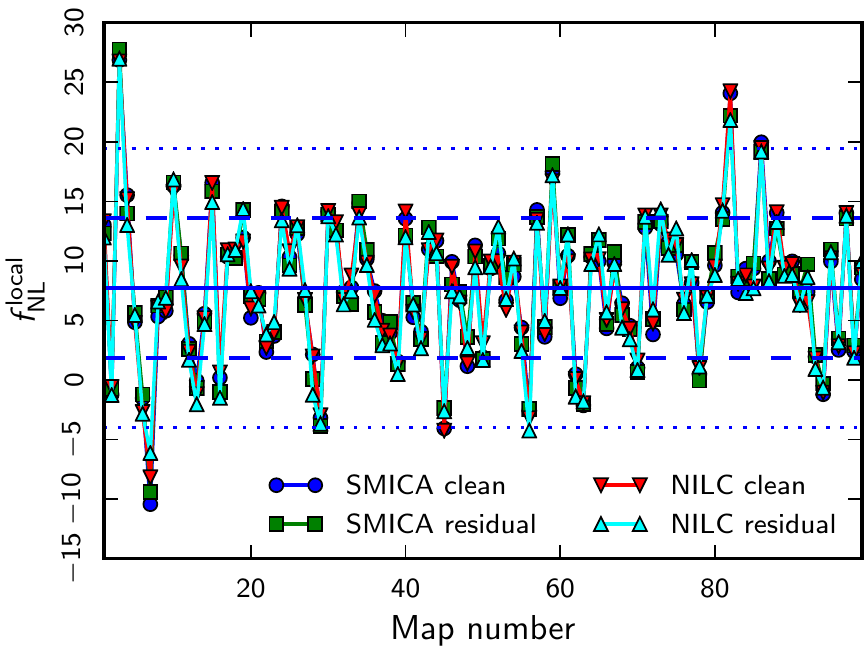}
\caption{The measured $f_{\rm NL}^{\rm local}$ for the first 99 
maps in the lensed FFP6 simulation sample used for 
the foreground studies presented in Sect.~\ref{sec:Valid_FGresid}. 
We show measurements from \SMICA\ and \NILC\ processed maps both
with and without foreground residuals. 
The horizontal solid line is the average value of the \SMICA\ clean maps, and the
dashed and dotted horizontal lines correspond to $1\sigma$ and $2\sigma$
deviations, respectively.
The plot clearly shows the very 
small impact of including residuals, and the very good consistency 
between the two component separation pipelines.
}
\label{Fig_ffp6}
\end{figure}
%%%%%%%%%%%%%%%%%%%%%%%%%%%%%%%%%%%%%%%%%%%%%%%%%%%%

In Sect.~\ref{sec:Results} we applied our bispectrum estimators to {\it Planck} 
data filtered through four different component separation
methods: \SMICA, \NILC, \SEVEM\ and \CR\ (for a detailed
description of component separation techniques used for {\it Planck}
see \cite{planck2013-p06}). 
The resulting set of $f_{\rm NL}$ measurements
shows very good internal consistency both between different
estimators (as expected from our MC validation tests of
bispectrum pipelines described in Sect.~\ref{sec:Validation}) and between
different foreground-cleaned maps. This already makes it clear that
foreground residuals in the data are very well under control, and
their impact on the final $f_{\rm NL}$ results is only at the level of
a small fraction of the measured error bars. In this Section we
further investigate this issue, and validate our previous findings on
data by running extensive tests in which we compare simulated data sets
with and without foreground residuals from two different component
separation pipelines, \SMICA\ and \NILC. The goal is to provide a MC-based assessment of the expected $f_{\rm NL}$ systematic error
from residual foreground contamination.

For each component separation pipeline, we consider two sets of
simulations. One set includes realistic {\it Planck} noise and beam,
is masked and inpainted in the same way as we do for real data, and
is processed through \SMICA\ and \NILC\, but it does {\em not} contain
any foreground component. The other set is obtained by adding to the
first one a number of diffuse foreground residuals: thermal
and spinning dust components; free-free and synchrotron emission;
kinetic and thermal SZ; CO lines and correlated CIB. These
residuals have been evaluated by applying the component separation
pipelines to accurate synthetic {\it Planck} datasets including
foreground emission according to the PSM~\citep{delabrouille2012},
and are of course dependent on the cleaning method adopted. The simple
procedure of adding foreground residuals to the initially clean
simulations is made possible because we consider only linear
component separation methods for our analysis.
Linearity is in general an important requirement for
foreground cleaning algorithms aiming at producing maps suitable for
NG analyses. All maps in both samples are lensed using the {\tt LensPix}
algorithm. We analyse both sets using different bispectrum estimators
(modal, KSW, binned) for cross-validation purposes. 

The presence of
residual foreground components in the data can have two main effects
on the measured $f_{\rm NL}$. The first is to introduce a bias in the
$f_{\rm NL}$ measurements due to the correlation between the
foreground and primordial 3-point function for a given shape. The
second is to increase the error bars while leaving the bispectrum
estimator asymptotically unbiased. This is a consequence of accidental
correlations between primordial CMB anisotropies and foreground
emission. Of course these ``CMB-CMB-foregrounds" bispectrum terms average to zero but
they do not cancel in the bispectrum variance 6-point function. An
interesting point to note is that a large foreground three-point
function will tend to produce a {\em negative} bias in the local
bispectrum measurements. That is because foreground emission 
produces a positive skewness of the CMB temperature distribution
(``excess of photons"), and a positive skewness is in turn
related to a negative $f_{\rm NL}^{\rm local}$ \footnote{While not rigorous, this argument captures the leading effect since Galactic foregrounds predominantly contaminate large scales. In principle, positively skewed, small scale foreground residuals  ($\ell>60$), or the negatively skewed SZ effect, can contribute positive bias. Our simulations with foreground residuals demonstrate that these contributions are subdominant.  }. A large negative
$f_{\rm NL}^{\rm local}$ is thus a signature of significant foreground
contamination in the map. This is indeed what we observe if we
consider raw frequency maps with a small Galactic cut, which is why
our frequency-by-frequency analysis in Sect.~\ref{sec:maskdep} was
performed using a very large mask. For more details regarding effects of
foreground contamination on primordial NG measurements in the context
of the {\it WMAP} analysis see \citealt{2008PhRvL.100r1301Y} and \citealt{2010JCAP...01..028S}.

In our test we built maps contaminated with foreground residuals by simply 
adding residual components to the clean maps. That means that the difference
$(f^{\rm residual}_{\rm NL} - f^{\rm clean}_{\rm NL})_i$ for the $i$-th realization 
in the simulated sample exactly quantifies the change in $f_{\rm NL}$ due to
foregrounds on that realization. In order to assess the level of residual 
foreground contamination on primordial and ISW-lensing shapes, first of all we consider the 
difference between average values and standard deviations of $\fnl$ measured from the 
two map samples for each shape. As shown in Table~\ref{Tab_fgresid} we find that neither the
average nor the standard deviation shows a significant change
between the two datasets. That means that foreground residuals are
clearly subdominant, as they do not bias the estimator for any shape
and they do not increase the variance through spurious correlations
with the CMB primordial signal. 

We also consider the difference $f^{\rm residual}_{\rm NL} - f^{\rm clean}_{\rm NL}$ 
on a map-by-map basis and compute its standard
deviation. This is used as an estimate of the expected bias on a
single realization due to the presence of residuals. As already
expected from the negligible change in the standard deviations of the
two samples, the variance of the map-by-map differences is also very
small: Table~\ref{Tab_fgresid} again shows that it is at most about
${\sigma_{f_{\rm NL}}/6}$ for any given shape, where $\sigma_{f_{\rm  NL}}$ 
is the $\fnl$ standard deviation for that shape. As an
example, in Fig.~\ref{Fig_ffp6} we show the measured values of
$f_{\rm NL}^{\rm local}$ for the first 99 maps in both the \SMICA\ and
\NILC\ samples, comparing results with and without residuals. It is
evident also from this plot that the change in central value due to
including residuals is very small. The very good agreement between the
two component separation pipelines is also worth notice.

From the study shown here and from the comparison between different
component separation methods in Sect.~\ref{sec:Results}, we can thus
conclude that the combination of foreground-cleaned maps and 
$f_{\rm  sky} = 0.73$ sky coverage we adopt in this work provide a very
robust choice for $\fnl$ studies.

\subsection{Impact of HFI time-ordered information processing on NG constraints}

Our validation tests are designed to determine whether instrumental or data processing systematics could lead to a spurious detection of primordial NG. Our analysis protocol passes these tests which allows us to rule out this concern with confidence. 
These validation tests do not exclude the possibility that the extensive processing of HFI time-ordered information (TOI) could somehow \textit{remove} non-Gaussian signals present on the sky and thus negatively impact \Planck's ability to detect them. If this were true it would weaken the bounds on primordial NG reported here.  HFI TOI processing includes glitch fitting, gain drift and correction (ADC non-linearity),  4K cooler line removal, telegraph noise step correction, and TOD filtering to correct for the bolometer time constant (\citealt{planck2013-p03,planck2013-p03c,planck2013-p03e}).

The following facts constitute strong evidence that HFI TOI processing does not remove non-Gaussian signals: 
\begin{enumerate}
\item
\Planck's $2.6 \sigma$ ISW-lensing bispectrum measurement is consistent with expectations from the LCDM model, and has the right skew-$C_\ell$ shape. Like NG of local type, this is a squeezed bispectrum template.
\item
When frequency maps are combined to maximize the CIB signal, \Planck\ finds a $25\sigma$ detection of the nearly squeezed CIB-lensing bispectrum, consistent with the CIB 2-point correlations inferred by \Planck\ \citep{planck2013-pip56}.
\item
\Planck\ detects residual point sources in the \SMICA\ maps at $5\sigma$, seen as a bispectrum of equilateral 
shape at high $\ell$  consistent with the  expected skew-spectrum shape (Fig.~2) on the angular scales  where residual infrared sources and  far infrared background are expected to be the dominant contaminants of the power spectrum according  to the foreground residuals in FFP6 simulations of \SMICA\ maps (Fig.~E.3 in \citealt{planck2013-p06}).
\item 
The trispectrum signal imprinted by \Planck's motion with respect to the CMB rest frame is seen at a sensible level and in a plausible direction \citep{planck2013-pipaberration}.
\item
\Planck\ detects the trispectrum signal due to lensing by  large scale structure  with  high significance  and in accordance with LCDM predictions based on the \Planck\ parameter likelihood.
\item
Fig.~\ref{Fig_lmaxdep} shows that we reproduce the {\it WMAP}-9 local NG results on those larger angular scales where the {\it WMAP}-9 data are cosmic variance limited.
\end{enumerate}
We conclude that all the forms of NG that should have been seen by \Planck\ have been seen, in quantitative agreement with the expected level across the entire range of angular scales probed. While it cannot be excluded with absolute certainty that some unknown aspect of HFI TOI processing could have affected some unknown form of NG, the presence of these non-Gaussian features in the resulting maps (in addition to signals such as the extracted map of y-distortion, and galactic and extragalactic foregrounds), gives us confidence in the force of the bounds on primordial NG described in this paper.

%%%%%%%%%%%%%%% Section 9 %%%%%%%%%%%%%%%
\section{Implications for early Universe physics}
\label{sec:Implications}

The NG analyses performed in this paper show that \textit{Planck} data are consistent with Gaussian primordial fluctuations. The standard models of single-field slow-roll inflation have therefore survived the most stringent tests of Gaussianity performed to date. On the other hand, the NG constraints obtained on different primordial bispectrum shapes (e.g., local, equilateral and orthogonal), after properly accounting for various contaminants, severely limit various classes of mechanisms for the generation of the primordial perturbations proposed as alternatives to the standard models of inflation.  

In the following subsections, unless explicitly stated otherwise, we translate limits on NG to limits on parameters of the theories by constructing a posterior assuming the following: the sampling distribution is Gaussian (which is supported by Gaussian simulations); the likelihood is approximated by the sampling distribution but centred on the NG estimate (see \citealt{2009ApJS..184..264E}); uniform or Jeffreys' prior (where stated), over ranges which are physically meaningful, or as otherwise stated.  Where two parameters are involved, the posterior is marginalized to give one-dimensional constraints on the parameter of interest.

%%%%%%%%%%%%%%%%%%%%%%%%%%%%%%%%%%%%%%%%%%%%%%%%%%%%

\subsection{General single-field models of inflation}

\noindent{\it DBI models}: The constraints on $f_{\rm NL}^{\rm equil}$ provide constraints on the sound speed with which the inflaton fluctuations can propagate during inflation. For example, for DBI models of inflation~(\citealt{2004PhRvD..70j3505S,2004PhRvD..70l3505A}), where the inflaton field features a non-standard kinetic term,  the predicted value of the nonlinearity parameter is $f_{\rm NL}^{\rm DBI}=-(35/108) (c_\mathrm{s}^{-2}-1)$~(\citealt{2004PhRvD..70j3505S,2004PhRvD..70l3505A,2007JCAP...01..002C}). Although very similar to the equilateral shape, we have obtained constraints directly on the theoretical (nonseparable) predicted shape~(Eq.~\ref{dbiBis})). The constraint $f_{\rm NL}^{\rm DBI}=11 \pm 69$ at $68\%$ CL (see Eq.~(\ref{fnlequilfamily})) implies 
\begin{equation}
c_\mathrm{s}^{\rm DBI} \geq 0.07 \quad \quad  \text{95$\%$ CL}\, . 
\end{equation}
The DBI class contains two possibilities based on string constructions. In {\sl ultraviolet} (UV) DBI models, the inflaton field moves under a quadratic potential from the UV side of a warped background to the infrared side. It is known that this case is already at odds with observations, if theoretical internal consistency of the model and constraints on power spectra and primordial NG are taken into account \citep{2007PhRvD..75l3508B,2007JCAP...07..002L,2007JCAP...05..004B,2007PhRvD..76j3517P}. Our results strongly limit the relativistic r\' egime of these models even without applying the theoretical consistency constraints. 

It is therefore interesting to look at {\sl infrared} (IR) DBI models~(\citealt{2005PhRvD..71f3506C,2005JHEP...08..045C}) where the inflaton field moves from the IR to the UV side, and the inflaton potential is $V(\phi)=V_0-\frac{1}{2} \beta H^2 \phi^2$, with a wide range $0.1 < \beta < 10^9$ allowed in principle. In previous work \citep{2008PhRvD..77b3527B}  a 95\% CL limit of $\beta < 3.7$ was obtained using {\it WMAP}. In a minimal version of IR DBI, where stringy effects are neglected and the usual field theory computation of the primordial curvature perturbation holds, one finds~\citep{2005PhRvD..72l3518C,2007JCAP...01..002C} $c_\mathrm{s} \simeq (\beta N/3)^{-1}$, $n_\mathrm{s}-1=-4/N$, where $N$ is the number of $e$-folds; further, primordial NG of the equilateral type is generated with an amplitude $f^\mathrm{DBI}_\mathrm{NL}=-(35/108)\, [(\beta^2\, N^2/9)-1]$. For this model, the range $N \geq 60$ is compatible with {\it Planck}'s $3\sigma$ limits on $n_\mathrm{s}$ \citep{planck2013-p17}. Marginalizing over $60 \leq N \leq 90$, we find 
\begin{equation}
\beta \leq 0.7 \quad \quad  \text{95$\%$ CL}\, ,
\end{equation}
dramatically restricting the allowed parameter space of this model.

\medskip

\noindent{\it Power-law $k$-inflation}: These models~(\citealt{1999PhLB..458..209A,2007JCAP...01..002C}) predict $f_{\rm NL}^\mathrm{equil} = -170/(81 \gamma)$, where $n_\mathrm{s}-1=- 3\gamma$, $c_\mathrm{s}^2 \simeq \gamma /8$. Assuming a prior of $0 < \gamma < 2/3$, our constraint $f_\mathrm{NL}^\mathrm{equil} = -42 \pm 75$ at $68\%$ CL (see Table~\ref{tab:fNLsmicah}) leads to the limit $\gamma \geq 0.05$ at 95\% CL. On the other hand, {\it Planck}'s constraint on $n_\mathrm{s}-1$ yields the limit $0.01 \leq \gamma \leq 0.02$ \citep{planck2013-p17}. These conflicting limits severely constrain this class of models.

\medskip

\noindent{\em Flat Models and higher derivative interactions:}  
Flat NG can characterize inflationary models which arise from independent interaction terms different from the  $(\dot{\pi})^3$ and $\dot{\pi} (\nabla \pi)^2$ discussed in Sect.~\ref{SectionII} (see also Sect.~\ref{Implications_EFT}). An example is given by models of inflation based on a Galilean symmetry~(\citealt{2011JCAP...02..006C}), where one of the inflaton cubic interaction terms allowed by the Galilean symmetry, $M_3[ \ddot{\pi} (\partial_i \partial_j \pi)^2/a^4-2H \dot{\pi} \ddot{\pi}^2+3 H^3 \dot{\pi}^3]$, contributes to the flat bispectrum with 
an amplitude $f^{\rm flat}_{\rm NL}=(35/256) (M_3 H)/(\epsilon M_{\rm Pl}^2)$~(\citealt{2011JCAP...02..006C}). Here, $\pi$ is the relevant inflaton scalar degree of freedom, $\epsilon$  the usual slow-roll parameter and $a$ the scale factor and $H$ the Hubble parameter during inflation. Our constraint $\fnl^{\rm flat}=37 \pm 77$ at $68\%$ CL (see Table~\ref{tab:fnlnonstandard}) leads to $(M_3 H)/(\epsilon M_{\rm Pl}^2)=270 \pm 563$ at $68\%$ CL. 
These interaction terms are similar to those arising in general inflaton field models that include extrinsic curvature terms, e.g., parameterized in the Effective Field Theory approach as $M^2 \dot{\pi} (\partial_{ij} \pi)^2/a^4$~(\citealt{2010JCAP...08..008B}), which contribute to a flat bispectrum with an amplitude $f^{\rm flat}_{\rm NL}= (50/108)\, M^2/(c_\mathrm{s}^2 \epsilon M_{\rm Pl}^2)$. In this case, we obtain 
$M^2/(c_\mathrm{s}^2 \epsilon M_{\rm Pl}^2)= 80 \pm 166$ at $68\%$ CL.

%%%%%%%%%%%%%%%%%%%%%%%%%%%%%%%%%%%%%%%%%%%%%%%%%%%%

\subsection{Implications for Effective Field Theory of Inflation}
 \label{Implications_EFT}

The effective field theory approach to inflation~(\citealt{2008PhRvD..77l3541W,2008JHEP...03..014C}) provides a general way to scan the NG parameter space of inflationary perturbations. For example, one can expand the Lagrangian of the dynamically relevant degrees of freedom into the dominant operators satisfying some underlying symmetries.  We will focus on general single-field models parametrized by the following operators (up to cubic order) 
 \begin{eqnarray}
 S&=&\int d^4x \sqrt{-g} \left[ -\frac{M^2_{\rm Pl} \dot{H}}{c_\mathrm{s}^2} \left( \dot{\pi}^2-c_\mathrm{s}^2 \frac{(\partial_i \pi)^2}{a^2} \right) 
 \right. \\   
 &-& \left. M_{\rm Pl}^2 \dot{H} (1-c_\mathrm{s}^{-2}) \dot{\pi} \frac{(\partial_i \pi)^2}{a^2}  
 + \left( M_{\rm Pl}^2 \dot{H}  (1-c_\mathrm{s}^{-2})  -\frac{4}{3} M_3^4 \right) \dot{\pi}^3 \right] \nonumber 
\end{eqnarray}
where $\pi$ is the scalar degree of freedom ($\zeta=- H \pi$). The measurements on $f_{\rm NL}^{\rm equil}$ and $f_{\rm NL}^{\rm ortho}$ can be used to constrain the magnitude of the inflaton interaction terms 
 $\dot{\pi} (\partial_i \pi)^2$ and $(\dot{\pi})^3$ which give respectively $f_{\rm NL}^{\rm EFT1}=-(85/324)(c_\mathrm{s}^{-2}-1)$ and $f_{\rm NL}^{\rm EFT2}=-(10/243)(c_\mathrm{s}^{-2}-1) \[\tilde{c}_3+(3/2) c_\mathrm{s}^2\]$~(\citealt{2010JCAP...01..028S}, see also ~\citealt{2007JCAP...01..002C,2010AdAst2010E..72C}). These two operators give rise to shapes that peak in the equilateral configuration that are, nevertheless, slightly different, so that the total NG signal will be a linear combination of the two, possibly leading also to 
an orthogonal shape.  There are two relevant NG parameters, $c_\mathrm{s}$, the sound speed of the the inflaton fluctuations, and $M_3$ which characterizes the amplitude of the other operator $\dot{\pi}^3$. Following~\cite{2010JCAP...01..028S} we will focus on the dimensionless parameter ${\tilde c}_3 (c_\mathrm{s}^{-2}-1)=2 M_3^4 c_\mathrm{s}^2 /(\dot{H} M_{\rm Pl}^2)$. For example, DBI inflationary models corresponds to 
$\tilde{c}_3= 3(1-c_\mathrm{s}^2)/2$, while the non-interacting model (vanishing NG) correspond to $c_\mathrm{s}=1$ and $M_3=0$ (or ${\tilde c}_3 (c_\mathrm{s}^{-2}-1)=0$). 

%%%%%%%%%%%%%%%%%%%%%%%%%%%%%%%%%%%%%%%%%%%%%%%%%%%%
\begin{figure}[!t]
\includegraphics[width=\hsize]{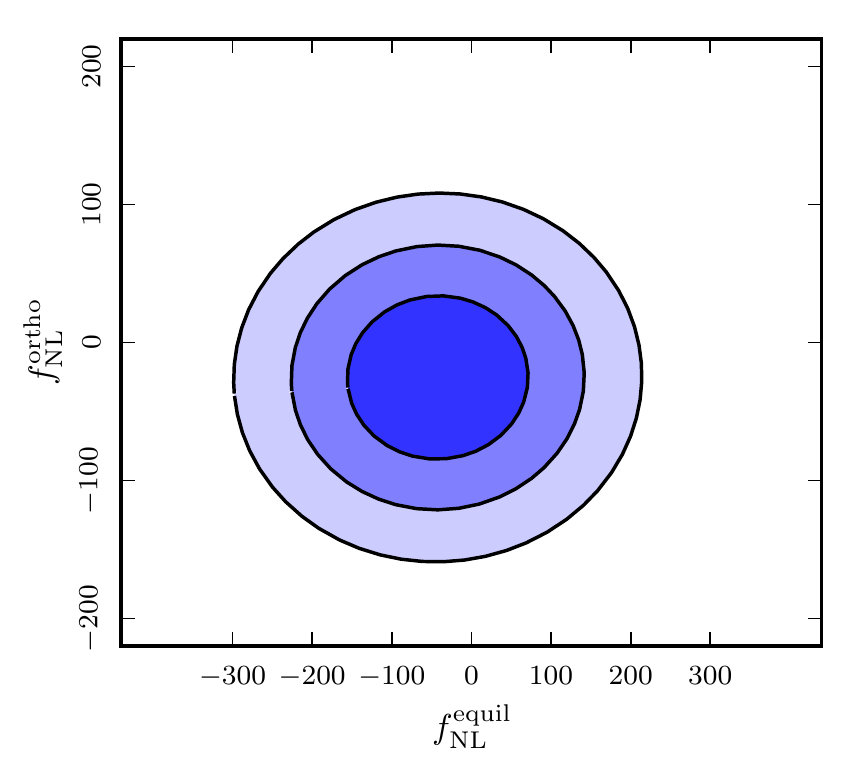}
\caption{$68\%$, $95\%$, and $99.7\%$ confidence regions in the parameter space $(f_\mathrm{NL}^\mathrm{equil}, f_\mathrm{NL}^\mathrm{ortho})$, defined by thresholding $\chi^2$ as described in the text.}
\label{fig:eq_ort}
\end{figure}
%%%%%%%%%%%%%%%%%%%%%%%%%%%%%%%%%%%%%%%%%%%%%%%%%%%%

%%%%%%%%%%%%%%%%%%%%%%%%%%%%%%%%%%%%%%%%%%%%%%%%%%%%
\begin{figure}[!t]
\includegraphics[width=\hsize]{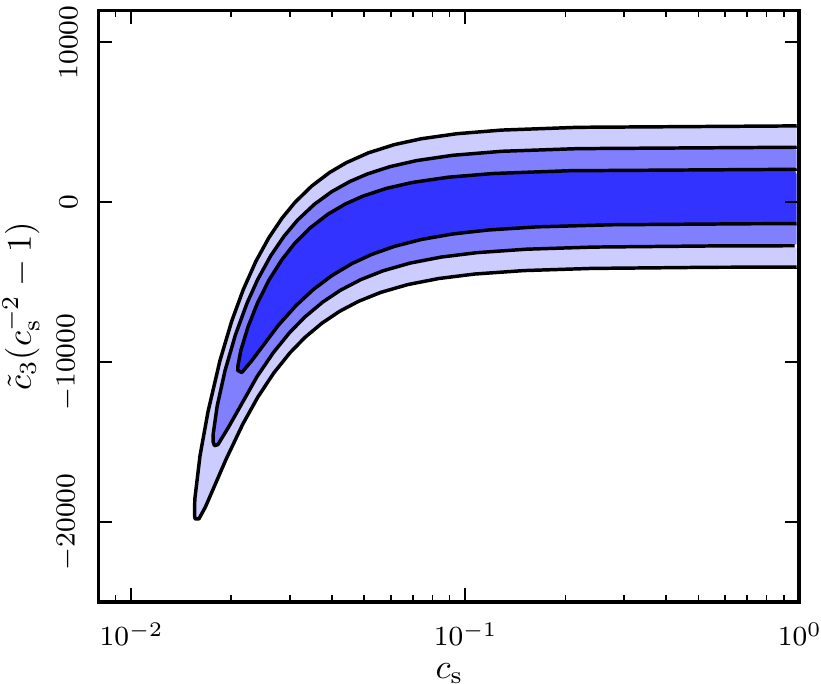}
\caption{$68\%$, $95\%$, and $99.7\%$ confidence regions in the single-field inflation parameter space $(c_\mathrm{s}, \tilde{c}_3)$, obtained from Fig.~\ref{fig:eq_ort} via the change of variables in Eq.~(\ref{meanfNL}).}
\label{fig:cs_c3}
\end{figure}
%%%%%%%%%%%%%%%%%%%%%%%%%%%%%%%%%%%%%%%%%%%%%%%%%%%%

The mean values of the estimators for equilateral and orthogonal NG amplitudes are given in terms of $c_\mathrm{s}$ and $\tilde{c}_3$ by
\begin{eqnarray}
\label{meanfNL}
f^{\rm equil}_{\rm NL} &=&\frac{1-c_\mathrm{s}^2}{c_\mathrm{s}^2} (-0.275 +  0.0780 A) \nonumber \\
f^{\rm ortho}_{\rm NL} &=&\frac{1-c_\mathrm{s}^2}{c_\mathrm{s}^2} (0.0159 - 0.0167 A)
\end{eqnarray}
where $A=-(c_\mathrm{s}^2+(2/3) \tilde{c}_3)$, and the coefficients are computed from the Fisher correlation matrix between the equilateral and orthogonal template bispectra and the theoretical bispectra arising from the two operators $\dot{\pi} (\nabla \pi)^2$ and $\dot{\pi}^3$. Given our constraints on $f^{\rm equil}_{\rm NL}$ and $f^{\rm ortho}_{\rm NL}$, and the covariance matrix $\tens{C}$ of the joint estimators, we can define a $\chi^2$ statistic given by $\chi^2(\tilde{c}_3,c_\mathrm{s})={\vec v}^T(\tilde{c}_3,c_\mathrm{s}) \tens{C}^{-1} {\vec v}(\tilde{c}_3,c_\mathrm{s})$, where the vector ${\vec v}$ is given by $v^i(\tilde{c}_3,c_\mathrm{s})=f^i(\tilde{c}_3,c_\mathrm{s})-f^i_{P}$.  $f^i_{P}$, where $i$=\{equilateral, orthogonal\}, are the joint estimates of the equilateral and orthogonal $f_{\rm NL}$ measured by \Planck\, and $f^i(\tilde{c}_3,c_\mathrm{s})$ is given by Eq.~(\ref{meanfNL}).  Figure~\ref{fig:eq_ort} shows the $68 \%, 95 \%$, and $99.7\%$ confidence regions for $f^{\rm equil}_{\rm NL}$ and $f^{\rm ortho}_{\rm NL}$, obtained by requiring $\chi^2 \leq 2.28, 5.99$, and $11.62$ respectively, as appropriate for a $\chi^2$ variable with two degrees of freedom. The corresponding confidence regions in the $(\tilde{c}_3,c_\mathrm{s})$ parameter space are shown in Fig.~\ref{fig:cs_c3}.
After marginalizing over $\tilde{c}_3$ we find the following conservative bound on the inflaton sound-speed 
\begin{equation}
c_\mathrm{s} \geq 0.02 \quad \quad  \text{95$\%$ CL}\, .
\end{equation}

\noindent Note that we have also looked explicitly for the non-separable shapes in Sect.~\ref{nonsep}, in particular the two effective field theory shapes and the DBI inflation shape (see Eqs.~(\ref{EFT1Bis},~\ref{EFT2Bis},~\ref{dbiBis})) .  

%%%%%%%%%%%%%%%%%%%%%%%%%%%%%%%%%%%%%%%%%%%%%%%%%%%%

\subsection{Multi-field models}

\noindent{\it Curvaton models:}
\textit{Planck} NG constraints have interesting implications for the simplest adiabatic curvaton models. They predict~\citep{2004PhRvD..69d3503B,2004PhRvL..93w1301B}
\begin{equation}
f_\mathrm{NL}^\mathrm{local} = \frac{5}{4r_{\rm D}} - \frac{5 r_{\rm D}}{6} - \frac{5}{3}\, ,
\end{equation}
for a quadratic potential of the curvaton field~(\citealt{2002PhLB..524....5L,2003PhRvD..67b3503L,2005PhRvL..95l1302L,2006JCAP...09..008M,2006PhRvD..74j3003S}), where $r_{\rm D}=[3\rho_{\rm curvaton}/(3 \rho_{\rm curvaton}+4\rho_{\rm radiation})]_{\rm D}$ is the ``curvaton decay fraction''  evaluated at the epoch of the curvaton decay in the sudden decay approximation. Assuming a prior $0<r_\mathrm{D}<1$,  given our constraint $f_{\rm NL}^{\rm local}=2.7 \pm 5.8$  at 68\% CL, we obtain 
\begin{equation}
r_{\rm D} \geq 0.15 \quad \quad  \text{95$\%$ CL}\, . 
\end{equation}
In \citet{planck2013-p17} a limit on $r_{\rm D}$ is derived from the constraints on isocurvature perturbations under the assumption that there is some residual isocurvature fluctuations in the curvaton field. For this restricted case, they find $r_{\rm D} > 0.98$  ($95 \%$ CL), compatible with the constraint obtained here. 

\medskip

\noindent{\it Quasi-single field inflation}:
It is beyond the scope of this paper to perform a general multi-field analysis employing the local NG constraints. However, we have performed a detailed analysis for the quasi-single field models (see Eq.~(\ref{qsiBis})). Quasi-single field (QSF) inflation models \citep{2010JCAP...04..027C, 2010PhRvD..81f3511C, Baumann:2011nk} are a natural consequence of inflation model-building in string theory and supergravity (see Sect.~\ref{sec:qsf}). In addition to the inflaton field, these models have extra fields with masses of order the Hubble parameter, which are stabilized by supersymmetry. A distinctive observational signature of these massive fields is a one-parameter family of large NG whose squeezed limits interpolate between the local and the equilateral shape. Therefore, by measuring the precise momentum-dependence of the squeezed configurations in the NG, in principle, we are directly measuring the parameters of the theory naturally determined by the fundamental principle of supersymmetry.
These models produce a bispectrum (Eq.~\eqref{qsiBis})  depending on two parameters $\nu$, $\fnl^{\rm QSI}$, with a shape that interpolates between the local shape, where $\nu=1.5$ and the equilateral shape, where $\nu = 0$.  

\begin{figure}
\includegraphics[width=\hsize]{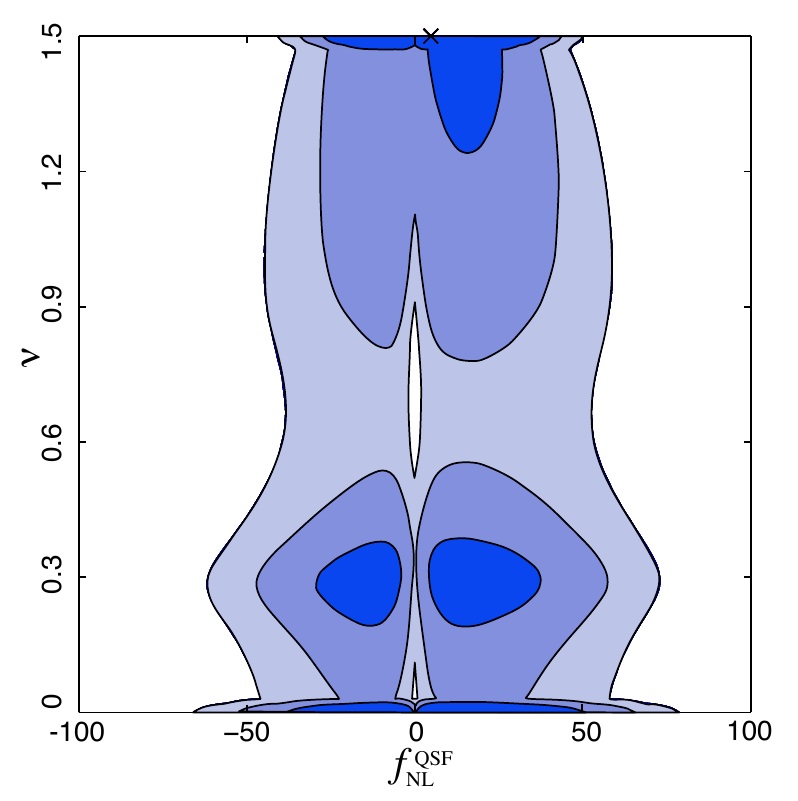}
\caption{
$68\%$, $95\%$, and $99.7\%$ confidence intervals for $\nu$ and 
$f_\mathrm{NL}^{\rm QSI}$ for quasi-single field inflation. The best fit value of 
$\nu=1.5$, $f_\mathrm{NL}^{\rm QSI} = 4.75$ is marked with an X.  The contours were 
calculated using MC methods by creating $2\times10^9$ simulations using the 
$\beta$ covariance matrix around this best fit model.
}
\label{QSF}
\end{figure}

Results are shown in Fig.~\ref{QSF} (see Sect.~\ref{qsfresults} for details of the analyses). The best fit value corresponds to 
$\nu=1.5$, $f_\mathrm{NL} = 4.79$  which would imply, within this scenario, that the extra 
 field different from the inflaton has a mass $m \ll H$.  However, the figure shows that there is no preferred value for $\nu$ with all values allowed at $3\sigma$.  
 
%%%%%%%%%%%%%%%%%%%%%%%%%%%%%%%%%%%%%%%%%%%%%%%%%%%%
%%%%%%%%%%%%%%%%%%%%%%%%%%%%%%%%%%%%%%%%%%%%%%%%%%%%

%%%%%%%%%%%%%%%%%%%%%%%%%%%%%%%%%%%%%%%%%%%%%%%%%%%%

\subsection{Non-standard inflation models}

\noindent{\it Constraints on excited initial states}:
Results from Sect.~\ref{STS} constrain a variety of models with flattened bispectra, often in combination with a non-trivial squeezed limit. The most notable examples are bispectra produced in excited initial state models (non-Bunch-Davies vacua), which can be generated by strong disturbances away from background slow-roll evolution or additional trans-Planckian physics \citep{2007JCAP...01..002C, 2008JCAP...05..001H,Meerburg:2009ys,Agullo:2011xv}. The  constraints we have obtained are summarized in Table~\ref{tab:fnlnonstandard}, and cover four representative cases (see Eq.~(\ref{NBD2Bis},~\ref{NBD3Bis})) in the literature.  We find no strong evidence for these flattened bispectra in the \textit{Planck} data after subtraction of the ISW-lensing signal, with which all these models have some correlation. This is consistent with an earlier constraint on the NBD model from {\it WMAP} \citep{2010arXiv1006.1642F}. However, this investigation is limited by the present resolution of the polynomial modal estimator ($n_{\rm max}=600$), so more strongly flattened models are not excluded by this analysis. 

\medskip 
\noindent{\it Directional dependence motivated by  fields}:
Directionally-dependent bispectra (Eq.~(\ref{vectorBis})), motivated by inflation with gauge fields, have also been constrained (see Table~\ref{tab:fnlnonstandard}).
For example, models with a kinetic term of the vector field(s) $\mathcal{L}=-I^2(\phi) F^2/4$, where $F^2$ is the strength of the gauge field, and $I(\phi)$ is a function of the inflaton field which, with an appropriate time dependence (see, e.g.,~\citealt{1992ApJ...391L...1R}), can generate vector fields during inflation. For these models the $L=0,2$ modes in the bispectrum are excited with 
$f_{\rm NL}^\mathrm{L}=X_L (|g_*|/0.1)\, (N_{k_{3}}/60)$, where $X_{L=0}=(80/3)$ and $X_{L=2}=-(10/6)$, respectively~\citep{Barnaby:2012tk,Bartolo:2012sd,Shiraishi:2013vja}. 
Here $g_*$ is the amplitude of a quadrupolar anisotropy in the power spectrum (see, e.g.,~\cite{2007PhRvD..75h3502A}) and $N$ is the number of $e$-folds before the end of inflation when the relevant scales exit the horizon. 
These modes therefore relate the bispectrum amplitude to the level of statistical anisotropy in the power spectrum. 
Marginalizing over a prior $50 \leq N \leq 70$ assuming uniform priors on $g_*$, our constraints in Table~\ref{tab:fnlnonstandard} lead to the limits 
$-0.05 < g_* < 0.05$ and $-0.36 < g_* < 0.36$ from the $L=0$, $L=2$ modes respectively ($95 \%$ CL). Note, however, that in the current analysis only a modest correlation was possible with the shape corresponding to the $L=2$ mode. These results apply to all models where curvature perturbations are sourced by a $I^2(\phi) F^2$ term (see references in ~\citealt{Shiraishi:2013vja}).

\medskip

\noindent{\it Feature models}:
Non-scale-invariant oscillatory bispectrum shapes can be generated by sharp or periodic features in the inflaton potential, with particular recent interest on axion monodromy models motivated by string theory (see Sect.~\ref{New}).   We have undertaken a survey of simple feature models (Eq.~\eqref{featureBis}) which have a periodicity accessible to the polynomial modal estimator (wavenumbers $K=k_1+k_2+k_3 \gtrsim 0.01$), a two-parameter space spanned by  $K$ and a phase $\phi$.    There are interesting best fit models for the \Planck\ CMB bispectrum around $K=0.01875$, $\phi=0$ (that is, with a large-$\ell$ bispectrum periodicity around $\Delta\ell=260$), with results shown in Table~\ref{tab:fnlfeature}.  We note important caveats on the statistics of parameter surveys like this in Sect.~\ref{featureres}; given the large numbers of feature models studied, we have to apply a higher threshold for statistical significance as shown for a survey of 200 Gaussian simulations.  This feature survey takes forward earlier results for the {\it WMAP} data \citep{2010arXiv1006.1642F}, with the apparent fit reflecting the signal observed in the \Planck\ CMB reconstruction (see Fig.~\ref{fig:recondetail}).   Most attention on feature models has been motivated by the simplest single-field case for which there are correlated signals predicted in the bispectrum and power spectrum (see e.g.,~\citealt{2007JCAP...06..023C,Adshead:2011jq}).   In this case, regions with small $k\sim 0.001$ are favoured for producing an observable bispectrum\footnote{\cite{planck2013-p11} confirms an anomaly in the power spectrum at $20 \lesssim \ell \lesssim 40$ first noted in {\it WMAP}, which leads to an improvement in likelihood when fitted with a feature in the inflationary potential \citep{peiris2003}. Unfortunately, the best-fit parameters in this case do not lead to an observable bispectrum~\citep{2011PhRvD..84d3519A}.}, given existing {\it WMAP} power spectrum constraints on these models.  Periodicities $\Delta\ell \lesssim 20$ are anticipated (see~\citealt{Adshead:2011jq}) which are not accessible to the present modal bispectrum estimator analysis, but which are discussed in \citep{planck2013-p17}.  Conversely, the  \Planck\ feature model survey using the power spectrum  \citep{planck2013-p17} does cover bispectrum scales and parameters investigated in this paper.  An analysis of the Bayesian posterior probability~\citep{planck2013-p17} does not appear to provide evidence favoring parameters associated with the current best-fit bispectrum model. More detailed analysis using the specific bispectrum envelope for the single-field feature solution is being undertaken.    

\medskip

\noindent{\it Resonance models}:
We have also investigated resonance models  of Eq.~\eqref{resonantBis} such as axion monodromy and enfolded resonance models  of Eq.~\eqref{resNBDBis}, in which non-Bunch-Davies vacua are excited by sharp features. This limited analysis focuses on periodicities associated with the best-fit feature model and the results are described in Tables~\ref{tab:fnlresonant}  and \ref{tab:fnlresNBD}.   No significant signal was found in this domain for either of these two models.   However, we note that the logarithmic dependence of the bispectrum creates challenges at low $k$, as we must ensure important features do not fall below the modal resolution.  This restricts the present survey range, which will be extended in future.   Again, we note that most attention on these models has focused on higher $\ell$-periodicities than those accessible to the present modal estimator (see e.g., \citealt{2011JCAP...01..017F,Peiris:2013opa}), for the same reason as the feature models.   A  resonance model survey using the \Planck\ power spectrum has been undertaken and the results can be found in~\cite{planck2013-p17}; however, this also currently excludes high frequencies that have received attention in the literature.

\medskip

\noindent{\it Warm inflation}:
This model, where dissipative effects are important, predicts $f_\mathrm{NL}^\mathrm{warm} = -15 \ln \left(1 + r_\mathrm{d}/14 \right) - 5/2$
\citep{2007JCAP...04..007M} where the dissipation parameter $r_\mathrm{d} = \Gamma/(3H)$ must be large for strong dissipation. The limit from {\it WMAP} is $r_\mathrm{d} \leq 2.8 \times 10^4$ \citep{2007JCAP...04..007M}. Assuming a prior $0 \leq \log_{10} r_\mathrm{d} \leq 4$, our constraint $f_\mathrm{NL}^\mathrm{warmS} = 4 \pm 33$ at 68\% CL (see Sect.~\ref{warmres}) yields a limit on the dissipation parameter of $\log_{10} r_\mathrm{d} \leq 2.6$ ($95 \%$ CL), improving the previous limit by nearly two orders of magnitude. The strongly-dissipative regime with $r_\mathrm{d} \gtrsim 2.5$ still remains viable; however, the {\it Planck} constraint puts the model in a regime which can lead to an overproduction of gravitinos (see~\cite{2008JCAP...01..027H} and references within).

\subsection{Alternatives to inflation}
Perhaps the most striking example is given by the ekpyrotic/cyclic models~(for a review, see~\citealt{2010AdAst2010E..67L}) proposed as alternative to inflationary models. Typically they predict a local NG $|f_{\rm NL}^{\rm local}| > 10$. In particular, the so-called ``ekpyrotic conversion'' mechanism (in which isocurvature perturbations are converted into curvature perturbations during the ekpyrotic phase) yields $f_{\rm NL}^\mathrm{local} = -(5/12)\, c_1^2$, where $c_1$ is a parameter in the potential, requiring $10 \gtrsim c_1  \gtrsim 20$ for compatibility with power spectrum constraints. This case was $\sim 4\sigma$ discrepant with {\it WMAP} data, and with {\it Planck} it is decisively ruled out given our bounds $f_{\rm NL}^{\rm local}=2.7 \pm 5.8$  at $68\%$ CL (see Table~\ref{tab:fNLsmicah}) yielding $c_1 \leq 4.2$ at 95\% CL. The predictions for the local bispectrum from other ekpyrotic models (based on the so called ``kinetic conversion'' taking place after the ekpyrotic phase) yield $f_{\rm NL}^\mathrm{local} = (3/2) \, \kappa_3 \sqrt{\epsilon} \pm 5$, where  values $-1 < \kappa_3 <5$ and $\epsilon \sim 100$ are natural~\citep{2010AdAst2010E..67L}. Taking $\epsilon \sim 100$, in this case we obtain $-0.9 < \kappa_3 <  0.6$  and $-0.2 < \kappa_3 <  1.3$ at 95\% CL, for the plus and minus sign in $f_{\rm NL}^\mathrm{local}$ respectively, significantly restricting the viable parameter space of this model.

%%%%%%%%%%%%%%%%%%%%%%%%%%%%%%%%%%%%%%%%%%%%%%%%%%%%

\subsection{Implications of CMB trispectrum results}
The non-detection of local-type $\fnl$ by \planck\ raises the immediate
question of whether there might still be  a large trispectrum. In this first
analysis we have focused on the local shape $\taunl$. Most inflation models
predict $\taunl \sim \mathcal{O}(\fnl^2)$ \citep{2006PhRvD..74l3519B,
2012JCAP...09..001E}, and hence given our tight $\fnl$ limits, would be predicted
to be very small. This is easily consistent with our conservative upper
limit $\taunl < 2800$, and also with the significantly smaller signals
found in the modulation dipole and quadrupole. Our upper limit also
restricts the freedom of curvaton-like models with quadratic terms that are nearly uncorrelated with the curvature
perturbation, which could in principle have $\fnl\sim0$ but a large
trispectrum \citep{2010AdAst2010E..76B}.

%%%%%%%%%%%%%%% Section 10 %%%%%%%%%%%%%%%
\section{Conclusions}
\label{sec:Conc}

In this paper we have derived constraints on primordial NG, using the
CMB maps derived from the \textit{Planck} nominal mission data.  Using three optimal bispectrum estimators
-- KSW, binned, and modal -- we obtained consistent values for the primordial local,
equilateral, and orthogonal bispectrum amplitudes, quoting as our
final result $\fnllocal=2.7 \pm 5.8$, 
$f_\textrm{NL}^\textrm{equil}=-42 \pm 75$, and 
$f_\textrm{NL}^\textrm{ortho}= - 25 \pm 39$ ($68\,\%$ CL statistical).  
Hence there is no evidence for primordial NG of one
of these shapes. We did, however, measure the ISW-lensing bispectrum expected in
the \LCDM\ scenario, as well as a contribution from diffuse point
sources, and these contributions are clearly seen in the form of the associated skew-$C_\ell$.  Indeed, the detection of ISW-lensing and Poisson point source 
skew-$C_\ell$ with the right shapes and at expected levels is good evidence that the data processing is not destroying non-gaussianity in the maps, and gives
confidence that the non-detections of primordial non-gaussianity are robust.
These results have been confirmed by measurements using the
wavelet bispectrum, and Minkowski functional estimators, and demonstrated to be stable for the four different component 
separation techniques \SMICA, \NILC, \SEVEM, and \CR, showing their
robustness against foreground residuals. They have also passed
an extensive suite of tests studying the dependence on the maximum multipole
number and the mask, consistency checks between frequency channels, 
and several null tests.
In addition, we have summarized in this paper an extensive validation 
campaign for the three optimal bispectrum estimators on Gaussian and 
non-Gaussian simulations.

Extending our analysis beyond estimates of individual shape amplitudes, we presented 
model-independent, three-dimensional reconstructions of the \textit{Planck} 
CMB bispectrum using the modal and binned bispectrum estimators.
These results were also shown to be fully consistent between the different component
separation techniques even for the full bispectrum, and contained no 
significant NG signals of a type not captured by our standard templates at
low multipole values. At high multipoles, some indications of unidentified NG signals were found, as also evidenced by
the results from the skew-$C_\ell$ estimator. Further study will be required to ascertain whether these indications are 
due to foreground residuals, beams, data processing, or a more interesting signal.

Using the modal decomposition, we have presented constraints on key
primordial NG scenarios, including general single-field models with non-separable shapes, excited initial states (non-Bunch-Davies vacua), and directionally-dependent vector models. We have also undertaken an initial survey of scale-dependent feature and resonance models. 

Moving beyond three-point correlations,  we have obtained limits from the \textit{Planck} data on the
amplitude of the local four-point function. Our trispectrum reconstruction yielded a signal consistent with zero
except for an anomalously large octopole.
Frequency dependence indicated that this was unlikely to be primordial, but
allowing the signal to be primordial we placed an upper limit
$\tau_{\mathrm{NL}} < 2800$ ($95\%$ CL).

We have discussed the impact of these results on the inflationary model space, and derived bispectrum constraints on a selection of specific inflationary mechanisms, including both general single-field inflationary models (extensions to the standard single-field slow-roll case) and multi-field models. 
Our results led to a lower bound on the speed of sound, $c_\mathrm{s} \geq 0.02$ ($95\%$ CL), in an effective field theory approach to inflationary models which includes models with non-standard kinetic terms. Strong constraints on other scenarios such as IR DBI, $k$-inflation, inflationary models involving gauge fields, and warm inflation have been obtained. Within the class of multi-field models, our measurements limit the curvaton decay fraction to $r_{\rm D} \geq 0.15$ ($95\%$ CL). 
Ekpyrotic/cyclic scenarios were shown to be under pressure from the \textit{Planck} data: we robustly ruled out the so-called ``ekpyrotic conversion'' mechanism, and found that the parameter space of the ``kinetic conversion'' mechanism is significantly limited.

With these results, the paradigm of standard single-field slow-roll inflation has survived its most stringent tests to-date.

%%%%%%%%%%%%%%% acknowledgements %%%%%%%%%%%%%%%

\begin{acknowledgements}

The development of \Planck\ has been supported by: ESA; CNES and CNRS/INSU-IN2P3-INP (France); ASI, CNR, and INAF (Italy); NASA and DoE (USA); STFC and UKSA (UK); CSIC, MICINN, JA and RES (Spain); Tekes, AoF and CSC (Finland); DLR and MPG (Germany); CSA (Canada); DTU Space (Denmark); SER/SSO (Switzerland); RCN (Norway); SFI (Ireland); FCT/MCTES (Portugal); and PRACE (EU). A description of the Planck Collaboration and a list of its members, including the technical or scientific activities in which they have been involved, can be found at \url{http://www.sciops.esa.int/index.php?project=planck&page=Planck_Collaboration}.
The primary platform for analysis with the modal and KSW estimators was the
COSMOS supercomputer, part of the DiRAC HPC Facility supported by STFC and the UK
Large Facilities Capital Fund. We gratefully acknowledge IN2P3 Computer Center (http://cc.in2p3.fr) for
providing a significant amount of the computing resources and services needed for the analysis with the binned 
bispectrum estimator. This research used resources of the National Energy Research Scientific Computing Center, which is supported
by the Office of Science of the U.S. Department of Energy under Contract No. DE-AC02-05CH11231.
We acknowledge the use of resources from the Norewegian national super computing facilities NOTUR.
We also acknowledge the IAP magique3 computer facilities as well as the curvaton computer at the Department of Physics of the
University of Illinois (Urbana, IL).  
We further acknowledge the computer resources and technical assistance
provided by the Spanish Supercomputing Network nodes at Universidad de
Cantabria and Universidad Polit\'ecnica de Madrid as well as 
the support provided by the Advanced Computing and e-Science team at IFCA.

\end{acknowledgements}

%%%%%%%%%%%%%%% References %%%%%%%%%%%%%%%
\bibliographystyle{aa_arxiv}
\bibliography{Planck_bib,P09a_bib}

%%%%%%%%%%%%%%% Appendix %%%%%%%%%%%%%%%
\appendix

\section{Full Focal Plane Simulations}\label{sec:FFP6}

The purpose of the Full Focal Plane (FFP) simulations is to provide a complete, coherent realization of the full \Planck\ mission (both HFI and LFI) using the best available estimates of its characteristics, including the satellite pointing, the individual detector beams, band passes and noise properties, and the various data flags. Each FFP data set consists of three parts: a fiducial observation of an input sky (including our best estimates of all of the foreground components and a realization of the CMB sky corresponding to a chosen cosmology), together with separate MC realization sets of the CMB and noise. The maps made from these simulated data are used both to validate and verify the analysis algorithms and their implementations and to quantify uncertainties in and remove biases from the analysis of the real data. The FFP simulations have been used for validating map-making, component separation (for both CMB and diffuse foreground recovery), power spectra estimation, cosmological parameter
estimation and measurements of primordial NG and lensing. The complete specification of any FFP simulation consists of the definition of the inputs (mission characteristics and sky), the outputs (time-ordered data and maps), and the processes used to generate the latter from the former. For this data release we have used the sixth FFP simulation suite, henceforth referred to as FFP6.

\subsection{Mission Characteristics}
\label{sec:mission_characteristics}

The goal of FFP6 is to replicate the 2013 \Planck\ data release (DR1) as closely as possible. The satellite pointing incorporates the PTCOR6 wobble corrections used by HFI~\citep{planck2013-p03}, and the focal plane geometry of each instrument is provided by the corresponding DPC in the form of a reduced instrument model (RIMO) (\citealt{planck2013-p02} and~\citealt{planck2013-p03}).  The overall data flags are a combination of the pointing and detector flags.  For both instruments, satellite repointing maneuvers and planet crossings are flagged. For pairs of detectors in the same horn, each sample flagged in one detector is flagged in the other member of the pair.

In addition to the geometric detector offsets, the RIMOs for both instruments provide a representation of the detector bandpasses.  For LFI, the RIMO also provides a parameterized noise model (white noise level, knee frequency and power law index) for each detector~\citep{planck2013-p02}.  LFI beams were measured from Jupiter scans, fitted using Gaussian circular and elliptical approximations, converted into a polarized beam model, and smeared to account for the satellite motion~\citep{planck2013-p02d}.  For HFI, the noise power spectral densities (PSDs) are averaged and smoothed versions of the raw ring-by-ring spectra determined by the HFI pipeline. This processing produced the per detector mean noise spectrum over the duration of the nominal mission, calibrated to convert from pre-processed Time Ordered Information (TOI) units to thermodynamic $K$~\citep{planck2013-p03}. HFI beam maps were synthesized from the elliptical beam parameters and Gauss-Hermite coefficients from the Mars observation~\citep{planck2013-p03c}.

\subsection{Input Sky}
\label{sec:input_sky}

The FFP6 input sky is generated using the \Planck\ Sky Model (PSM)~\citep{delabrouille2012} and includes the CMB together with the current best estimate foreground templates for the cosmic infrared background, CO line emission, free-free, synchrotron, spinning dust, thermal dust, kinetic and thermal SZ, and radio and infrared point sources. For simulation purposes, point sources are considered strong if their flux is more than $100$ mJy at any of $30$, $100$, $300$ or $1000$ GHz; strong point sources are provided in catalogs, whilst weak sources are mapped. All maps are provided in thermodynamic $K$ at HEALPix resolution 2048~\citep{gorski2005} and smoothed with a 4 arcminute beam, while catalogues are in Jy. The CMB is generated as a lensed sum of a scalar, tensor and non-Gaussian realization with particular values of the tensor to scalar ratio ($r$) and NG parameter ($f_{\rm NL}$). For every detector, each foreground component and catalogue of strong point sources also takes the appropriate band-passed from the RIMO and applies it.

\subsection{FFP6 Outputs}
\label{sec:ffp6_outputs}

The FFP6 simulation consists of maps of the fiducial sky together with sets of MC realizations of the CMB and of the noise, each of which is produced using a distinct processing pipeline; full details on these can be found in the Explanatory Supplement~\citep{planck2013-p28}.

The fiducial sky maps are made from explicit time-ordered data sets which are generated using the Level S software suite~\citep{reinecke2006} for every detector and for every component; this results in 770 time-ordered datasets occupying 17TB of disk space. For every component we then make maps of the data at each frequency using the {\tt Madam/TOAST} map-maker \citep{keihanen2010} for all the detectors, all of the detector quadruplets and all of the unpolarized detector singlets, and using both the full data and the two half-ring data subsets. In addition we make maps of the sum of all of the components' time streams with simulated noise added. HFI data are mapped at HEALPix resolution 2048 and LFI data at 1024; this results in 2370 maps occupying 0.4TB of disk space. We also produced hit maps, binned maps and the 3x3 block diagonal white-noise covariance matrices and their inverses.

The CMB MC provides a set of 1000 temperature-only maps of the full data combining all of the detectors at each frequency. These are made directly in the map domain by calculating the effective beam at each frequency using the {\tt FEBeCoP} software~\citep{mitra2010} and applying it to each of 1000 realizations of the CMB sky drawn from the best-fit {\it WMAP} spectrum and generated using the HEALPix {\tt synfast} tool~\citep{gorski2005}; this results in 9,000 maps occupying 1.2TB of
disk space.

The MC noise maps are generated using {\tt Madam/TOAST}, which has the ability to simulate the noise time stream on the fly (bypassing the otherwise cripplingly expensive writing and reading of time-ordered data objects). For every detector combination at each frequency, and for both the full data and half-ring data subsets, we generate 1000 noise realizations; this results in 222,000 maps occupying 33TB of disk space.

The bulk of the FFP simulations were performed at the US Department of Energy's National Energy Research Scientific Computing Center (NERSC); the LFI noise MC for the various detector quadruplets were produced at the Finnish Supercomputing Center (CSC); in total the generation of FFP6 required some 15 million CPU-hours and 50TB of disk space. All of the temperature data in the FFP6 simulation set are available for public use at NERSC.

\subsection{Validation and Exploitation}\label{sec:ffp6_valid}

Specific tests for assessing our confidence in the claims were conducted for all the main results in the \Planck\
2013 release, namely CMB reconstruction through foreground cleaning, likelihood to cosmology, lensing and primordial
NG; these are discussed in detail in the corresponding papers (\citealt{planck2013-p06, planck2013-p08, planck2013-p11,planck2013-p12, planck2013-p09} and this work).
The validation was conducted with the necessary accuracy for measuring second-order cosmological effects like in the
case of primordial NG as well as lensing. In the former case, blind tests on an FFP set with a non-zero
and detectable $f_{\rm NL}^{\rm local}=20.4075$ value in the CMB component were performed and the correct value (within the error bars) was detected after
foreground cleaning. In the case of lensing, null detection tests were conducted successfully after the process of foreground cleaning on maps that did not
contain this signal as input.

Residual systematics which were not taken into account belong to the low-$\ell$ and high-$\ell$ 
regimes for LFI and HFI, respectively. Although the latter were not translated directly into spurious 
NG effects, their effect was quantified. For LFI, sidelobe corrections have been quantified to 
a level which is comparable or below $0.05\%$ in the beam transfer function on the range of $\ell$ 
between $200$ and $1500$, being increasingly sub-dominant with respect to noise at high $\ell$s. 
A number of HFI systematic effects and their treatment in the data processing are not included 
in the simulations. On large angular scales, the most important effects are the Analogue Digital 
Correction (ADC) non-linearities (which principally manifest themselves as an apparent gain drift), 
Zodiacal light emission, and far sidelobe pick-up of the Galaxy. On small angular scales the most important 
effects are cosmic ray hits (flagging of those from real data was taken into account into the FFP6 processing) 
and the 4He-JT (4K) cooler lines.  The processing steps used to remove these effects and assessments of the 
residuals are described in \citet{planck2013-p03,planck2013-p03f,planck2013-p03e,planck2013-pip88}.

\Planck\ will be updating its suite of simulations for the forthcoming release including polarization.
The instrumental characteristics described above will be updated, and particular care is being taken regarding the ability to simulate and control the main systematic effects affecting the polarised signal.

\section{Expected level of agreement from bispectrum estimators with correlated weights}
\label{sec:AA}

The estimator cross-validation work presented in Sect.~\ref{sec:Sec_valid_est} was based 
on comparing results from
different estimators using sets (typically 50 to 100 simulations in
size) of both Gaussian and non-Gaussian simulations. We started from
idealized maps (e.g., full-sky, noiseless, Gaussian
simulations) and then went on to include an increasing number of
realistic features at each additional validation step. This allowed
better testing  and characterization of the response of different pipelines and
bispectrum decompositions to various potential spurious effects in the
data.  As a preliminary step, we derived a general formula describing
the expected level of agreement between estimators with different but
strongly correlated weights, with the simplifying assumption of full-sky  
measurements and homogeneous noise. This theoretical expectation, summarized by 
Eq.~(\ref{eqn:expectation}),  was then
used as a benchmark against which to assess the quality of the
results.  
The aim of this appendix is to describe in detail how we obtained Eq.~(\ref{eqn:expectation}). 

Let us consider the idealized case of full-sky,
noiseless CMB measurements (note that the following conclusions also work
for homogeneous noise, because the pure cubic $f_{\rm NL}$ estimators without linear term 
corrections are still optimal) . Under these assumptions, the $f_{\rm NL}$
estimator for a given CMB shape $B_{\ell_1 \ell_2 \ell_3}$ can be
written simply as (see Sect.~\ref{sec:se} for details):
\begin{equation}\label{eqn:estideal}
\hat{f}_{\rm NL} = \frac{1}{F}\sum_{\ell_1 \leq \ell_2 \leq \ell_3} \sum_{m_1 m_2 m_3} \frac{B_{\ell_1 \ell_2 \ell_3}^{\rm th} a_{\ell_1 m_1} a_{\ell_2 m_2} a_{\ell_3 m_3}}{g_{\ell_1 \ell_2 \ell_3} C_{\ell_1} C_{\ell_2} C_{\ell_3}},
\end{equation}
where $B^{\th}_{\ell_1 \ell_2 \ell_3}$ is the angle-averaged bispectrum
template for a given theoretical shape, $a_{\ell_1 m_1} a_{\ell_2
  m_2} a_{\ell_3 m_3}$ is a bispectrum estimate constructed from the data,
and $F$ is the normalization of the estimator, provided by 
the Fisher matrix number 
\begin{equation}
F=\sum_{\ell_1 \leq \ell_2 \leq \ell_3} \frac{(B^{\rm th}_{\ell_1 \ell_2 \ell_3})^2} {g_{\ell_1 \ell_2 \ell_3}C_{\ell_1}C_{\ell_2}C_{\ell_3}}.
\end{equation}
As explained in Sect.~\ref{sec:se}, the different
$f_{\rm NL}$ estimation techniques implemented in this work can be
seen as separate implementations of the optimal estimator of
Eq.~(\ref{eqn:estideal}).  Each implementation is based on expanding the
angular bispectrum as a sum of basis templates defined in different
domains: for example in our analyses we built templates out of
products of wavelets at different scales, cubic
combinations of 1-dimensional polynomials and plane waves (what we call ``modal estimator" in the main text), 
and $\ell$-binning of the bispectrum (what we called "binned estimator" in the main text). 
Our initial theoretical templates in this work
are the local, equilateral and orthogonal separable cases used in
the KSW and skew-$C_\ell$ estimators. In this sense the KSW/skew-$C_\ell$ estimator provides an
``exact'' estimate of $f_{\rm NL}$ for this choice of shapes, while
the other pipelines provide an approximate result that approaches KSW
measurements as the expansions get more accurate. The differences
between results from different pipelines became smaller and smaller as
the approximate expanded templates converge to the starting one
(e.g., by increasing the number of $\ell$-bins or the number of
wavelets and polynomial triplets). The degree of convergence can
be measured through the correlation coefficient $r$ between the
initial bispectrum and its reconstructed expanded version. The
correlation coefficient is, as usual, defined as
\begin{equation}\label{eqn:corr}
 r \equiv \frac{  
\sum_{\ell_1 \leq \ell_2 \leq \ell_3} \frac{ B^{\rm th}_{\ell_1 \ell_2 \ell_3} B^{\rm exp}_{\ell_1 \ell_2 \ell_3}}{g_{\ell_1 \ell_2 \ell_3} C_{\ell_1} C_{\ell_2} C_{\ell_3}}}
{\sqrt{\sum_{\ell_1 \leq \ell_2 \leq \ell_3} \frac{ (B^{\rm th}_{\ell_1 \ell_2 \ell_3})^2}{g_{\ell_1 \ell_2 \ell_3} C_{\ell_1} C_{\ell_2} C_{\ell_3}}
\sum_{\ell_1 \leq \ell_2 \leq \ell_3} \frac{ (B^{\rm exp}_{\ell_1 \ell_2 \ell_3})^2}{g_{\ell_1 \ell_2 \ell_3} C_{\ell_1} C_{\ell_2} C_{\ell_3}}} } \; ,
\end{equation}
where the label ``th'' denotes the initial bispectrum shape to fit to
the data, and ``exp'' is the approximate expanded one. The correlator 
between shapes naturally defines a scalar
product in bispectrum space, and in the following the operation of
correlating two shapes will be denoted by the symbol $\langle \, ,
\rangle$, so that the definition above would read $r = {\langle
  B^{\rm th}, B^{\rm exp} \rangle / \sqrt{\langle B^{\rm exp},B^{\rm exp} \rangle
    \langle B^{\rm th},B^{\rm th} \rangle}}$.
 
Whichever basis and separation scheme we have chosen, let us call
$\mathcal{R}_n(\ell_1, \ell_2, \ell_3)$ the $n$-th template in the
adopted bispectrum expansion, and $\alpha_n$ the coefficients of the
expansion, so that we can write:
\begin{equation}
 B^{\rm exp}_{\ell_1 \ell_2 \ell_3} = \sum_{n=0}^{N_{\rm exp}} \alpha_n \mathcal{R}_n(\ell_1, \ell_2, \ell_3) \; .
\label{eqn:Bexpalpha}
\end{equation}
From now on we will also call $\mathcal{R}_n(\ell_1, \ell_2, \ell_3)$
the ``modes'' of our expansion, $N_{\rm exp}$ is the number of modes we
are using to approximate the starting template in order to obtain a
correlation coefficient $r$. 
We will also assume that
the modes form an orthonormal basis, that is:
\begin{equation}
\langle \mathcal{R}_{n_1}, \mathcal{R}_{n_2} \rangle = \delta_{n_1}^{n_2} \; ,
\end{equation}
where $\delta_{n_1}^{n_2}$ is the Kronecker delta symbol. The
orthogonality assumption does not imply loss of generality since any
starting set of modes can always be rotated and orthogonalized. We now
consider an expansion with a number of coefficients $N_{\rm th} > N_{\rm exp}$
that perfectly reproduces the initial bispectrum (i.e., $r=1$), and is
characterized by the same modes and coefficients as the previous one
up to $N_{\rm exp}$. So we can write
\begin{equation}
 B^{\th}_{\ell_1 \ell_2 \ell_3} = \sum_{n=0}^{N_{\th}} \alpha_n \mathcal{R}_n(\ell_1, \ell_2, \ell_3) \; .
\end{equation}
We now build two optimal estimators of $f_{\rm NL}$ for the shape
$B_{\th}$: an ``exact'' estimator and an ``approximate'' one. In the
exact estimator we fit the actual $B_{\th}$ shape to the data and
obtain the estimate $\hat{f}_{\th}$, while in the approximate estimator
we fit the expanded shape $B_{\exp}$ to obtain $\hat{f}_{\exp}$.  We
want to understand by how much the ``exact'' and ``approximate''
$f_{\rm NL}$ measurements are expected to differ, as a function of the
correlation coefficient $r$ between the weights $B_{\th}$ and $B_{\exp}$
that appear in the two estimators.

For each mode template $\mathcal{R}_n(\ell_1, \ell_2, \ell_3)$ we can
build an optimal estimator (following Eq.~(\ref{eqn:estideal}))
in order to fit the mode to the data and get a corresponding amplitude
estimate $\beta_n$. In other words, the observed bispectrum can then
be reconstructed as in Eq.~(\ref{eqn:Bexpalpha}), but with coefficients
$\beta_n$ instead of $\alpha_n$. Due to the orthonormality of the 
$\mathcal{R}_n$, the $\beta$ coefficients have unit variance.
It is then possible to show~(\citealt{2010PhRvD..82b3502F}) that the
$f_{\rm NL}$ estimate for a given $B_{\th}$ with expansion parameters
$\alpha_n$ is given by
\begin{equation}
\hat{f}_{\rm NL} = \frac{1}{F}\sum_n \alpha_n \beta_n \; .
\end{equation}
In the light of all the above, the exact and approximate estimators are:
\begin{eqnarray}
\hat{f}^{\th}_{\rm NL} & = & \frac{1}{F_{\th}}\sum_{n=0}^{N_{\th}} \alpha_n \beta_n  \;; \\
\hat{f}^{\exp}_{\rm NL} & = & \frac{1}{F_{\exp}}\sum_{n=0}^{N_{\exp}} \alpha_n \beta_n \; .
\end{eqnarray}
Thanks to the orthonormality properties of the modes, we can easily
relate the Fisher matrix normalization $F$ to the expansion
coefficients $\alpha$:
\begin{eqnarray}
F & \equiv & \sum_{\ell_1 \leq \ell_2 \leq \ell_3} \frac{B^2_{\ell_1 \ell_2 \ell_3}} {g_{\ell_1 \ell_2 \ell_3}C_{\ell_1}C_{\ell_2}C_{\ell_3}} \nonumber\\
& = & \sum_{\ell_1 \leq \ell_2 \leq \ell_3} \frac{\sum_{n_1=0}^{N} \alpha_{n_1} \mathcal{R}_{n_1}(\ell_1,\ell_2,\ell_3) \sum_{n_2=0}^{N} \alpha_{n_2} \mathcal{R}_{n_2}(\ell_1,\ell_2,\ell_3)} {g_{\ell_1 \ell_2 \ell_3}C_{\ell_1}C_{\ell_2}C_{\ell_3}} \nonumber\\
  & = & \sum_{n_1=0}^{N} \alpha_{n_1}\sum_{n_2=0}^{N} \alpha_{n_2}
        \langle  \mathcal{R}_{n_1},\mathcal{R}_{n_2} \rangle \nonumber \\
  & = & \sum_{n_1=0}^{N} \sum_{n_2=0}^{N} \alpha_{n_1}\alpha_{n_2} \delta_{n_1}^{n_2} \nonumber \\ 
  & = & \sum_{n=0}^{N} \alpha^2_{n} \label{eqn:alpha2norm} \; .
\end{eqnarray}
In analogous fashion we can derive an expression for the correlation
coefficient $r$ between the two estimators we
are comparing. It is easy to check that the following holds:
\begin{equation}\label{eqn:alpha2corr}
r^2 = \frac{\sum_{n=0}^{N_{\exp}} \alpha^2_n}{\sum_{n=0}^{N_{\th}} \alpha^2_n} \; ,
\end{equation}
and using the equation just derived above we can also write $r^2 =
{F_{\exp}/F_{\th}}$, i.e., the square of the correlation coefficients
between the estimators equals the ratio of the normalizations.

Armed with this preliminary notation we can now calculate the expected
scatter between the exact and approximate $f_{\rm NL}$ measurement
when we apply the two estimators to the same set of data. In order to
quantify it we will consider the variable $\delta_{f_{\rm NL}} \equiv
\hat{f}^{\th}_{\rm NL} - \hat{f}^{\exp}_{\rm NL}$, and calculate its standard
deviation. We find
\begin{eqnarray}
\sigma^2_{\delta f_{\rm NL}} & \equiv & \left\langle  \left( \frac{1}{F_{\th}} \sum_{n=0}^{N_{\th}} \alpha_n \beta_n - \frac{1}{F_{\exp}} \sum_{n=0}^{N_{\exp}} \alpha_n \beta_n \right)^2 \right\rangle \nonumber \\
                             & = & \frac{1}{F^2_{\exp}} \left\langle  \left(r^2 \sum_{n=0}^{N_{\th}} \alpha_n \beta_n - \sum_{n=0}^{N_{\exp}} \alpha_n \beta_n \right)^2 \right\rangle \nonumber \, , \\
\end{eqnarray}
where we made use of Eq.~(\ref{eqn:alpha2corr}). 
It is easy to show that orthonormality of the $\mathcal{R}$ templates implies 
no correlation of the amplitudes $\beta$, i.e., $\langle \beta_p \beta_q = \delta_p^q\rangle$. A straightforward calculation then yields:
\begin{equation}\label{eqn:deltafnl1}
\sigma^2_{\delta f_{\rm NL}} = \frac{1}{F^2_{\exp}} \sum_{n=0}^{N_{\exp}} \alpha_n^2 -2r^2 \sum_{n=0}^{N_{{\exp}}} \alpha^2_n + r^4\sum_{n=0}^{N_{{\th}}} \alpha^2_n \; .
\end{equation}
This, together with Eqs.~(\ref{eqn:alpha2norm},~\ref{eqn:alpha2corr})  finally gives:
\begin{equation}\label{eqn:expectationA}
\sigma_{\delta f_{\rm NL}} = \Delta_{\th} \frac{\sqrt{1-r^2}}{r} \; ,
\end{equation}
where $\Delta_{\th}$ is the standard deviation of the exact
estimator. Eq. ~(\ref{eqn:expectationA}) is the starting point of our validation tests.  
It provides an estimate of the expected scatter between $f_{\rm
  NL}$ estimators with correlated weights, as a function of the
$f_{\rm NL}$ error bars and of the correlation coefficient $r$. Note
that this formula has been obtained under the simplifying assumptions
of Gaussianity, full-sky coverage and homogeneous noise. For
applications dealing with more realistic cases we might expect the
scatter to become {\em larger} than this expectation, while remaining
qualitatively consistent.

In our tests we started from sets of simulated maps and compared
$f_{\rm NL}$ results from different pipelines and shapes on a
map-by-map basis. In our validation tests the correlation levels between templates 
 in different expansions were varying between $r \sim 0.95$ and $r \sim 0.99$, depending 
on the estimators and the shapes under study. Using the formula above, this corresponds to an
expected scatter $ 0.15 \, \Delta  \leq \sigma \leq 0.3 \, \Delta$, where $\Delta$ is
the $f_{\rm NL}$ standard deviation for the shape under study. 

In Sect.~\ref{sec:Sec_valid_est} we presented several applications to simulated
data sets, showing that the recovered results are fully consistent with
these expectations, and thus the different pipelines are fully
consistent with each other.

\section{Targeted Bispectrum Constraints}
\label{sec:AB}

\begin{table*}[h!]
\begingroup
\newdimen\tblskip \tblskip=5pt
\caption{Results from a limited $\fnl$ survey of resonance models of Eq.~\eqref{resonantBis} with $0.25\le k_{\rm c}\le0.5$ using the \SMICA\ component-separated map.   These models have  a large-$\ell$ periodicity similar to the feature models in Table~\ref{tab:fnlfeature}.}
\label{tab:fnlresonant}
\medskip
\nointerlineskip
\vskip -6mm
\footnotesize
\setbox\tablebox=\vbox{
   \newdimen\digitwidth 
   \setbox0=\hbox{\rm 0} 
   \digitwidth=\wd0 
   \catcode`*=\active 
   \def*{\kern\digitwidth}
   \newdimen\signwidth 
   \setbox0=\hbox{+} 
   \signwidth=\wd0 
   \catcode`!=\active 
   \def!{\kern\signwidth}
   \newdimen\dotwidth 
   \setbox0=\hbox{.} 
   \dotwidth=\wd0 
   \catcode`;=\active 
   \def;{\kern\dotwidth}
    \halign{\hbox to 1in{#\leaderfil}\tabskip 1.0em&
            \hfil#\hfil&
            \hfil#\hfil&
            \hfil#\hfil&
            \hfil#\hfil&
            \hfil#\hfil&
            \hfil#\hfil\tabskip 0pt\cr
    \noalign{\doubleline\vskip 2pt}
 \omit&\multispan6\hfil $\fnl\pm\Delta\fnl$\hfil\cr
\omit&\multispan6\hrulefill\cr   
Wavenumber $k_{\rm c}$\hfill & $\phi=0$ &$\phi=\pi/5 $ & $\phi=2\pi/5$ & $\phi=3\pi/5$ & $\phi=4\pi/5$ & $\phi=\pi$ \cr
 %\omit   Wavenumber  \hfil& $\fnl\pm\Delta\fnl$ &$\fnl\pm\Delta\fnl$ & $\fnl\pm\Delta\fnl$ & $\fnl\pm\Delta\fnl $ & $\fnl\pm\Delta\fnl $ & $\fnl\pm\Delta\fnl $ \cr
     \noalign{\vskip 4pt\hrule\vskip 6pt}
$0.25$ & $-16 \pm 57 $ & $!*6 \pm 63 $ & $*19 \pm 67 $ & $!31 \pm 69 $ & $!38 \pm 68 $ & $*-6 \pm 60 $ \cr
$0.30$ & $-66 \pm 73$ & $-57 \pm 74 $ & $-44 \pm 73$ & $-26 \pm 72 $ & $*-7 \pm 71 $ & $-65 \pm 73 $ \cr
$0.40$ & $!*5\pm 57 $ & $!40 \pm 66 $ & $!55 \pm 71 $ & $!63 \pm 73 $ & $!63 \pm 71 $ & $!22 \pm 61 $ \cr
$0.45$ & $!25 \pm 56$ & $!34 \pm 59 $ & $!36 \pm 62 $ & $!34 \pm 67 $ & $!27 \pm 69 $ & $!30 \pm 56 $ \cr
$0.50$ & $*-2\pm 65 $ & $-13 \pm 72$ & $-16 \pm 69$ & $-16 \pm 60$ & $-14 \pm 55 $ & $*-7 \pm 71 $ \cr
\noalign{\vskip 3pt\hrule\vskip 4pt}
}}
\endPlancktablewide

\endgroup
\end{table*}
%%%%%%%%%%%%%%%%%%%%%%%%%%%%%%%%%%%%%%%%%%%%%%%%%%%%

 %%%%%%%%%%%%%%%%%%%%%%%%%%%%%%%%%%%%%%%%%%%%%%%%%%%%
\begin{table*}[!t]
\begingroup
\newdimen\tblskip \tblskip=5pt
\caption{Results from a limited $\fnl$ survey of non-Bunch-Davies feature models (or enfolded resonance models) of Eq.~\eqref{resNBDBis} with $4\le k_{\rm c}\le 12$, again overlapping in periodicity with the feature model survey.    }
\label{tab:fnlresNBD}
\medskip
\nointerlineskip
\vskip -6mm
\footnotesize
\setbox\tablebox=\vbox{
   \newdimen\digitwidth 
   \setbox0=\hbox{\rm 0} 
   \digitwidth=\wd0 
   \catcode`*=\active 
   \def*{\kern\digitwidth}
   \newdimen\signwidth 
   \setbox0=\hbox{+} 
   \signwidth=\wd0 
   \catcode`!=\active 
   \def!{\kern\signwidth}
   \newdimen\dotwidth 
   \setbox0=\hbox{.} 
   \dotwidth=\wd0 
   \catcode`;=\active 
   \def;{\kern\dotwidth}
    \halign{\hbox to 1.2in{#\leaderfil}\tabskip 1.0em&
            \hfil#&
            \hfil#&
            \hfil#&
            \hfil#\tabskip 0pt\cr
    \noalign{\doubleline\vskip 2pt}
    \omit&\multispan4\hfil $\fnl\pm\Delta\fnl ~~(\sigma)$\hfil\cr
\omit&\multispan4\hrulefill\cr
Wavenumber $k_{\rm c}$\hfill & $\phi=0$ \hfil &$\phi=\pi/4 $ \hfil & $\phi=\pi/2$ \hfil & $\phi=3\pi/4$ \hfil \cr
%\omit    Wavenumber  \hfil& $\fnl\pm\Delta\fnl ~~(\sigma)$ &$\fnl\pm\Delta\fnl ~~(\sigma)$ & $\fnl\pm\Delta\fnl ~ ~(\sigma)$ & $\fnl\pm\Delta\fnl  ~~(\sigma)$ \cr
     \noalign{\vskip 4pt\hrule\vskip 6pt}
$*4$ & *$11 \pm 146 ~~(!0.1)$ & **$2 \pm 145 ~~(!0.0)$ & **$-7 \pm 143 ~~(!0.0)$ & *$-15 \pm 142 ~~(-0.1)$ \cr
$*6$ & *$52 \pm 202 ~~(!0.3)$ & *$63 \pm 203 ~~(!0.3)$ & !*$72 \pm 201 ~~(!0.4)$ & !*$80 \pm 197 ~~(!0.4)$ \cr
$*8$ & $100 \pm 190 ~~(!0.5)$ & $130 \pm 189 ~~(!0.7)$ & !$158 \pm 189 ~~(!0.8)$ & !$183 \pm 190 ~~(!1.0)$ \cr
$10$ & $188 \pm 241 ~~(!0.8)$ & $210 \pm 242 ~~(!0.9)$ & !$230 \pm 242 ~~(!1.0)$ & !$248 \pm 243 ~~(!1.0)$ \cr
$12$ & $180 \pm 307 ~~(!0.6)$ & $171 \pm 310 ~~(!0.6)$ & !$158 \pm 312 ~~(!0.5)$ & !$142 \pm 314 ~~(!0.5)$ \cr
\noalign{\vskip 3pt\hrule\vskip 4pt}
}}
\endPlancktablewide
\endgroup
\end{table*}
%%%%%%%%%%%%%%%%%%%%%%%%%%%%%%%%%%%%%%%%%%%%%%%%%%%%

%%%%%%%%%%%%%%%%%%%%%%%%%%%%%%%%%%%%%%%%%%%%%%%%%%%%
\begin{table*}[!h]
\begin{center}
\begingroup
\newdimen\tblskip \tblskip=5pt
\caption{Summary of results from the modal estimator survey of primordial models for the main non-standard bispectrum shapes.   This is an extended version of Table~\ref{tab:fnlnonstandard}, but with results from \SMICA, \NILC\ and \SEVEM.  }
\label{tab:modalmapmethods}
\medskip
\nointerlineskip
\vskip -6mm
\footnotesize
\setbox\tablebox=\vbox{
   \newdimen\digitwidth 
   \setbox0=\hbox{\rm 0} 
   \digitwidth=\wd0 
   \catcode`*=\active 
   \def*{\kern\digitwidth}
   \newdimen\signwidth 
   \setbox0=\hbox{+} 
   \signwidth=\wd0 
   \catcode`!=\active 
   \def!{\kern\signwidth}
   \newdimen\dotwidth 
   \setbox0=\hbox{.} 
   \dotwidth=\wd0 
   \catcode`;=\active 
   \def;{\kern\dotwidth}
\begin{tabular}{ l  c  c  c  c  c  c }
\hline
\hline
\noalign{\vskip 4pt}
\omit& $F_{\rm NL}$ & $F_{\rm NL}$-Clean & StDev & Fisher & $\sigma$ & $\sigma$-clean\\
\noalign{\vskip 4pt}
\hline
\noalign{\vskip 4pt}
\qquad \SMICA\ \\
Const (see text)  & !*$26$;** & !*$14$;** & !*$44$;** & !*$42$;** & !**$0.6$* & !**$0.3$*\\
EFT1 shape \eqref{EFT1Bis}  & !*$13$;** & !**$8$;** & !*$73$;** & !*$70$;** & !**$0.2$* & !**$0.1$*\\
EFT2 shape \eqref{EFT1Bis}  & !*$27$;** & !*$19$;** & !*$57$;** & !*$54$;** & !**$0.5$* & !**$0.3$*\\
DBI inflation \eqref{dbiBis} & !*$17$;** & !*$11$;** & !*$69$;** & !*$67$;** & !**$0.2$* & !**$0.2$*\\
Ghost inflation (see text)  & *$-27$;** & *$-24$;** & !*$88$;** & !*$87$;** & **$-0.3$* & **$-0.3$*\\
Flat model \eqref{flatBis} & !*$70$;** & !*$37$;** & !*$77$;** & !*$71$;** & !**$0.9$* & !**$0.5$*\\
NBD (see text)  & !$178$;** & !$155$;** & !*$78$;** & !*$76$;** & !**$2.2$* & !**$2.0$*\\
NBD1 flattened  \eqref{NBD2Bis} & !*$31$;** & !*$19$;** & !*$13$;** & !*$12$;** & !**$2.4$* & !**$1.4$*\\
NBD2 squeezed \eqref{NBD2Bis} & !**$0.8$* & !**$0.2$* & !**$0.4$* & !**$0.5$* & !**$1.8$* & !**$0.5$*\\
NBD3 non-canonical \eqref{NBD3Bis}  & !*$13$;** & !**$9.6$* & !**$9.7$* & !**$9.0$* & !**$1.3$* & !**$1.0$*\\
Vector model $L=1$ \eqref{vectorBis} & *$-18$;** & **$-4.6$* &  !*$47$;** & !*$45$;** & **$-0.4$* & **$-0.1$*\\
Vector model $L=2$ \eqref{vectorBis} & !**$2.8$* & **$-0.4$* & !**$2.9$* & !**$2.8$* & !**$1.0$* & **$-0.1$*\\
WarmS inflation (see text)  & **$-8$;** & !**$1$;** & !*$33$;** & !*$33$;** & **$-0.2$* & !**$0.04$\\
%\hline
%\end{tabular}
%\begin{tabular}{ l  c  c  c  c  c  c }
\noalign{\vskip 4pt}
\qquad\SEVEM\ &&&\\
Const (see text)  & !*$23$;** & !*$11$;** & !*$44$;** & !*$42$;** & !**$0.5$* & !**$0.2$*\\
EFT1 shape \eqref{EFT1Bis}  & !**$9$;** & **$-1$;** & !*$72$;** & !*$71$;** & !**$0.1$* & **$-0.02$\\
EFT2 shape \eqref{EFT1Bis}  & !*$21$;** & !*$14$;** & !*$56$;** & !*$54$;** & !**$0.4$* & !**$0.2$*\\
DBI inflation \eqref{dbiBis} & !*$13$;** & !**$7$;** & !*$68$;** & !*$67$;** & !**$0.2$* & !**$0.1$*\\
Ghost inflation  & *$-24$;** & *$-21$;** & !*$88$;** & !*$88$;** & **$-0.3$* & **$-0.2$*\\
Flat model \eqref{flatBis} & !*$63$;** & !*$31$;** & !*$76$;** & !*$72$;** & !**$0.8$* & !**$0.4$*\\
NBD (see text)  & $159$;** & $137$;** & !*$78$;** & !*$76$;** & !**$2.0$* & !**$1.8$*\\
NBD1 flattened  \eqref{NBD2Bis} & !*$30$;** & !*$18$;** & !*$12$;** & !*$12$;** & !**$2.4$* & !**$1.4$*\\
NBD2 squeezed \eqref{NBD2Bis} & !**$0.9$* & !**$0.4$* & !**$0.4$* & !**$0.5$* & !**$2.1$* & !**$0.8$*\\
NBD3 non-canonical \eqref{NBD3Bis}  & !*$12$;** & !**$9$;** & !*$10$;** & !**$9$;** & !**$1.2$* & !**$0.9$*\\
Vector model $L=1$ \eqref{vectorBis} & *$-15$;** & **$-2$;** & !*$47$;** & !*$45$;** & **$-0.3$* & **$-0.04$\\
Vector model $L=2$ \eqref{vectorBis} & !*$3.5$* & !**$0.3$* & !**$2.7$* & !**$2.8$* & !**$1.3$* & !**$0.1$*\\
WarmS inflation  & *$-11$;** & **$-2$;** & !*$33$;** & !*$33$;** & **$-0.3$* & **$-0.1$*\\
%\hline
%\end{tabular}
%\begin{tabular}{ l  c  c  c  c  c  c }
\noalign{\vskip 4pt}
\qquad \NILC\ &&&\\
Const (see text)  & !*$37$;**& !*$25$;**& !*$44$;**& !*$42$;**& !**$0.8$* & !**$0.6$*\\
EFT1 shape \eqref{EFT1Bis}  & !*$20$;**& **$-4$;**& !*$72$;**& !*$70$;**& !**$0.3$* & **$-0.05$\\
EFT2 shape \eqref{EFT1Bis}  & !*$39$;**& !*$32$;**& !*$56$;**& !*$54$;**& !**$0.7$* & !**$0.6$*\\
DBI inflation \eqref{dbiBis} & !*$26$;**& !*$20$;**& !*$69$;**& !*$67$;**& !**$0.4$* & !**$0.3$*\\
Ghost inflation   & *$-36$;**& *$-33$;**& !*$88$;**& !*$87$;**& **$-0.4$* & **$-0.4$*\\
Flat model \eqref{flatBis} & !$100$;**& !*$68$;**& !*$76$;**& !*$71$;**& !**$1.3$* & !**$0.9$*\\
NBD (see text)  & !$189$;**& !$165$;**& !*$78$;**& !*$76$;**& !**$2.4$* & !**$2.1$*\\
NBD1 flattened  \eqref{NBD2Bis} & !*$35$;**& !*$22$;**& !*$13$;**& !*$12$;**& !**$2.7$* & !**$1.7$*\\
NBD2 squeezed \eqref{NBD2Bis} & !**$0.8$ & !**$0.3$ & !**$0.4$ & !**$0.5$ & !**$1.9$* & !**$0.6$*\\
NBD3 non-canonical \eqref{NBD3Bis}  & !*$17$;**& !*$13$;**& !**$9$;**& !**$9$;**& !**$1.8$* & !**$1.4$*\\
Vector model $L=1$ \eqref{vectorBis} & *$-41$;**& *$-28$;**& !*$46$;**& !*$45$;**& **$-0.9$* & **$-0.6$*\\
Vector model $L=2$ \eqref{vectorBis} & !**$3.8$* & !**$0.6$* & !**$2.6$* & !**$2.8$* & !**$1.4$* & !**$0.2$*\\
WarmS inflation  & *$-23$;**& *$-14$;**& !*$32$;**& !*$33$;**& **$-0.7$* & $**-0.4$*\\
\hline
\end{tabular}
}
\endPlancktablewide
\endgroup
\end{center}
\end{table*}
%%%%%%%%%%%%%%%%%%%%%%%%%%%%%%%%%%%%%%%%%%%%%%%%%%%%
%
%
%
%
%%%%%%%%%%%%%%%%%%%%%%%%%%%%%%%%%%%%%%%%%%%%%%%%%%%%
\begin{table*}[!t]
\begingroup
\newdimen\tblskip \tblskip=5pt
\caption{Cross-validation of best fit feature model results for the \SMICA, \NILC\ and \SEVEM\ foreground-cleaned maps.   Results are only presented for feature models with better than a $2.5\sigma$ result on the full domain (see Table~\ref{tab:fnlfeature}).}
\label{tab:fnlfeaturemethods}
\medskip
\nointerlineskip
\vskip -6mm
\footnotesize
\setbox\tablebox=\vbox{
   \newdimen\digitwidth 
   \setbox0=\hbox{\rm 0} 
   \digitwidth=\wd0 
   \catcode`*=\active 
   \def*{\kern\digitwidth}
   \newdimen\signwidth 
   \setbox0=\hbox{+} 
   \signwidth=\wd0 
   \catcode`!=\active 
   \def!{\kern\signwidth}
   \newdimen\dotwidth 
   \setbox0=\hbox{.} 
   \dotwidth=\wd0 
   \catcode`;=\active 
   \def;{\kern\dotwidth}
    \halign{\hbox to 1.5in{#\leaderfil}\tabskip 1.0em&
            \hfil#&
            \hfil#&
              \hfil#\tabskip 0pt\cr
    \noalign{\doubleline\vskip 2pt}
\omit&\multispan3\hfil $fnl\pm\Delta\fnl ~~(\sigma)$\hfil\cr
\omit&\multispan3\hrulefill\cr
Wavenumber; Phase\hfill & \NILC\ \hfil & \SEVEM\ \hfil & \SMICA\ \hfil \hfil \cr
%    Feature model  \hfil& $\fnl\pm\Delta\fnl ~~(\sigma)$ &$\fnl\pm\Delta\fnl ~~(\sigma)$ & $\fnl\pm\Delta\fnl  ~~(\sigma)$ \cr
     \noalign{\vskip 4pt\hrule\vskip 6pt}
$k_{\rm c}=0.01125;\,\phi=0$ & $458 \pm 169 ~~(!2.7)$ & $409 \pm 169 ~~(!2.4)$ &  $434 \pm 170 ~~(!2.6)$ \cr
$k_{\rm c}=0.01750;\,\phi=0$ & $-337 \pm 131 ~~(-2.6)$ & $-328 \pm 128 ~~(-2.6)$ &  $-335 \pm 137 ~~(-2.4)$ \cr
$k_{\rm c}=0.01750;\,\phi=3\pi/4$ & $368 \pm 124 ~~(!3.0)$ & $348 \pm 121 ~~(!2.9)$ &  $366 \pm 126 ~~(!2.9)$ \cr
$k_{\rm c}=0.01875;\,\phi=0$ & $-359 \pm 118 ~~(-3.1)$ & $-366 \pm 115 ~~(-3.2)$ & $-348 \pm 118 ~~(-3.0)$ \cr
$k_{\rm c}=0.01875;\,\phi=\pi/4$ & $-339 \pm 117 ~~(-2.9)$ & $-328 \pm 115 ~~(-2.9)$ &  $-323 \pm 120 ~~(-2.7)$ \cr
$k_{\rm c}=0.02000;\,\phi=\pi/4$ & $-305\pm 118 ~~(-2.6)$ & $-334 \pm 118 ~~(-2.8)$ & $-298 \pm 119 ~~(-2.5)$ \cr
\noalign{\vskip 3pt\hrule\vskip 4pt}
}}
\endPlancktablewide
\endgroup
\end{table*}
%%%%%%%%%%%%%%%%%%%%%%%%%%%%%%%%%%%%%%%%%%%%%%%%%%%%
%
%

%%%%%%%%%%%%%%%%%%%%%%%%%%%%%%%%%%%%%%%%%%%%%%%%%%%%
\begin{table*}[!t]
\begingroup
\newdimen\tblskip \tblskip=5pt
\caption{Comparison of $\fnl$ results for the hybrid polynomial and Fourier modes for a variety of non-separable and feature models.  }
\label{tab:polysvswaves}
\medskip
\nointerlineskip
\vskip -6mm
\footnotesize
\setbox\tablebox=\vbox{
   \newdimen\digitwidth 
   \setbox0=\hbox{\rm 0} 
   \digitwidth=\wd0 
   \catcode`*=\active 
   \def*{\kern\digitwidth}
   \newdimen\signwidth 
   \setbox0=\hbox{+} 
   \signwidth=\wd0 
   \catcode`!=\active 
   \def!{\kern\signwidth}
   \newdimen\dotwidth 
   \setbox0=\hbox{.} 
   \dotwidth=\wd0 
   \catcode`;=\active 
   \def;{\kern\dotwidth}
    \halign{\hbox to 1.5in{#\leaderfil}\tabskip 1.0em&
            \hfil#&
              \hfil#\tabskip 0pt\cr
    \noalign{\doubleline\vskip 2pt}
    \omit&\multispan2\hfil $\fnl\pm\Delta\fnl ~~(\sigma)$\hfil\cr
\omit&\multispan2\hrulefill\cr
        \hfil Model\hfill & Polynomial\hfil &  Fourier ISW  \hfil  \cr
%    Model  \hfil& $\fnl\pm\Delta\fnl ~~~~(\sigma)\hfil$  & $\fnl\pm\Delta\fnl  ~~~~(\sigma)$\hfil \cr
     \noalign{\vskip 4pt\hrule\vskip 6pt}
Const (see text)  &!*$14;**	\pm *44	;** **(!0.3*)$&$	!*10;** \pm *44;**	**(!0.2)$\cr
EFT1 shape \eqref{EFT1Bis}  & !**$8;**	\pm 	*73;**	**(!0.1*	)$&!**$6;**	\pm *73;**	**(!0.1)$\cr
EFT2 shape \eqref{EFT1Bis}  & !*$19;**	\pm 	*57;**	**(!0.3*	)$&!*$13;**	\pm 	*57;**	**(!0.2)$\cr
DBI inflation \eqref{dbiBis} & !*$12;** 		\pm *69	;****(!0.2*	)$&**$-0.3*	\pm *70;**	**(!0.0)$\cr
Ghost inflation (see text)  &  *$-24;**		\pm *88	;****(-0.3*)$&*$	-48	;**\pm 	*89;**	**(-0.5)$\cr
Flat model \eqref{flatBis} &     !*$37	;**  \pm 	*77;**	**(!0.5*)$&!*$	38;**	\pm *76;**	**(!0.5)$\cr
NBD (see text)  &                 !$155;**	\pm 	*78;**	**(!2.0*	)$&!$116;**	\pm 	*92	;** **(!1.3)$\cr
NBD1 flattened  \eqref{NBD2Bis} & !*$19;**	\pm 	*13;**	**(!1.4*)$&!**$	4;**	\pm 	*19;**	**(!0.2)$\cr
NBD2 squeezed \eqref{NBD2Bis} & !**$0.25	\pm 	**0.45	**(!0.5*)$&**$	-0.3*		\pm **0.5*	**(-0.5)$\cr
NBD3 non-canonical \eqref{NBD3Bis}  & !*$10;**	\pm 	*10;**	**(!1.0*)$&!**$	4;**		\pm *11;**	**(!0.3)$\cr
Vector model $L=1$ \eqref{vectorBis} & **$-5;** 	\pm 	*47;**	**(-0.1*)$&**$	-24	;**\pm 	*50;**	**(-0.5)$\cr
Vector model $L=2$ \eqref{vectorBis} & **$-0.4*	\pm **2.8*	**(-0.1*)$&**$	-1.0*	\pm **3.2*	**(-0.3	)$\cr
WarmS inflation (see text)  & !**$1;**	\pm 	*33;**	**(!0.04)$&*$	-16;** 	\pm 	*41;**	**(-0.4	)$\cr
Feature $k_{\rm c}=0.015$ & $-313;** \pm 144;** **(!2.1*)$&  $-264;** \pm 161;** **(!1.6)$\cr
Feature $k_{\rm c}=0.020$ & $-155;** \pm 110;** **(-1.4*)$&  $-167;** \pm 122;** **(-1.4)$\cr
Feature $k_{\rm c}=0.025$ & !$106;** \pm *93;** **(!1.1*)$&  !$110;** \pm *98;** **(!1.1)$\cr
Feature $k_{\rm c}=0.030$ & !*$56;** \pm *89;** **(!0.6*)$&  !*$78;** \pm *96;** **(!0.8)$\cr
Feature $k_{\rm c}=0.035$ & !*$22;** \pm *82;** **(!0.3*)$&  !*$15;**  \pm *84;** **(!0.2)$\cr
Feature $k_{\rm c}=0.040$ & **$-0.9* \pm *68;** **(-0.01)$&  !**$4;** \pm *69;** **(!0.1)$\cr
Feature $k_{\rm c}=0.050$ & **$-0.2* \pm *63;**  **(!0.00)$&  !*$15;** \pm *66;** **(!0.2)$\cr
Feature $k_{\rm c}=0.080$ & !*$16;** \pm  *64;** **(!0.2*)$&  !*$21;** \pm *64;** **(!0.3)$\cr
\noalign{\vskip 3pt\hrule\vskip 4pt}
}}
\endPlancktablewide
\endgroup
\end{table*}
%%%%%%%%%%%%%%%%%%%%%%%%%%%%%%%%%%%%%%%%%%%%%%%%%%%%

This Appendix provides further tables of results for primordial models, extending those given in Sect.~\ref{STS},  notably for resonance models, while it also gives additional validation checks for the modal bispectrum estimator, beyond the extensive tests described already in the paper for the standard bispectrum shapes.   In particular, using each of the \SMICA\, \NILC\ and \SEVEM\ maps, we will quote results for the main paradigms for non-standard bispectrum models, including comparisons for the feature model results.  Remarkably consistent results are again obtained using the different foreground-separation methodologies and using different modal basis functions.

%%%%%%%%%%%%%%%%%%%%%%%%%%%%%%%%%%%%%%%%%%%%%%%%%%%%

\subsection{Additional results for resonance models and enfolded resonance models}

As described in Sect.~\ref{STS}, we have surveyed resonance models (Eq.~\eqref{resonantBis}) in the region of most interest for feature models, that is, with comparable periodicities to those with described in Table~\ref{tab:fnlfeature} near the best-fit feature model.   The results from this initial survey for the \SMICA\ component-separated map  produced no significant signal, with the results Table~\ref{tab:fnlresonant}.  We note that while the feature and resonance models proved similar for the {\it WMAP} analysis \citep{2010arXiv1006.1642F}, wavelengths are much more differentiated for \Planck\ and so it can be difficult to resolve the shortest wavelengths at low $\ell$.   This is the key limitation on surveys with the present estimator and will be circumvented in future, by using specific separable templates to improve the overall resolution.   Just as local and ISW templates can be incorporated into the analysis, so can targeted feature modes.    
Note that consistent results were obtained using the \NILC\ and \SEVEM\ maps, though we only show results here for feature models with an envelope (see discussion below with Table~\ref{tab:fnlfeaturemethods}).

The enfolded resonant  (or non-Bunch-Davies feature) models (Eq.~\eqref{resNBDBis}) were also surveyed for these periodicities and also yielded no significant signal; see Table~\ref{tab:fnlresNBD}. These are interesting shapes which hold out the prospect of exhibiting both oscillatory and flattened features observed in the \Planck\ bispectrum reconstruction, see Fig.~\ref{fig:recondetail}.    Due to similar resolution restrictions, only relatively large multipole periodicities ($\ell > 100$) have been surveyed in the present work, again searching around the periodicities exhibited for feature models.

%%%%%%%%%%%%%%%%%%%%%%%%%%%%%%%%%%%%%%%%%%%%%%%%%%%%

\subsection{Comparison of $f_{\rm NL}$ results from \SMICA, \NILC\ and \SEVEM\ foreground-cleaned maps}

\noindent  Tables~\ref{tab:modalmapmethods} and~\ref{tab:fnlfeaturemethods} compare bispectrum results extracted from \Planck\ maps created using the three different component-separation techniques, \SMICA, \NILC\ and \SEVEM.   In addition to the models discussed in Sect.~\ref{STS}, we have also included the constant model which is defined by $B^{\rm const}(\kall) = 6A ^2\fnl^{\rm const}/ (k_1k_2k_3)^2$.  
The abbreviations EFT denotes the effective field theory single field shapes, NBD the non-Bunch-Davies (excited initial state) models, vector models are gauge field inflation with directional dependence,  along with DBI, Ghost and Warm inflation, also described previously.  The expression ``cleaned'' refers to removal of the predicted ISW-lensing signal and the measured point source signal, unless stated otherwise. 

We note  that there is good consistency between the different foreground-separation techniques for all models, whether equilateral, flattened, or squeezed as shown in Table~\ref{tab:modalmapmethods}.  For the  non-scaling case, differences for the best-fit feature models were below $1/3\sigma$ confirming the interesting results discussed in  Sect.~\ref{Sec_modalest},  see Table~ \ref{tab:fnlfeaturemethods}.

%%%%%%%%%%%%%%%%%%%%%%%%%%%%%%%%%%%%%%%%%%%%%%%%%%%%

\subsection{Comparison of $f_{\rm NL}$ results from ISW Fourier basis with hybrid polynomials}

As described in Sect.~\ref{Sec_modalest}, the modal bispectrum estimator can flexibly incorporate any suitable  basis functions with which to expand the bispectrum and separably filter CMB maps 
\citep{2010PhRvD..82b3502F}.   For the \Planck\ analysis, we have evolved two sets of basis functions to fulfil three basic criteria: first, to provide a complete basis for the bispectrum up to a given $\ell$-resolution, secondly, to represent accurately primordial models of interest and, thirdly, to incorporate the CMB ISW-lensing signal, which with diffuse point sources provides a significant secondary signal which must be subtracted.   The first basis functions are $n_{\rm max} = 600$ polynomials (closely related to Legendre polynomials) which are supplemented with the Sachs-Wolfe local mode in order to  represent more accurately the squeezed limit; enhanced orthogonality is preconditioned by choosing these from a larger set of polynomials.   The second basis functions are $n_{\rm max} = 300$ Fourier modes, augmented with the same SW local mode, together with the the separable ISW-lensing modes.    Both modal expansions proved useful, providing important validation and cross-checks, however, the twofold resolution improvement from the polynomials meant that most quantitative results employed this basis.    This improved resolution was particularly important in probing flattened models on the edge of the tetrapyd.   

In Fig.~\ref{fig:recondetail}, we show a direct comparison between the modal reconstruction of the 3D bispectrum using the polynomial and Fourier mode expansions.    The basic features of the two reconstructions are in good agreement, confirming a central feature which changes sign at low $\ell$ and a flattened signal beyond as discussed previously in Sect.~\ref{Sec_bisprec}.   Notably the polynomial basis, with double the resolution, preserves the large-scale features observed in the Fourier basis.  

In Table~\ref{tab:polysvswaves}, results from both basis expansions are shown for a variety of non-separable models.   These demonstrate consistent results where the Fourier basis had sufficient resolution, as indicated by the ratio of the variance  reflecting the Fisher ratio (i.e., above 90\% correlation as indicated by the results in Appendix~\ref{sec:AA}).   Note that we independently determine the estimator correlation between the exact solution and primordial decomposition and then at late times with the CMB decomposition; we use a polynomial basis as the overall benchmark here.     This analysis also includes several feature models (phase $\phi=0$) showing good agreement from the Fourier basis while the Fisher estimates remain accurate.  Again, the hybrid Fourier basis degrades in accuracy towards $k=0.02$ as it reaches its resolution limit, when the variance disparity rises towards 10\%.  With $n_{\rm max}=600$ modes and $\ell_{\rm max}=2000$, the polynomial basis retains a good correlation for all primordial feature models for $k>0.01$. 
The accuracy and robustness of the feature model results have been verified using $\ell_{\rm max}=1500$ for the polynomial expansion, for example, obtaining $3.1\sigma$ with the \texttt{SMICA} map for the best-fit model ($K= 0.01875, \;\phi=0$).

%\newpage

\raggedright
\end{document}